\documentclass[trackchanges,twocolumn]{aastex7}

\usepackage{xcolor}
\usepackage{booktabs}
\usepackage{amsmath}
\definecolor{lavender}{rgb}{.9,.9,.7}
\definecolor{gold}{rgb}{.99,.84,.0}
\definecolor{khaki}{rgb}{.94,.90,.55}
\definecolor{asparagus}{rgb}{0.53, 0.66, 0.42}
\usepackage{ulem} 
\usepackage{array}

\newcommand{\uvic}{Department of Physics and Astronomy, University of Victoria, Victoria, BC V8P 5C2, Canada}
\newcommand{\nrchaa}{National Research Council of Canada, Herzberg Astronomy \& Astrophysics Research Centre, Victoria, BC, V9E 2E7, Canada}
\newcommand{\csa}{Canadian Space Agency, 6767 Route de l’A\'eroport, Saint-Hubert, QC J3Y 8Y9, Canada}
\newcommand{\noirlab}{NSF’s NOIRLab, 950 N. Cherry Avenue, Tucson, AZ 85719, USA}

\begin{document}

\title{The Next Generation Virgo Cluster Survey (NGVS). XXIX. Demography of Nuclear Star Clusters}

\correspondingauthor{Max Kurzner}
\email{mkurz21@gmail.com}

\author[0000-0001-5208-4697]{Max M. Kurzner}
\affiliation{\uvic}
\email{mkurz21@gmail.com}

\author[0000-0003-1184-8114]{Patrick Côté}
\affiliation{\nrchaa}
\email{Patrick.Cote@nrc-cnrc.gc.ca}

\author[0000-0002-8224-1128]{Laura Ferrarese}
\affiliation{\nrchaa}
\email{Laura.Ferrarese@nrc-cnrc.gc.ca}

\author[0000-0002-1685-4284]{Chelsea Spengler}
\affiliation{\uvic}
\email{g.spengleri@gmail.com}

\author[0000-0002-2073-2781]{Eric W. Peng}
\affiliation{\noirlab}
\email{eric.peng@noirlab.edu}

\author[0000-0003-4945-0056]{Rub\'en S\'anchez-Janssen}
\affiliation{UK Astronomy Technology Centre, Royal Observatory, Blackford Hill, Edinburgh EH9 3HJ, UK}
\email{ruben.sanchez-janssen@stfc.ac.uk}

\author[0000-0002-0363-4266]{Joel C. Roediger}
\affiliation{\csa}
\email{joel.roediger@asc-csa.gc.ca}

\author[0000-0003-4672-8497]{Kaixiang Wang}
\affiliation{Department of Astronomy, Peking University, Beijing, China Kavli Institute for Astronomy and Astrophysics, Peking University, Beijing, China}
\email{kkwang.astro@gmail.com}

\author[0000-0002-4718-3428]{Chengze Liu}
\affiliation{State Key Laboratory of Dark Matter Physics, Shanghai Key Laboratory for Particle Physics and Cosmology, School of Physics and Astronomy \& Tsung-Dao Lee Institute, Shanghai Jiao Tong University, Shanghai 200240, China}
\email{czliu@sjtu.edu.cn}

\author[0000-0002-5049-4390]{Sungsoon Lim}
\affiliation{Division of Science Education, Kangwon National University, Chuncheon 24341, Republic of Korea}
\affiliation{Kangwon National University Research Institute for Mathematical Sciences, Chuncheon 24341, Republic of Korea}
\email{sslim00@gmail.com}

\author[0000-0003-3009-4928]{Matthew A. Taylor}
\affiliation{University of Calgary, 2500 University Drive NW, Calgary Alberta T2N 1N4, Canada}
\email{matthew.taylor2@ucalgary.ca}

\author[0009-0008-9732-2407]{Eleanore B Todd}
\affil{Department of Physics and Astronomy, University of Victoria, Victoria, BC V8P 5C2, Canada}
\email{blonsbrough@uvic.ca}

\author[0000-0001-9427-3373]{Patrick R. Durrell}
\affiliation{Department of Physics, Astronomy, GIS, Geology and Environmental Sciences, Youngstown State University, One Tressel Way, Youngstown, OH 44555 USA}
\email{prdurrell@ysu.edu}

\author[0000-0001-6443-5570]{Elisa Toloba}
\affiliation{Department of Physics and Astronomy, University of the Pacific, Stockton, CA, USA}
\email{etoloba@pacific.edu}

\author[0000-0001-8867-4234]{Puragra Guhathakurta}
\affiliation{UCO/Lick Observatory, Department of Astronomy and Astrophysics, University of California Santa Cruz, Santa Cruz, CA, USA}
\email{raja@ucolick.org}

\author[0000-0002-7089-8616]{J. Christopher Mihos}
\affiliation{Department of Astronomy, Case Western Reserve University, Cleveland, OH 44106, USA}
\email{mihos@case.edu}

\author[0000-0002-5213-3548]{John P. Blakeslee}
\affiliation{\noirlab}
\email{john.blakeslee@noirlab.edu}

\author[0000-0003-1632-2541]{Hong-Xin Zhang}
\affiliation{Department of Astronomy, University of Science and Technology of China, Hefei, Anhui 230026, China,School of Astronomy and Space Science}
\affiliation{University of Science and Technology of China, Hefei, Anhui 230026, China}
\email{hzhang18@ustc.edu.cn}

\begin{abstract}

Nuclear star clusters (NSCs) provide insights into processes shaping galaxy cores. We use deep $u^*giz$ CFHT imaging from the Next Generation Virgo Cluster Survey (NGVS) to study NSCs in Virgo cluster galaxies. Our parent sample consists of 3680 certain, likely or possible cluster members spanning more than six orders of magnitude in stellar mass. Using a self-consistent classification framework, we identify 1144 candidate nucleated galaxies --- 804 high-confidence NSCs and 340 lower-confidence candidates. This is the largest and most uniform NSC catalog for any environment studied to date.
We explore the nucleation fraction, $f_{\rm nuc}$, as a function of stellar mass, morphology and environment. Consistent with previous studies, we find stellar mass a key factor driving nucleation probability. However, host-galaxy morphology is also important, with the peak of the $f_{\rm nuc}$ distribution occurring at progressively lower masses as one moves from later to earlier morphological types. We interpret this shift as evidence that the more extended structures of late-type galaxies reduce the efficiency of GC inspiral at fixed stellar mass, while gas-rich secular processes and central star formation may contribute increasingly to NSC growth at higher masses. Environment plays a smaller role, with only a slight enhancement in high-density regions.
We characterize the nonlinear relationship between NSC stellar mass and host-galaxy mass, and assess the role of selection effects. We conclude that NSC formation is best explained by a hybrid scenario, with inspiraling star clusters and {\it in situ} star formation contributing, and their relative importance dictated primarily by galaxy mass.

\end{abstract}



\section{Introduction}
\label{sec:introduction}

Nuclear star clusters (NSCs) are among the densest stellar systems in the Universe \citep{Boker:2002aa,Walcher:2005aa,Neumayer2020}, occupying the centers of galaxies across a wide range of mass, size and morphology \citep{Boker:2002aa, Cote:2006aa, Turner:2012aa, Georgiev:2014aa,Poulain+2021,Ashok+2023,Hoyer+2023,Lambert+2024,Khim+2024}. With typical half-light radii of only a few parsecs \citep{Boker:2004aa, Cote:2006aa, Georgiev:2016aa} and stellar masses spanning $\sim10^4$--$10^9\,M_\odot$ \citep{Walcher:2005aa, Rossa:2006aa, Georgiev:2016aa, Neumayer2020,Poulain+2025a,Ashok+2023}, NSCs represent an extreme endpoint of star formation and accretion. Understanding how they form and evolve is therefore central to broader questions of galaxy assembly, central mass buildup, and the interplay between baryonic and dark matter in galaxy cores \citep{Ferrarese:2006aa, Wehner:2006aa, Graham2009, Antonini:2015aa}.

The formation and growth of NSCs are thought to be complex, stochastic processes. It is attributed to two primary, and not necessarily mutually exclusive, mechanisms: a {\it dissipationless} process in which star clusters (such as globular clusters, or GCs) fall to the center of the galaxy due to dynamical friction \citep[e.g.,][]{Tremaine:1975aa,Oh:2000aa,Lotz:2001aa,Capuzzo-Dolcetta:2008aa,Antonini:2012aa,Gnedin:2014aa}; and a {\it dissipative} process that funnels gas into the galactic center to form the stellar content of an NSC  {\it in situ} due to disk instabilities \citep[e.g.,][]{Milosavljevic:2004aa,Bekki:2007aa,Schinnerer:2008aa} or galaxy mergers \citep{Mihos:1994aa,Bekki:2015aa,Roman+2023,Cao+2025,Poulain+2025b}. Recent theoretical and numerical work increasingly supports a hybrid picture, in which both star cluster mergers and  {\it in situ} star formation contribute meaningfully to NSC growth, with their relative importance varying across host galaxy mass and evolutionary history \citep{Pfeffer:2014aa,Antonini:2015aa,Guillard:2016aa,vanDonkelaar+2023,Jo2026}. Some observations of nucleated early-type galaxies in clusters indicate that GC infall may be the dominant mode of NSC growth at low galaxy masses, although some gas infall is still required to match the properties of NSCs \citep{Turner:2012aa,den-Brok:2014aa,Spengler:2017aa,Poulain+2025b}. Similarly, NSCs in late-type galaxies show evidence of younger and multiple stellar populations requiring episodes of dissipative formation \citep{Rossa:2006aa,Walcher:2006aa,Carson:2015aa,Ashok+2023}, but perhaps still with a contribution from GC infall \citep{Hartmann:2011aa}. 

While the growth mechanisms of NSCs show clear dependencies on galaxy mass and morphology, environment is also likely at play. For example, in denser environments, galaxies host more massive GC systems \citep{Peng:2008aa, Harris2013,Hudson2014,Forbes2018}, suggesting more favorable conditions for the formation of compact star clusters (perhaps including NSCs).This also implies that dissipationless NSC formation might be more easily facilitated in higher-density environments, given that more GCs are available. Galaxy interactions will also be more frequent in denser environments, so NSCs may experience additional periods of growth during galaxy mergers, which have been shown to trigger bursts of star formation in the galactic center \citep{Mihos:1994aa,Bekki:2015aa,Roman+2023,Cao+2025,Poulain+2025b}. Additionally, the increased density of the intergalactic medium in denser environments may suppress supernova-driven outflows from escaping the galaxy, retaining star-forming gas for the NSC \citep{Babul:1992aa}. These considerations suggest that both the efficiency of NSC formation and the likelihood that a galaxy hosts an NSC at all may depend on environment. 

This connection between NSCs and environment was previously explored by \citet[][hereafter SJ+19]{Sanchez-Janssen2019} who used data from the Next Generation Virgo Cluster Survey \citep[NGVS;][]{Ferrarese:2012aa, Ferrarese2020} to study NSCs in the core of the Virgo Cluster. Following a comparison with literature samples for other environments, they found that, at fixed host galaxy mass, NSCs were most common in the Coma Cluster, less common in the Virgo and Fornax clusters, and least common in the Local Group. This variation was attributed to a dependence on total halo mass, with the most massive clusters having the highest fraction of NSC-hosting galaxies.

SJ+19 also characterized the NSC-galaxy mass relation. Using a sample of low-mass nucleated galaxies from the NGVS, they found that NSCs in low-mass galaxies become substantially more massive relative to their hosts (up to 0.3\% of the host mass at $M_\star \sim 10^6 M_\odot$, compared to 0.03\% at higher galaxy masses), even though NSC abundance tapers off below galaxy stellar masses of $\sim10^9\,M_\odot$. They further noted that the shape of this relation mirrors the stellar mass-dark matter halo mass relation inferred from abundance-matching \citep[e.g.,][]{Behroozi2013, Moster2013}. This suggests that NSC growth may be regulated by the depth of the host galaxy's dark matter potential well, rather than by stellar mass alone \citep[see also][]{Ferrarese:2006aa, Antonini:2015aa}. However, in the absence of comparably large and homogeneous samples of low-mass galaxies in clusters beyond Virgo, it remains unclear whether this scaling is universal or is itself modulated by environment.

This NGVS paper is the latest in a series of papers focused on compact stellar systems in the Virgo cluster. Apart from the study of SJ+19, previous investigations have explored the connection between nucleation and galaxy shapes \citep{Sanchez2019b} and characterized NSC properties in high- and intermediate-mass galaxies by combining NGVS and HST data \citep{Spengler:2017aa}. Others have mapped out the distribution of ultra-compact dwarfs (UCDs) throughout the cluster \citep{Liu:2015aa,Liu+20}, while identifying and characterizing an evolutionary link between NSCs and UCDs \citep{Mihos2015,Wang2023}. The spatial distribution, and photometric and kinematic properties of GCs in the cluster have been described in \citet{Durrell2014,Zhang:2015aa,Ko2022,Lim2024,Lim2025}. 

This paper is structured as follows. In \S\ref{sec:data}, we describe the NGVS imaging, catalog parameters and other data products used in this study. In \S\ref{sec:results}, we present our results on the properties of NSCs --- including their luminosities, masses, and colors --- and their variation with host galaxy properties. In \S\ref{sec:discussion}, we compare our results for the Virgo Cluster to those for NSCs in other environments, discuss the agreement between the observations and current NSC formation models, explore NSC properties in some interesting galaxy subpopulations, and discuss the implications of our findings for NSC formation and evolution. We summarize our findings and conclude in~\S\ref{sec:conclusions}. 

All NSC classifications and measurements used in this paper, along with galaxy stellar masses, morphologies and structural parameters, are available in the final NGVS catalog \citep{Ferrarese+2026}.


\section{Sample and Analysis}
\label{sec:data}

\subsection{NGVS Imaging and Photometry}
\label{subsec:design}

In this study, we use imaging and data products from the Next Generation Virgo Cluster Survey \citep[NGVS;][]{Ferrarese2012} to assess the presence of NSCs in all 3680 Virgo Cluster galaxies in the NGVS catalog. The NGVS is a CFHT/MegaCam survey of the Virgo A and B subclusters covering 104 deg$^2$ in $u^{*}giz$, with image quality, depth, and low-surface-brightness sensitivity well matched to the detection of faint galaxies and compact central components. Full details of the observing strategy, dithering pattern, depth, and photometric calibration are given by \citet{Ferrarese2012} and \citet{Ferrarese2020}. For the present analysis, the key point is that the stacked images reach point-source depths of $u^{*},g,i,z = 26.3, 26.6, 25.8,$ and $24.8$ mag, respectively, and a $g$-band surface-brightness sensitivity of $\mu_g \sim 29$ mag arcsec$^{-2}$ \citep{Ferrarese:2016aa,Ferrarese2020}. 

\subsection{Image Processing and Data Products}
\label{subsec:image_analysis}

The NGVS reductions use the CFHT \textit{Elixir} system and the specialized \textit{Elixir-LSB} processing described by \citet{Ferrarese2012}. This pipeline models and removes large-scale sky and instrumental backgrounds, preserving the low-surface-brightness structure needed for Virgo dwarf-galaxy work. Residual limitations from chip gaps, saturated-star halos, Galactic cirrus, and local background variations are discussed in detail by \citet{Ferrarese2020}. Those effects are relevant for source detection, but injection--recovery tests and visual inspection show that the effective detection limit is set primarily by image noise rather than by uncorrected background structure.

For NSC classification, we use the stacked NGVS images together with color composite thumbnails generated from the $u^{*}gi$ imaging. These products provide the visual context needed to identify compact central light concentrations and to distinguish likely nuclei from surrounding galaxy structure or nearby compact sources.


\begin{figure*}
    \plotone{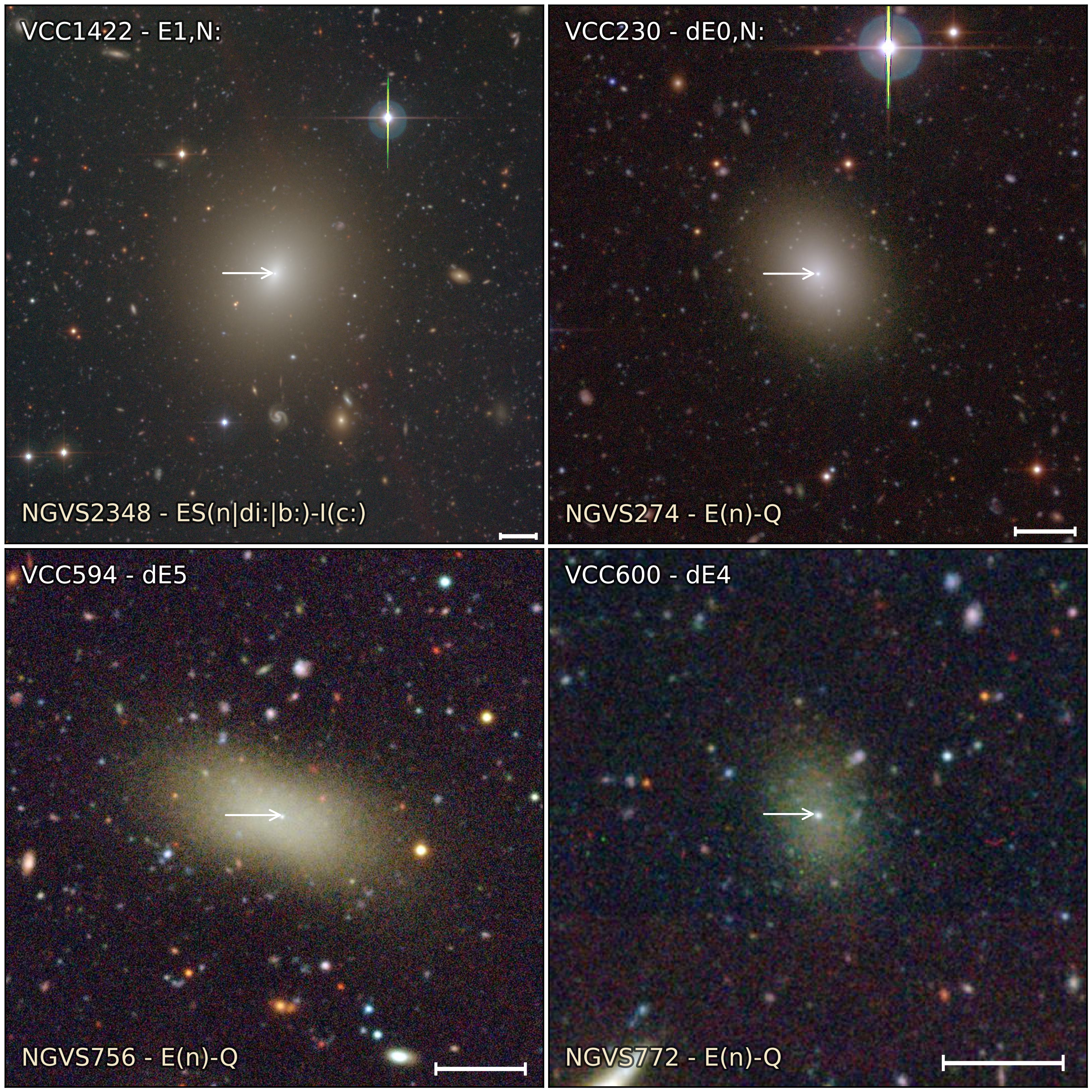}
    \caption{Color images for four Virgo galaxies that highlight differences between the NGVS and VCC in terms of galaxy classification and nuclear star cluster (NSC) identification. ID numbers and classifications from the two surveys are indicated in each panel. The galaxies are selected to span a range in $g$-band magnitude: 13.28 (VCC1422), 15.37 (VCC230), 17.00 (VCC594), and 19.04 (VCC600). Each of these galaxies is classified as containing an NSC in the NGVS, as indicated by the arrow. In the VCC, the existence of an NSC was uncertain for the first two galaxies, while the two fainter galaxies were classified as non-nucleated. The scale bar in each panel corresponds to $20\arcsec$.}
    \label{fig:mosaic}
\end{figure*}

\subsection{Supplementary Data}
\label{subsec:supp_data}

The analysis uses several ancillary NGVS and Virgo data products. Cluster membership probabilities are taken from \citet{Ferrarese2020}, who combine color, magnitude, structural parameters, spatial information, and available  redshifts to distinguish Virgo members from background sources. Host-galaxy morphologies are taken from \citet{Kurzner2025}, enabling NSC demographics to be examined as a function of galaxy structure and star-formation activity (\S\ref{subsec:nsc_nucfraction_morph}). Stellar masses are taken from \citet{Ferrarese+2026}, based on stellar-population fits to $u^{*}giz$ photometry (see \S\ref{subsec:SED}).

For galaxies with direct surface-brightness-fluctuation distances, we adopt the measurements of \citet{Cantiello:2018aa} and \citet{Cantiello2024}. For the remaining galaxies, we assume a Virgo distance of 16.5 Mpc \citep{Mei:2007aa,Blakeslee2009}, consistent with the ACS Virgo Cluster Survey distance scale \citep{Cote:2004aa}. We also use NGVS catalogs of candidate GCs \citep{Durrell2014,Lim2025}, UCDs \citep{Liu:2015aa,Liu+20,Wang2023,Wang+2026}, and large-scale low-surface-brightness structures from Burrell Schmidt imaging \citep{Mihos2005,Mihos2015,Mihos2017} to place NSC hosts in their broader compact-object and environmental context. 

Nearly all NSC classifications in this paper are based on NGVS imaging. The exception is the subset of 94 early-type galaxies brighter than $g \sim 15.6$ that overlap with the ACS Virgo Cluster Survey \citep[ACSVCS;][]{Cote:2004aa,Cote:2006aa}. For these galaxies, we adopt the ACSVCS nucleation classifications because the HST resolution is better suited to detecting faint nuclei against high central surface brightness profiles.\footnote{Six ACSVCS galaxies fall outside the NGVS footprint: VCC9, VCC21, VCC437, VCC751, VCC798, and VCC1913.} Independent NGVS classifications agree with the ACSVCS classifications for 77 of the 94 galaxies (82\%). When the two disagree, we adopt the ACSVCS classification but retain the NGVS-derived host and NSC measurements for internal consistency. Among the discrepant classifications, two galaxies (VCC 759 and VCC 1049) were classified as nucleated in the NGVS but non-nucleated (Type II) in the ACSVCS; as for all disagreements in the overlap sample, we adopt the ACSVCS classifications for these systems. Conversely, six galaxies classified as non-nucleated in the NGVS were classified as nucleated (Type Ia/Ib) in the ACSVCS.


As an additional check on the NGVS classifications for spiral galaxies, we compared our classifications with previous HST studies of nuclear clusters in spirals, drawing on the surveys of \citet{Carollo:1997aa,Carollo:1998aa},
\citet{Boker:2002aa,Boker:2004aa}, and \citet{Georgiev:2014aa}. The resulting NGVS overlap contains 21 spiral galaxies with previous HST classifications or nuclear-cluster measurements. Ten show direct agreement, with both the NGVS and previous HST work identifying a nucleus, while four additional systems show broad agreement at the uncertain or possible-nucleus level. Five galaxies with previous evidence for an NSC or nuclear candidate are assigned {\tt Nuc\_code}$=3$ in the NGVS because their central structures are sufficiently ambiguous or contaminated that they are excluded from the clean nucleated/non-nucleated comparison rather than treated as non-nucleated. Only two galaxies, IC~3298 (VCC~768) and NGC~4299 (VCC~491), show a direct classification tension, with the HST literature
supporting the presence of an NSC while the NGVS classification is non-nucleated ({\tt Nuc\_code}$=0$). In the case of NGC~4299, however, the nuclear-cluster fit reported by \citet{Boker:2002aa} was itself unconstrained. Overall, the late-type comparison shows broad consistency between the NGVS classifications and previous HST studies, while highlighting the greater difficulty of classifying nuclear structure in morphologically complex spiral galaxies from ground-based imaging.
\color{black}


\begin{figure*}
    \centering
    \includegraphics[width=7.0in]{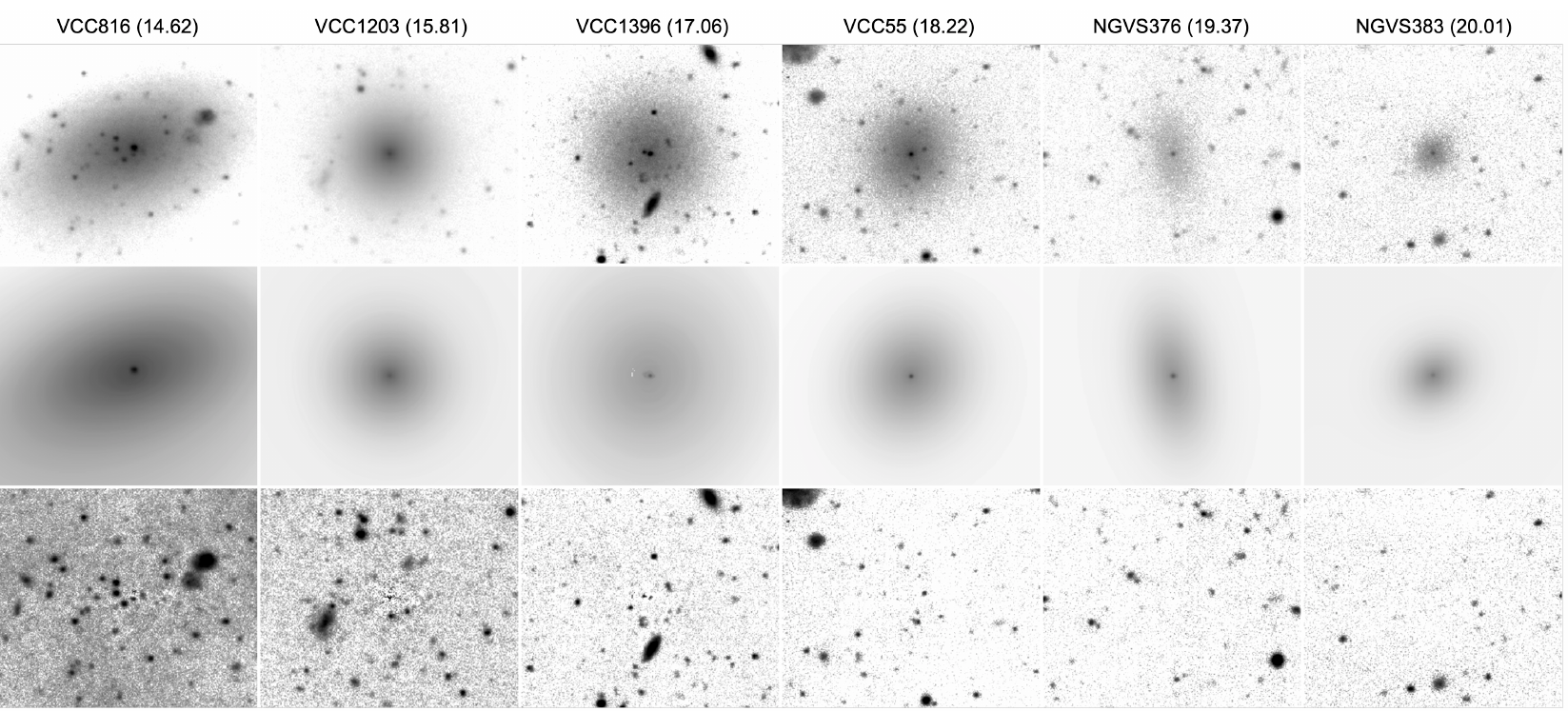}
    \caption{An illustration of model-fitting results for a sample of nucleated galaxies that span roughly two orders of magnitude in stellar mass. The selected galaxies are labeled at the top of the figure along with their $g$-band magnitudes. The top row shows the stacked $g$-band image for each galaxy. The middle row shows the best-fit 2D model, based either on a {\tt Typhon} or {\tt GALFIT} analysis: the three galaxies on the left are fitted using {\tt Typhon}, while the three on the right are fitted using {\tt GALFIT}. In the former approach, the nucleus and galaxy are both fitted with S\'ersic profiles; in the latter approach, we use a 2D S\'ersic model with a central point source. The lower row shows residual images for each galaxy.
    }
    \label{fig:nsc_models}
\end{figure*}

\subsection{Identification and Analysis of NSCs}
\label{subsec:ncs}

Figure~\ref{fig:mosaic} shows four representative nucleated galaxies in the NGVS, chosen to span a range of galaxy properties. VCC1422, in the upper left, has a magnitude of $g=13.28$ and a mean effective surface brightness of $\langle\mu_g\rangle_e = 22.15$~mag~arcsec$^{-2}$, while VCC600, in the lower right, has $g=19.04$ and $\langle\mu_g\rangle_e = 25.33 $~mag~arcsec$^{-2}$. All four galaxies were originally cataloged as likely or possible cluster members in the survey of \cite{Binggeli:1985aa} but in only two cases (VCC1422 and VCC230) was the presence of a possible NSC reported. In this section, we describe the methodology used to identify and analyze NSCs in the NGVS imaging, given the large diversity in galaxy properties.

\subsubsection{Model Fitting for Galaxies and NSCs}
\label{subsec:nsc_models}

In principle, the NGVS imaging allows NSCs to be detected down to a formal point-source limit of $g = 25.9$~mag (10$\sigma$), corresponding to $\log(M_{\rm NSC}/M_\odot) \approx 4.5$. However, the detection threshold will usually be higher given the enhancement in surface brightness due to the underlying galaxy. As a further complication, some NSCs are known to be displaced from the galaxy photocenter \citep[e.g.,][]{Binggeli2000, Lotz:2004aa, Cote:2006aa, Neumayer2020} while others may  appear slightly extended under the very best seeing conditions. Finally, the central regions of intermediate-mass galaxies often show complex, multi-component structures including bars, disks, or dust lanes \citep[see, e.g.,][]{Jerjen2000,Barazza2002,Turner:2012aa, Georgiev:2014aa,Ashok+2023}. These considerations suggest that a combination of quantitative profile fitting --- in either one or two dimensions, or both --- plus some degree of human inspection, is necessary to robustly identify and classify NSCs across the full sample. 

As described in \cite{Ferrarese2020}, NGVS galaxies are analyzed using two independent techniques. For high- and intermediate-luminosity galaxies, photometric and structural parameters are derived by fitting the 1D surface brightness profiles extracted with a custom-built isophotal analysis pipeline, {\tt Typhon}. This software package performs spline interpolation of isophotal parameters derived with the IRAF ELLIPSE task: i.e., surface brightness, ellipticity, major axis position angle, isophotal center, and deviations of the isophotes from pure ellipses. Derived parameters include integrated magnitudes and colors from 1D fitting of the $u^*giz$-band ELLIPSE models. 

For galaxies analyzed with {\tt Typhon}, the center for each galaxy is  defined as the isophotal center measured in the $g$-band and averaged between 1\arcsec~and one effective radius, where the effective radius is determined from the $g$-band curve of growth as described in \S5.1.3 of \citet{Ferrarese2020}.
Because the center is determined from isophotes spanning a wide radial range, it is not driven solely by the innermost light distribution. As a result, compact nuclear components—which typically contribute only a small fraction of the total galaxy luminosity—are not expected to significantly bias the derived center, even in cases where the nucleus is modestly offset from the photocenter. This definition is therefore well suited for identifying both centrally located and offset NSCs within a consistent framework. 

For systems without {\tt Typhon} modeling, single- or two-component S\'ersic models from {\tt GALFIT} \citep{Peng:2010aa} were used to fit the galaxies in two dimensions.\footnote{In the central $\simeq$ 4 deg${^2}$ of the cluster, every cataloged galaxy was fitted using both {\tt Typhon} and {\tt GALFIT} to assess possible systematics \citep{Ferrarese2020}. For the rest of the survey area, nearly all galaxies fainter than $g \approx 18$ were fitted exclusively with {\tt GALFIT} whereas all galaxies brighter than $g \approx 16$ (which are more spatially extended and often more structurally complex) were modeled with {\tt Typhon}. A comparison of parameters measured using the two techniques is presented in \S5.3 and Figures~28-29 of \citep{Ferrarese2020}. The two techniques generally show very good agreement, with the largest discrepancies found for the central parameters of structurally complex galaxies. As we discuss in \S\ref{subsec:nsc_samples}, any NSCs identified in such galaxies are usually classified with types {\tt 1*} or {\tt 2*}, meaning that these NSCs are only considered when computing nucleation fractions. }
The {\tt GALFIT} models consisted of an outer S\'ersic profile for the host galaxy and an inner PSF component for the NSC. Similar automated approaches have been widely used in the literature, typically relying on the detection of central luminosity excesses and parametric profile decomposition \citep[e.g.,][]{Boker:2002aa, Turner:2012aa, Georgiev:2014aa, den-Brok:2014aa, Pechetti2020, Neumayer2020, Ashok2023}. The position of the PSF component was constrained to lie within a small region (a few arcseconds in size) centered on the galaxy photocenter. While this approach nominally ensures that the fitted nucleus is consistent with a true central source, it may exclude offset NSCs that do not coincide with the underlying galaxy potential. Such cases are discussed in \S\ref{subsec:nsc_validation} below.

Representative results from this analysis are presented in Figure~\ref{fig:nsc_models} which shows six nucleated galaxies that span roughly two orders of magnitude in stellar mass. The first three galaxies (VCC816, VCC1203, VCC1396) were modeled using {\tt Typhon} while the next three galaxies (VCC55, NGVS376, NGVS383) were fitted using {\tt GALFIT}. For all galaxies, the top row shows the stacked $g$-band image, the middle row shows the best-fit model generated by {\tt Typhon} or {\tt GALFIT}, and the bottom row shows the residuals. These examples illustrate the quality and flexibility of the two-component decomposition across a range of luminosities. 


\begin{figure*}
    \centering
    \includegraphics[width=7.0in]{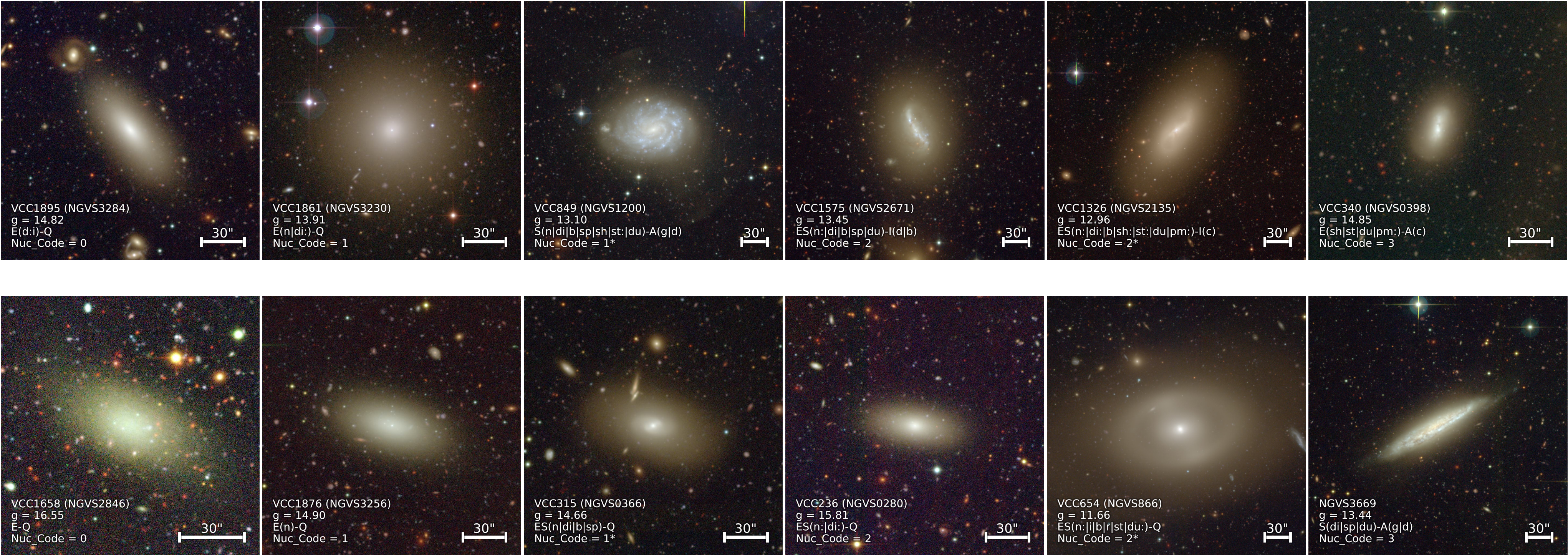}
    \caption{A selection of galaxies illustrating the various nuclear classification codes adopted in this analysis. From left to right, the panels show two examples of each of the following codes: {\tt Nuc\_code} = 0, 1, 1*, 2, 2* and 3 (see \S\ref{subsec:nsc_samples} for a description of the various classes). These classifications are adopted through a combination of one- and/or two-dimensional profile fitting, as well as a visual classification based on the multi-band images, model residuals, and unsharp masked images. Labels in each panel report the galaxy ID, $g$-band magnitude, morphological classification from \citet{Kurzner2025} and the adopted {\tt Nuc\_code}. Each panel includes a 30\arcsec\ scalebar.
    }
    \label{fig:manual_nsc_mosaic}
\end{figure*}

\subsubsection{NSC Validation and Classification}
\label{subsec:nsc_validation}

For the reasons described above, we categorize the nucleation status of every NGVS galaxy following a combination of model-fitting and visual classification. This latter step is needed to identify NSCs in complex galaxies, or offset NSCs missed by automated methods that rely on fixed isophotal centering and PSF-constrained fitting. This approach follows a long tradition of visually identifying NSCs in nearby galaxies \citep[e.g.,][]{Binggeli:1985aa, Cote:2006aa, Turner:2012aa, den-Brok:2014aa, Zanatta2021, Marleau2021, Hoyer2021,Su2022}, and is also well suited to faint and low‐surface‐brightness systems where possible NSCs can be difficult to identify and fit robustly in an automated manner. %

Fundamentally, we visually identify NSCs as central brightness ``excesses" in galaxies flagged during the  morphological classification process \citep{Kurzner2025}. Their presence is inferred from multi-band imaging, and confirmed through inspection of the $u^{*},g,i$ color composites, and unsharp masks when available \citep{MalinUnsharp}. This procedure enables the detection of compact and slightly extended nuclei, as well as higher and lower confidence sources.

Initial classifications were later reassessed during a visual inspection by three classifiers (MK, PC and LF) in order to reach internal consistency and minimize subjectivity. Any candidate NSCs missed in the automated analysis were re-fitted, if possible, using {\tt GALFIT} and the multi-band residuals inspected to determine whether the central light excess corresponded to a likely NSC. Following this step, we are left with a set of classifications and measurements that are used to characterize NSCs.

Anticipating the discussion in \S\ref{subsec:offset_nscs}, we pause to note that the classification of a source as an ``offset nucleus'' becomes increasingly ambiguous at large separations. Beyond some threshold --- comparable to a significant fraction of the galaxy effective radius --- a compact source may be more naturally interpreted as a massive GC or other compact stellar system rather than an offset NSC. This threshold is not purely geometric: brighter sources that dominate the central light distribution and remain morphologically associated with the host galaxy can plausibly be classified as offset nuclei at larger projected separations, whereas fainter sources at similar radii are more difficult to distinguish from ordinary GCs, foreground stars, or background objects. Our selection therefore implicitly defines a practical, luminosity-dependent upper limit on offset separation, motivated by source prominence, morphological association with the host galaxy, and consistency with expected dynamical timescales for orbital decay.

\subsubsection{Definition of NSC Samples}
\label{subsec:nsc_samples}

To facilitate our analyses and enable a comparison to previous NSC studies,
we define two samples that reflect different levels of classification confidence. The first sample consists of 740 high-confidence galaxies that exhibit unambiguous, centrally located, compact NSCs. These galaxies, designated in the catalog with a \texttt{Nuc\_code} of 1, contain NSCs that are point-like or barely resolved at NGVS resolution, and lie within $\sim 2\arcsec$ of the galaxy photocenter \citep{Ferrarese2020}. An additional 64 galaxies are deemed to contain high-confidence NSCs but exhibit complex central morphologies (i.e., inner bulges, bars, or nuclear disks) that make a clean separation between the NSC and its host galaxy difficult, and therefore the nuclear magnitudes cannot be measured reliably. These cases are flagged with {\tt Nuc\_codes} of 1$^\ast$. Together, the class 1 and 1* objects comprise a {\it high-purity} sample of 804 galaxies with the most secure NSC detections. This ``{\tt Sample A}'' forms a baseline for quantitative measurements of
nucleation fractions; the 740 {\tt Nuc\_code = 1} galaxies are also used to explore NSC properties and scaling relations..



\begin{table}[ht]
\centering
\caption{Nuclear Classification Codes }
~\\
\label{tab:nuc_code_counts}
\begin{tabular}{lccl }
\toprule 
{\tt Nuc\_code} & No. of Galaxies & Fraction & \\
\midrule
0 & 2505 & 0.680 \\
1 &  740 & 0.201 \\
1* &   64 & 0.017 \\
2 &  312 & 0.085 \\
2* &   28 & 0.008 \\
3 &   31 & 0.008 \\
\midrule
1, 1* (Sample A)  &  804 & 0.218 \\
2, 2* (Sample B)  &  340 & 0.092 \\
1, 1*, 2, 2* (Sample A+B)  & 1144 & 0.311 \\
\bottomrule
\end{tabular}
\end{table}

In addition to this sample, we identify 312 lower-confidence NSC candidates, designated with \texttt{Nuc\_code = 2}. These objects share some similarities with the high-purity sample but display one or more complicating factors such as shallow or diffuse NSC light profiles, partial blending with the main structure of the galaxy, unusually large brightnesses, or non-central positions. As with the previous sample, an additional 28 objects are assigned a {\tt Nuc\_code} of 2* due to complex inner structures that make an NSC fit unreliable.  These 340 lower-confidence classification objects define ``{\tt Sample~B}'', while the combined high- and lower-confidence sample ({\tt Sample~A+B}) contains 1144 galaxies with likely or possible NSCs, 1052 of which have fitted NSCs. This {\it high-completeness} sample is meant to be more inclusive, ensuring sensitivity to faint or structurally complex nuclei at the expense of some built-in classification uncertainty or contamination by interlopers. These two samples allow us to explore the impact of classification confidence on our derived sample properties and conclusions.

Finally, in a small number of cases, a \texttt{Nuc\_code} of 3 is used to flag galaxies for which a robust NSC classification is not possible due to central dust obscuration or other contaminants. These galaxies are excluded from the {\tt A} and {\tt B} samples used in this analysis, and omitted when estimating nucleation fractions (\S~\ref{subsec:nsc_nucfraction}). The distribution of these nuclear classification codes is summarized in Table~\ref{tab:nuc_code_counts} and a visual schematic is shown in Figure~\ref{fig:manual_nsc_mosaic}.

\subsection{SED Fitting and Mass Estimates}
\label{subsec:SED}

\subsubsection{Galaxy Masses}
\label{subsubsec:SED_galaxies}

For each galaxy in the NGVS catalog, a stellar mass is available from \citet{Ferrarese+2026} who used NGVS multi-band photometry to model galaxy spectral energy distributions (SEDs). Their analysis employed the Flexible Stellar Population Synthesis (FSPS) models of \cite{Conroy2009} to find a best-fit SED for each galaxy under the dual assumptions of an exponentially declining star formation history and a \cite{Chabrier2003} initial mass function. Although dustless models were adopted as the fiducial --- an oversimplification in the case of late-type, star-forming systems --- their SED analysis showed that dustless models were statistically preferred over the dusty models in nearly all cases. In any event, late-type systems are generally rare in the Virgo Cluster, with just $\sim$17.5\% of Virgo galaxies showing clear signs of active or intermediate-level star formation (see Table~2 of \citealt{Kurzner2025}). 

In the analysis of \citet{Ferrarese+2026}, FSPS SEDs were fitted to a grid of 50,000 synthetic models with metallicities in the $0.01 \le Z/Z_{\odot} \le 1.6$ range, star formation timescales $0.5 \le \tau \le 100$ Gyr$^{-1}$, and luminosity-weighted ages between 5 and 13 Gyr. The models are compared to the observed photometric measurements in the $u^*giz$ bands (the four filters for which measurements are available for nearly every member galaxy). Typical uncertainties on these stellar masses, based on a comparison with literature values, are $\sigma$(log $M_*$) $\simeq$ 0.15 dex.

\subsubsection{NSC Masses}
\label{subsubsec:SED_nscs}

To derive stellar masses for the NSCs, we used the variation in mass-to-light ratio, $\Upsilon_g$, with dereddened color measured directly from the NGVS galaxies with FSPS SED fits.\footnote{ We compute a Galactic extinction value for each object following \citet{Schlegel1998} and adopt extinction coefficients of
$A_g/E(B-V) = 3.560$, $A_{u^*}/E(B-V) = 4.594$, $A_r/E(B-V) = 2.464$, $A_i/E(B-V) = 1.813$ and $A_z/E(B-V) = 1.221$.} To capture the variation in $\Upsilon_g$ with color, we applied a non-parametric, Locally Weighted Scatterplot Smoothing (LOWESS) regression \citep{Cleveland:1979aa} to the relation between $\log(M_\ast/L_g)$ and the dereddened color $(g - i)_0$ for  a sample of 3568 galaxies with reliable photometry and SED-based stellar masses. The LOWESS fit reproduces the smooth, nonlinear behavior of the $M_\ast/L_g$  relation across the color range $0.20 \lesssim (g - i)_{0} \lesssim 1.57$. 


Of the 1052 galaxies in our sample having {\tt Nuc\_code} of 1 or 2, stellar masses were successfully computed for NSCs in 762 systems (72.4\% of the sample), with $\Upsilon_g$ values spanning the range 0.36 to 5.41 in solar units, with a median of 1.45. The remaining 290 galaxies
(27.6\%) were excluded due to the following issues: 213 lacked both $g$- and $i$-band photometry, 74 had NSC $(g-i)_0$ colors that fell outside the valid range for
the $\Upsilon_g$ calibration (0.2--1.6), and 3 yielded unphysical NSC-to-host stellar mass ratios. These cases correspond either to extremely blue or red systems or to
unreliable NSC--galaxy decompositions. For systems with valid photometry in only one of the $g$ or $i$ bands, mass estimates were computed, where possible, using the mean $M_\star/L$ appropriate to the available band.

This approach of calculating mass-to-light ratios from the dereddened ($g-i)_0$ colors, rather than full SED fitting, has some important advantages. First, the adopted relation improves on standard prescriptions
\citep[e.g.][]{Bell:2003aa, Taylor:2011aa} by being calibrated against the color dependence of $\Upsilon_g$ for the predominantly old, metal-poor stellar populations typical of Virgo cluster galaxies and NSCs. Moreover, it is optimized for the depth, image quality, and photometric system of the NGVS as it ensures NSC masses are based on the same color index for all NSCs, measured from two widely separated passbands having the highest-quality and most uniform data. As we will show in \S\ref{subsubsec:nsc_check}, the resultant NSC masses are in excellent agreement with those reported in SJ+19, who carried out full SED-fitting for NSCs in the core region. 

The sample of NSC masses spans more than five orders of magnitude, from $\log(M_{\mathrm{NSC}}/M_\odot)\approx 3.8$ to $8.9$, with a median value of $\sim5.6$. These parameters fall within the ranges reported in previous studies \citep[e.g.][]{Georgiev:2014aa, den-Brok:2014aa, Spengler:2017aa, Fahrion2021}.


\begin{figure*}
    \centering
    \includegraphics[width=\textwidth]{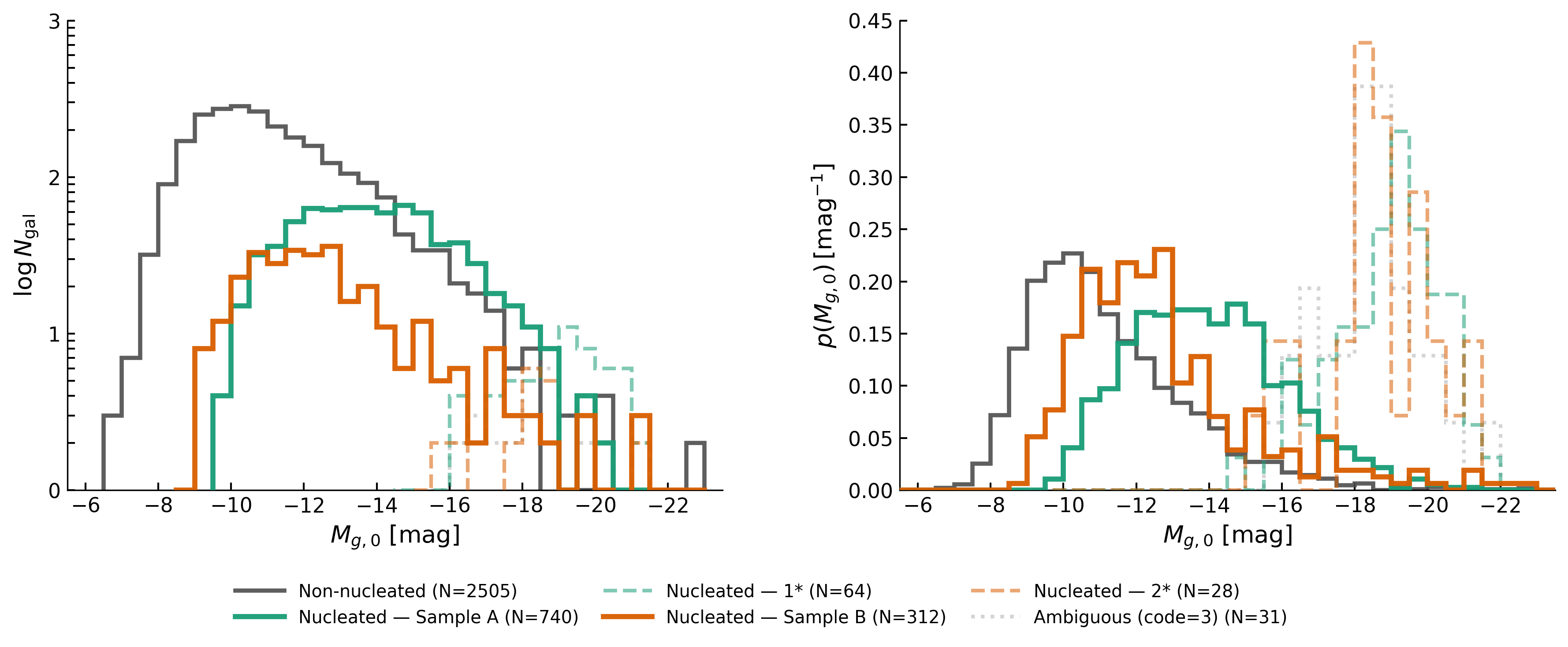}
    \caption{Absolute-magnitude distributions for galaxies separated by nucleation class. The left panel shows the raw number of galaxies per magnitude bin, while the right panel shows the corresponding normalized probability density, \(p(M_{g,0})\), such that each curve integrates to unity. Non-nucleated galaxies are shown in dark gray, Sample~A nucleated galaxies in green, and Sample~B nucleated galaxies in orange. The uncertain \(1^\ast\) and \(2^\ast\) classes are shown as dashed curves using the same colors as their parent samples, while ambiguous code~3 systems are shown as a muted dotted gray curve. See \S\ref{subsec:nsc_samples} for the definition of these classes. The strong overlap among the curves indicates that the luminosity distributions of the different nucleation samples are broadly similar, with modest differences visible at the bright end.}
    \label{fig:maghist}
\end{figure*}


\begin{figure*}
\centering

\includegraphics[width=\textwidth,trim=0 0 0 0,clip]{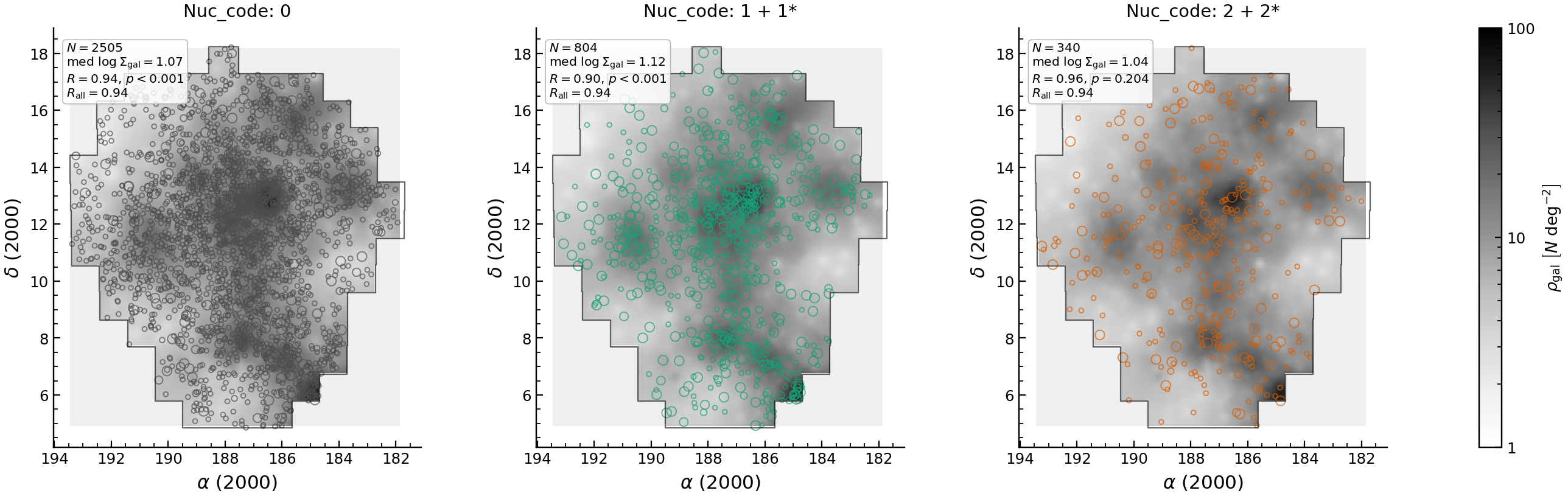}\\[0.8em]
\includegraphics[width=\textwidth,trim=0 0 0 0,clip]{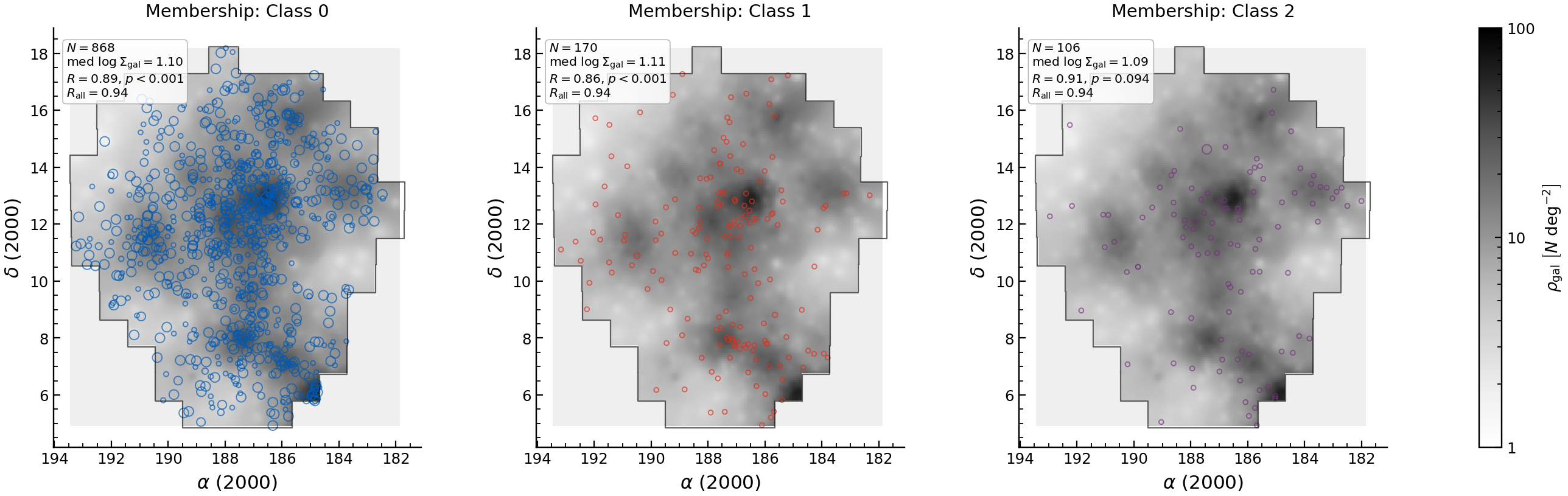}\\[0.8em]
\includegraphics[width=\textwidth,trim=0 0 0 0,clip]{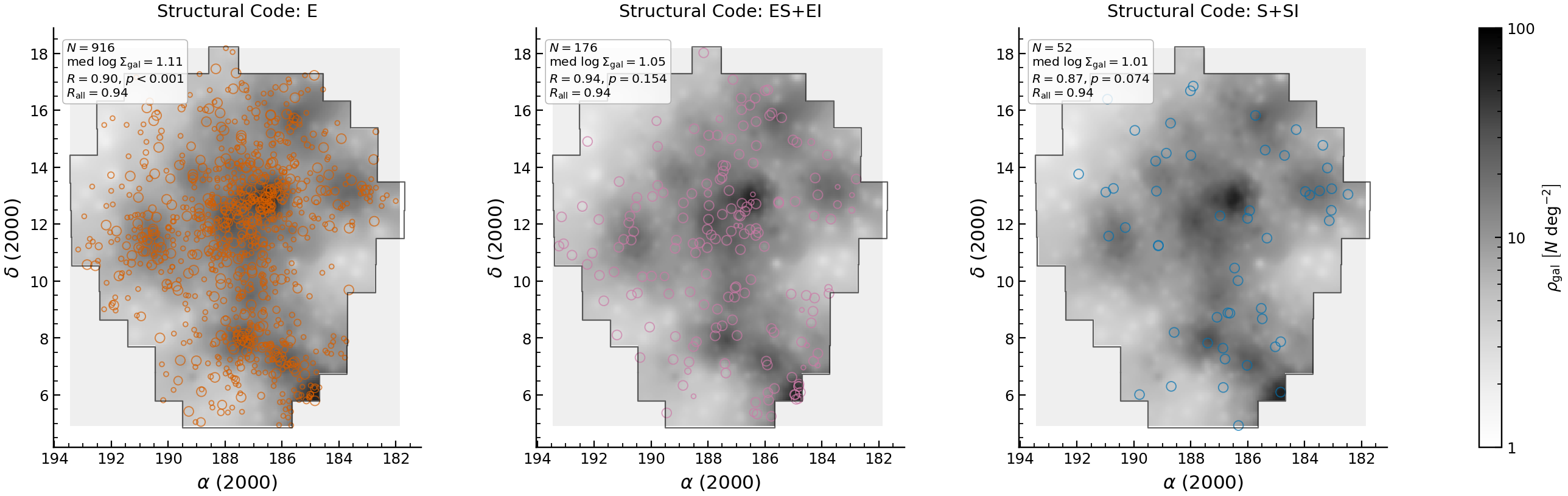}

\caption{Projected number density map of Virgo member galaxies across the NGVS footprint.  In all panels, symbol size scales with galaxy luminosity. The grayscale background  shows the local projected surface density $\Sigma_{\rm gal}$, estimated from the ten nearest bright galaxies ($M_g < -13.59$). For each subsample, the panel titles report the number of galaxies ($N$), the median projected density $\log \Sigma_{\rm gal}$, and the Clark--Evans nearest-neighbor 
clustering index, $R$ \citep{Clark1954}. (Top Row) Galaxies color-coded by nucleation class ({\tt Nuc\_code} $=0$, $1/1^\ast$, $2/2^\ast$). These correspond to non-nucleated galaxies (0); galaxies hosting centrally located, point-like NSCs (1/1$^\ast$); and galaxies with less certain central sources, including diffuse or shallow light profiles and/or sources offset from the photocenter (2/2$^\ast$). (Middle Row) Nucleated galaxies divided by NGVS membership class (0/1/2), corresponding to certain, likely and possible NGVS members.
(Bottom Row) Nucleated galaxies divided by structural code: ellipticals ({\tt E}), transition early-types ({\tt ES+EI}), and late-type systems ({\tt S+SI}). 
}
\label{fig:densmaps_combined}
\end{figure*}


\section{Results} 
\label{sec:results}


\subsection{General Sample Properties}
\label{subsec:sample_properties}

From the full NGVS catalog, we consider 3680 galaxies classified as certain, likely, or possible Virgo members (membership classes 0--2). Valid total $g$-band magnitudes are available for 3679 of these galaxies, spanning the range $8.4 \lesssim g \lesssim 24.3$ mag. Individual analyses below may use subsets of this parent sample depending on the availability and quality of the measurements required for a particular calculation; the relevant selection criteria are specified where appropriate.
\color{black}


\subsection{Magnitude Distribution}
\label{subsec:mag_dist}

Figure~\ref{fig:maghist} shows the distribution of galaxies as a function of absolute dereddened $g$-band magnitude, separated by nucleation class. The classes correspond to {\tt Nuc\_codes} of 0, 1, 1*, 2, 2* and 3, as defined in \S\ref{subsec:nsc_samples}. The left panel shows total counts of galaxies in each magnitude bin, while the right panel shows normalized probability densities, so that the area under each curve is unity.

Several conclusions may be drawn from these distributions. Non-nucleated galaxies ({\tt Nuc\_code} = 0) dominate the faint end of the sample, whereas nucleated systems ({\tt Nuc\_code} = 1 and~2) are most prevalent at intermediate magnitudes. In terms of absolute counts (left panel), the total number of nucleated galaxies reaches a broad maximum at
$-16 \lesssim M_{g,0} \lesssim -12$.
From the bright end of the sample ($M_{g,0}\sim -22$) down to the upper boundary of this plateau, the relative incidence of nucleated galaxies increases, consistent with the well-established trend that nucleation becomes more common toward intermediate-mass galaxies \citep[e.g.,][]{Cote:2006aa, Seth:2008aa, Turner:2012aa, Georgiev:2014aa, den-Brok:2014aa, Spengler:2017aa}.  

The normalized distributions in the right panel highlight differences in the shapes of the luminosity distributions independent of sample size. Because the curves are normalized to unit area, their relative heights reflect differences in the shapes of the luminosity distributions rather than differences in the number of galaxies in each class. In this representation, code~1 galaxies peak near $M_{g,0}\approx -13$, while the code~2 sample extends somewhat toward fainter magnitudes and exhibits a broader distribution. This difference likely reflects increasing uncertainty in NSC identification at low galaxy luminosities, where both surface-brightness limitations and stochastic star-formation features can complicate the classification of central components.

{\tt Nuc\_codes} 1* and 2* identify galaxies that are certainly (1*) or possibly (2*) nucleated, but for which the nuclear parameters could not be derived, often because the galaxy exhibits a complex morphology. These systems are strongly skewed toward the bright end of the distribution ($M_{g,0}\lesssim -17$) and likely trace the emergence of more complex central structures---such as bulges, nuclear disks, or bars---that become increasingly common in more massive galaxies \citep[e.g.,][]{Kormendy2004,Cote:2006aa,Fisher2010,Coelho2011,Turner:2012aa,Ashok+2023}. 

Galaxies with {\tt Nuc\_code = 3} are rare overall (31 objects in total) but also preferentially occupy the bright end of the luminosity function. Many of these systems exhibit central dust lanes, disks, or other morphological complexities that make the presence of a distinct NSC ambiguous. The nucleation fraction, $f_{\rm nuc}$, of Virgo galaxies as a function of galaxy mass and environment will be explored in more detail in \S\ref{subsec:nsc_nucfraction}.


\subsection{Spatial Distribution and Clustering}
\label{subsec:spatial_dist}

The spatial distribution of the nucleated galaxies is shown in Figure~\ref{fig:densmaps_combined} which maps the local projected number density, $\Sigma_{\rm gal}$, of galaxies across the NGVS footprint. The densities are indicated by the underlying grayscale image and have been computed using distances to the tenth nearest neighbor, considering only galaxies brighter than $M_{g,0} = -13.59$ mag (above which the survey is virtually complete; see \citealt{Ferrarese2020}). The highest projected densities correspond to the Virgo A subcluster, centered on M87, and the B subcluster, centered on M49, with progressively lower densities found in the cluster outskirts. Smaller peaks in projected density coincide with the Virgo C subcluster, in the vicinity of M60, the M, W  and W$^{\prime}$ Clouds as well as the Low Velocity Cloud \citep[e.g.,][]{Binggeli:1987aa, Hoffman1989, Mei:2007aa, Grossi2015, Cantiello2024}.

The top row of Figure~\ref{fig:densmaps_combined} separates galaxies by nucleation class. Non-nucleated galaxies ({\tt Nuc\_code} = 0) are distributed throughout the cluster footprint while galaxies with certain NSCs ({\tt Nuc\_code} = 1/1$^\ast$) appear to be preferentially concentrated toward the higher-density regions that define Virgo's main subclusters. Galaxies for which the presence of a nucleus is less certain, many of which may have offset nuclei (see Section~\ref{subsec:offset_nscs}), appear to follow an intermediate distribution, as would be expected if this class included galaxies from each of the two preceding classes. 

To quantify the degree of spatial clustering, we compute the Clark--Evans nearest-neighbor index \citep{Clark1954} $R$ for each subsample, defined as the ratio of the observed mean nearest-neighbor separation to that expected for a random distribution of points within the NGVS footprint. Values of $R < 1$ indicate spatial clustering, while $R \approx 1$ corresponds to a distribution consistent with complete spatial randomness. The expected nearest-neighbor separations are estimated using Monte Carlo realizations (1000 simulations) of random point distributions within the survey footprint; probabilities smaller than the simulation resolution are reported as $p<0.001$.

However, the more relevant comparison is not purely uniform spatial randomness, since Virgo galaxies themselves are clustered within the NGVS footprint. We therefore also compute the Clark--Evans statistic for the full NGVS parent sample of 3680 galaxies, irrespective of nuclear classification. Of these, 3678 lie within the adopted survey footprint used for the spatial-statistics calculation, yielding \(R_{\rm all}=0.94\) (\(p<0.001\)). This value provides a control against which the nucleation subsamples can be compared: if nucleation were assigned randomly among NGVS galaxies, then the nucleated subsamples would be expected to have \(R\) values similar to \(R_{\rm all}\), rather than necessarily approaching \(R=1\). 

With this control in mind, the secure nucleated sample ({\tt Nuc\_code}=1/1$^\ast$) is modestly more clustered than the full NGVS galaxy population, with \(R=0.90\) (\(p<0.001\)) and a median projected density of \(\log\Sigma_{\rm gal}=1.12\). Non-nucleated galaxies have \(R=0.94\) (\(p<0.001\)), essentially identical to the parent-sample value, and a slightly lower median density of \(\log\Sigma_{\rm gal}=1.07\). The uncertain-nucleus sample ({\tt Nuc\_code}=2/2$^\ast$) has \(R=0.96\) (\(p=0.204\)) and \(\log\Sigma_{\rm gal}=1.04\), indicating no statistically significant
excess clustering relative to a random distribution within the footprint and no evidence for stronger clustering than the parent NGVS population. These results support the visual impression that securely nucleated galaxies are somewhat more concentrated toward Virgo's dense substructures, while the uncertain central-source population is spatially broader.

The middle row shows the spatial distribution of nucleated galaxies by NGVS membership class. Certain members (membership class~0) span the full range of cluster environments, from the dense cores of Virgo A and B to the more diffuse outskirts, and are more clustered than the parent sample (\(R=0.89\), \(p<0.001\)). Likely members (class~1) show a similar or slightly stronger apparent concentration (\(R=0.86\), \(p<0.001\)), although the smaller sample size should be kept in mind. Possible members (class~2) have \(R=0.91\) (\(p=0.094\)), consistent with a weaker or statistically marginal clustering signal. Overall, the membership subsamples broadly trace the same large-scale Virgo substructure, supporting their inclusion as likely cluster-associated systems while also reflecting the greater uncertainty of the lower-confidence membership classes.

In the bottom row, nucleated galaxies are divided by NGVS structural code from \citet{Kurzner2025}. Nucleated {\tt E}-type systems dominate the sample and are more clustered than the full NGVS population, with \(R=0.90\) (\(p<0.001\)) and \(\log\Sigma_{\rm gal}=1.11\), consistent with the familiar morphology--density relation in clusters \citep{Dressler:1980aa, Postman1984, Binggeli:1987aa, Whitemore1993, Blanton2009, Cappellari2016}. Nucleated transitional systems ({\tt ES+EI}) have \(R=0.94\) (\(p=0.154\)), very similar to the parent NGVS value, and occupy a broader range of projected densities. Nucleated late-type systems ({\tt S+SI}) are much fewer in number, have the lowest median density (\(\log\Sigma_{\rm gal}=1.01\)), and show only marginal evidence for clustering (\(R=0.87\), \(p=0.074\)). These trends indicate that the apparent environmental dependence of nucleation is closely linked to the underlying morphology--density relation: secure nucleated early-type galaxies are preferentially found in dense Virgo substructures, while transitional and late-type nucleated systems are more broadly distributed across the cluster footprint.


\begin{figure*}
    \plotone{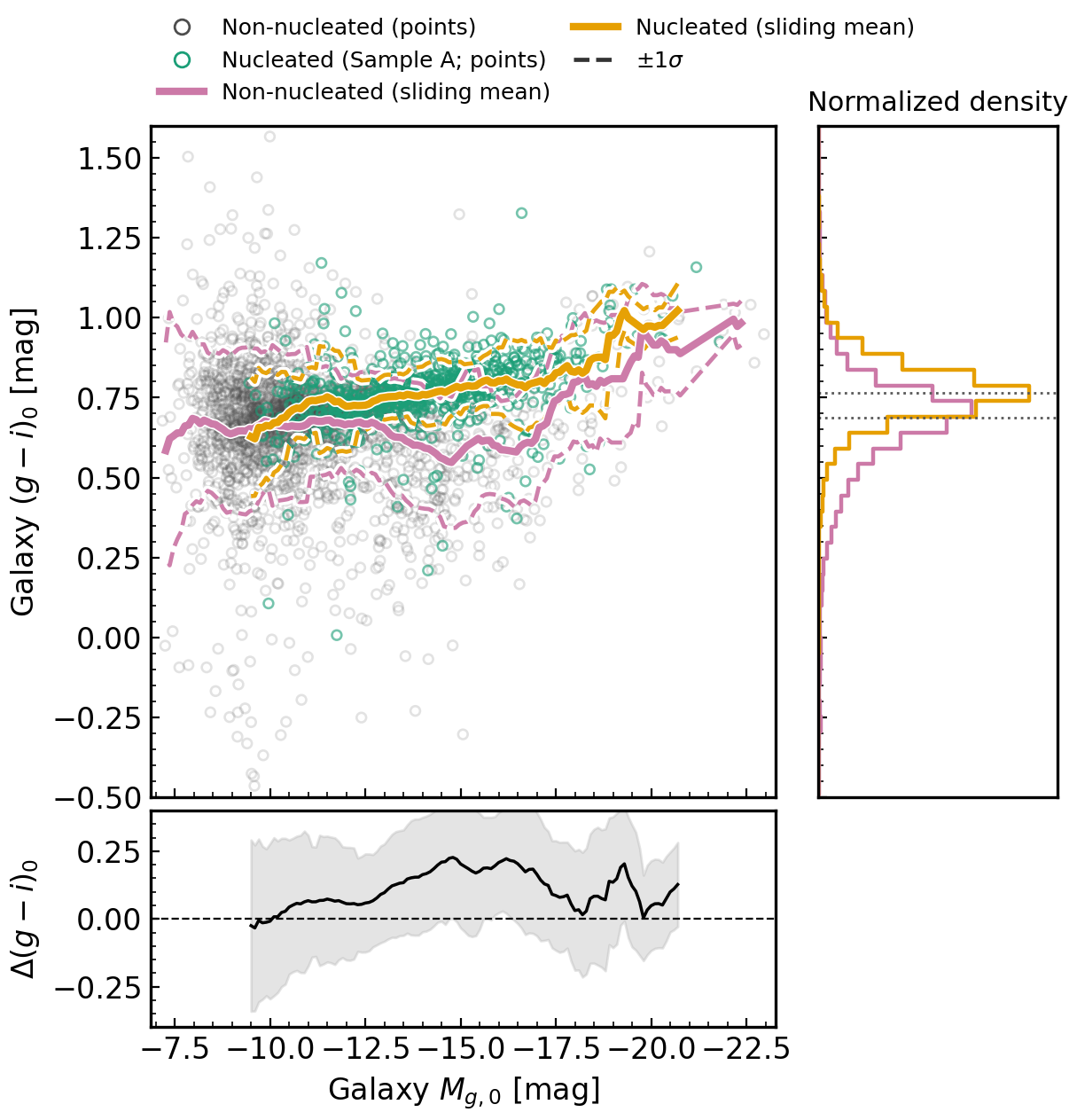}
    \caption{(Upper left) Dereddened galaxy color, ($g - i$)$_0$, as a function of extinction-corrected, absolute magnitude, $M_{g,0}$, for nucleated (green points) and non-nucleated (gray points) galaxies. The solid gold and lavender curves show the average color-magnitude relation for nucleated and non-nucleated galaxies, respectively, estimated from the mean of a sliding bin with a width of 1 mag. Dashed curves show the 1$\sigma$ spread for each population. (Upper right) Normalized color histograms of each population, following the same color coding as the previous panel.  Dotted lines show the median colors of the two groups, with nucleated galaxies, as a whole, appearing slightly redder than their non-nucleated counterparts by 0.077$\pm$0.003 mag. (Lower left) The color difference between nucleated and non-nucleated galaxies as a function of magnitude, using the average curves from the top left panel. The shaded region shows the 1$\sigma$ range. Nucleated galaxies are notably redder primarily in the magnitude range $-17 \lesssim M_{g,0} \lesssim -11$, where NSCs are most abundant.}
    \label{fig:nsc_cmd}
\end{figure*}

\color{black}


\subsection{Stellar Populations}
\label{subsec:stellar_pops_nsc}


\begin{figure}
    \centering
    \includegraphics[width=3.0in]{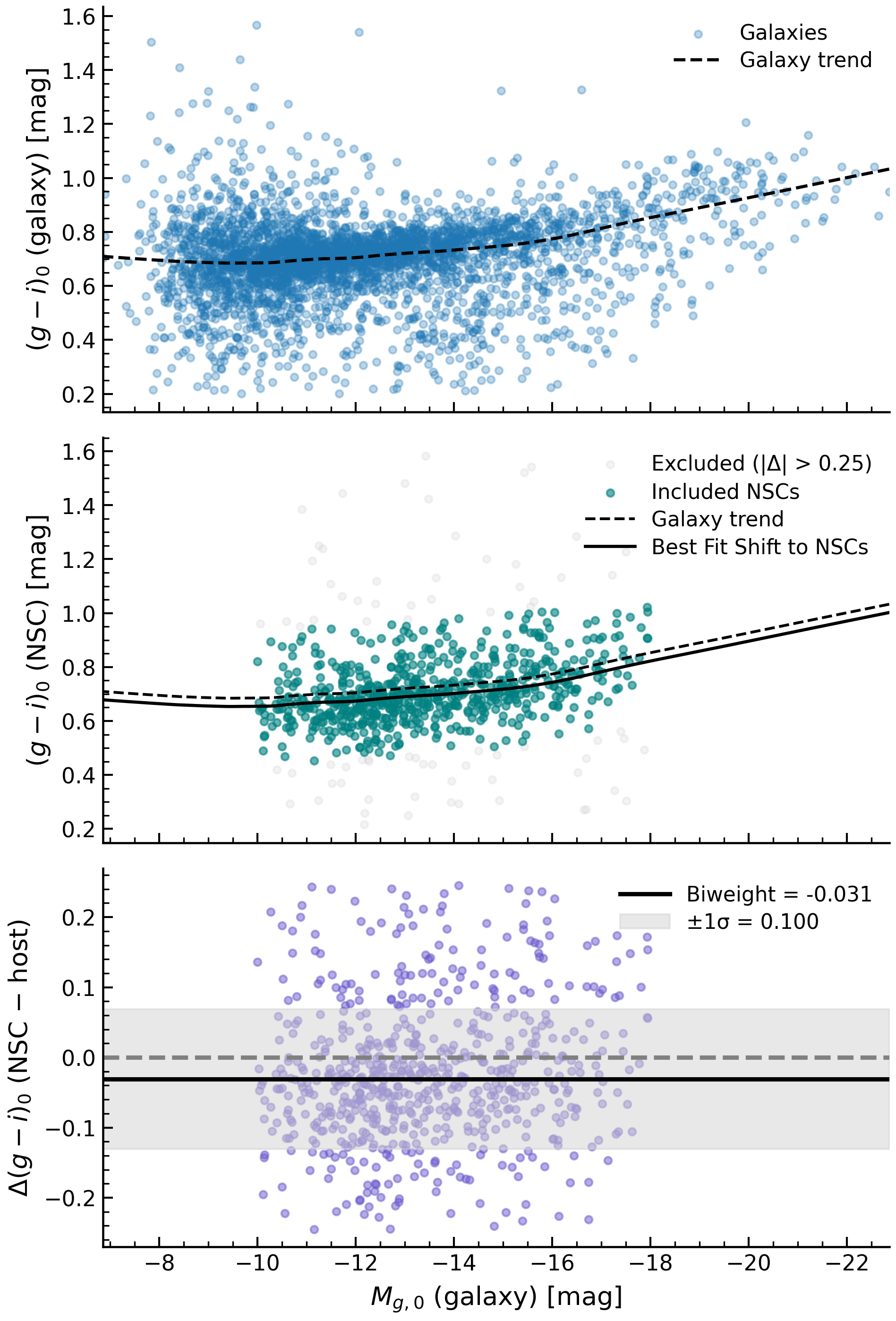}
    \caption{ (Upper) Color magnitude relation for Virgo cluster galaxies. Blue points show dereddened $(g-i)_0$ colors plotted as a function of absolute magnitude, $M_{g,0}$. The dashed black curve shows the LOWESS relation for these data. (Middle) Dereddened NSC color plotted against absolute magnitude of their host galaxies. The dashed black curve is the LOWESS relation shown in the preceding panel. (Bottom) Difference between NSC and host-galaxy color, $\Delta(g-i)_0$, plotted as a function of host-galaxy absolute magnitude, $M_{g,0}$. The dashed horizontal line marks $\Delta(g-i)_0=0$, while the solid black line shows the biweight offset of the NSC colors relative to the host-galaxy LOWESS relation. The grey shaded region shows the corresponding $1\sigma$ scatter, giving $\Delta(g-i)_0=-0.03\pm0.10$ mag.}
    \label{fig:nsc_col_vs_gal}
\end{figure}


\begin{figure}
\plotone{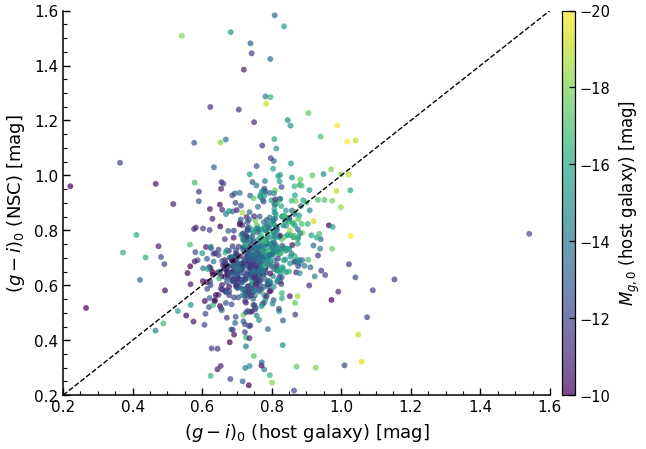}
    \caption{NSC $(g - i)_0$ colors plotted against those of their host galaxies. Each point is color-coded by host galaxy magnitude, with brighter galaxies shown in lighter colors and fainter galaxies shown in darker shades. The black dashed line indicates the one-to-one relation, illustrating that NSCs tend to have slightly bluer colors than their host galaxies across all colors and galaxy magnitudes. }
    \label{fig:nsc_vs_gal_colors}
\end{figure}

The main panel of Figure~\ref{fig:nsc_cmd} shows the color-magnitude relation (CMR) for Virgo galaxies from the NGVS. The CMR for Virgo's core region was presented and discussed in \cite{Roediger:2017aa}, so we refer the reader to that paper for details. Here, we are primarily interested in comparing the distribution of nucleated and non-nucleated galaxies in the color-magnitude plane. 

The gray points in this plot show 2505 non-nucleated galaxies, while the 740 galaxies belonging to Sample A with valid nuclear magnitudes (see Table~\ref{tab:nuc_code_counts}) are indicated by green points; recall that Sample A is the subset of galaxies for which the presence of an NSC can be established with a high degree of confidence. 
The solid gold and lavender curves trace the respective CMRs for nucleated and non-nucleated galaxies, estimated from the mean sliding bin with a width of 1 mag. Although the two galaxy populations are observed to follow broadly similar trends, the nucleated systems are systematically redder than their non-nucleated counterparts. Indeed, the nucleated galaxies are found to mostly reside on the cluster red sequence \citep{Roediger:2017aa}, and are redder across the full magnitude range where nucleated galaxies are common.  

This difference is highlighted in the bottom panel of the figure, which shows the average color difference between the nucleated and non-nucleated populations as a function of apparent magnitude. At the bright end of this distribution, where the nucleation fraction ``turns over" (see Section~\ref{subsec:nsc_nucfraction}), nucleated galaxies are typically redder by $\sim$ 0.05--0.1 mag in $(g - i)_0$, consistent with these galaxies having more evolved and metal-rich stellar populations. At the fainter end, the two populations are more similar in color although the overall number of nucleated galaxies at these magnitudes is small. In the panel on the right, we compare histograms of the nucleated and non-nucleated galaxies (in gold and lavender, respectively). At this global level, the two populations differ in $(g-i)_0$ color by $0.077\pm$0.003, in the sense that the nucleated galaxies are redder. 

A related question is how the colors of NSCs compare to those of their host galaxies. 
The upper panel of Figure~\ref{fig:nsc_col_vs_gal} shows the same CMR for Virgo galaxies, but with no distinction between nucleated and non-nucleated galaxies (blue symbols). The dotted black line shows a non-parametric LOWESS fit to the data. In the middle panel, we compare this relation to the NSCs themselves, based on the sample of 1052 galaxies for which reliable NSC photometry is available (combining Samples A and B; green symbols). The NSCs appear to follow the same trend as their hosts, including the gradual upturn to red colors at galaxy magnitudes above $M_g \sim -15$. The solid black curve in this panel shows this same relation, shifted to best match the NSC data. In this fit we only consider galaxies within the main locus of the NSC  ($-10 < M_{g} < -18$)  and within 0.25 magnitudes of the original host galaxy LOWESS fit to minimize scatter from outliers, leaving 657 systems after clipping.

This comparison should be interpreted with some caution, since broadband colors reflect a combination of stellar age and metallicity. Spectroscopic studies are needed to disentangle the relative importance of these effects. In this context, \citet{Lyu2025} recently compiled a large spectroscopic sample of nearby NSCs and found evidence that the stellar populations of NSCs vary systematically with host-galaxy mass: NSCs in galaxies with $\log(M_\star/M_\odot)<8.5$ are typically more metal-poor than their hosts, whereas those in galaxies with $\log(M_\star/M_\odot)\gtrsim9.5$ are generally more metal-rich. At intermediate masses, NSCs show a more mixed behavior, with comparable numbers being more metal-poor or more metal-rich than their host galaxies. These results suggest that the dominant NSC formation channel may vary with host-galaxy mass, and reinforce the need to interpret the NGVS color offsets as broad stellar-population indicators rather than as direct measures of age or metallicity alone.

The lower panel of Figure~\ref{fig:nsc_col_vs_gal} shows residuals between NSC and host galaxy colors, $\Delta(g-i)_0 = (g-i)_{0,{\rm NSC}} - (g-i)_{0,{\rm host}}$. The distribution of these residuals has a median of $-0.03$~mag and a standard deviation of $0.10$~mag across 657 systems. In other words, NSCs are, on average, slightly bluer than their hosts, consistent with previous findings that NSCs may host younger or more metal-poor stellar populations relative to their surrounding galaxy light \citep[e.g.,][]{Cote:2006aa, Paudel:2011aa, den-Brok:2014aa, Georgiev:2014aa, Pechetti2020}. That said, the large spread in $\Delta(g-i)_0$  highlights a clear diversity in the stellar population properties of NSCs and their hosts. This diversity explains at least some of the conflicting findings on color differences found in previous investigations \citep{Rossa:2006aa, Carson:2015aa, Spengler:2017aa, Ordenes-Briceno:2018aa, Fahrion2021}



\
Finally, in Figure~\ref{fig:nsc_vs_gal_colors} we compare the NSC colors to those of their host galaxies. We find a weak but statistically significant positive correlation: the Pearson correlation coefficient is $r=0.148$ ($p=3.95\times10^{-5}$), while the Spearman rank coefficient is $\rho=0.272$ ($p=1.98\times10^{-14}$). Thus, redder hosts tend to contain
redder NSCs, although with considerable object-to-object scatter. The
median color difference, $\Delta(g-i)_0=(g-i)_{0,\rm NSC}-(g-i)_{0,\rm host}$, is $-0.054$~mag, compared with a standard deviation of $0.192$~mag. The systematic offset toward bluer NSC colors is therefore modest relative to the broad distribution of color differences, consistent with substantial diversity in NSC stellar populations and/or evolutionary histories.
\color{black}


\subsection{Nucleation Fraction}
\label{subsec:nsc_nucfraction}
The fraction of galaxies hosting an NSC, $f_{\rm nuc}$, depends sensitively on host galaxy mass --- an established result that has been shown to hold across a diversity of environments and galaxy populations \citep[e.g.,][SJ+19]{Grant:2005aa,Cote:2006aa,Lisker:2007aa,Turner:2012aa,den-Brok:2014aa,Munoz:2015aa,Eigenthaler:2018aa}. Using the NGVS catalog, we now examine how nucleation fraction varies within the cluster and across the full range in galaxy properties.

\begin{figure*}
    \plotone{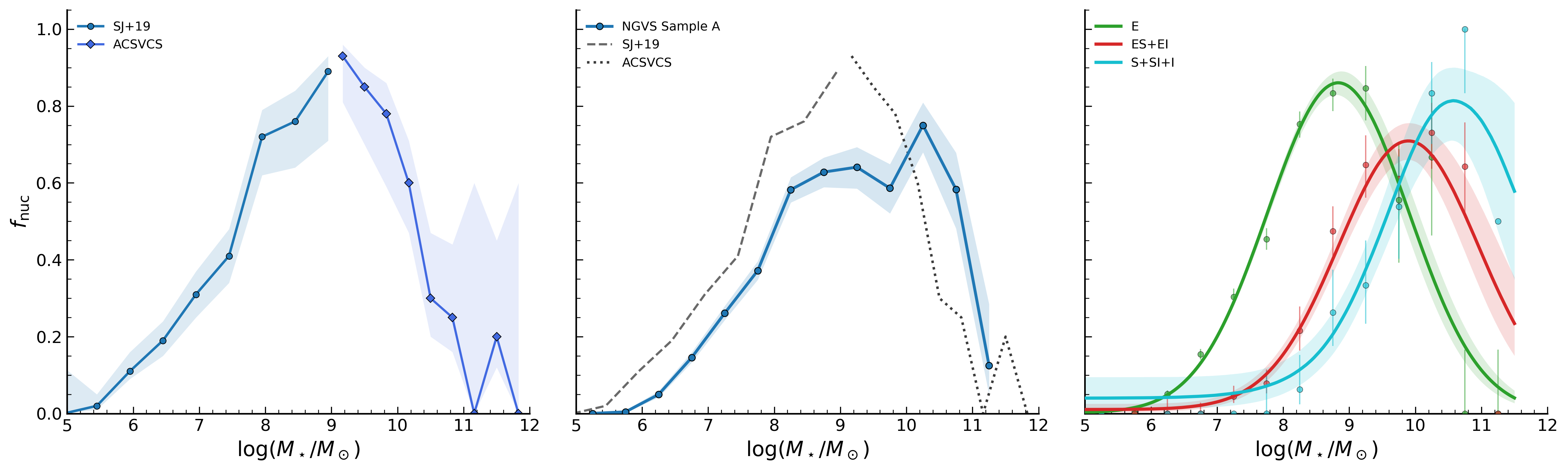}
    \caption{ 
    Nucleation fraction $f_{\rm nuc}$ in the Virgo cluster as a function of galaxy stellar mass. (Left) Published nucleation-fraction relations from the Virgo core sample of SJ+19 (circles) and the ACSVCS survey of high- and intermediate-mass early-type galaxies (diamonds). The shaded regions and error bars show their associated uncertainties. (Middle) The NGVS nucleation fraction for Virgo galaxies using the Sample~A nucleation definition (blue points and curve). The shaded blue region shows the approximately 68\% binomial confidence interval for the binned NGVS measurements, calculated using the Wilson score method. The published SJ+19 and ACSVCS relations from the left panel are overplotted directly for comparison as the gray dashed and dotted curves, respectively. (Right) The nucleation fraction using the Sample~A definition, separated into broad NGVS structural classes. The  green, red, and cyan curves correspond to {\tt E}, {\tt ES+EI}, and {\tt S+SI+I} galaxies, respectively. For each structural group, the colored points show binned nucleation fractions with approximately 68\% binomial confidence intervals calculated using the Wilson score method for visualization, while the solid curves show the median posterior relations obtained by fitting the nucleation state of each individual galaxy with a Bernoulli likelihood. The associated shaded regions show the 16th--84th percentile posterior credible intervals. 
    \color{black}
    }
\label{fig:fnuc_fig1}
\end{figure*}



\begin{figure*}
    \plotone{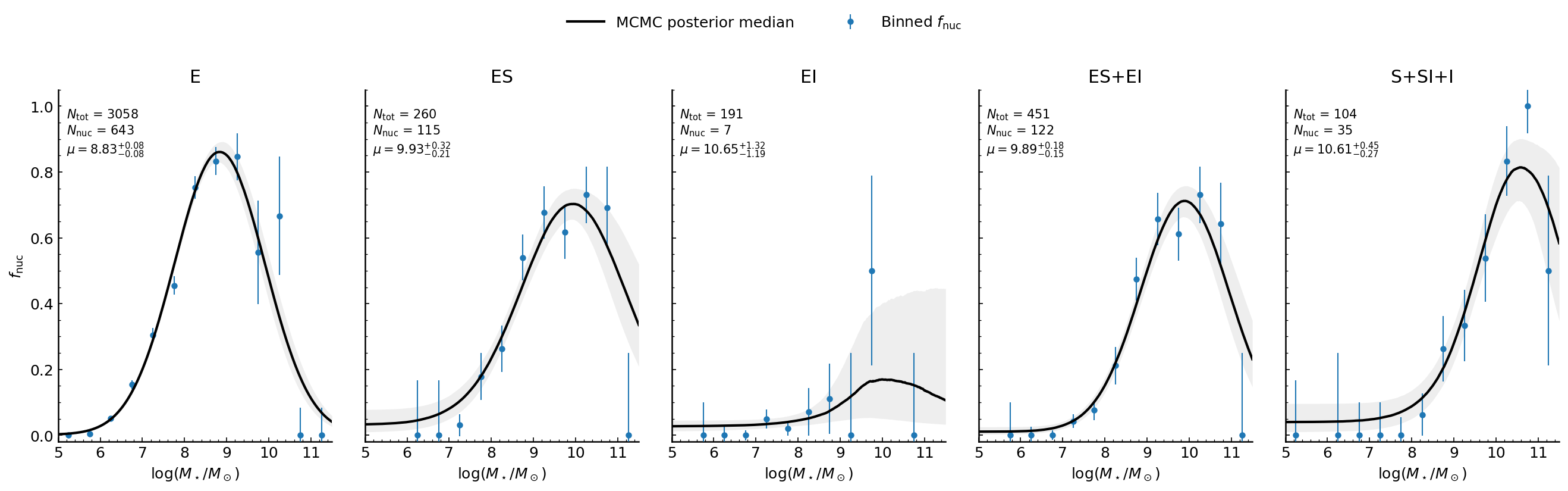}
    \caption{ Variation in $f_{\rm nuc}$ with stellar mass for individual NGVS structural classes. From left to right, the panels show results for {\tt E}, {\tt ES}, {\tt EI}, {\tt ES+EI}, and {\tt S+SI+I} structural codes. All samples use the Sample~A nucleation definition and are restricted to Virgo galaxies in {\tt Membership} classes 0--2. Blue points show the binned nucleation-fraction measurements with approximately 68\% binomial confidence intervals calculated using the Wilson score method and are included for visualization only. The solid black curves show the median posterior relations for the Gaussian-form nucleation model obtained by fitting the nucleation state of each individual galaxy with a Bernoulli likelihood, while the gray shaded regions show the corresponding 16th--84th percentile posterior credible intervals. The inset text in each panel lists the total number of galaxies, the number of nucleated systems, and the posterior constraint on the characteristic peak location $\mu$.} 
\label{fig:fnuc_fig2}
\end{figure*}


\begin{figure*}
  \centering

  \includegraphics[width=16cm]{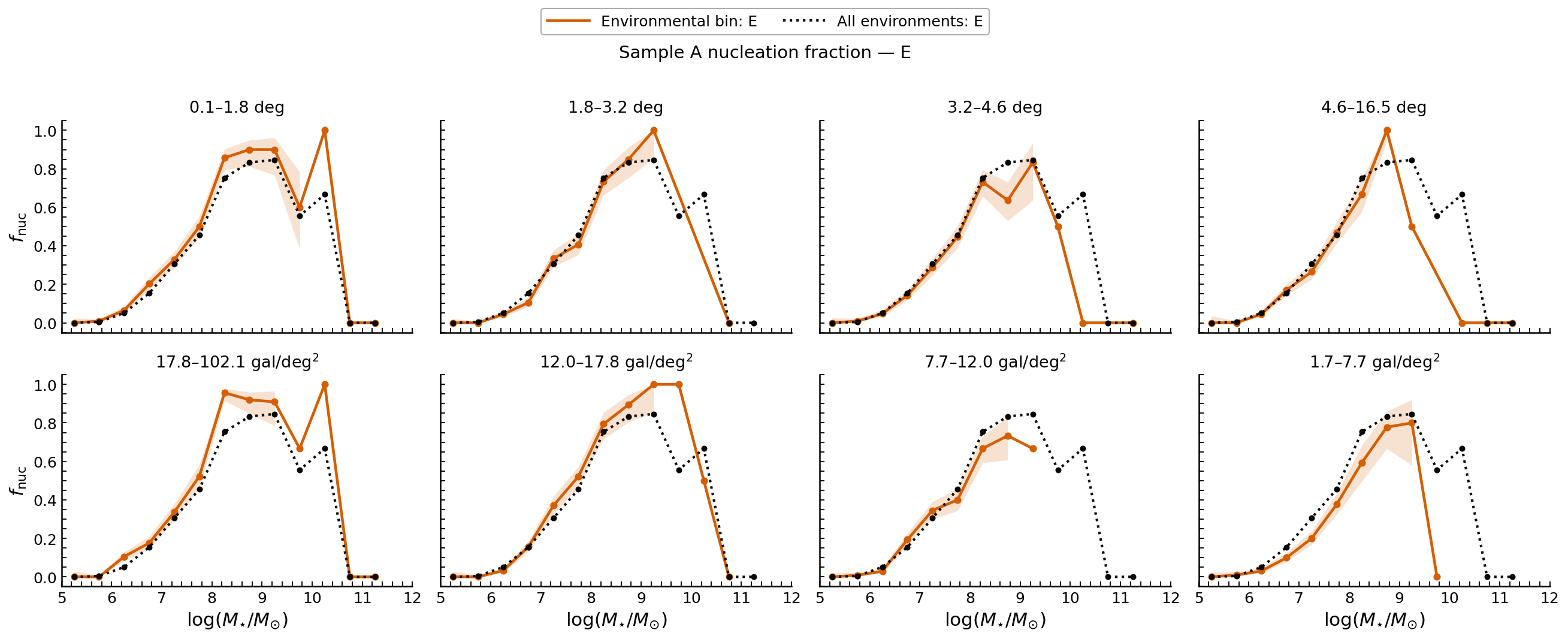}\par\vspace{0.6ex}
  \includegraphics[width=16cm]{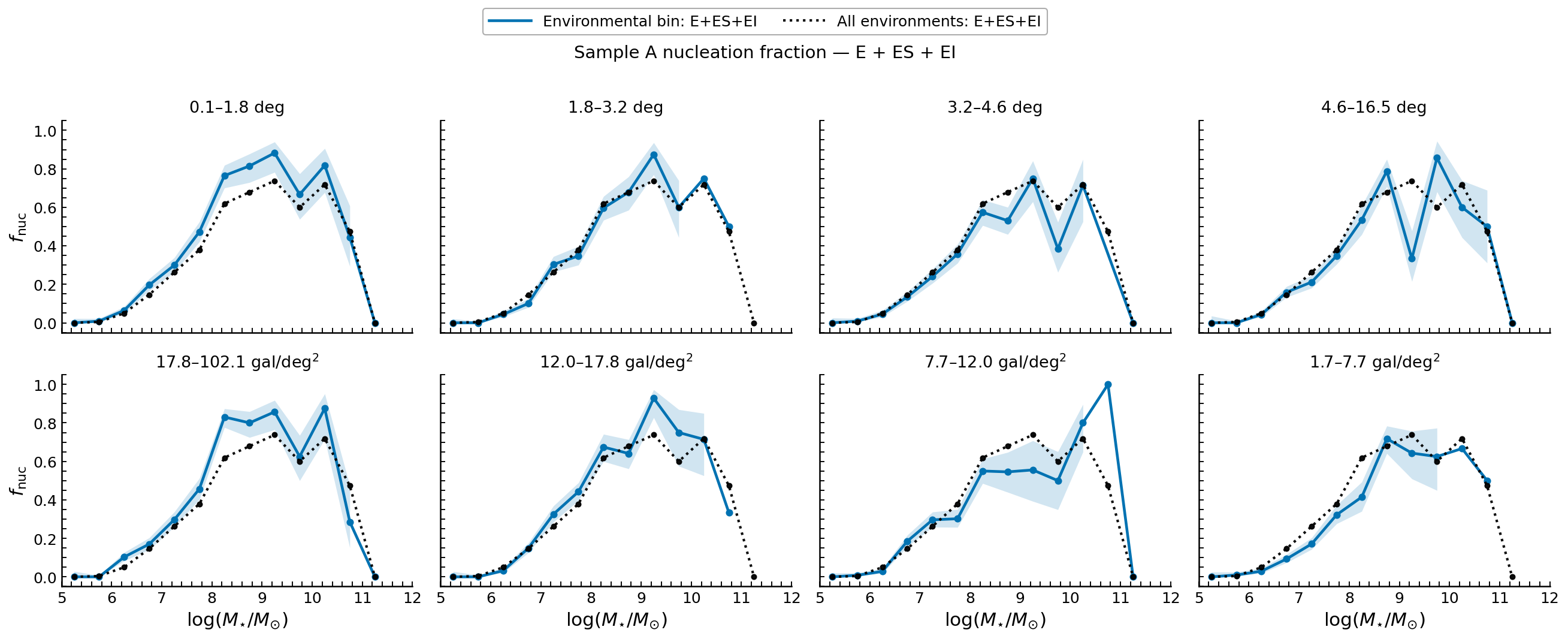}\par\vspace{0.6ex}
  \includegraphics[width=16cm]{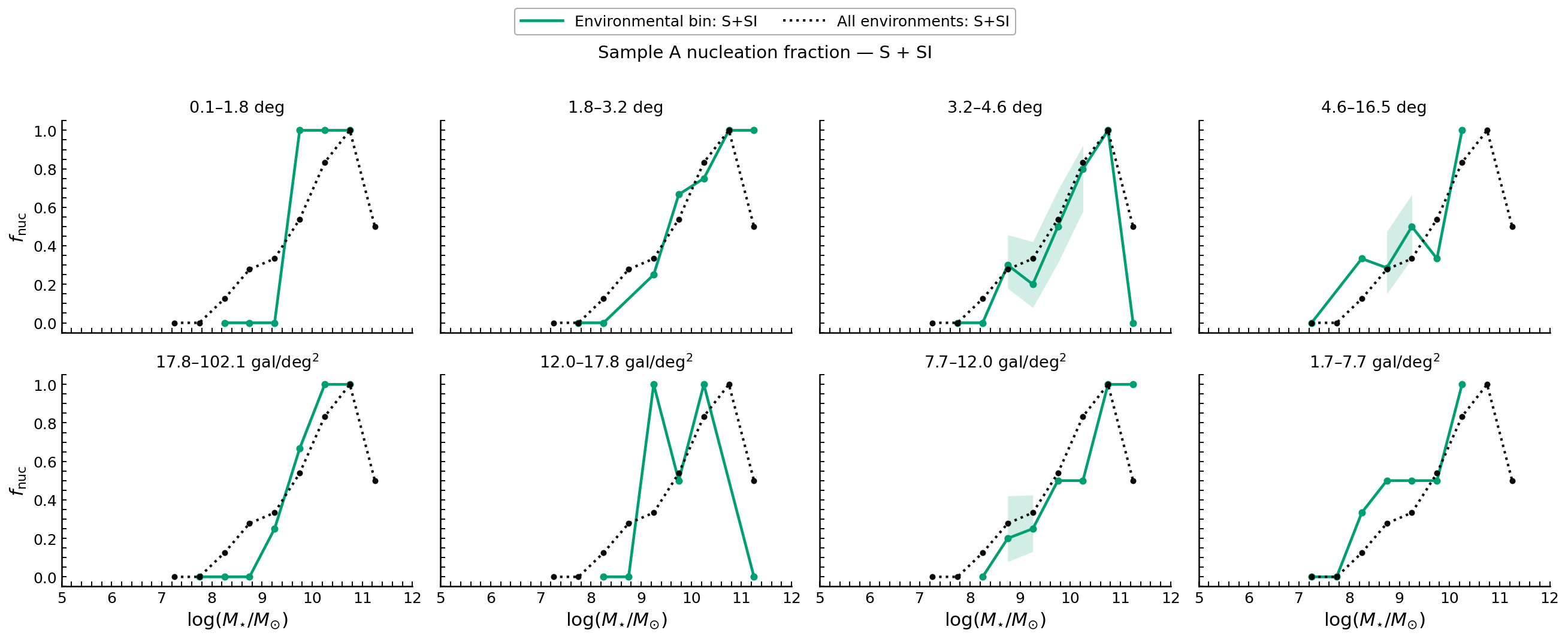}

    \caption{ Environment-binned nucleation fractions for Virgo galaxies using the Sample~A nucleation definition as a function of stellar mass, shown for three structural-morphology selections (top to bottom): {\tt E} only, {\tt E+ES+EI}, and {\tt S+SI}. For each morphology selection, the upper row shows bins in projected distance from M87, while the lower row shows bins in local projected surface density, with density decreasing from left to right. Colored curves show the nucleation fraction within the environmental bin indicated in each panel, with shaded approximately 68\% binomial confidence intervals calculated using the Wilson score method where the number of galaxies is sufficient. The dotted black curve shows the corresponding nucleation-fraction relation for the same morphology selection using galaxies from all environments. The black reference curve is therefore identical among the environmental panels within a given morphology selection, but differs among the three morphology selections. \color{black}}
  \label{fig:nsc_frac_env_binned_by_morph}
\end{figure*}


\begin{figure*}
  \includegraphics[width=17cm]{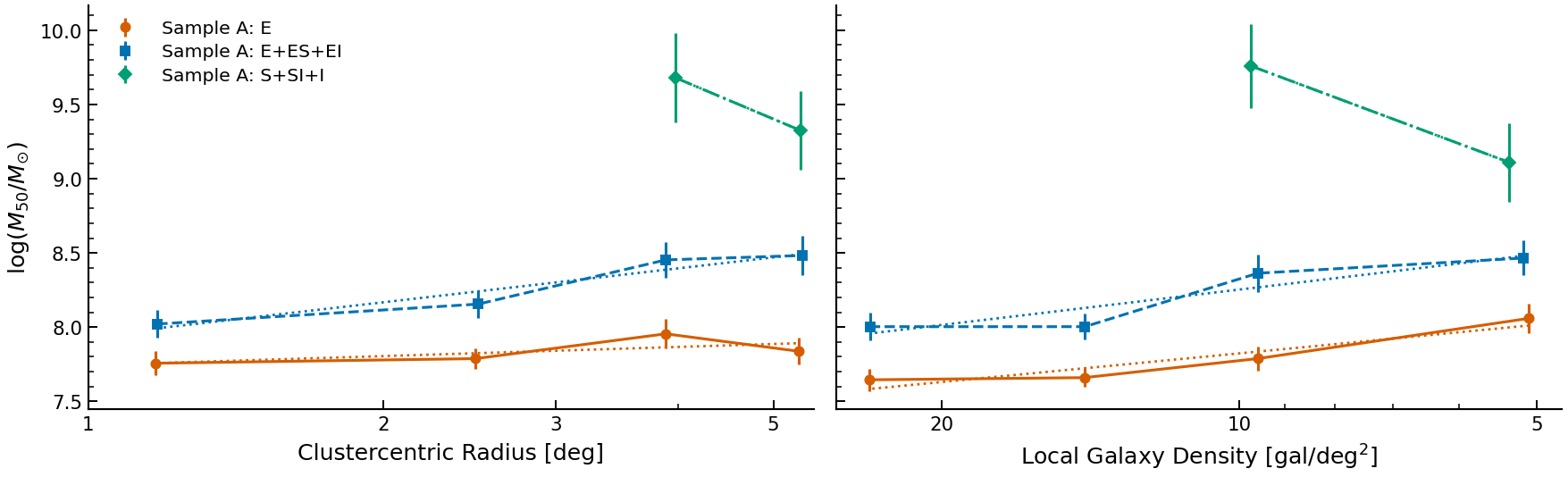}\par\vspace{0.6ex}
  \includegraphics[width=17cm]{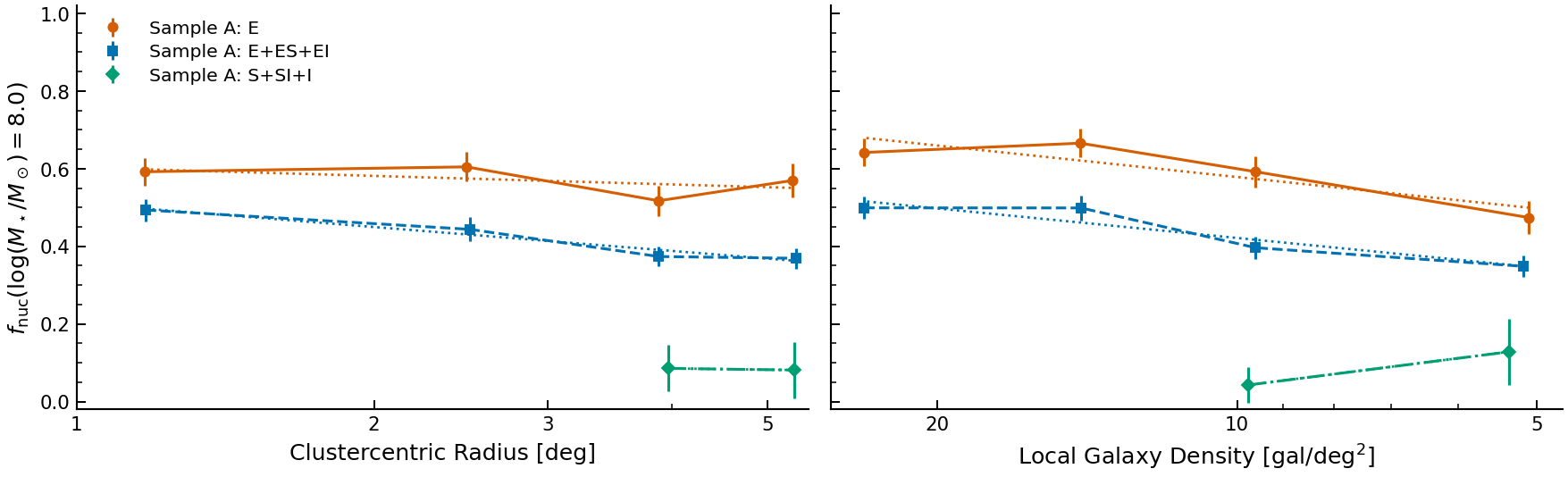}\par\vspace{0.6ex}

  \caption{Variation in nucleation fraction across the Virgo cluster.
(Upper) Stellar mass at which 50\% of galaxies are nucleated, $M_{50}$,
plotted against distance from M87, on the left, and local galaxy density,
on the right. Results are presented for different morphology groupings as
shown in the legend. Within each environmental bin, the nucleation state
of each individual galaxy is fitted with a logistic model, from which
$M_{50}$ is derived. Fits are shown only for bins containing at least 30
galaxies, including at least five nucleated and five non-nucleated systems.
Dotted lines show the best-fit linear relations for each population
grouping. (Lower) Same as above, except showing the nucleation probability
predicted by the fitted relation at a fixed stellar mass of
$\log(M_\star/M_\odot)=8.0$.
}
  \label{fig:fnuc_environment}
\end{figure*}

\begin{table*}[t]
\centering
\caption{ 
Posterior constraints on the nucleation-fraction model for different NGVS structural samples. The model is fitted directly to the individual galaxy nucleation classifications using a Bernoulli likelihood. Reported parameter values are posterior medians with 16th--84th percentile credible intervals.
}
\label{tab:nucfrac_model_params}

\setlength{\tabcolsep}{5pt}
\renewcommand{\arraystretch}{1.15}

\begin{tabular}{lrrrrrr}
\toprule
Group
& $N_{\rm tot}$
& $N_{\rm nuc}$
& $A$
& $\mu$
& $\sigma$
& $C$ \\
\midrule

{\tt E}
& 3056
& 641
& $0.86_{-0.03}^{+0.03}$
& $8.84_{-0.08}^{+0.08}$
& $1.07_{-0.04}^{+0.04}$
& $0.0015_{-0.0011}^{+0.0024}$ \\

{\tt ES}
& 257
& 114
& $0.68_{-0.06}^{+0.06}$
& $9.95_{-0.22}^{+0.34}$
& $1.24_{-0.18}^{+0.27}$
& $0.033_{-0.024}^{+0.046}$ \\

{\tt EI}
& 186
& 7
& $0.30_{-0.23}^{+0.37}$
& $10.67_{-1.20}^{+1.33}$
& $1.11_{-0.72}^{+0.93}$
& $0.027_{-0.015}^{+0.017}$ \\

{\tt ES+EI}
& 443
& 121
& $0.70_{-0.05}^{+0.05}$
& $9.90_{-0.15}^{+0.19}$
& $1.06_{-0.10}^{+0.13}$
& $0.011_{-0.008}^{+0.015}$ \\

{\tt E+ES+EI}
& 3499
& 762
& $0.76_{-0.03}^{+0.02}$
& $9.12_{-0.07}^{+0.08}$
& $1.24_{-0.04}^{+0.04}$
& $0.0014_{-0.0010}^{+0.0022}$ \\

{\tt S+SI+I}
& 104
& 35
& $0.80_{-0.11}^{+0.09}$
& $10.61_{-0.27}^{+0.46}$
& $1.04_{-0.23}^{+0.34}$
& $0.040_{-0.029}^{+0.054}$ \\

\bottomrule
\end{tabular}

\vspace{1mm}

\begin{minipage}{0.96\textwidth}
\footnotesize
\emph{Note.}
The nucleation probability is modeled as
$p_{\rm nuc}(x)=C+A\exp[-(x-\mu)^2/(2\sigma^2)]$, where
$x=\log(M_\star/M_\odot)$.
The fits use individual-galaxy nucleation states rather than binned
nucleation fractions. The NGVS samples are restricted to Virgo members
in {\tt Membership} classes 0--2, with
$5.0\leq\log(M_\star/M_\odot)\leq11.5$.
Following the {\tt Sample~A} nucleation definition, galaxies with
{\tt Nuc\_code}$=1$ or $1^\ast$ are treated as nucleated, while
galaxies with {\tt Nuc\_code}$=0$, $0^\ast$, $2$, or $2^\ast$ are
included as non-nucleated; systems with {\tt Nuc\_code}$=3$ or
$3^\ast$ are excluded. The binned values shown in
Figures~\ref{fig:fnuc_fig1} and \ref{fig:fnuc_fig2} are used only for
visualization and do not enter the likelihood. Because only seven
{\tt EI} galaxies are nucleated, the corresponding model parameters
are substantially less well constrained than those of the other
structural samples. \color{black}
\end{minipage}

\end{table*}


\subsubsection{Dependence on Morphology}
\label{subsec:nsc_nucfraction_morph}

Figure~\ref{fig:fnuc_fig1} shows the nucleation fraction as a function of galaxy stellar mass. The left panel compares the published relations from the Virgo core sample of SJ+19 and the ACSVCS \citep{Cote:2006aa}. Both of these samples are dominated by early-type galaxies: ACSVCS explicitly targeted early-type Virgo members, while the SJ+19 sample was drawn from the Virgo core region, where early-type systems are preferentially found. These literature-based relations therefore provide a useful reference for the behavior expected in early-type dominated samples.

The middle panel of Figure~\ref{fig:fnuc_fig1} shows the NGVS nucleation fraction for the 3603 galaxies entering the Sample~A nucleation-fraction analysis, compared with the SJ+19 and ACSVCS relations (dotted black lines). The overall trend remains the same: $f_{\rm nuc}$ rises with stellar mass, peaks near $\log(M_\star/M_\odot)\sim 9$--10, and then declines again at the highest masses. However, the NGVS relation shows broader structure than is seen in the literature comparison samples. This difference is expected because the NGVS sample spans the full Virgo footprint and includes a much wider range of galaxy morphologies than the core-dominated and early-type-dominated literature samples.


The right panel of Figure~\ref{fig:fnuc_fig1} shows that this broad and multi-peaked behavior is driven by the mixture of host morphologies in the NGVS sample. The data points in this panel show results for three 
morphologically distinct populations: {\tt E} (green), {\tt ES+EI} (red) and {\tt S+SI+I} (cyan). We have fitted the data for these samples using a Bernoulli likelihood, assuming the nucleation probability can be described by a Gaussian distribution. Posterior distributions for the model parameters are sampled using a
Markov chain Monte Carlo (MCMC) method implemented with the {\tt emcee}
ensemble sampler \citep{Foreman-Mackey:2013aa}, yielding direct constraints on the characteristic peak location, $\mu$, and width, $\sigma$, for each structural subsample. The resulting posterior constraints are listed in Table~\ref{tab:nucfrac_model_params}.

The groups show a clear progression in the stellar mass at which the nucleation fraction peaks. For the {\tt E} population, the posterior median peak is $\mu = 8.84^{+0.08}_{-0.08}$, while the combined transition early-type population ({\tt ES+EI}) peaks at $\mu = 9.90^{+0.19}_{-0.15}$. The late-type systems ({\tt S+SI+I}) peak at  higher masses still, $\mu = 10.61^{+0.46}_{-0.27}$. These results show that stellar mass is the primary driver of nucleation fraction, but host morphology introduces a clear secondary dependence. The apparent dip in the full-sample relation near $\log(M_\star/M_\odot)\sim10$ (see the middle panel of this figure) can likewise be understood as a consequence of this morphological mixture. The {\tt E} nucleation fraction is already declining at these masses, while the {\tt ES+EI} relation peaks at higher stellar mass and the {\tt S+SI+I} population peaks at still higher masses.
\color{black}

This morphology dependence is explored in more detail in Figure~\ref{fig:fnuc_fig2}, which shows the individual structural classes and groupings {\tt E}, {\tt ES}, {\tt EI}, {\tt ES+EI}, and {\tt S+SI+I}. The {\tt E} galaxies exhibit a pronounced and well-constrained peak, at $\mu=8.84^{+0.08}_{-0.08}$, while the {\tt ES} and {\tt ES+EI} populations show similar behavior shifted to higher stellar mass, with $\mu=9.95^{+0.34}_{-0.22}$ and $\mu=9.90^{+0.19}_{-0.15}$, respectively. The {\tt EI} subsample contains only seven nucleated galaxies, so its inferred peak location, $\mu=10.67^{+1.33}_{-1.20}$, is weakly constrained. The latest-type systems ({\tt S+SI+I}) again favor a peak at high stellar mass, $\mu=10.61^{+0.46}_{-0.27}$, but with larger uncertainties owing to their smaller sample size.  Figures~\ref{fig:fnuc_fig1} and \ref{fig:fnuc_fig2} together show that the global Virgo nucleation-fraction relation is not universal across host morphologies: the familiar early-type trend is recovered for {\tt E} galaxies, while the other structural populations generally favor higher characteristic stellar masses. 


\subsubsection{Dependence on Environment}
\label{subsec:nsc_nucfraction_env}


We next consider how the nucleation-fraction relation varies with environment. Figure~\ref{fig:nsc_frac_env_binned_by_morph} shows $f_{\rm nuc}$ as a function of stellar mass in bins of projected clustercentric distance from M87 and local projected galaxy density (upper and lower rows, respectively). Results are shown separately for three broad morphology selections: {\tt E} only (orange),
{\tt E+ES+EI} (blue), and {\tt S+SI} (green). Within each morphology selection, the colored curve shows the relation measured in the environmental bin indicated in that panel, while the dotted black curve shows the corresponding relation for galaxies of the same morphology selection drawn from all environments. Thus, the dotted black reference curve is identical across the environmental panels within a given morphology selection, but differs among the three morphology selections. Across all subsamples, the overall mass dependence remains similar from bin to bin, with the dominant feature still being the rise to a broad maximum near intermediate masses followed by a decline at higher masses.
\color{black}

The first point to note from Figure~\ref{fig:nsc_frac_env_binned_by_morph} is that the environmental variations are modest compared to the much stronger dependencies on stellar mass and morphology. Across all three morphology groupings, the dominant feature remains the same broad rise to a maximum at intermediate stellar masses followed by a decline toward higher masses. 
That said, there are discernible shifts with environment. For the {\tt E} sample, the environmental bins largely track the fiducial relation, with typical deviations from the overall relation of only $\langle |\Delta f_{\rm nuc}| \rangle \lesssim 0.1$ across the intermediate-mass regime. Here $\Delta f_{\rm nuc}$ denotes the difference between the nucleation fraction measured in a given environmental bin and the reference relation for the full sample at the same stellar mass. The dispersion among the environmental curves at fixed mass is similarly modest, with typical ranges of $\sim0.18$--0.21 between the lowest- and highest-density bins.

The {\tt E+ES+EI} sample shows somewhat larger environmental variation, with intermediate-mass ranges of $\sim0.22$--0.27 and maximum departures approaching $\sim0.4$--0.5 in a few bins. By contrast, the later-type {\tt S+SI} sample shows apparently larger fluctuations in some cases, but these are based on very few usable mass bins and are therefore much less well constrained. The strongest environmental differences thus appear at intermediate stellar masses, where the nucleation fraction is both most readily measured and intrinsically highest. At the lowest and highest masses, environmental trends are harder to assess because the occupation fraction becomes small and the measurements noisier.

To quantify these relatively subtle environmental variations,
Figure~\ref{fig:fnuc_environment} shows two derived quantities for each
environmental bin: (1) the stellar mass, $M_{50}$, at which
$f_{\rm nuc}=0.5$ (upper panels); and (2) the nucleation fraction evaluated
at a fixed stellar mass of $\log(M_\star/M_\odot)=8.0$ (lower panels).
For these derived fits, we use the broader late-type grouping
{\tt S+SI+I}, rather than the {\tt S+SI} selection shown in
Figure~\ref{fig:nsc_frac_env_binned_by_morph}, in order to retain
sufficient galaxy numbers for the fits in the individual environmental
bins. In the {\tt E} sample, $M_{50}$ varies only modestly with projected
clustercentric radius, spanning
$\log(M_{50}/M_\odot)\simeq7.75$--$7.95$, a total range of
$\sim0.20$ dex. The dependence on local density is somewhat stronger:
$M_{50}$ decreases from $\log(M_{50}/M_\odot)\simeq8.07$ in the
lowest-density bin to $\simeq7.65$ in the highest-density bin, a shift of
$\sim0.42$~dex. This is consistent with the visual impression from
Figure~\ref{fig:nsc_frac_env_binned_by_morph} that the early-type relation
is relatively stable across the cluster. The {\tt E+ES+EI} sample shows
somewhat larger variation, with $M_{50}$ spanning $\sim0.46$ dex across
the radial bins and $\sim0.47$ dex across the density bins. The
{\tt S+SI+I} sample exhibits the highest $M_{50}$ values overall,
reflecting the fact that later-type galaxies must typically reach higher
stellar masses before becoming comparably likely to host NSCs, although
again we note that these measurements are based on far fewer usable
environmental bins and are therefore less secure.

The lower panels of Figure~\ref{fig:fnuc_environment} show that the
fixed-mass nucleation fraction at $\log(M_\star/M_\odot)=8.0$ is highest
for the {\tt E} sample, intermediate for {\tt E+ES+EI}, and lowest for
{\tt S+SI+I} across the environmental bins for which the fits are
sufficiently constrained. For the {\tt E} sample, $f_{\rm nuc}(8)$
changes by only $\Delta f_{\rm nuc}\simeq-0.03$ from the innermost to
the outermost radial bin, while increasing by
$\Delta f_{\rm nuc}\simeq+0.17$ between the lowest- and highest-density
bins. The {\tt E+ES+EI} sample shows a decrease of
$\Delta f_{\rm nuc}\simeq-0.13$ from the innermost to the outermost
radial bin and an increase of $\Delta f_{\rm nuc}\simeq+0.15$ between
the lowest- and highest-density bins.

Figures~\ref{fig:fnuc_fig1}--\ref{fig:fnuc_environment} show that the nucleation fraction across Virgo is governed primarily by stellar mass: it reaches a broad maximum near $\log(M_\star/M_\odot)\sim9$ and declines
toward both lower- and higher-mass systems. This behavior is consistent with previous work in Virgo and other nearby environments \citep[e.g.,][]{Cote:2006aa,Turner:2012aa,Eigenthaler:2018aa}. At fixed stellar mass, galaxy morphology introduces a clear second-order dependence, with the peak of the $f_{\rm nuc}$ relation shifting towards higher masses for progressively later morphological types. Environmental trends are also present within the cluster, but they are subtler than the mass and morphology dependencies and are best understood as modest shifts in the normalization and characteristic mass scale of the underlying relation. These findings show that {\it interpreting a global nucleation fraction curve for any single environment is complicated since the relative number of galaxies belonging to each morphological type itself varies as a function of galaxy mass.}


\subsection{Nuclear Star Cluster Masses}\label{subsec:nsc_massfn}
\label{subsec:nsc_masses}

\subsubsection{Comparison to \citet{Sanchez-Janssen2019}} 
\label{subsubsec:nsc_check}


As an external check on the NSC stellar mass measurements, we compare our NSC masses with the published estimates from SJ+19. Recall from \S\ref{subsubsec:SED_nscs} that the NSCs in this paper are computed from mass-to-light ratios derived from the NSC $(g-i)$ colors while those in SJ+19 are based on full $u^*griz$ SED-fitting for NSCs in the central 3.7 deg$^2$ of the cluster. 

Of the 109 NSCs with published stellar masses in SJ+19, 96 have mass estimates in our sample. Defining $\Delta\log M_{\rm NSC} =
\log M_{\rm NSC,this\ work}-\log M_{\rm NSC,SJ+19}$, we find a median offset of only $+0.010$ dex, with a 95\% bootstrap confidence interval of $-0.011$ to $+0.047$ dex. The residuals have an NMAD of $0.110$ dex and a standard deviation of $0.165$ dex, comparable to the typical $0.15$--$0.20$ dex uncertainties of our NSC mass estimates, while both the Pearson and Spearman correlation coefficients are $0.965$.

Only one of the 96 matched nuclei differs by more than 0.5 dex. This is VCC~923, for which our inferred NSC mass is higher by approximately 0.80 dex. An inspection of the underlying measurements indicates that this discrepancy is driven by the derived nuclear luminosity rather than by the colour-based mass-to-light ratio. Excluding this object changes the median offset only from $+0.010$ to $+0.008$ dex and the NMAD from $0.110$ to $0.109$ dex. We therefore find no evidence for a significant systematic offset in our NSC stellar masses and only modest object-to-object scatter relative to the independent SJ+19 measurements.
\color{black}

\begin{figure}
    \plotone{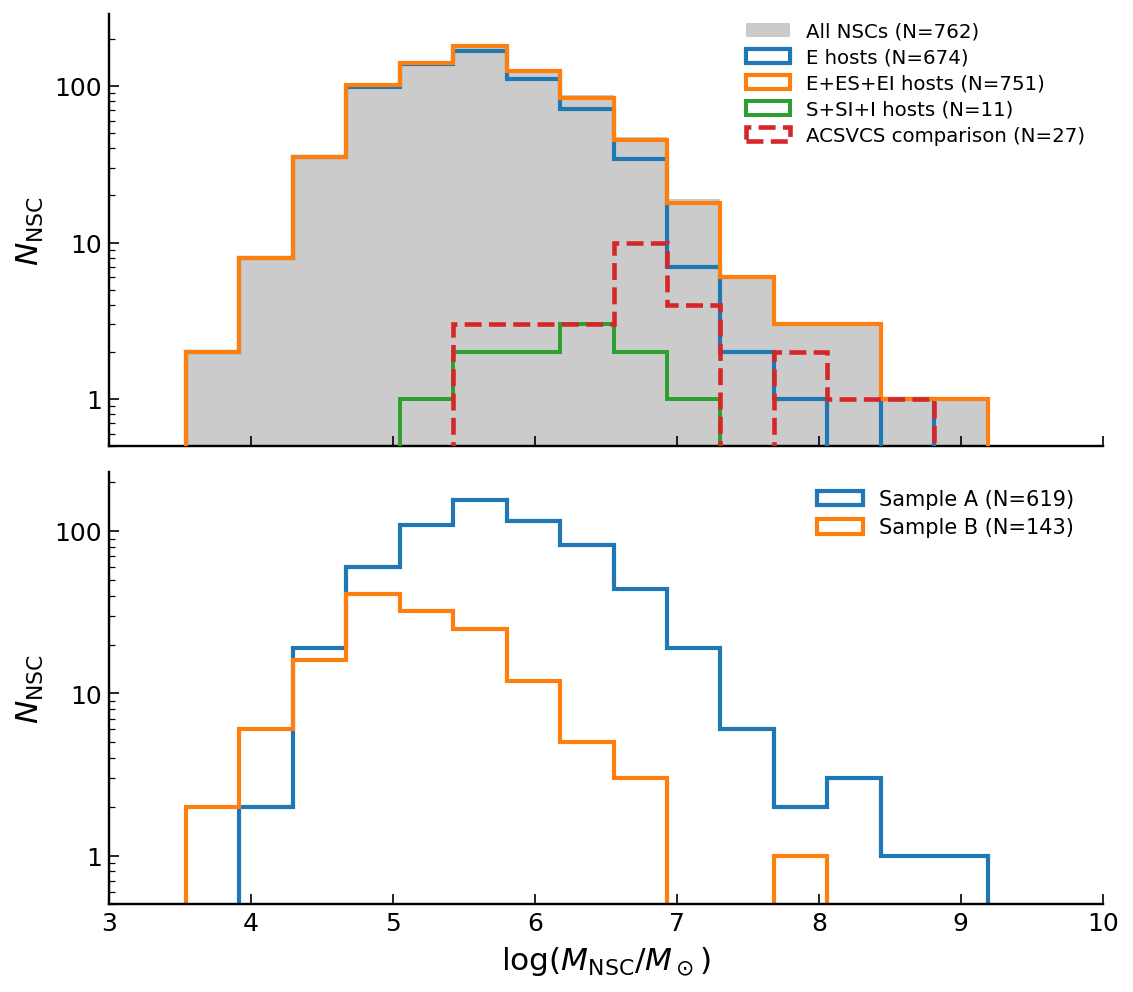} 
    \caption{(Upper) Distribution of NSC masses for all nucleated galaxies (Sample A+B; see \S\ref{subsec:SED} for details). (Lower) The same distribution, but divided into the Sample A (blue) and Sample B (orange) subsets. Lower-confidence nucleated systems (typically belonging to Sample B) tend to be found at the low-mass end of the distribution where NSCs are rarer, and more difficult to detect and classify robustly.}
    \label{fig:nsc_mass_hist}
\end{figure}


\begin{figure*}
\plotone{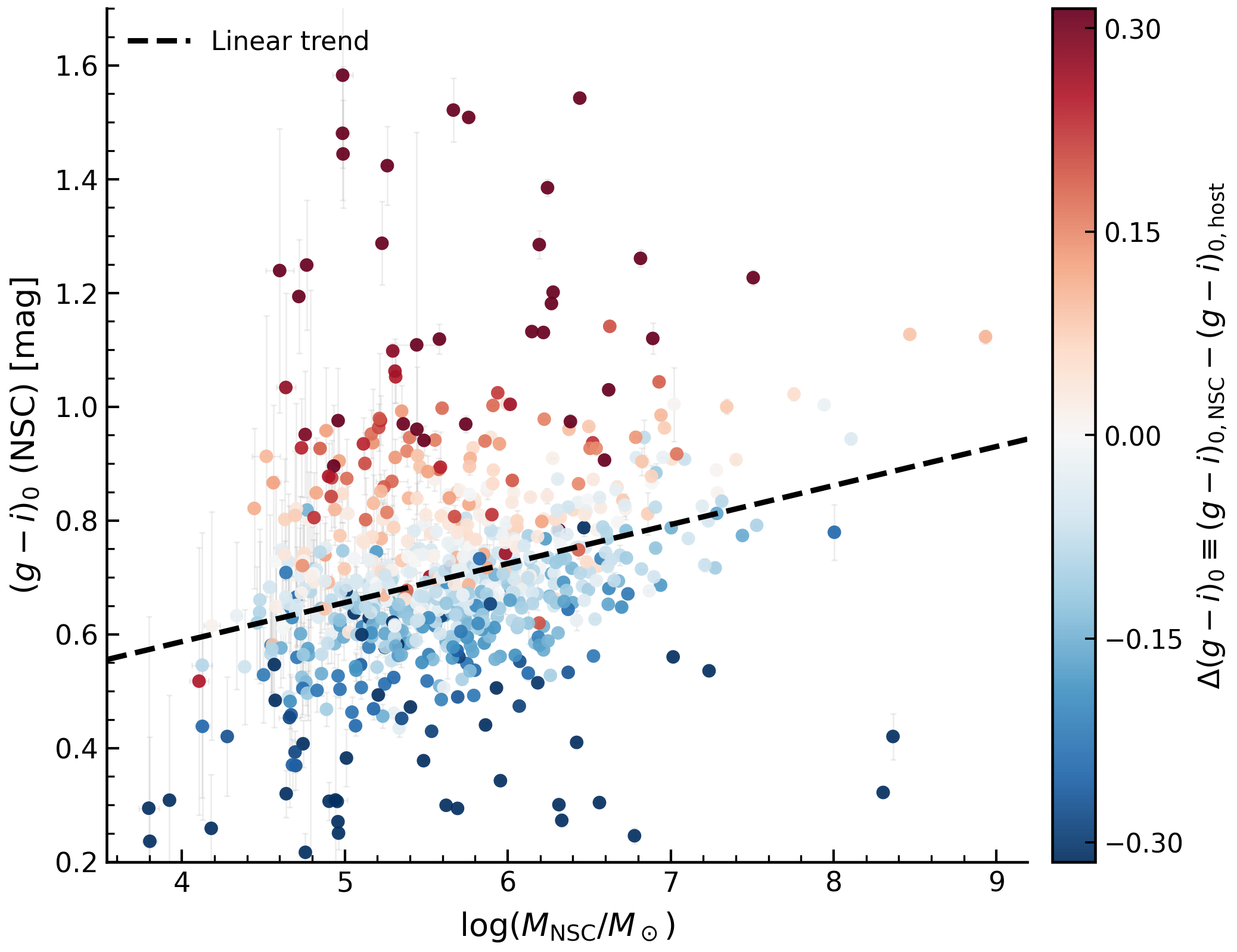
}
    \caption{NSC color, $(g-i)_0$, as a function of NSC stellar mass for Sample~A galaxies. Points are color-coded by the color offset relative to the host galaxy, $\Delta(g-i)_0 = (g-i)_{0,\rm NSC}-(g-i)_{0,\rm host}$, with blue and red points indicating NSCs that are respectively bluer and redder than their hosts. The dashed line shows the best-fit linear trend, selected using the Akaike Information Criterion \citep{Akaike1974}. The substantial scatter at fixed NSC mass reflects the diversity of NSC stellar populations.} 
    \label{fig:nsc_vs_gal_masses}
\end{figure*}

\subsubsection{Distribution of Masses}

The upper panel of Figure~\ref{fig:nsc_mass_hist} shows the NSC mass function (for the combined A and B samples) to be broad and roughly log-normal in form, with a peak at $\log(M_{\rm NSC}/M_\odot) \simeq 5.67$.  The high-mass tail reaching to $\log(M_{\mathrm{NSC}}/M_\odot)\gtrsim 8$ may indicate the presence of genuinely massive NSCs in a handful of galaxies, or large errors in the measured masses for some of the brightest and most structurally complex galaxies. The colored histograms in this panel show the division of NSC masses by host galaxy morphology, as indicated in the legend. This breakdown demonstrates that most of the NSCs in this high-mass tail are associated with galaxies having transitional morphologies ({\tt ES} and, to a lesser extent, {\tt EI} systems). The inner regions of these galaxies are often structurally complex and show evidence for star formation associated with dissipational processes.
\color{black}

The lower panel of Figure~\ref{fig:nsc_mass_hist} shows the result of splitting the sample into our high-confidence (Sample~A) and low-confidence (Sample~B) subsamples. Most high- and intermediate-mass NSCs reside in Sample~A, whereas Sample~B contributes disproportionately at the low-mass end. This behavior is not unexpected: low-mass NSCs are intrinsically fainter, more susceptible to confusion with background sources and therefore less likely to be confidently identified as {\it bona fide} NSCs during visual classification.

The comparison between the NGVS and literature-based samples highlights the importance of survey depth and host galaxy selection when interpreting the NSC mass function. Because the ACSVCS sample is drawn from relatively massive early-type galaxies, it preferentially probes the high-mass end of the NSC distribution. In contrast, the NGVS extends to much lower stellar masses and includes a substantially larger population of dwarf galaxies, revealing a broader NSC mass distribution that peaks at lower masses. We will return to this point when discussing the dependence of NSC mass on host galaxy properties.

\subsubsection{Masses and Stellar Populations} 

Figure~\ref{fig:nsc_vs_gal_masses} shows that, for Sample A, NSC colors correlate weakly with stellar mass, in the sense that more massive NSCs tend to be redder, although with substantial scatter at fixed $M_{\rm NSC}$. The dashed line shows the best-fit linear trend, selected over alternative functional forms using the Akaike Information Criterion \citep{Akaike1974}. The broad dispersion about this relation indicates that NSCs do not follow a single, tight stellar-population sequence, and likely reflects a range of formation and enrichment histories. We return to the possible origins of this diversity in Section~\ref{sec:discussion}. \S\ref{subsec:nsc_samples}

The points in Figure~\ref{fig:nsc_vs_gal_masses} are color-coded by their NSC--host color offset, $\Delta(g-i)_0$, as defined in \S\ref{subsec:stellar_pops_nsc}. This comparison shows that, at fixed NSC mass, nuclei span a wide range of colors relative to their host galaxies. Many NSCs are bluer than their hosts, particularly among low- and intermediate-mass nuclei, while others are comparable to or redder than the surrounding galaxy light. This diversity suggests that NSC color is not determined by host color alone, but instead reflects a combination of stellar mass, local enrichment history, and the relative contributions of different formation channels.

The tendency for many NSCs to be bluer than their hosts may indicate that they are marginally younger and/or more metal-poor than the surrounding galaxy light \citep[e.g.,][]{Paudel:2011aa, Georgiev:2014aa, Spengler:2017aa,Fahrion2021}. At the same time, the large scatter in $\Delta(g-i)_0$ suggests that no single stellar-population pathway dominates across the full sample. This is consistent with scenarios in which NSCs assemble through a combination of {\it in situ} star formation and the inspiral of GCs \citep[e.g.,][]{Tremaine:1975aa, CapuzzoDolcetta1993, Lotz:2001aa, Milosavljevic:2004aa, Hartmann:2011aa, Antonini:2012aa, Guillard:2016aa}.

\begin{figure*}
    \centering
    \includegraphics[width=0.95\textwidth]{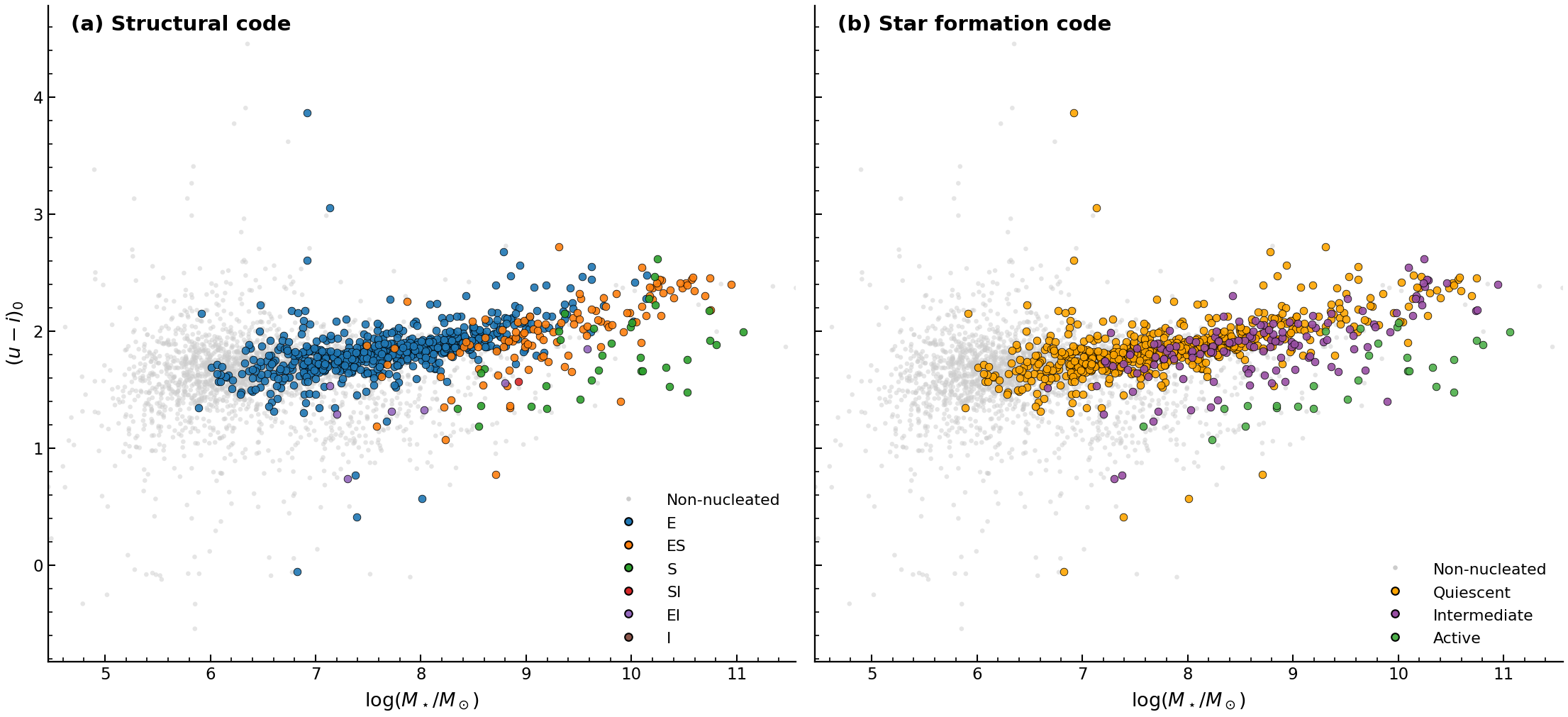}
    \caption{Plots of galaxy $(u-i)_0$ color versus stellar mass for different morphological types. (Left) Nucleated galaxies belonging to Sample A are color-coded by NGVS structural code from \citet{Kurzner2025}. (Right) As in the preceding panel but with Sample A galaxies color-coded by NGVS star formation code \citep{Kurzner2025}. In both panels, non-nucleated galaxies are shown as faint gray points to illustrate the full galaxy population. Nucleated galaxies are found mainly on the red sequence and have ({\tt E} and {\tt ES}) structural codes and {\tt Q}-type star formation codes (with a lesser contribution by {\tt I}-types). At the high-mass end of the population, {\tt S}-type galaxies with {\tt A-}type star formation codes begin to dominate the sample of nucleated galaxies. }
    \label{fig:cmd_nsc}
\end{figure*}

Further insights are available from the color--mass relation shown in Figure~\ref{fig:cmd_nsc}, which serves as a coarse proxy for stellar population age and chemical enrichment \citep{Bell:2003aa,Boselli2006}. In this figure, nucleated galaxies from Sample~A are shown in the context of the full galaxy population, with non-nucleated systems plotted as faint gray points. In the left panel, where galaxies are color-coded by structural type, nucleated systems are found predominantly along the red sequence and are largely associated with early-type morphologies ({\tt E}). Transitional systems ({\tt ES} and {\tt EI}) span a broader range in color at fixed stellar mass, indicating a mix of star formation histories, while late-type galaxies contribute more significantly at higher masses, where they extend toward slightly bluer colors. The right panel, which shows the same galaxies color-coded by star formation code, reinforces this picture. Nucleated galaxies are primarily associated with quiescent ({\tt Q}) systems, consistent with their concentration along the red sequence, with a smaller contribution from intermediate ({\tt I}) types. At higher stellar masses, however, actively star-forming ({\tt A}) galaxies become more prevalent among the nucleated population, indicating that NSC formation can also occur in systems with ongoing star formation.

Taken together, these trends suggest that NSCs are most commonly found in galaxies that have undergone significant chemical enrichment and are evolving along the red sequence, but are not restricted to such systems \citep[e.g.,][]{Georgiev:2014aa, Neumayer2020}. In particular, the presence of nucleated galaxies among both quiescent and star-forming populations indicates that NSC formation can occur across a range of star formation states.

This result does not uniquely distinguish between formation channels. For example, GC inspiral can operate in both quiescent and star-forming galaxies, while ongoing star formation in gas-rich systems may additionally contribute to NSC growth \citep[e.g.,][]{Milosavljevic:2004aa, Hartmann:2011aa, Antonini:2012aa, Guillard:2016aa}. The coexistence of nucleated galaxies in both regimes therefore supports a picture in which multiple formation pathways contribute, with their relative importance likely varying with galaxy mass and evolutionary state.


\begin{figure*}
    \centering
     \includegraphics[width=0.80\textwidth]{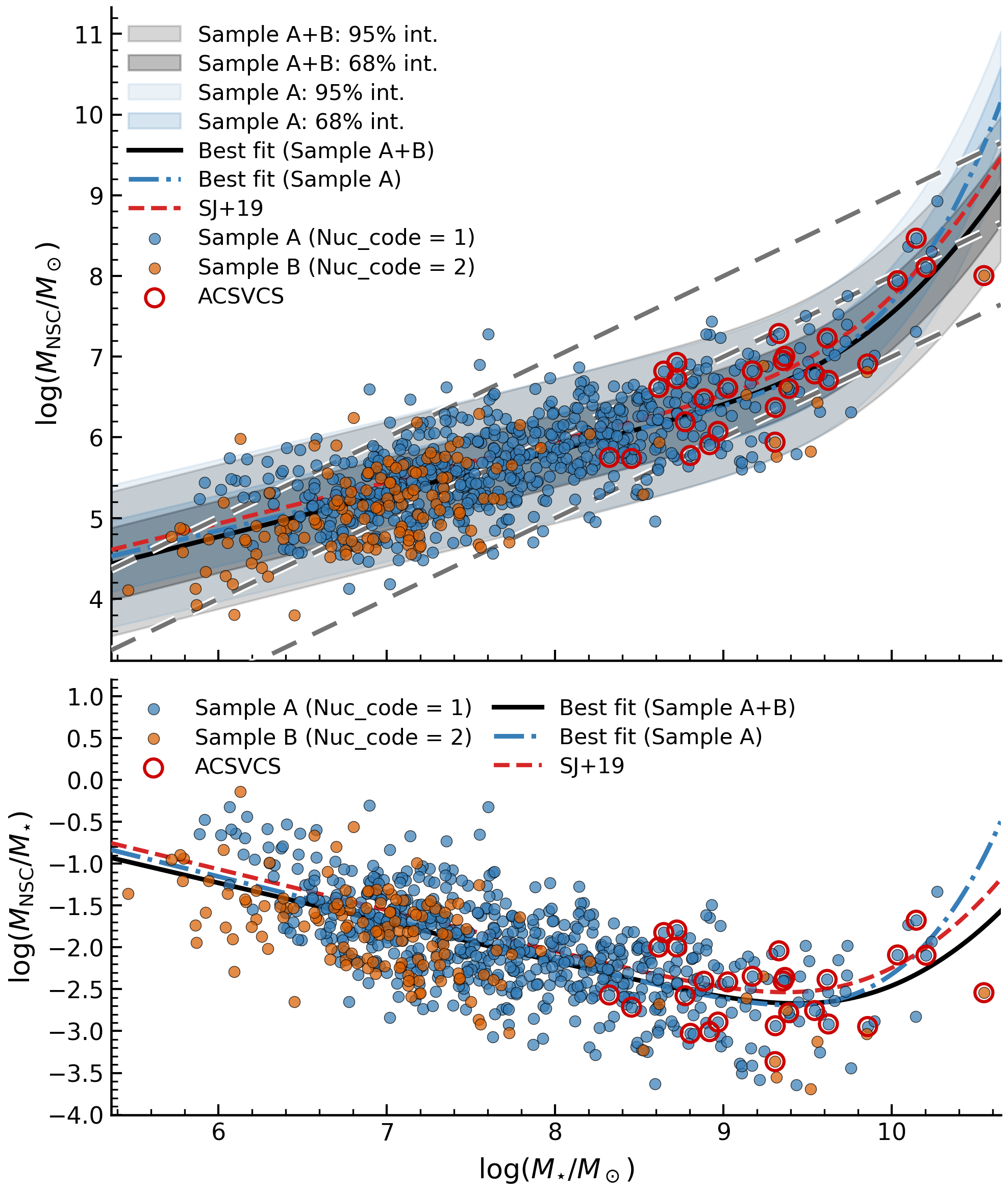}
    \caption{
        (Upper) NSC mass plotted against galaxy stellar mass. Blue points refer to galaxies in Sample~A, while orange points show those in Sample~B. The black curve indicates the best-fit relation from this work for the full sample, and the blue dash-dotted curve shows the fit to Sample~A alone. The red dashed curve shows the relation for the Virgo core region from \citet{Sanchez-Janssen2019}, while open red circles indicate galaxies from the ACSVCS \citep{Cote:2004aa}. Shaded bands around the blue and black curves represent the 68\% and 95\% total residual-scatter intervals for each sample. For reference, the dashed gray lines indicate constant mass fractions of $M_{\mathrm{NSC}}/M_\star $ = 0.001, 0.01 and 0.1.
        (Lower) The NSC-to-galaxy stellar mass ratio, $\Delta_{\rm NSC} = \log(M_{\mathrm{NSC}}/M_\star)$, as a function of host galaxy stellar mass. The curves are derived from the best-fit relations in the upper panel (i.e., not independently refit). This representation highlights the non-linear dependence of NSC mass fraction on galaxy mass, with a minimum near $\log(M_\star/M_\odot) \sim 9.8$ and an increase toward both lower and higher masses, although the high-mass behavior remains poorly constrained.}
    \label{fig:nsc_mass_relation_data}
\end{figure*}

\subsubsection{$M_{\rm NSC}$--$M_{\star}$ Relation and NSC Mass Fractions} 

To put the NSC masses in context, we now compare them to the SED-based stellar masses 
of their host galaxies. Figure~\ref{fig:nsc_mass_relation_data} shows the $M_{\rm NSC}$--$M_{\star}$ relation for Samples A and B, using blue and orange symbols, respectively. A well-defined relation is apparent over more than five orders of magnitude in galaxy stellar mass; this is consistent with earlier results, based on a smaller sample of galaxies, found in the Virgo core region (SJ+19). To guide the eye, 
the dashed gray lines in this figure indicate constant mass fractions of $M_{\mathrm{NSC}}/M_\star $ = 0.001, 0.01 and 0.1. On average, the data roughly follow the $M_{\mathrm{NSC}}/M_\star $ = 0.01 relation, but it is clear that no single mass fraction can match the full dataset. Most notably, there is an apparent upturn at $M \gtrsim 10^{10}M_{\odot}$, with a shallower relation exhibited by galaxies below this inflection point. 

As first noted by SJ+19, the shape of the relation in Figure~\ref{fig:nsc_mass_relation_data} is  reminiscent of the stellar-to-halo mass relation (SHMR). We follow SJ+19 in fitting a functional form inspired by the SHMR \citep[e.g.][]{Behroozi2013, Moster2013}: 

\begin{eqnarray}
\log M_{\rm NSC}
 &=& \log M_{0,{\rm NSC}}
     - \frac{1}{2}
     + \beta\,\log\!\left(\frac{M_{\star}}{M_{0,{\rm gal}}}\right) \nonumber\\
 &&\quad
     + \frac{\left(\dfrac{M_{\star}}{M_{0,{\rm gal}}}\right)^{\delta}}
            {1 + \left(\dfrac{M_{\star}}{M_{0,{\rm gal}}}\right)^{-\gamma}} \, .
\label{eq:nsc_mass_relation}
\end{eqnarray}
This five-parameter functional form  reproduces the observed curvature in the $M_{\rm NSC}$--$M_{\star}$ relation, including the steepening (flattening) at high (low) galaxy masses. For this choice of parameterization: 
\begin{itemize}
    \item $\log M_{0,{\rm NSC}}$ and $\log M_{0,{\rm gal}}$ are characteristic 
    NSC and galaxy masses that define the turnover scale of the relation;
    \item $\beta$ sets the power-law slope at the low-mass end 
    ($M_{\rm NSC}\propto M_{\star}^{\beta}$);
    \item $\delta$ sets the asymptotic high-mass slope, controlling how rapidly the NSC mass relation steepens above $M_{0,{\rm gal}}$;
    \item $\gamma$ regulates the smoothness and symmetry of the transition between the low-mass and high-mass power-law regimes.
\end{itemize}


We fix $\gamma \equiv 1$ because the limited number of galaxies at the high-mass end provides insufficient leverage to constrain this parameter robustly. This was also the choice adopted by SJ+19, who found $\gamma$ to be largely unconstrained for the same reason. Fixing $\gamma$ reduces the dimensionality of the fit and facilitates a direct comparison with their results. The four remaining parameters of the mean relation --- $\log M_{0,{\rm NSC}}$, $\log M_{0,{\rm gal}}$, $\beta$, and $\delta$ --- are inferred simultaneously with the residual-variance parameter $\ln V$.

We fitted the relation separately to Sample~A and to the combined Sample~A+B population. The resulting samples contain 617 and 760 galaxies, respectively. For galaxy $i$, with $x_i=\log(M_{\star,i}/M_\odot)$ and $y_i=\log(M_{{\rm NSC},i}/M_\odot)$, we adopt a Gaussian likelihood,

\begin{equation}
\ln \mathcal{L}
=
-\frac{1}{2}\sum_i
\left[
\frac{\left(y_i-f(x_i;\boldsymbol{\theta})\right)^2}{V}
+\ln(2\pi V)
\right],
\label{eq:nsc_mass_likelihood}
\end{equation}
where $f(x_i;\boldsymbol{\theta})$ is the relation defined in
Equation~\ref{eq:nsc_mass_relation} and $V=\exp(\ln V)$. No additional
per-object measurement-error term is included in this likelihood, so
$\sqrt{V}$ represents the total residual scatter about the fitted
relation rather than a measurement-error-corrected intrinsic scatter.

We adopt bounded uniform priors of
$5<\log M_{0,{\rm NSC}}<9$,
$6<\log M_{0,{\rm gal}}<11$,
$0<\beta<10$,
$0<\delta<10$, and
$-5<\ln V<2$. We sample the posterior using \texttt{emcee} \citep{Foreman-Mackey:2013aa} with 64 walkers for 10,000 steps, discarding the first 2,000 steps as burn-in and retaining every tenth sample. Convergence is assessed using the integrated autocorrelation time; for both fits, the post-burn-in chain length exceeds 90 times the largest estimated autocorrelation time. Parameter estimates are reported as the medians of the marginalized posterior distributions, with uncertainties given by the 16th and 84th percentiles.

\color{black}
The upper panel of Figure~\ref{fig:nsc_mass_relation_data} shows the best-fit $M_{\rm NSC}$--$M_{\star}$ relations for both Sample A (dotted-dashed blue curve) and Sample A+B combined (solid black curve). The dashed red curve shows the best-fit relation from SJ+19. Overall, our results are in good agreement with the earlier findings from the core region, aside from a difference in normalization: i.e., the global values for the cluster point to average NSC masses, for a given galaxy mass, that are 1.3--1.9$\times$ lower than what is found in the cluster core over the range $6 \le \log({M_*/M_{\odot}}) \le 9$.

The low-mass slope, $\beta$, is consistent within the quoted uncertainties, indicating that NSCs in the Virgo core and throughout the cluster scale with host mass at a similar rate. The characteristic masses $(M_{0,{\rm NSC}}, M_{0,{\rm gal}})$ and high-mass curvature term, $\delta$, show mild changes relative to SJ+19, which most likely reflect a combination of sample differences and our nucleation classification encompassing both high- and low-probability nuclei. The constraint on $\delta$ is sensitive to the sparse sampling at the high-mass end. In particular, removing VCC~1692, the highest-mass Sample~B galaxy at $\log(M_{\star}/M_\odot)=10.55$, changes the Sample~A+B posterior from $\delta=0.39^{+0.16}_{-0.15}$ to $\delta=1.09^{+0.59}_{-0.37}$ and substantially broadens the credible interval. Removing the next two highest-mass Sample~B galaxies produces comparatively little additional change, indicating that VCC~1692 provides substantial leverage on the inferred high-mass behavior. The high-mass curvature should thus be interpreted with caution. The residual scatter remains substantial
($\sim0.4$\,dex), underscoring the large object-to-object variation in NSC masses at fixed host mass.
\color{black} 

In the lower panel of Figure~\ref{fig:nsc_mass_relation_data}, we plot these same data but now showing the variation of logarithmic mass fraction
\begin{eqnarray}
\Delta_{NSC}  \equiv \log(M_{\rm NSC} / M_{\star,{\rm host}}).
\label{eq:nsc_mass_fraction}
\end{eqnarray}
The three curves shown in the upper panel have now been recast in terms of this ratio. The overall trends can be summarized as follows. First, the NSC -- host galaxy mass fraction, $\Delta$, is clearly coupled --- nonlinearly --- to host galaxy mass, with a progressive increase as one approaches the regime of the lowest-mass dwarfs. Second, an apparent minimum in $\Delta$ is found at a galaxy stellar mass of $\log(M_{\mathrm{gal},\star}/M_\odot) = 9.87 \pm 0.11$ --- close to where the NSC nucleation fraction reaches its maximum value (see, e.g., Figure~\ref{fig:fnuc_fig1}). Third, the NSC mass fraction increases above this point, although the behavior in this high-mass regime is poorly constrained by the NGVS data alone, which is sparse. Finally, although these general trends are clear, the relation exhibits significant residual scatter at all masses, which may point to a diversity of NSC formation pathways and/or additional variables that control the growth of NSCs across the galaxy mass spectrum. We shall return to these issues in \S\ref{sec:discussion}.



\begin{deluxetable*}{lccc} 
\tablecaption{Posterior Constraints on the NSC-to-galaxy Stellar Mass Relation. For the NGVS fits, $\sqrt{V}$ denotes the total residual scatter in dex.
\label{tab:nsc_massfit_params}}
\tablehead{
\colhead{Parameter} &
\colhead{Sample A} &
\colhead{Sample A+B} &
\colhead{SJ+19}
}
\startdata
$\log M_{0,\mathrm{NSC}}$      & $7.25^{+0.12}_{-0.13}$ & $7.32^{+0.11}_{-0.11}$ & $7.29^{+0.11}_{-0.13}$ \\
$\log M_{0,\mathrm{gal}}$      & $9.80^{+0.12}_{-0.13}$ & $9.88^{+0.11}_{-0.11}$ & $9.73^{+0.10}_{-0.12}$ \\
$\beta$                        & $0.50^{+0.03}_{-0.03}$ & $0.53^{+0.02}_{-0.02}$ & $0.46^{+0.03}_{-0.04}$ \\
$\delta$                       & $0.64^{+0.34}_{-0.23}$ & $0.39^{+0.15}_{-0.15}$ & $0.43^{+0.23}_{-0.17}$ \\
$\ln V$                        & $-1.64^{+0.06}_{-0.06}$ & $-1.61^{+0.05}_{-0.05}$ & $-1.83^{+0.09}_{-0.09}$ \\
$\sqrt{V}$ (dex)               & $0.44$ & $0.45$ & $0.40$ \\
\enddata
\end{deluxetable*}



\begin{figure*}
    \centering
     \includegraphics[width=0.70\textwidth]{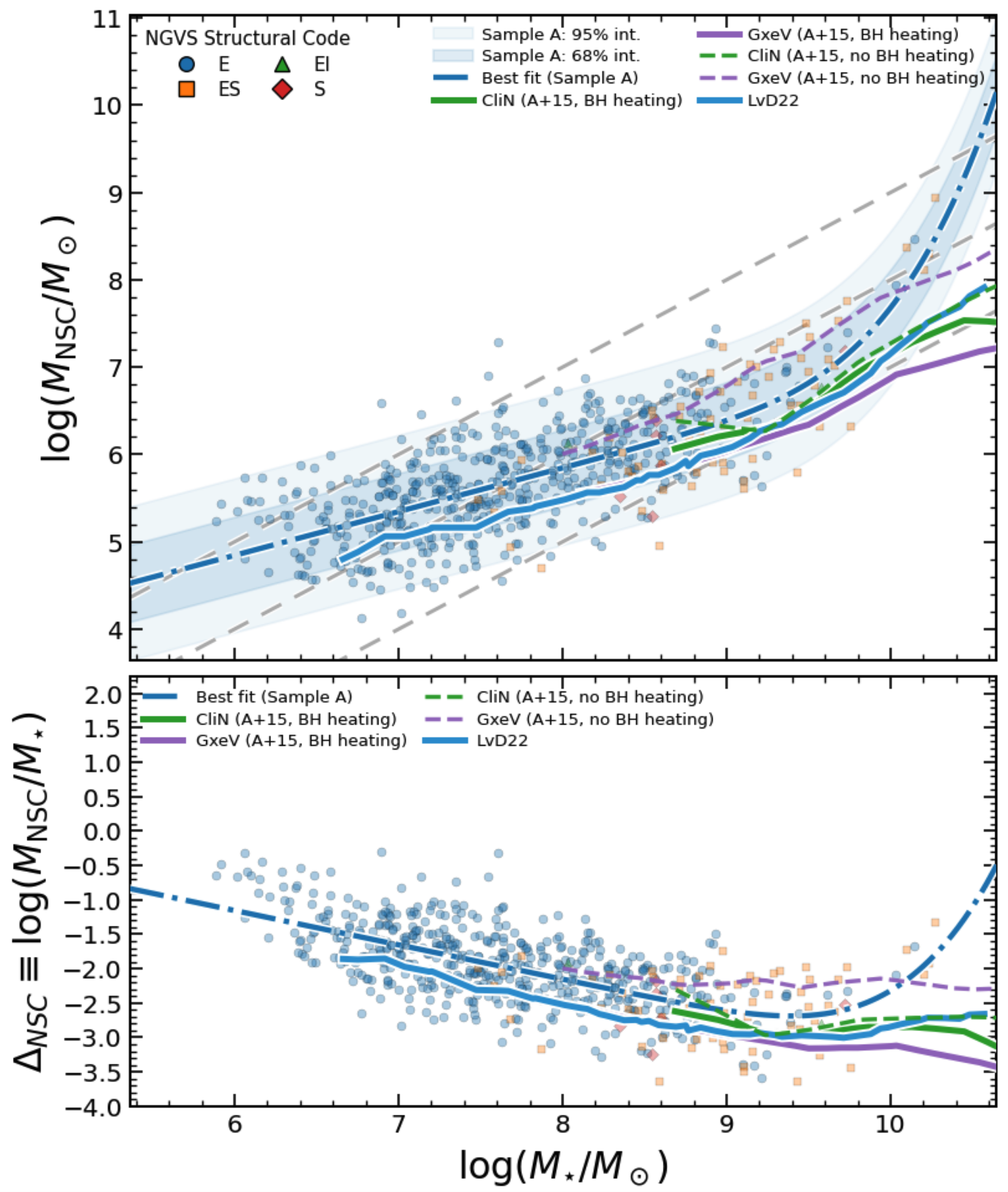} \caption{ NSC mass-host galaxy relation (top) and NSC-to-galaxy mass ratio (bottom) for nucleated NGVS galaxies in Sample~A. Galaxy morphologies are indicated using different symbols, as shown in the legend. The posterior-median relation for Sample~A is shown by the dot-dash blue curve, with the shaded regions indicating the corresponding 68\% and 95\% residual-scatter intervals. We also include predictions from several NSC formation models: the {\tt CliN} and {\tt GxeV} models of \citet{Antonini:2015aa} (with and without black hole heating) and the GC infall model of \citet{Leaman2022} ({\tt LvD22}). \color{black}
}
    \label{fig:nsc_mass_fraction_theory}
\end{figure*}


\begin{figure*}
    \centering
     \includegraphics[width=0.80\textwidth]{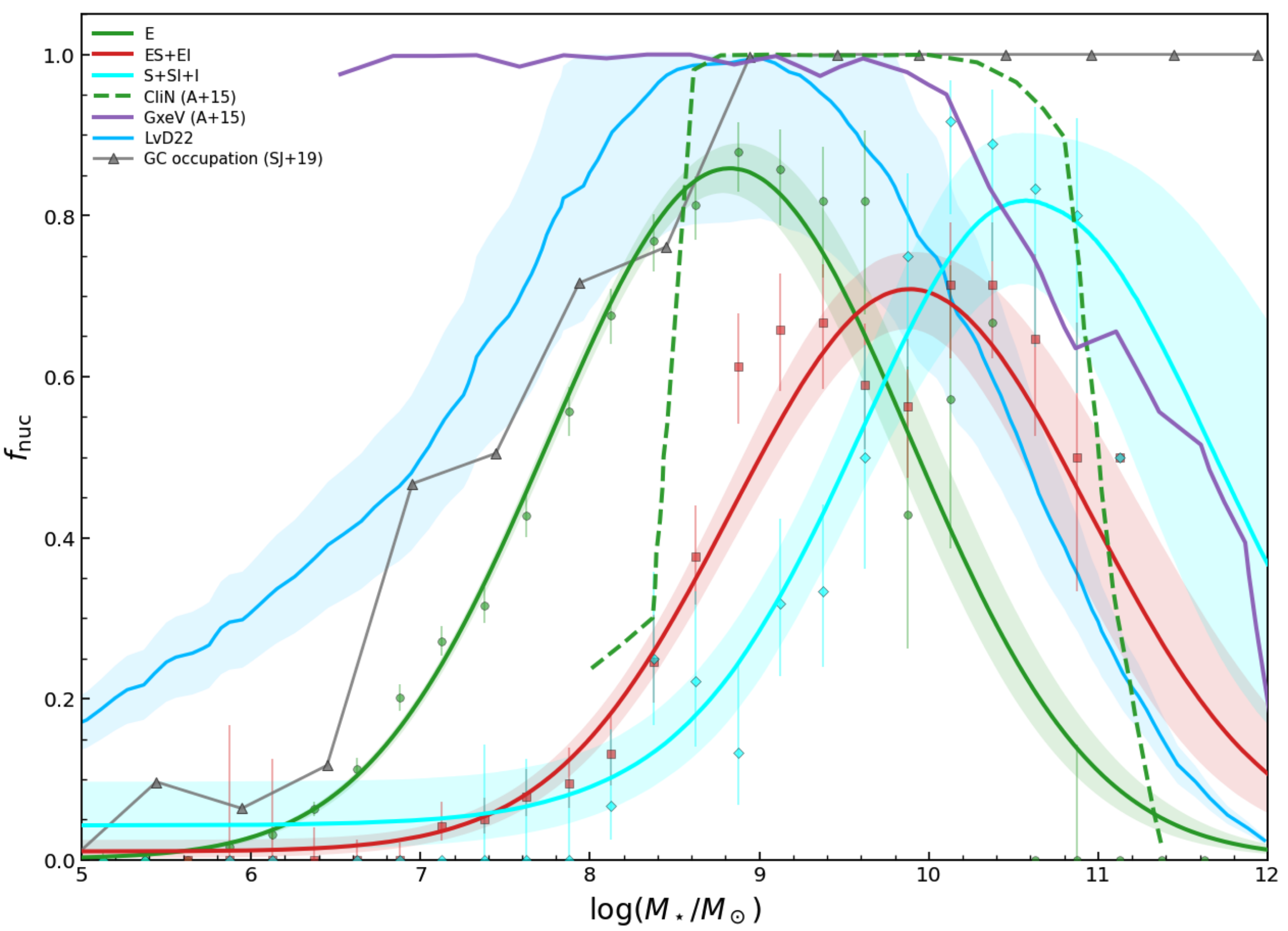}
     \caption{Nucleation fraction, $f_{\rm nuc}$, as a function of host-galaxy stellar mass for Virgo galaxies, shown separately by structural morphology. The green points show the nucleation fraction for galaxies with {\tt E}-type morphologies using the Sample~A nucleation definition, computed in sliding stellar-mass windows; vertical error bars show approximately 68\% binomial confidence intervals calculated using the Wilson score method. The solid green curve shows the posterior-median nucleation-fraction relation obtained by fitting the individual galaxy nucleation states with the Bernoulli likelihood model described in \S\ref{subsec:nsc_nucfraction_morph}, while the green shaded region shows the corresponding 16th--84th percentile posterior interval. The red and cyan points, curves, and shaded regions show the corresponding measurements and posterior constraints for the broader morphological groupings {\tt ES+EI} and {\tt S+SI+I}, respectively. We also show the median relation from \citet[][{\tt LvD22}]{Leaman2022} and model predictions from \citet{Antonini:2015aa} ({\tt CliN} and {\tt GxeV}). Gray triangles denote the GC occupation fraction in Virgo galaxies from \citet{Sanchez-Janssen2019}.} 
     \label{fig:nsc_morph_theory}
\end{figure*}


\subsection{NSCs and Globular Clusters}
\label{subsec:globulars}

We now attempt to place these results into the broader context of GC populations and their role in NSC formation and evolution. By jointly considering the NSC mass function, the NSC occupation fraction, and the stellar mass fraction contained in NSCs as complementary diagnostics, we aim to explore the connections between NSCs and GCs \citep[e.g.,][]{Hoyer+2023,Khim+2024}. 

To guide the comparison, we consider two sets of theoretical models. The semi-analytic models of \citet[][hereafter A+15]{Antonini:2015aa} bracket two limiting NSC formation channels: (1) pure GC inspiral ({\tt CliN}), in which NSCs assemble exclusively through dynamical-friction-driven GC decay and merging, and (2) hybrid models ({\tt GxeV}) that additionally include centrally concentrated {\it in situ} star formation triggered by gas inflow.  Solid and dashed curves show their results with and without the added effect of black hole heating. \color{black} Meanwhile, the models of \citet[][hereafter LvD22]{Leaman2022} explicitly follow the orbital evolution, tidal stripping, and disruption of GC populations within realistic galaxy potentials, providing a complementary framework in which the efficiency of NSC assembly is regulated by cluster survival and dynamical evolution.

Figure~\ref{fig:nsc_mass_fraction_theory} compares the NGVS NSC mass relation and corresponding stellar mass fractions with predictions from these models, with galaxy structural codes indicated as in the legend. The data show that NSCs become increasingly prominent toward lower galaxy masses, reaching mass fractions of several percent in the lowest-mass systems, in broad agreement with expectations from GC inspiral models. At higher masses, however, the relation flattens and begins to diverge from simple GC model predictions, consistent with reduced cluster-migration efficiency or perhaps the increasing importance of competing processes such as central star formation and/or disruption by black holes.

These GC models reproduce the overall shape of the relation, including the presence of a (subtle) inflection, but differ in detail. The LvD22 models capture the curvature and high-mass suppression reasonably well, reflecting the increasing importance of GC disruption and tidal effects in massive galaxies. At the same time, the normalization is imperfect, and the significant scatter and strong morphology dependence seen in the NGVS data may not be fully reproduced. This may indicate that additional formation channels, including centrally concentrated {\it in situ} star formation, contribute to shaping the observed NSC population, although the present comparison does not uniquely
distinguish among these possibilities. Similar arguments apply to the A15 models, which span a range in normalization depending on the importance of black hole heating. These models are only available for the high-mass half of the NGVS sample, though, and do not capture the apparent upturn above $\log M_{\star}/M_{\odot} \sim 9.5$.
\color{black}

Motivated by these trends, we now turn to the nucleation fraction itself, which provides a direct measure of the incidence of NSC formation across galaxy mass and morphology. Figure~\ref{fig:nsc_morph_theory} shows the NGVS nucleation fraction as a function of host-galaxy stellar mass for Sample~A, separated by host-galaxy morphological class, and compared with the GC occupation fraction in the Virgo core from SJ+19, defined as the fraction of galaxies that host at least one GC above the survey detection limit, as well as predictions from the same theoretical models detailed above.


Overall, the NGVS $f_{\rm nuc}$ relations are in only fair qualitative agreement with the A+15 models. The observed curves rise with increasing host mass and peak at roughly the expected stellar mass, indicating that these models capture the basic large-scale behavior of the nucleation fraction. The agreement is nevertheless imperfect. There are differences at the low-mass end of the models, where the observed NGVS relations fall below the A15 {\tt GxeV} model predictions, and at the highest masses, where the models do not match the observed nucleation fractions. These discrepancies likely reflect the fact that the NGVS spans a broader range of host morphologies than the predominantly early-type systems for which such models are most naturally applicable. 
Indeed, the early-type subsamples most closely resemble the A+15
predictions, whereas the late-type galaxies are the most discrepant.
Neither the {\tt CliN} nor {\tt GxeV} model reproduces the
morphology-dependent NGVS relations in detail, particularly the shift of
the late-type ({\tt S+SI+I}) nucleation-fraction peak toward higher stellar
masses. We therefore do not regard this comparison as providing a clear
preference for either model, or as a clean diagnostic of the relative
importance of GC inspiral and dissipational/{\it in situ} NSC growth.
\color{black}

The LvD22 models reproduce the overall shape of the observed relations, including the turnover and decline toward both higher and lower galaxy masses. At high masses, the low nucleation rate reflects the reduced efficiency of GC inspiral in massive systems where clusters are increasingly disrupted or unable to reach the center. The models were not designed to reproduce the normalization or the strong morphology dependence seen in the NGVS data. The early-type relations are broadly the most similar to the LvD22 predictions, but even there the agreement is imperfect, particularly at low masses where the predicted nucleation fractions significantly exceed the observations (i.e., by factors of three or more below $\log(M_{\ast}/M_{\odot}) \simeq 7$). The observed relation for late-type galaxies is even more discrepant.  However, there are a number of assumptions implicit in these calculations including GC formation distance, the galaxy size-mass relation, the efficiency of clustered star formation, the assumed form of the GC mass function and the density structure of the host galaxy. Fine tuning of these inputs may be needed to match the observations, including the apparently low nucleation fractions observed for the faintest galaxies (c.f., \S\ref{subsec:discussion_selection_effects}).
\color{black}

The GC occupation fraction provides an important complementary constraint because it traces the reservoir of clusters from which NSCs may assemble. As Figure~\ref{fig:nsc_morph_theory} shows, it exhibits a broadly similar mass dependence to the NSC nucleation fraction and is generally consistent with the LvD22 curves. This agreement is particularly suggestive because GCs are common in the same mass regime in which NSCs are most prevalent, while the GC occupation fraction declines toward lower galaxy masses in tandem with the drop in $f_{\rm nuc}$. These trends strengthen the case that GCs and NSCs are physically connected. At the same time, the decline in NSC incidence above $\log M_{\star}/M_{\odot} \sim 9$ points to additional regulating processes in massive galaxies, such as longer dynamical-friction timescales, more efficient GC disruption, or the destruction of pre-existing NSCs during mergers.


\subsection{Connection to UCDs}
\label{subsec:ucds}

Ultra-compact dwarf galaxies (UCDs) are among the densest stellar systems known, occupying a parameter space between classical dwarfs and massive GCs in both size and luminosity \citep{Hilker+99,Drinkwater+00,Drinkwater:2003aa}. It was suggested early on that some UCDs may be the remnants of nucleated dwarfs that have undergone severe tidal stripping or ``threshing'' \citep{Bekki2001,Bekki+03}. Many studies during the past two decades have presented evidence in support of this scenario \cite[e.g.,][]{Pfeffer:2013aa,Seth:2014aa,Liu:2015aa,Zhang:2015aa,Ahn2017,Ahn2018,Liu+20,Mayes2021b,Taylor+25}. Targeted studies of ultra diffuse galaxies (UDGs) have revealed the presence of NSCs in a number of these galaxies, whose LSB nature may be related to tidal stripping. This is another potential link between NSCs, UCDs and UDGs \citep{Lambert+2024,Khim+2024,Kurzner2025}.

\begin{figure*}
    \centering
     \includegraphics[width=0.80\textwidth]{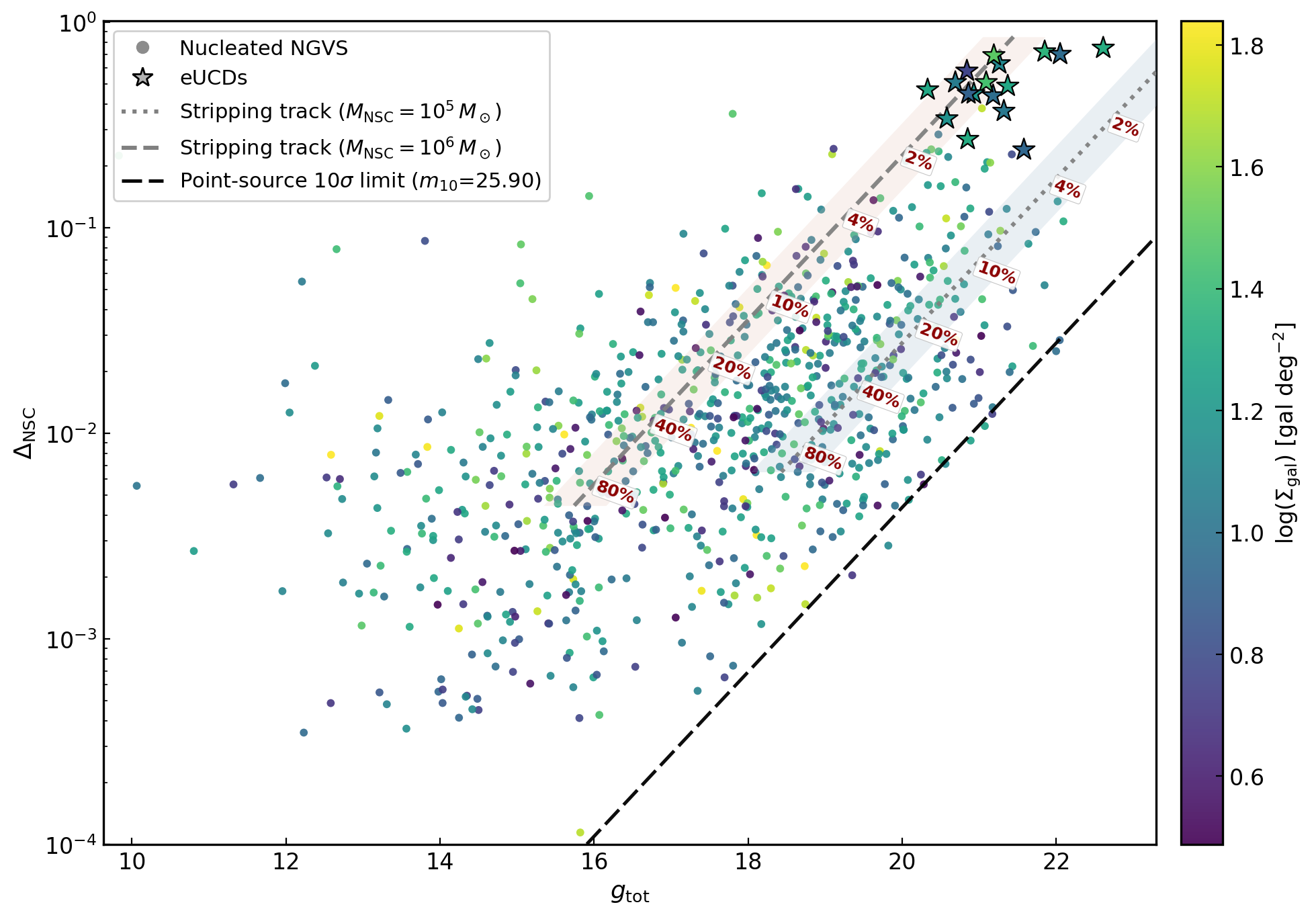}
    \caption{NSC mass fraction, $\Delta_{\rm NSC}$, plotted against total $g$-band magnitude for nucleated galaxies in Sample~A (filled circles). UCDs with detectable envelopes (eUCDs) identified by \citet{Wang2023,Wang+2026} are shown as filled stars. In both cases, symbols are color-coded by local density to highlight possible trends with environment. The NGVS $g$-band detection limit for point sources is shown by the dashed black curve. The grey dotted and dashed curves show the expected evolution of a tidally stripped galaxy containing an NSC of mass $\log(M_{\ast}/M_{\odot}) = 5$ and 6, respectively, while the shaded regions indicate the corresponding ranges spanned by the stripping models. Red labels indicate the percentage of mass remaining along the indicated stripping tracks.}
    \label{fig:nsc_ucd}
\end{figure*}

We can examine the connection between NSCs and UCDs using our updated catalog of nucleated galaxies.  Figure~\ref{fig:nsc_ucd} plots the NSC-to-galaxy mass ratio, $\Delta_{\rm NSC}$, against $g$-band magnitude for Sample~A galaxies (filled circles). Filled stars show the sample of UCDs with envelopes (eUCDs), taken from the study of \citet{Wang+2026}. Both samples are color-coded by the local density as shown by the color-bar at the right of the figure. The black line corresponds to the (10$\sigma$) point-source limit for the NGVS $g$-band imaging. While this does not translate to a firm detection limit on NSCs, it does highlight the importance of selection effects for both populations: i.e., galaxies containing NSCs fainter than this would be classified as non-nucleated, while UCDs fainter than $g \sim 20$ would tend to be classified as GC candidates.

The dotted and dashed lines show the expected evolution of two hypothetical nucleated galaxies containing NSCs with masses of
$\log(M_{\rm NSC}/M_{\odot})=6$ and 5, respectively, as the galaxies are stripped away. This evolution is indicated by the red labels showing the surviving percentage of stellar mass, leaving an increasingly prominent NSC/UCD. Despite our earlier warning about the importance of selection effects in this diagram, this comparison shows the eUCDs lie on the extension of the relation defined by nucleated galaxies, consistent with a tidal stripping origin. That said, there does not appear to be a strong trend with local density, as one might expect if the most severely threshed objects are found in high-density regions. 

\color{black}


\subsection{Galaxies with Possible Offset NSCs}
\label{subsec:offset_nscs}

Galaxies hosting offset nuclei --- in which the photometric nucleus is measurably displaced from the geometric center of the host galaxy  --- may provide insights into the early assembly of NSCs. For Virgo galaxies, offsets reported in the past are typically of order $\sim$1\arcsec, corresponding to $\sim$80~pc, with some systems exhibiting even larger displacements that represent a significant fraction of the host galaxy scale radius \citep[e.g.,][]{Binggeli2000,Lisker:2007aa,Sanchez-Janssen2019}. Such offsets are easily resolvable in ground-based imaging surveys as they exceed typical uncertainties involved in centroiding the galaxy light. In practice, though, the issue is complicated by the often uncertain nature of the nuclei with the largest displacements as {\it bona fide} NSCs (rather than interloping stars, background galaxies or even massive GCs).

Offset NSCs in Virgo and Local Volume dwarf galaxies \citep[e.g.,][]{Binggeli2000, Lisker:2007aa, Georgiev:2014aa, Sanchez-Janssen2019} have usually been interpreted as transient configurations rather than long-lived equilibrium states. Proposed formation pathways include incomplete orbital decay of massive star clusters via dynamical friction, localized {\it in situ} star formation away from the potential minimum, and dynamical perturbations induced by tidal interactions or minor mergers \citep[e.g.,][]{Antonini:2012aa, Gnedin:2014aa, Neumayer2020,Roman+2023,Cao+2025,Poulain+2025b}. Offset NSCs may thus provide direct observational constraints on the timescales and physical mechanisms governing NSC assembly.

Within our Sample~A, we identify 79 systems as the most promising candidates for {\it bona fide} offset NSCs. 
\footnote{These are a subset of 172 galaxies with candidate offset nuclei in the NGVS catalog, some of which are uncertain due to low-S/N detections, small offsets, unusual NSC-galaxy mass ratios, or other factors.  This assignment is inevitably subjective, and is typically used for those galaxies that show a bright, point-like source offset by $\lesssim$1-2 arcseconds from the photocenter. \color{black}} We compare these galaxies against a mass-matched control sample constructed from centrally nucleated galaxies, comprising 395 offset--control pairs drawn from 289 unique control galaxies. By construction, the two samples have nearly identical stellar mass distributions, with median $\log(M_\star/M_\odot)=7.68$ for the offset sample and $7.66$ for the control sample ($p_{\rm KS}=0.99$).



\begin{figure}
    \plotone{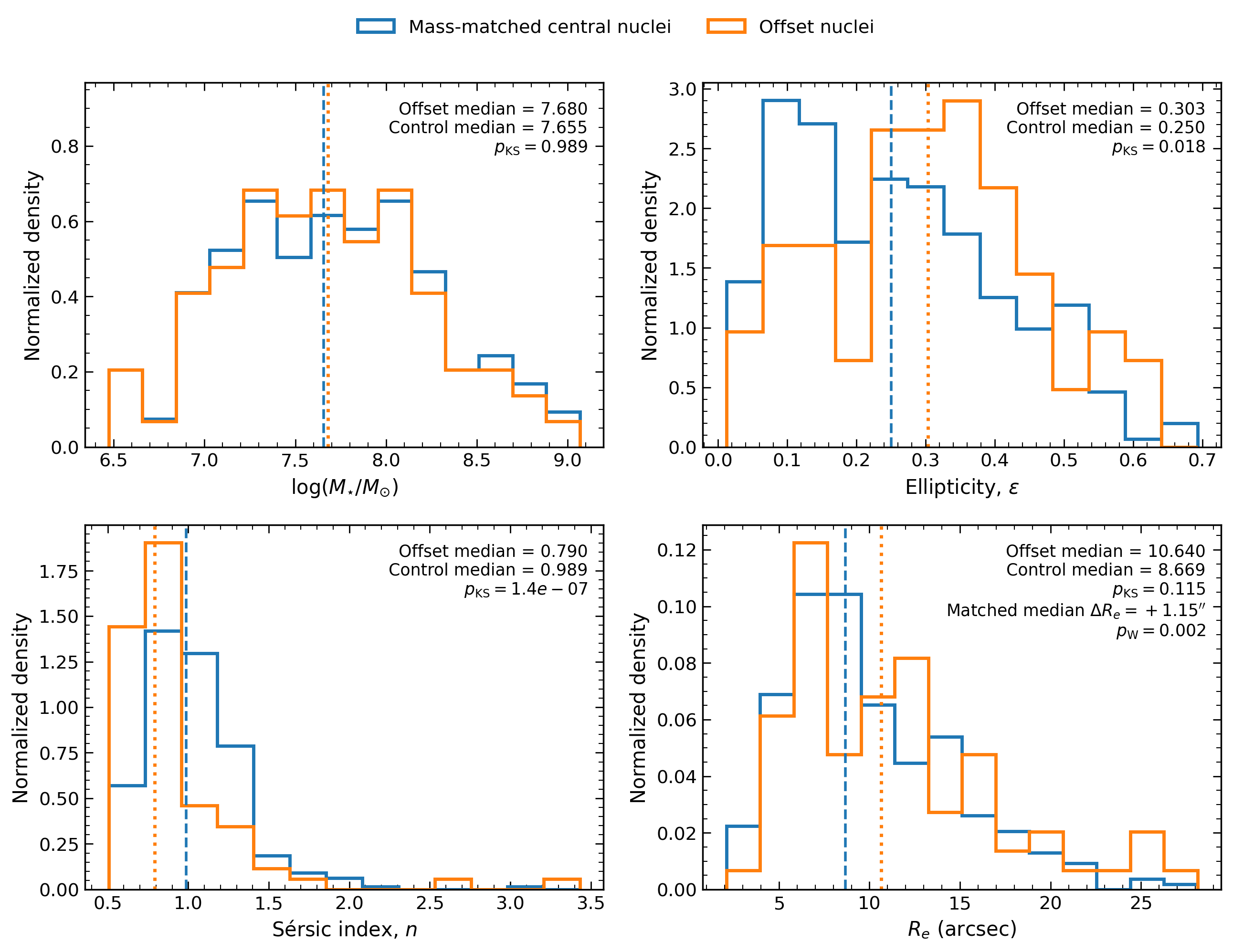} 
    \caption{ Comparison of structural properties for the 79 candidate offset-NSC hosts and 289 unique centrally nucleated galaxies used as their mass-matched control population. The upper-left panel demonstrates that the stellar-mass distributions are nearly identical. The remaining
    panels compare host-galaxy ellipticity, Sérsic index, and effective radius. Vertical dotted and dashed lines indicate the medians of the offset and control samples, respectively, and the corresponding two-sample Kolmogorov--Smirnov probabilities are given in each panel. Offset-NSC hosts have modestly higher ellipticities and substantially lower Sérsic indices. Although the pooled effective-radius distributions overlap, a galaxy-by-galaxy comparison with the median of each object's mass-matched controls gives a median $\Delta R_e=+1.15\arcsec$ and a Wilcoxon signed-rank probability of $p=0.0019$, indicating that offset hosts tend to be somewhat larger at fixed stellar mass. \color{black}}
    \label{fig:offset_structural}
\end{figure}

Comparing structural and environmental properties, we find that offset NSC hosts have systematically lower Sérsic indices than the control sample galaxies (median $n=0.79$ vs.\ $0.99$, $p_{\rm KS}=1.4\times10^{-7}$), as well as modestly higher ellipticities (median $\epsilon=0.303$ vs.\ $0.250$, $p_{\rm KS}=0.018$). These differences are illustrated in Figure~\ref{fig:offset_structural}. The Sérsic-index difference is particularly robust when each offset galaxy is compared directly with the median of its own mass-matched controls: 58 of 77 offset hosts have lower $n$, with a median paired difference of $\Delta n=-0.17$ and a Wilcoxon signed-rank probability of $p=1.8\times10^{-4}$.

The effective radii show a similar, though more modest, trend. The  $R_e$ distributions overlap substantially (median $R_e=10.64\arcsec$ for the offset hosts and $8.67\arcsec$ for the unique control galaxies; $p_{\rm KS}=0.115$), but the explicitly matched comparison reveals that offset hosts are typically larger at fixed stellar mass. The median paired difference is $\Delta R_e=+1.15\arcsec$ ($\simeq0.04$ dex), with a Wilcoxon signed-rank probability of $p=0.0019$. This result remains significant after removing the most extreme size differences and is not driven by any single galaxy. Offset hosts also show some evidence of inhabiting lower-density environments, with median local density $\Sigma_{\rm gal}=10.1$ compared with $13.3~{\rm gal~deg^{-2}}$ for the pooled control sample ($p_{\rm KS}=0.004$), but this difference is less significant when the galaxy-by-galaxy mass matching is explicitly taken into account.



\color{black}

In terms of stellar populations and morphology, the offset sample contains a mix of systems spanning both the red sequence and the blue cloud, with a modest but significant shift toward bluer colors relative to the control sample. Transitional and later-type morphologies ({\tt ES, EI, S}) are enhanced among the offset hosts compared to the controls, and there is a corresponding increase in galaxies with intermediate or active {\tt I} or {\tt A} star-formation classifications, whereas the control sample is dominated by ({\tt Q}) systems.

We also compare the properties of the nuclei themselves in the offset and control samples. Within the uncertainties, the candidate offset NSCs show broadly similar magnitudes and colors to their centrally located counterparts at fixed galaxy stellar mass, with no strong evidence for systematic differences. This suggests that the primary distinction between the two samples lies in the dynamical state of the nuclei rather than their stellar populations, although the relatively small number of offset systems and the larger measurement uncertainties for displaced nuclei limit the sensitivity of this comparison.

Some of these differences can be subtle, but they support an
interpretation in which offset NSCs preferentially reside in galaxies that are less centrally concentrated, modestly more flattened, and somewhat more extended at fixed stellar mass. There is also some evidence that they favor lower-density environments, although this environmental difference is less robust in the explicitly mass-matched comparison. Together with the enhancement of transitional and star-forming systems, these differences may suggest that offset NSCs represent a transient phase in NSC assembly, in which NSCs are dynamically young (e.g., still undergoing orbital decay of infalling clusters and/or oscillating about the center of the potential well), recently formed, or weakly perturbed by tides or minor interactions \citep{Roman+2023,Cao+2025,Poulain+2025b}.
\color{black}

\subsection{Galaxies with Blue Cores}
\label{subsec:blue_cores}

Early-type galaxies with blue cores provide direct evidence that recent or ongoing star formation can contribute to the build-up of nuclear components \citep[e.g.,][]{Lisker:2007aa, Paudel:2011aa, Ulrich2017}. From an analysis of SDSS imaging, \citet{Lisker+06} identified 23 Virgo dwarf ellipticals (dEs) having blue central colors and noted that an additional 14 objects previously classified as blue compact dwarfs showed similar properties. These galaxies, they argued, would appear like ordinary dEs $\sim 1$ Gyr after the last episode of central star formation. The nucleation properties of such objects may therefore shed light on the origin of NSCs in low-mass galaxies.

We identify a sample of 97 low-mass galaxies that show evidence for a blue core in the NGVS imaging. This sample includes 28 of the 37 galaxies noted by \citet{Lisker+06}; the other nine galaxies either fall outside the NGVS footprint or were classified differently in the NGVS. The first point to note is that these galaxies are not drawn uniformly from all types. Rather, the population is dominated by galaxies with early-type structural codes: 61 ($63\%$) are classified as {\tt E}, 26 ($27\%$) as {\tt ES}, and 10 (10\%) as {\tt EI}. Nearly all (94/97 $\simeq$ 97\%) are assigned an intermediate-type star formation code ({\tt I}), as might be expected given their blue central colors.

To compare the blue-core population to a stellar-mass-matched control sample, we consider a sample of 388 mass-matched control galaxies spanning the range $\log(M_\star/M_\odot)\approx 6.6$--$9.7$, with a median stellar mass of $\log(M_\star/M_\odot)=7.91\pm0.71$. The corresponding blue-core sample has a very similar median stellar mass, $\log(M_\star/M_\odot)=8.04\pm0.74$. Relative to the control sample, the blue-core galaxies are preferentially transitional, disky systems ({\tt ES}) with intermediate star-formation properties. In the color--magnitude diagram, only about half of the galaxies fall along the red sequence, indicating a clear enhancement in the fraction of systems displaced toward the star-forming region of color--magnitude space. In terms of internal structure, the blue-core galaxies tend to have slightly higher S\'ersic indices and smaller effective radii. The blue-cored galaxies also have a slightly more extended distribution in the cluster than the control sample.

The nucleation fraction of the blue-core galaxies is $f_{\rm nuc}=33\pm5\%$, comparable to that of the mass-matched controls ($38\pm2\%$) as well as the nucleation fraction for the full sample of galaxies in the same mass range ($34\pm1\%$). The median NSC mass for the blue-core galaxies is $\log(M_{\rm NSC}/M_\odot)=5.61$, compared to 6.00 for the controls. The logarithmic NSC-to-host mass ratio is likewise lower for the blue-core galaxies than for the controls ($\Delta_{\rm NSC}=-2.48$ versus $-2.14$), implying that their nuclei are undermassive by a factor of $\sim 2$ relative to those of the control sample.

These results suggest that the blue-core phenomenon is associated less with unusually massive or dynamically mature nuclei than with recent or ongoing central star formation superimposed on relatively modest nuclear mass concentrations. In this picture, central star formation may eventually enhance the visibility of the NSC or may contribute to its growth, but does not by itself imply that the NSC has already reached the masses seen in more evolved systems. The predominance of {\tt E}, {\tt ES}, and {\tt EI} morphologies among the blue-core population further suggests that this process is especially relevant in galaxies that are structurally early-type but still retain (or have recently acquired) enough gas to sustain central star formation.
Overall, the properties of the blue-core population support a picture of NSC assembly in which centrally concentrated star formation can contribute to NSC growth alongside other channels such as star-cluster migration. 


\subsection{Connection to AGNs}
\label{subsec:agns}

Up to this point, we have bypassed the issue of AGN and their possible coincidence with NSCs. This relationship will be explored in an upcoming paper in the NGVS series, so we present only a brief overview here, noting at the outset that it is now well established that AGN and NSCs can indeed coexist (e.g., \citealt{Seth:2008aa}, \citealt{Baldassare+2022}, \citealt{Ashok+2023}, \citealt{Hoyer+2024}, \citealt{Dullo+2024}). It is thus natural to ask: how many galaxies in the NGVS sample contain AGNs and what fraction of those galaxies also show evidence for an NSC?

To answer these questions, we turn to three studies of AGN in the Virgo cluster. 
First, \citet{Decarli+2007} used optical spectroscopy for 213 late-type Virgo galaxies with dynamical masses above 10$^{8.5}M_{\odot}$ to identify 77 systems exhibiting signs of nuclear activity (i.e., Seyfert galaxies, LINERS and transition objects).
Second, \citet{Edler+2023} performed a radio survey of the cluster (132 deg$^2$) at 120-168 MHz with LOFAR to search for radio emitters, including AGNs, as part of the VICTORIA (VIrgo Cluster multi-Telescope Observations in Radio of Interacting galaxies and AGN) project.
Third, \citet{Hou+2024} carried out an X-ray search for AGNs in Virgo using source catalogs from the first all-sky scan from eROSITA, finding 50 X-ray sources down to a 0.2-2.3 keV luminosity of $\sim$10$^{39}$~erg~s$^{-1}$.
%


Cross matching these samples to the NGVS, we find 70 unique
 AGN hosts, including 36 high-confidence nucleated galaxies
({\tt Nuc\_code}=1/1$^\ast$), 10 lower-confidence NSC candidates
({\tt Nuc\_code}=2/2$^\ast$), 9 non-nucleated galaxies
({\tt Nuc\_code}=0), and 15 systems with ambiguous or otherwise
uncertain nuclear classifications. This exercise makes clear that AGN and NSC candidates coexist in a substantial number of Virgo galaxies, as expected.


The raw nucleation fraction among the securely classified AGN hosts is
fairly high ($36/45=0.80$), but the literature AGN samples are strongly
weighted toward relatively massive galaxies. The AGN surveys also have
heterogeneous selection functions and a morphology distribution that
differs substantially from that of the full NGVS population.

Nevertheless, a one-to-one comparison with non-AGN galaxies matched in stellar mass and morphology shows no measurable enhancement or suppression of nucleation probability. We find no evidence that AGN incidence is independently associated with NSC occupation in the present sample. That said, AGN emission may contribute to the unresolved nuclear light in individual systems, but separating stellar and non-stellar components would require data beyond the scope of this study. A more complete examination of the link between AGN and NSCs will be presented in a future paper.
\color{black}

\section{Discussion}
\label{sec:discussion}



\begin{figure*}
    \centering
    \includegraphics[width=0.98\textwidth]{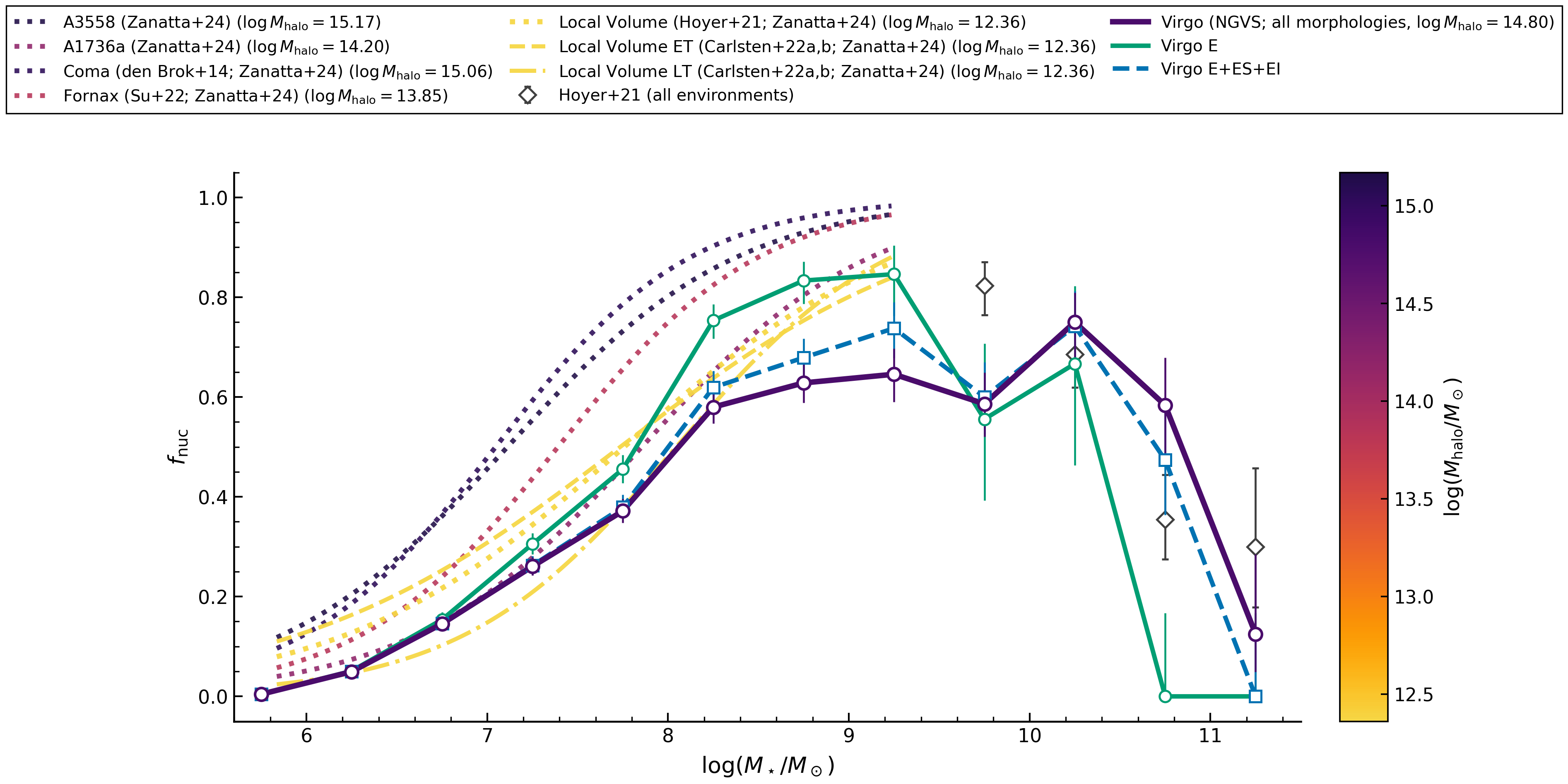}
    \caption{{Nucleation fraction, $f_{\rm nuc}$, for Virgo galaxies compared with literature measurements in other nearby environments. NGVS results are shown for the full Virgo sample by the solid purple curve, for {\tt E} galaxies by the solid green curve, and for the combined {\tt E+ES+EI} sample by the dashed blue curve. The colored dotted, dashed, and dot-dashed curves show the Bayesian logistic-regression fits compiled by \citet{Zanatta2024} for the Shapley clusters A3558 and A1736a, Coma, Fornax, and Local Volume samples, with curve colors indicating host halo mass. The gray diamonds show an independent high-mass comparison constructed by combining the Local Volume, Coma, Fornax, and Virgo galaxy catalogs of \citet{Hoyer2021}; the points give the nucleation fraction in 0.5-dex stellar-mass bins, with approximately 68\% Wilson binomial confidence intervals, and bins containing fewer than seven classified galaxies are omitted. Because the literature samples have different morphological selections, stellar-mass ranges, and spatial coverages, the comparison is intended to illustrate broad differences among environments rather than to define a controlled environmental sequence, as discussed in the text.}}
    \label{fig:nsc_env_comparison}
\end{figure*}


\begin{figure*}
    \centering
     \includegraphics[width=1.00\textwidth]{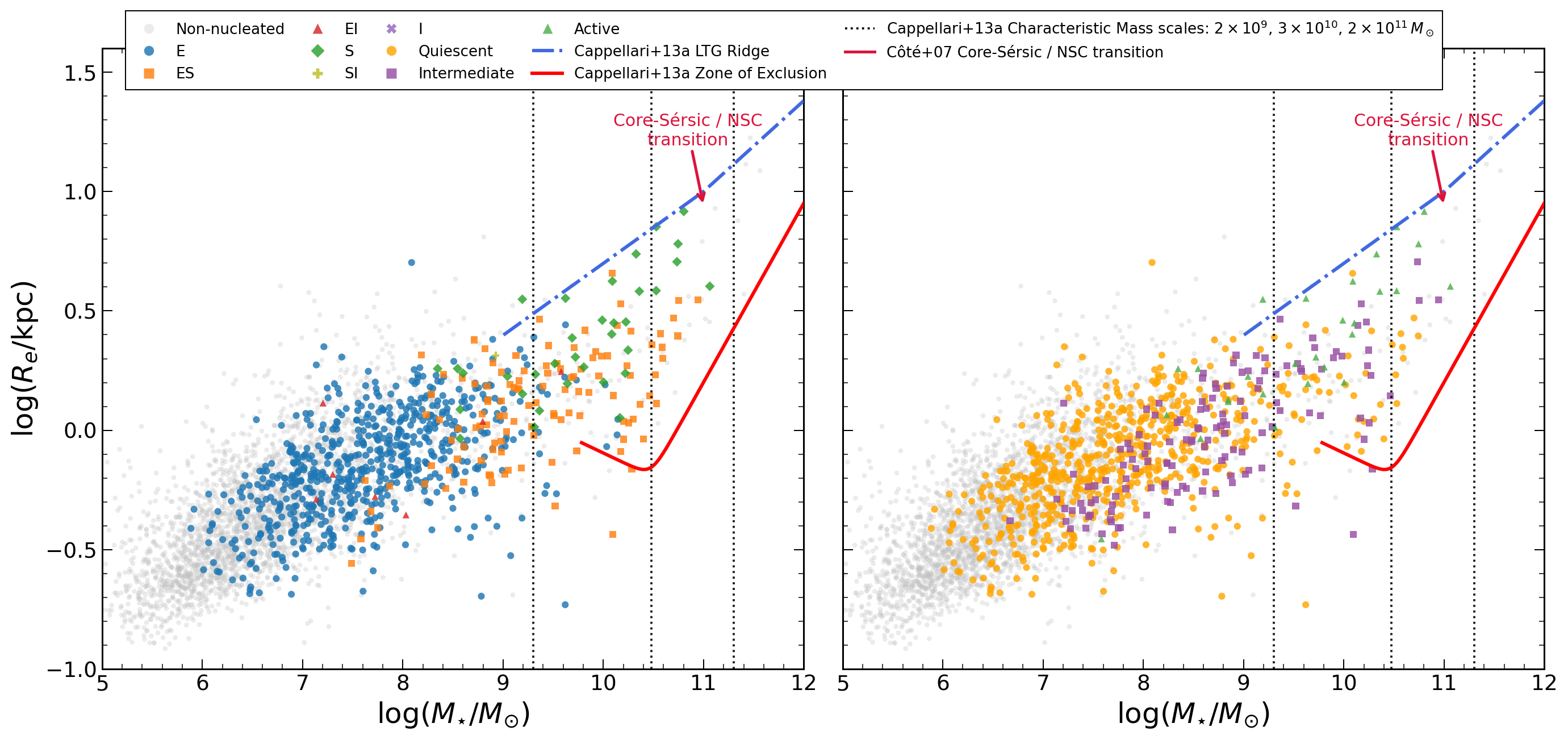}
    \caption{Mass–size diagram for Virgo galaxies, colored by structural (left) and star formation (right) classifications. Galaxies are shown in the $\log(R_e/{\rm kpc})$–$\log(M_\star/M_\odot)$ plane with non-nucleated systems plotted in gray and nucleated galaxies color-coded either by structural class ({\tt E, ES, EI, S, SI, I}; left panel) or by star formation class (quiescent, intermediate, active; right panel). Overlaid are several reference relations and characteristic mass scales from Figure~14 of \citet{Cappellari2013b} --- the late-type galaxy size–mass relation (blue dot-dashed line), the ``zone of exclusion" (red curve), and vertical dotted lines marking three key stellar masses at $2\times10^9$, $3\times10^{10}$, and $2\times10^{11}\,M_\odot$ (see text).
    The distribution of NGVS galaxies illustrates how early-type systems dominate the intermediate-mass regime and follow a relatively tight sequence, while progressively later-type galaxies have larger effective radii at fixed mass. In terms of star formation, quiescent galaxies align with the red sequence and coincide with the region of highest nucleation efficiency, whereas actively star-forming systems populate lower-mass and more extended regions. The annotated core–Sérsic/NSC transition highlights the regime where central stellar ``excesses" give way to depleted cores in the most massive galaxies. }
    \label{fig:size_mass}
\end{figure*}

\begin{figure*}
    \centering
    \includegraphics[width=1.00\textwidth]{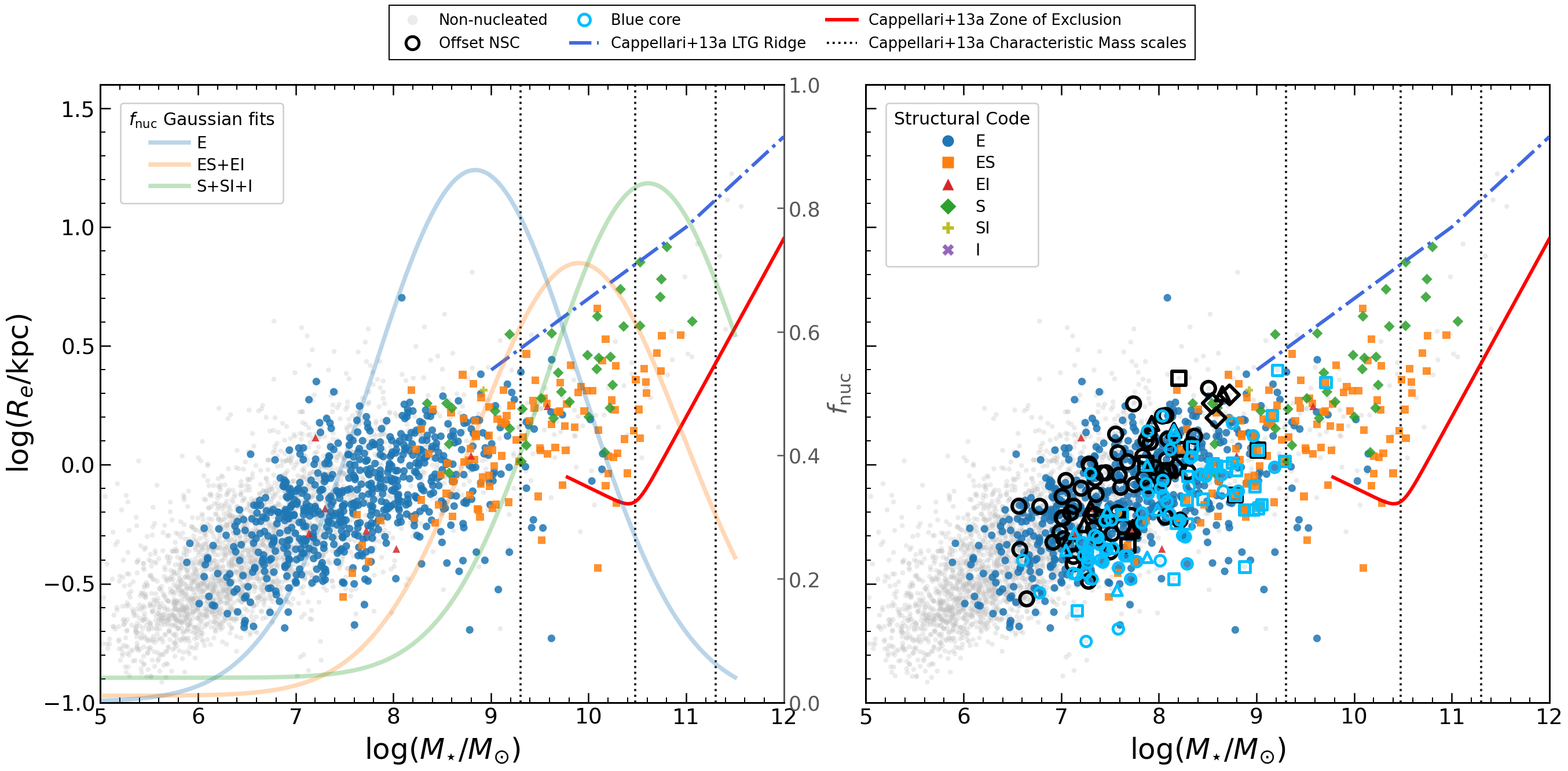}
    \caption{{Mass--size} diagram highlighting key galaxy subsamples and the dependence of the nucleation fraction on stellar mass. {\bf (Left)} The same {mass--size diagram} as in Figure~\ref{fig:size_mass} with overlaid Gaussian models of $f_{\rm nuc}$ for different structural groupings ({\tt E, ES+EI, S+SI+I}), {using the posterior-median parameters from the final individual-galaxy Bernoulli-MCMC fits listed in Table~\ref{tab:nucfrac_model_params}. The curves are shown over the stellar-mass range used for the fits, $5 \leq \log(M_\star/M_\odot) \leq 11.5$.} The vertical reference lines mark the characteristic mass scales identified by \citet{Cappellari2013b}, which delineate transitions in galaxy structure. The $f_{\rm nuc}$ curves show where nucleation is most efficient, with {morphology-dependent peak locations of $\mu=8.84$, $9.90$, and $10.61$ for {\tt E}, {\tt ES+EI}, and {\tt S+SI+I}, respectively.} {\bf (Right)} The {mass--size diagram} with specific galaxy populations of note, such as galaxies with offset nuclei (black open symbols) and blue-core galaxies (cyan open symbols), highlighted against all nucleated galaxies color-coded according to their morphological classification. The fact that galaxies with offset nuclei as well as nucleated blue-core galaxies are preferentially found in the mass range where $f_{\rm nuc}$ is maximized suggests that they trace transitional phases in NSC growth and dynamical evolution, where galaxies are sufficiently massive to support central buildup but not yet fully dominated by processes that suppress or disrupt NSCs.}
    \label{fig:size_mass_fnuc}
\end{figure*}


\begin{figure*}
    \centering
     \includegraphics[width=1.00\textwidth]{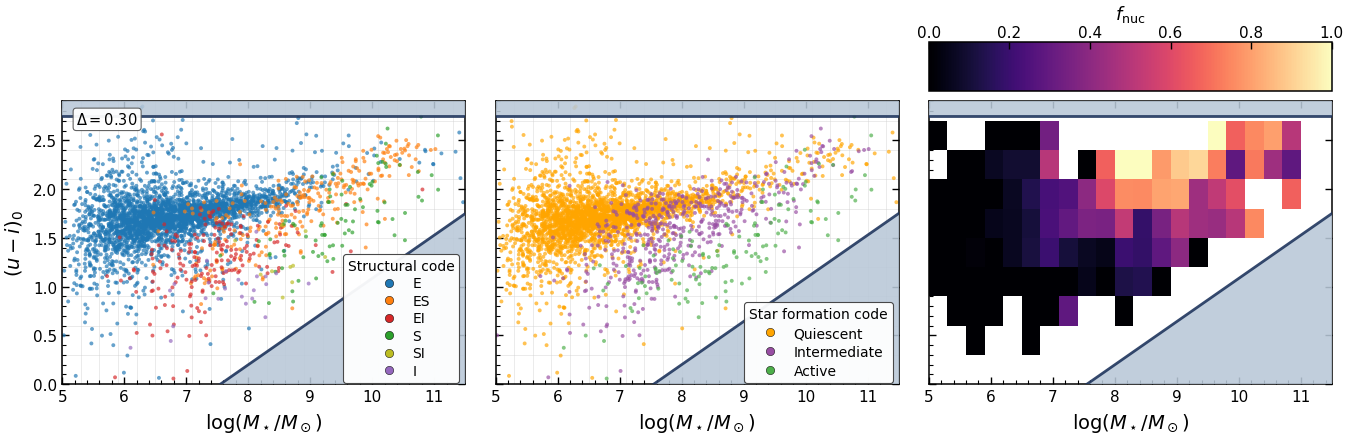}
    \caption{Color--mass nucleation maps for Virgo galaxies using fixed-grid binning with $\Delta \log M = \Delta (u-i)_0 = 0.30$. The left and middle panels show the $(u-i)_0$--$\log(M_\star/M_\odot)$ distribution color-coded by NGVS structural class ({\tt E, ES, EI, S, SI, I}) and star-formation class (quiescent, intermediate, active), respectively. The  right panel shows the resulting nucleation fraction $f_{\rm nuc}$ in the same bins. Shaded regions mark portions of color--mass space excluded from the analysis because they lie outside the main galaxy locus. The highest nucleation fractions are concentrated along the red sequence at intermediate-to-high stellar masses, while lower values are found among bluer and lower-mass systems. The star-formation-coded panels help illustrate that the high-$f_{\rm nuc}$ locus is dominated primarily by quiescent galaxies, with active systems contributing mainly at bluer colors and lower masses.}
    \label{fig:nuc_maps}
\end{figure*}

\subsection{Recap of NSC Properties and Trends}
\label{subsec:discussion_properties}

Unlike many past NSC surveys, which tended to focus on galaxies of a specific morphological class, the NGVS makes it possible to analyze NSCs in galaxies across wide ranges in stellar mass and morphology. Early studies proceeded within more limited regimes, beginning with observations of bulge-dominated, early-type spirals \citep{Carollo:1997aa,Carollo:1998aa}, followed by systematic investigations of nuclei in late-type, bulgeless disks \citep{Boker:2002aa,Boker:2004aa} and then early-type galaxies of high- and intermediate-mass \citep{Cote:2004aa,Jordan:2007aa,Turner:2012aa}. Having NSC data for only a few morphological types, or in narrow mass ranges, makes it difficult to disentangle the relative roles of stellar mass, morphology, and environment in a cohesive~way.  

The results presented in \S\ref{sec:results} reveal a number of important NSC properties and trends that we briefly summarize here. Nucleated galaxies are more spatially clustered than the full galaxy sample (\S\ref{subsec:spatial_dist}), as first noted by \citet{Binggeli:1987aa} and \citet{Ferguson:1989aa}. The variation in nucleation fraction, $f_{\rm nuc}$ across the cluster is, however, fairly modest (\S\ref{subsec:nsc_nucfraction_env}). 

NSCs have optical colors that are, on average, slightly bluer than their host galaxies (\S\ref{subsec:stellar_pops_nsc}; \citealt{Cote:2006aa,Turner:2012aa,Georgiev:2014aa}), consistent with either younger stellar populations and/or lower metallicities. While this may suggest delayed or ongoing central star formation in some systems \citep{Lotz:2004aa,Lisker+06,Paudel2010,Paudel:2011aa,Spengler:2017aa}, the substantial scatter in this relation likely reflects a diversity of stellar population histories.

NSC masses and mass-ratio scaling relations --- which are known to vary smoothly and continuously with galaxy mass \citep[][SJ+19]{Cote:2006aa,Graham:2012aa,Scott:2013aa,Neumayer2020} --- are now seen to extend over roughly five decades in galaxy mass range and show little dependence on morphology (\S\ref{subsec:nsc_masses}). By contrast, $f_{\rm nuc}$ itself is found to depend strongly on both stellar mass and morphology, with a peak height and location varying systematically across structural codes (\S\ref{subsec:nsc_nucfraction_morph}). Further insights are available from galaxy types of special interest, such as galaxies with extended blue cores (\S\ref{subsec:blue_cores}), offset nuclei (\S\ref{subsec:offset_nscs}), and ultra-compact dwarfs (\S\ref{subsec:ucds}), which appear to highlight distinct stages of NSC formation and/or evolution. 

Consistent with previous studies, we find galaxy stellar mass to be the primary factor governing the likelihood that a galaxy hosts an NSC. The nucleation fraction, $f_{\rm nuc}$, has a Gaussian-like form, rising toward progressively lower stellar masses, reaching a broad maximum near $\log(M_\star/M_\odot)\sim9$, before declining at the lowest galaxy masses probed in the survey. An important finding from this work is that stellar mass alone does not uniquely determine a galaxy's nucleation probability. At fixed stellar mass, there is clear evidence for a systematic variation with host morphology. For instance, at $\log(M_\star/M_\odot)\sim10$, the nucleation fraction differs by more than a factor of two between early-type and late-type systems, with transitional morphologies ({\tt ES}/{\tt EI}) exhibiting intermediate behavior. This morphology-dependent trend remains even if all galaxies with {\tt Nuc\_code = 3} are conservatively treated as nucleated, indicating that it is not driven simply by missed nuclei in dusty or otherwise complex systems (see \S\ref{subsec:nsc_nucfraction_morph}).

These results demonstrate that the commonly reported nucleation-fraction relation is not universal, but instead reflects a composite behavior of morphologically distinct galaxy populations. Asymmetries observed at the high-mass end of the global nucleation relation arise naturally from the superposition of these populations. Although the observed nucleation fraction curve exhibits a broadly Gaussian-like form, both the height and location of the peak vary systematically with galaxy morphology (see Table~\ref{tab:nucfrac_model_params}).

\subsection{Variations with Environment}
\label{subsec:discussion_environment}

Given this dependence on both galaxy mass and morphology --- as well as the modest variation in $f_{\rm nuc}$ with radius and local density at fixed morphological class (see \S\ref{subsec:nsc_nucfraction_env} and Figures~\ref{fig:nsc_frac_env_binned_by_morph} and \ref{fig:fnuc_environment}) --- we reconsider the reported variations in $f_{\rm nuc}$ between field, group, and cluster environments. SJ+19 first noted that the nucleation fraction, $f_{\rm nuc}$, in the Virgo core is lower than that found for the Coma cluster, comparable to that for the Fornax cluster, and higher than that in the Local Group (albeit based on a sparse sample). This behavior is suggestive of an environmental trend, where the normalization of the $f_{\rm nuc}$ relation increases with density, perhaps due to an evolution of galaxy structure and size in rich environments \citep[e.g.,][]{Leaman2022}.

Figure~\ref{fig:nsc_env_comparison} revisits this comparison using a broader compilation of group and cluster environments from \citet{Zanatta2024}, spanning a wide range of halo masses. The NGVS measurements are shown for the full Virgo sample, for galaxies with {\tt E}-type morphologies, and for the combined early-type/transition sample ({\tt E+ES+EI}). For comparison at the high-mass end, we also show nucleation fractions derived from the combined Local Volume, Coma, Fornax, and Virgo galaxy catalogs compiled by \citet{Hoyer2021}. These measurements are shown in 0.5-dex stellar-mass bins and provide a broader literature reference than any single cluster sample. The combined Hoyer+21 measurements show a clear decline in nucleation fraction toward high stellar masses, from $f_{\rm nuc}\simeq0.82$ at $\log(M_\star/M_\odot)=9.75$ to $f_{\rm nuc}\simeq0.30$ at $\log(M_\star/M_\odot)=11.25$, although the statistical uncertainties increase as the number of galaxies decreases.

The NGVS measurements show the same general behavior: the nucleation fraction rises from low stellar masses toward a maximum at intermediate masses and then declines (and becomes increasingly uncertain) at the highest masses. The agreement in overall form with the combined Hoyer+21 high-mass measurements supports the conclusion that the decline in $f_{\rm nuc}$ at high galaxy mass is not unique to the NGVS Virgo sample. At the same time, differences between the individual NGVS morphology curves demonstrate that the precise value of $f_{\rm nuc}$ at fixed stellar mass depends on the structural composition of the galaxy sample.

The separate NGVS morphology curves illustrate why this distinction matters. The full Virgo relation, the {\tt E}-only relation, and the combined {\tt E+ES+EI} relation are similar in shape but differ in detail, showing that the measured nucleation fraction at fixed stellar mass depends on the structural mix of the sample. This is important when comparing to the literature relations compiled by \citet{Zanatta2024}, since those measurements are not derived from uniformly selected galaxy populations. Differences in the relative numbers of early-type, transition, and late-type galaxies can therefore contribute to apparent differences in $f_{\rm nuc}$ between environments.

Some of the previously reported environmental variation in $f_{\rm nuc}$ may thus be driven by changes in the underlying galaxy population, or by differences in radial sampling, rather than by halo mass alone, reflecting the well-known morphology--density relation. At fixed stellar mass and morphology, any residual dependence of $f_{\rm nuc}$ on environment appears to be weak. We conclude that environment plays a secondary role in shaping the nucleation fraction, acting primarily through its indirect influence on galaxy structure and morphology.

A caveat to the cross-environment comparison in Figure~\ref{fig:nsc_env_comparison} is that the literature samples considered were not selected in an identical manner. Several of the cluster measurements are based specifically on early-type or quiescent dwarf galaxies and/or on HST pointings toward cluster cores. For example, the Coma measurements include early-type dwarf galaxies from the HST/ACS Coma Cluster Survey, where \citet{den-Brok:2014aa} analyzed 200 early-type dwarfs with $16.0<m_{\rm F814W}<22.6$, as well as deeper HST observations of faint quiescent dwarfs in two fields in the Coma core \citep{Zanatta2021}. The Shapley cluster measurements are based on deep HST/ACS imaging of dwarf galaxies in A3558 and A1736a, with A3558 probing the central cluster environment \citep{Zanatta2024}. The Fornax measurements of \citet{Su2022} are broader in stellar mass and include 557 galaxies with available structural decompositions, but are explicitly separated into the Fornax main cluster and the Fornax A group. By contrast, the Local Volume comparison samples are not cluster-core surveys: \citet{Hoyer2021} studied nucleation across a broad Local Volume galaxy sample, while the Exploration of Local VolumE Satellites (ELVES)-based measurements focus on low-mass early-type satellites of Milky Way-like and small-group hosts \citep{Carlsten2022b}, drawn from the broader ELVES satellite survey \citep{Carlsten2022a}. The additional high-mass measurements \citep{Hoyer2021} shown in Figure~\ref{fig:nsc_env_comparison} combine galaxies from several environments, including Virgo, Fornax, Coma, and the Local Volume, and should therefore be interpreted as a general high-mass literature comparison rather than as an additional measurement for a single environment.

\subsection{NSC Formation Across Mass and Morphology}
\label{subsec:discussion_formation}

Figure~\ref{fig:size_mass} shows the distribution of nucleated and non-nucleated galaxies in the size--mass plane, color-coded by structural and star formation codes from \citet[][left and right panels, respectively]{Kurzner2025}. The distribution of galaxies with structural properties from the NGVS closely follows that found in the ATLAS$^{\rm 3D}$ survey \citep{Cappellari+2011} which showed that the size--mass diagram encodes the dominant evolutionary pathways of galaxies, including the transition between dissipational growth and merger-driven assembly \citep{Cappellari2013b,Cappellari2013a}. Figure~\ref{fig:size_mass} extends the ATLAS$^{\rm 3D}$ framework to the full sample of Virgo galaxies  --- across all galaxy types and over six decades in mass --- including the separate nucleated and non-nucleated populations. Several well-established structural features of this plane are highlighted here for reference, as they define the physical regimes within which NSC formation and survival are expected to operate.

First, the transition between ``core-S\'ersic" and ``S\'ersic" galaxies \citep{Graham2004} occurs at $M _B \sim -19.75$ \citep{Cote:2007}, which is roughly equivalent to $\log(M_\star/M_\odot)\sim10.7$. This transition reflects a change in the innermost structure of galaxies --- systems on the low-mass side of this boundary tend to have steeply rising brightness profiles, thought to be shaped by dissipative processes, while higher-mass galaxies show depleted central ``cores". Such cores are believed to arise in dissipationless (dry) mergers and dynamical scouring associated with supermassive black hole binaries \cite[see, e.g.,][]{Faber1997,Milosavljevic+2001}. 

Second, \citet{Cappellari2013b} showed (see their Figure 14) that the sequence of late-type galaxies falls below a distinct ridge in the size--mass plane, corresponding to systems that grow primarily through the retention of gas and secular processes. In such galaxies, ongoing star formation and gas inflows can build up central mass concentrations, including NSCs \citep{Antonini:2015aa,Neumayer2020}.

Third, the distribution of high-mass galaxies is bounded by a ``zone of exclusion'' \citep{Cappellari2013b} that reflects physical limits on how compact a galaxy can be at a given stellar mass \citep{Burstein+1997,Cappellari2013b}. This boundary is thought to arise from a combination of dynamical and feedback processes that regulate galaxy structure across cosmic time. Above $M_\ast \sim 10^9M_{\odot}$, nearly every NGVS galaxy is contained within these upper and lower boundaries, with a distribution of morphological types that is remarkably consistent with the one presented in Figures~9 and~14 of \citet{Cappellari2013b}.

The characteristic masses labeled in Figure~\ref{fig:size_mass} correspond to the distinct structural and evolutionary regimes identified by \citet{Cappellari2013b}. At $M_\star \approx 2\times10^9M_\odot$, the lower size envelope defined by ``normal" early-type galaxies in the ATLAS3D sample merges with the sequence of low-mass galaxies, which is dominated by early morphological types. \color{black} At $M_\star \approx 3\times10^{10}M_\odot$, early-type galaxies reach their minimum size and maximum stellar density, marking a transition in the slope of the mass--size relation toward larger radii at higher masses. Above $M_\star \approx 2\times10^{11}M_\odot$, the galaxy population is dominated by massive, round slow rotators with cored central profiles \citep{Faber1997,Graham2004,Cote:2007}, while spirals and fast rotators become increasingly rare. The distribution of Virgo galaxies from the NGVS aligns with these trends: {\tt ES}-type systems dominate the intermediate-mass regime and trace a relatively tight sequence, while later-type galaxies ({\tt S, SI, I}) occupy larger effective radii at fixed mass. In terms of star formation, quiescent galaxies align with the region of highest nucleation efficiency, whereas actively star-forming systems preferentially occupy lower-mass and more extended regions.

Together, these features define a set of structural regimes that are expected to influence NSC formation through different physical mechanisms. In particular, low- and intermediate-mass galaxies lie in a regime where both dissipational processes, such as gas inflows and central star formation, and non-dissipational processes, such as GC inspiral through dynamical friction \citep[e.g.,][]{Tremaine:1975aa,CapuzzoDolcetta1993,Antonini:2012aa}, can operate efficiently. By contrast, the highest-mass systems are increasingly dominated by merger-driven evolution and core formation, where NSCs may be disrupted or suppressed. Thus, Figure~\ref{fig:size_mass} shows that nucleation is not simply a function of stellar mass, but occurs within specific regions of the size--mass plane where galaxy structure allows NSCs to form and survive.

Within this framework, nucleated galaxies do not populate the size--mass plane uniformly. Transitional and late-type systems lie mainly along the dissipational sequence, while {\tt E} galaxies span both the lower-mass regime where nucleation is efficient and the higher-mass regime where NSCs are suppressed or destroyed. This structural segregation provides a natural explanation for the morphology dependence of the $f_{\rm nuc}$ relation.

Building directly on this structural framework, Figure~\ref{fig:size_mass_fnuc} links these regimes to the observed nucleation properties by overplotting the best-fitting $f_{\rm nuc}$ relations for different morphological classes. The figure shows that the peaks in $f_{\rm nuc}$ occur in different parts of the size--mass plane for different morphologies: compact early-type systems peak at lower stellar masses, whereas more extended late-type systems peak at higher stellar masses. In this sense, Figure~\ref{fig:size_mass_fnuc} connects the empirical morphology dependence of nucleation to the underlying structural sequence of Virgo galaxies. These regimes reflect the changing balance between dissipational processes, such as gas inflows and central star formation, and non-dissipational mechanisms, such as GC infall in lower-mass, quenched systems.

The shift of the nucleation-fraction peak toward higher stellar masses in {\tt S}/{\tt SI} galaxies likely reflects a change in the balance of NSC formation channels rather than a simple delay of the same process seen in early-type galaxies. At fixed stellar mass, these systems are generally more extended than early-type galaxies, which lengthens dynamical-friction timescales and makes GC inspiral less efficient. As a result, a larger stellar mass, and therefore a larger available cluster population, may be required before inspiral can contribute appreciably to NSC assembly. At the same time, late-type galaxies retain gas and disk structure, which may allow secular inflows and central star formation to contribute to NSC growth. The higher-mass peak in {\tt S}/{\tt SI} galaxies therefore most likely marks the regime where dissipational growth and a sufficient cluster reservoir both become available, while these galaxies have not yet entered the massive, merger-dominated regime where NSCs are more likely to be disrupted or replaced by depleted cores.

In the right panel of Figure~\ref{fig:size_mass_fnuc}, we show the distribution of two notable populations discussed in \S\ref{subsec:offset_nscs} (i.e., galaxies with possible offset NSCs) and \S\ref{subsec:blue_cores} (i.e., early-type transitional galaxies with blue cores). Both populations span a range in mass, but are mostly found in the range $7 \lesssim \log(M_\ast/M_{\odot}) \lesssim 9$. The blue core galaxies are slightly more compact than most galaxies in this regime, likely the result of recent or ongoing central star formation. It is likely that both populations represent transitional phases in the growth of NSCs through dissipational and non-dissipational processes, as we will discuss in the next section. 

A complementary view of NSC demographics is provided by Figure~\ref{fig:nuc_maps}, which shows how the nucleation fraction varies across the color--mass diagram. We measure $f_{\rm nuc}$ in fixed cells of size $\Delta\log M_\star = \Delta(u-i)_0 = 0.30$. This binning scale provides a compromise between resolving local structure in the color--mass plane and maintaining adequate number statistics in each cell. The three panels show, from left to right, the galaxy distribution color-coded by structural classification, the same distribution color-coded by star-formation classification, and the resulting binned nucleation fraction. Shaded regions mark portions of the color--mass plane excluded from the analysis because they fall outside the adopted color range or contain too few galaxies for a robust nucleation-fraction estimate.

The highest values of $f_{\rm nuc}$ are either concentrated along the red sequence at intermediate stellar masses, or among high-mass, star-forming galaxies --- as expected from \S\ref{subsec:nsc_nucfraction_morph}. Lower $f_{\rm nuc}$ values are found elsewhere in the CMD, including among low-mass galaxies (both quiescent and star-forming) and among the most massive systems. This behavior is consistent with the interpretation from the size–mass plane: nucleation is most efficient in galaxies that have assembled sufficient mass to support central growth, but have not yet entered regimes where dynamical processes, such as mergers and core scouring by supermassive black hole binaries, can either destroy NSCs or suppress their formation. The region where $f_{\rm nuc}$ reaches its maximum corresponds to the mass range where both dissipational processes and GC infall are expected to play a role in central galaxy evolution. Among the lowest mass galaxies, quenching has depleted the reservoir of available gas, and the dwindling number of GCs limits any NSC growth from dynamical friction. These trends provide a natural bridge to the physical mechanisms driving NSC formation. In the next section, we quantify the efficiency of dynamical friction and assess the extent to which GC inspiral can account for the observed nucleation properties across the NGVS sample. 

\subsection{Dynamical Friction, Globular Cluster Infall and {\it In Situ} Star Formation}
\label{subsec:discussion_df}

\begin{figure}
    \centering
     \includegraphics[width=0.50\textwidth]{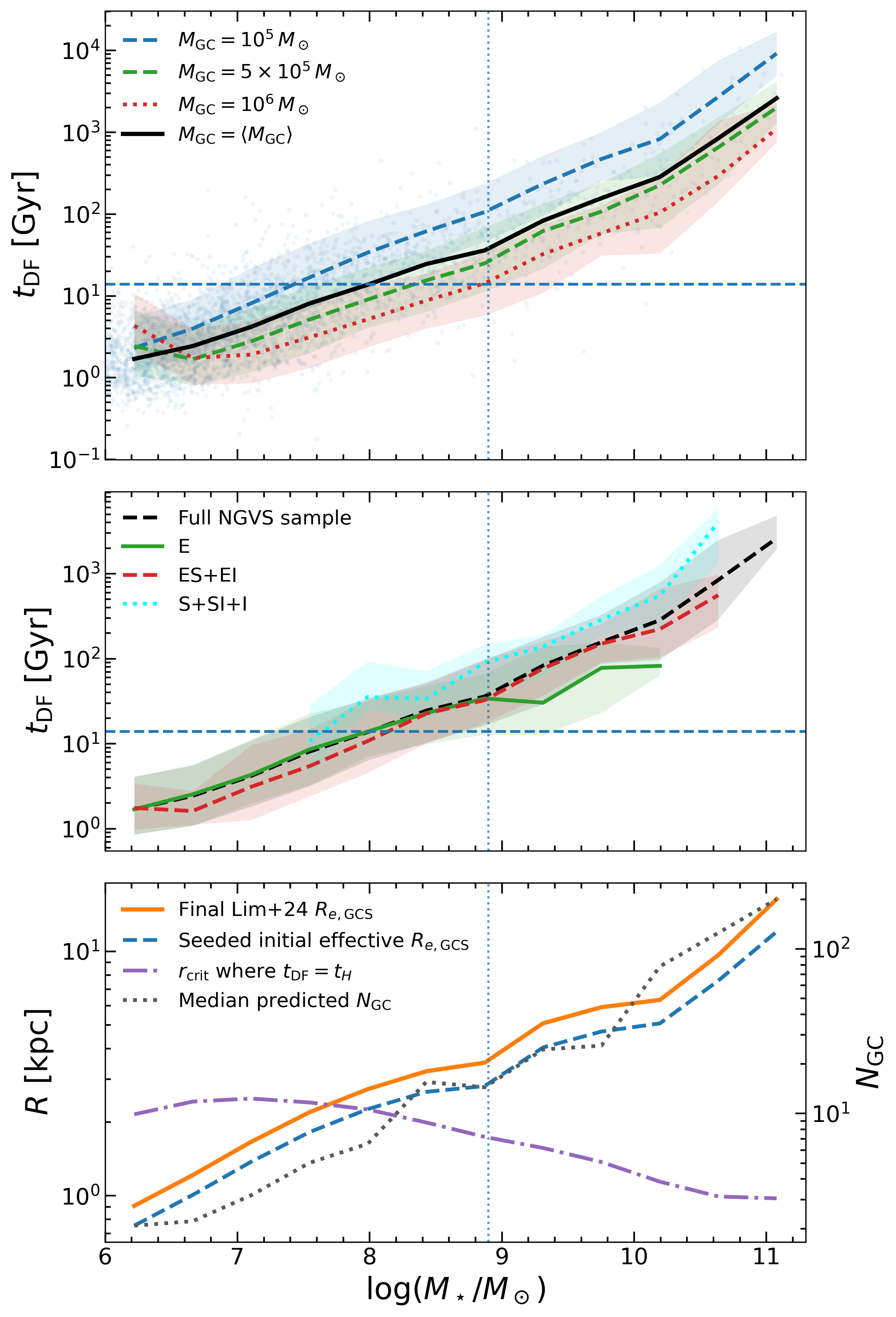} \caption{  Dynamical-friction timescale, $t_{\rm DF}$, for inspiraling GCs in NGVS galaxies computed using Monte Carlo realizations of GC systems evolved over a Hubble time. The seeded initial GC distributions are described by a two-component S\'ersic profile: the outer component is anchored to the observed GC-system scaling relations of \citet{Lim2024,Lim2025}, with its radial scale evaluated for each galaxy using the individually measured host-galaxy effective radius, while an additional compact inner component allows for a centrally concentrated population that can be preferentially depleted by dynamical friction. {\it (Top)} $t_{\rm DF}$ as a function of galaxy stellar mass for GCs of different masses. Curves show median timescales from Monte Carlo sampling, with shaded regions indicating the 16th--84th percentile range. The horizontal line marks a Hubble time, and the vertical line indicates $\log(M_\star/M_\odot)=8.9$, near the peak in nucleation fraction. {\it (Middle)} Morphology-dependent $t_{\rm DF}$ tracks computed using the mean GC mass--stellar mass relation of \citet{DornanHarris2026}. Morphology-specific curves are shown only for stellar-mass bins containing at least five galaxies. {\it (Bottom)} Characteristic GC-system scales as a function of galaxy stellar mass. The orange curve shows the outer GC-system effective radius prescribed by the \citet{Lim2024} relation, while the blue curve shows the effective radius of the complete seeded two-component initial GC distribution. The purple curve indicates the maximum radius within which a GC of the characteristic mass can inspiral within a Hubble time ($r_{\rm crit}$). For comparison, the black dotted curve shows the median predicted GC population size (right-hand axis; \citealt{DornanHarris2026}). 
}
    \label{fig:df_timescales_sersic}
\end{figure}

\begin{figure}
    \centering
     \includegraphics[width=0.50\textwidth]{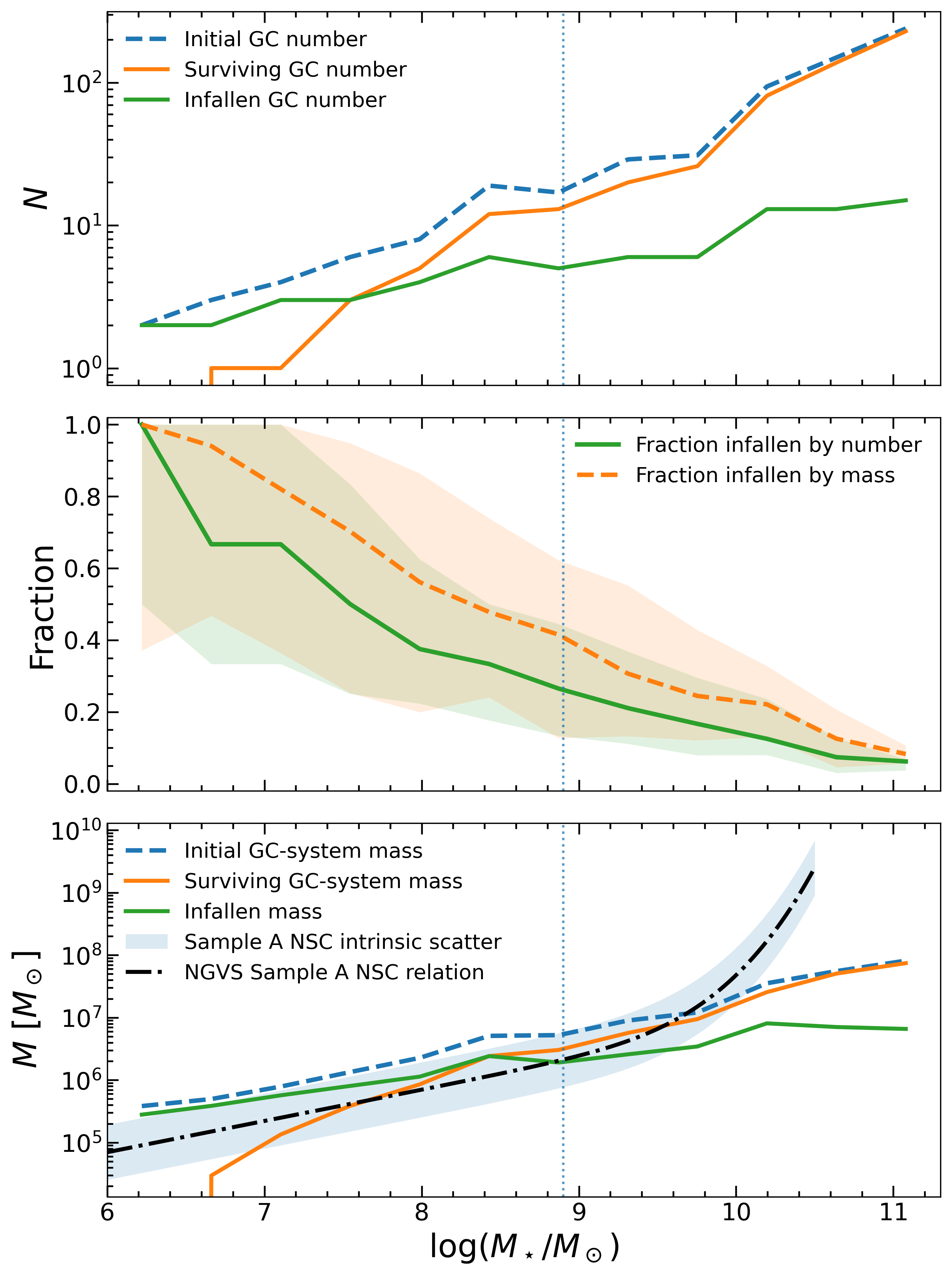}
    \caption{Toy model of GC population evolution due to dynamical friction, using the same  prescriptions as Figure~\ref{fig:df_timescales_sersic}, but evolving the full initialized GC population with individual cluster masses drawn from the adopted GC mass function. {\it (Top)} Evolution of GC number with host-galaxy stellar mass. The dashed blue curve shows the initialized GC population, the solid orange curve the surviving population after a Hubble time, and the green curve the number of clusters that reach the galaxy center. GC number is conserved for each modeled galaxy, such that $N_{\rm init}=N_{\rm surv}+N_{\rm infall}$. The curves shown here are medians evaluated independently within stellar-mass bins and therefore need not satisfy this equality exactly when the three median values are combined. {\it (Middle)} Fraction of the initialized GC population that reaches the galaxy center within a Hubble time, shown by number (green) and by mass (orange dashed), with shaded regions indicating the 16th--84th percentile range among galaxies in each stellar-mass bin. {\it (Bottom)} Evolution of GC-system mass. The dashed blue and solid orange curves show the initialized and surviving GC-system masses, respectively, while the green curve shows the stellar mass deposited at the galaxy center by inspiraling clusters. Mass is likewise conserved for each modeled galaxy, such that $M_{\rm init}=M_{\rm surv}+M_{\rm infall}$. The black dash-dotted curve shows the observed Sample~A NSC mass--host galaxy relation, while the blue shaded region indicates the corresponding total residual scatter about the fitted relation. Because the high-mass behavior of this empirical relation is poorly constrained, it is shown only over $6.0\leq\log(M_\star/M_\odot)\leq10.5$. The comparison illustrates the decreasing ability of GC inspiral alone to account for NSC growth as galaxy mass and GC-system extent increase. \color{black}
    }
    \label{fig:gc_population_sersic}
\end{figure}

We now connect the empirical trends in the NGVS to the physical mechanisms responsible for NSC formation. The observations support a hybrid picture in which both GC inspiral and {\it in situ} star formation contribute, with their relative importance governed primarily by host galaxy mass, and secondarily by host structure and morphology \citep{Turner:2012aa, Antonini:2015aa, Fahrion2022,Leaman2022,Gray+2025,Ashok+2023}. Our goal in this section is to provide a rough evaluation of the efficiency of GC infall using the sample of NGVS galaxies.

At low masses, NSC formation is expected to be dominated by GC inspiral driven by dynamical friction \citep{Chandrasekhar1943,Tremaine:1975aa}. In our model, the inspiral timescale is computed as
\begin{equation}
t_{\rm DF}=
\frac{2.64\times10^{2}}{\ln\Lambda}
\left(\frac{r_i}{2\,{\rm kpc}}\right)^2
\left(\frac{v_c}{250\,{\rm km\,s^{-1}}}\right)
\left(\frac{10^6\,M_\odot}{M_{\rm GC}}\right)\ {\rm Gyr},
\end{equation}
where the Coulomb logarithm is
\begin{equation}
\ln\Lambda=
\ln\left[\frac{b_{\max}v_c^2}{G(M_{\rm GC}+m_\star)}\right].
\end{equation}
We adopt $m_\star=0.5\,M_\odot$, set $b_{\max}=R_e$ and take $v_c=\sqrt{2}\,\sigma$. We make this choice in order to set an approximate velocity scale within the central few effective radii, where the stellar component is expected to dominate over the dark matter halo. We therefore treat the resulting $t_{\rm DF}$ values as order-of-magnitude estimates rather than precise orbital decay times.

The host-galaxy velocity scale is estimated from a virial approximation,
\begin{equation}
\sigma^2=\frac{G\,M_{\rm dyn}}{K(n)\,R_e},
\end{equation}
where we adopt $M_{\rm dyn}=2M_\star$ as a simple stellar-dominated inner-galaxy approximation rather than a full halo mass estimate. The expression
\begin{equation}
K(n)=8.87-0.831\,n+0.0241\,n^2, 
\end{equation} accounts for structural non-homology in Sérsic profiles \citep{Bertin2002,Cappellari2006}.  

The inspiral timescale is defined for an individual GC of mass $M_{\rm GC}$ initialized at radius $r_i$, rather than for the GC system as a whole. In Figure~\ref{fig:df_timescales_sersic}, we first use this expression as a diagnostic by evaluating inspiral times for representative fixed GC masses, $M_{\rm GC}=10^5$, $5\times10^5$, and $10^6\,M_\odot$, as well as for a host-mass-dependent mean GC mass, $\langle M_{\rm GC}\rangle$, predicted from  empirical GC-population scaling relations. For each host galaxy, initial radii are drawn from the seeded GC spatial profile, and the resulting timescale distributions are summarized by their median values in bins of host-galaxy stellar mass. This diagnostic figure therefore illustrates the mass and radius dependence of dynamical friction, rather than representing the full GC mass spectrum of each galaxy.

GCs are initialized with radii drawn from a two-component S\'ersic
surface-density model representing the seeded initial GC system, which is
subsequently evolved under dynamical friction. Each component follows
\color{black}

\begin{equation}
\Sigma(R)=\Sigma_e\,
\exp\left\{-b_n\left[\left(\frac{R}{R_{e,\mathrm{GCS}}}\right)^{1/n}-1\right]\right\},
\end{equation}
where $R_{e,\mathrm{GCS}}$ denotes the effective radius of the S\'ersic
component being sampled and $b_n$ is defined such that
$R_{e,\mathrm{GCS}}$ encloses half of its projected number distribution.
Radial sampling is based on the corresponding enclosed fraction,

\begin{equation}
F(<R)=
P\!\left(
2n,\;
b_n\left(\frac{R}{R_{e,\mathrm{GCS}}}\right)^{1/n}
\right),
\end{equation}
where $P$ is the regularized lower incomplete gamma function.

For each NGVS galaxy, we assign the present-day outer GC-system effective
radius using the all-GC-system size relation of \citet{Lim2024},
\begin{equation}
\log\left(\frac{R_{e,\mathrm{GCS}}}{\mathrm{kpc}}\right)
=
0.51
+
1.22\log\left(\frac{R_e}{\mathrm{kpc}}\right),
\end{equation}
evaluated at the individually measured host-galaxy effective radius,
$R_e$. Thus, at fixed stellar mass, the observed differences in galaxy
size among the structural classes propagate directly into the adopted
GC-system size and hence into the distribution of initial GC radii,
$r_i$. The same measured $R_e$ is also used as $b_{\max}$ in the
Coulomb logarithm above. We retain the GC-system S\'ersic-index
prescription based on \citet{Lim2025}, so that the observed
galaxy-by-galaxy structural differences are propagated consistently
through both the GC-system radial scale and the dynamical-friction
calculation.
\color{black}

The total GC population is assigned using the empirical $N_{\rm GC}$--$M_\star$ relation of \citet{DornanHarris2026}, with a high-mass extension above the transition mass. The characteristic GC mass is approximated as a piecewise function of host stellar mass, varying from $2.05\times10^5\,M_\odot$ at $\log(M_\star/M_\odot)\leq 6.0$, to $3.70\times10^5\,M_\odot$ at $\log(M_\star/M_\odot)\geq 10.2$, with linear interpolation in between.

The resulting dynamical friction timescales are shown in Figure~\ref{fig:df_timescales_sersic}. Although dynamical friction is efficient in low-mass galaxies, the timescale increases rapidly with stellar mass due to the combined growth of galaxy size and velocity dispersion, yielding $t_{\rm DF} \propto v_c r_i^2$. Because the GC systems are usually more spatially extended than the underlying light profiles \citep[see][and references therein]{Lim2024}, many GCs originate at large radii, further increasing the effective inspiral timescale. As a result, for galaxies with $\log(M_\star/M_\odot) \gtrsim 9$, typical inspiral timescales exceed a Hubble time for all but the most massive GCs \citep[see also][]{Turner:2012aa, Antonini:2012aa}. This transition helps to explain the observed rise and fall of the nucleation fraction with stellar mass. It is also consistent with the independent modeling of \citet{Fahrion2022}, who found that lower mass NSCs are expected to form predominantly through GC accretion, whereas more massive NSCs require an increasing contribution from {\it in situ} star formation.

The morphology-dependent tracks in the middle panel show that galaxy
structure introduces an additional dependence in the dynamical-friction
timescale at fixed stellar mass. Over the well-populated range
$9 \lesssim \log(M_\star/M_\odot) \lesssim 10.2$, the late-type systems
({\tt S+SI+I}) are systematically more extended than the {\tt E}
population at comparable stellar mass, leading to correspondingly larger
GC-system effective radii. Because $t_{\rm DF}\propto v_c r_i^2$, the
quadratic dependence on the initial GC radius dominates over the lower
velocity scales inferred for the more extended late-type galaxies. As a result, the median $t_{\rm DF}$ values for the {\tt S+SI+I} population are roughly a factor of $\sim4$--$7$ longer than those of the {\tt E} population across this mass range. Thus, while stellar mass remains the
primary driver of the overall increase in $t_{\rm DF}$, morphology-dependent galaxy structure can substantially modulate the efficiency of GC inspiral at fixed stellar mass.

The bottom panel of Figure~\ref{fig:df_timescales_sersic} highlights the
interplay between the spatial extent of the GC system and the available
GC population. The predicted number of GCs increases strongly with host
stellar mass, while the characteristic GC-system radius also grows and
increasingly exceeds $r_{\rm crit}$, the maximum radius from which a GC
of the characteristic mass can reach the galaxy center within a Hubble
time. Thus, although more massive galaxies contain larger GC populations,
an increasing fraction of those clusters reside at radii where dynamical
friction is inefficient. Conversely, low-mass galaxies have shorter
inspiral times but substantially smaller GC populations. The competition
between GC supply and inspiral efficiency therefore produces a regime
at intermediate stellar masses in which the absolute mass delivered by
inspiraling clusters becomes substantial, even though the fraction of
clusters able to reach the center is already declining with increasing
host mass.

In the population-evolution model shown in
Figure~\ref{fig:gc_population_sersic}, we move beyond the fixed-mass
diagnostics of Figure~\ref{fig:df_timescales_sersic} by assigning each
cluster a mass drawn from the adopted log-normal GC mass function and
evolving the full GC population under dynamical friction. This model
should be interpreted as a controlled feasibility experiment rather than
a uniquely constrained reconstruction of the initial GC system. Clusters
are initialized with a spatial distribution drawn from the same
two-component Sérsic profile described above. The outer component is
anchored to the observed large-scale GC-system structure using the
\citet{Lim2024,Lim2025} scaling relations, while the compact inner
component provides a centrally concentrated population that can be
preferentially depleted by dynamical friction.

An additional inner component is included to test whether a plausible
centrally concentrated reservoir of clusters could supply material to the nucleus through dynamical friction. This component is not directly
constrained by present-day GC profiles; rather, it parametrizes a possible initial population that would have been preferentially depleted by orbital decay. In the realizations shown here, 22\% of the initialized clusters are assigned to this compact component, with an effective radius $R_{e,\rm inner}=0.28\,R_{e,\rm outer}$. Its S\'ersic index is set to $n_{\rm inner}=n_{\rm outer}+2$, restricted to the range $2.5\leq n_{\rm inner}\leq8$. The initialized GC population is normalized to 1.20 times the empirical $N_{\rm GC}$ prediction, allowing for clusters subsequently removed from the surviving GC system by orbital decay. Individual clusters are then drawn from this two-component spatial distribution, assigned masses from the adopted log-normal GC mass function, and evolved independently according to their dynamical-friction timescales. We do not model tidal mass loss or disruption of individual GCs during their inspiral, so each cluster retains its initialized mass throughout the calculation. We track both the number of clusters that reach the galaxy center within a Hubble time and the total stellar mass delivered by those clusters.
\color{black}

With this setup, the outer GC system remains tied to the observed
large-scale GC structure of each host galaxy, while the inner component
allows us to test whether dynamical friction acting on a centrally enhanced initial cluster distribution can reproduce the observed NSC trends. Because the outer GC-system radius is evaluated from each galaxy's measured $R_e$, the model preserves the morphology-dependent structural differences identified above while allowing the initially concentrated component to be preferentially depleted through orbital decay. This forward-modeling approach avoids assuming that the present-day GC system was identical to the initial condition, but it should be regarded as illustrative rather than as a uniquely constrained formation model.

The outcomes of this toy model illustrate how the efficiency of GC
inspiral changes with host-galaxy stellar mass. The top panel of
Figure~\ref{fig:gc_population_sersic} compares the initialized GC
population with the subsets that remain at large radii or reach the galaxy center within a Hubble time. At low stellar masses, a comparatively large fraction of the available clusters can undergo orbital decay, whereas at higher masses the increasing spatial extent of the GC system causes a progressively smaller fraction of clusters to reach the center. The middle panel shows the corresponding infall fractions by both number and mass, highlighting the decline in dynamical-friction efficiency toward higher stellar masses. The bottom panel translates these trends into the total GC mass delivered to the central regions for comparison with the observed NSC
mass--host galaxy relation.

The mass delivered to the galaxy center by inspiraling GCs can be compared directly with the observed NSC mass--host galaxy relation in the bottom panel of Figure~\ref{fig:gc_population_sersic}. At the lowest stellar masses, dynamical friction is efficient but the small number of available GCs limits the total mass that can be transported to the center. The predicted infall mass becomes appreciable at intermediate masses, where the combination of a larger GC population and still-effective orbital decay allows GC inspiral to contribute substantially to NSC growth. Toward higher stellar masses, however, the increasing dynamical-friction timescales and larger characteristic GC-system radii reduce the fraction of clusters able to reach the center. The median infall mass consequently begins to level off at a few $\times10^6\,M_\odot$, while the observed NSC mass relation continues to rise. This divergence indicates a transition to a regime in which GC inspiral alone becomes increasingly insufficient to account for NSC growth, requiring a larger contribution from additional processes. 

In summary, these simple calculations support a physically motivated
picture in which the efficiency of GC-driven NSC growth depends on both
the supply of clusters and the dynamical-friction timescale. At low stellar masses, dynamical friction is efficient but the available GC population is small. At intermediate masses, the combination of a larger GC population and still-efficient orbital decay provides favorable conditions for NSC growth. At higher masses, increasing galaxy and GC-system sizes lead to longer inspiral times, so that an increasing fraction of clusters are unable to reach the central regions within a Hubble time. The morphology-dependent differences in galaxy size further modulate this process at fixed stellar mass, with more extended late-type systems experiencing less efficient GC inspiral. Together, these trends are qualitatively consistent with the observed peak in nucleation fraction at intermediate stellar masses and support
an increasing contribution from additional growth channels, including
{\it in situ} star formation, toward higher galaxy masses.
\color{black}

\subsection{Detection Thresholds and Selection Effects}
\label{subsec:discussion_selection_effects}


\begin{figure*}
    \centering
     \includegraphics[width=0.80\textwidth]{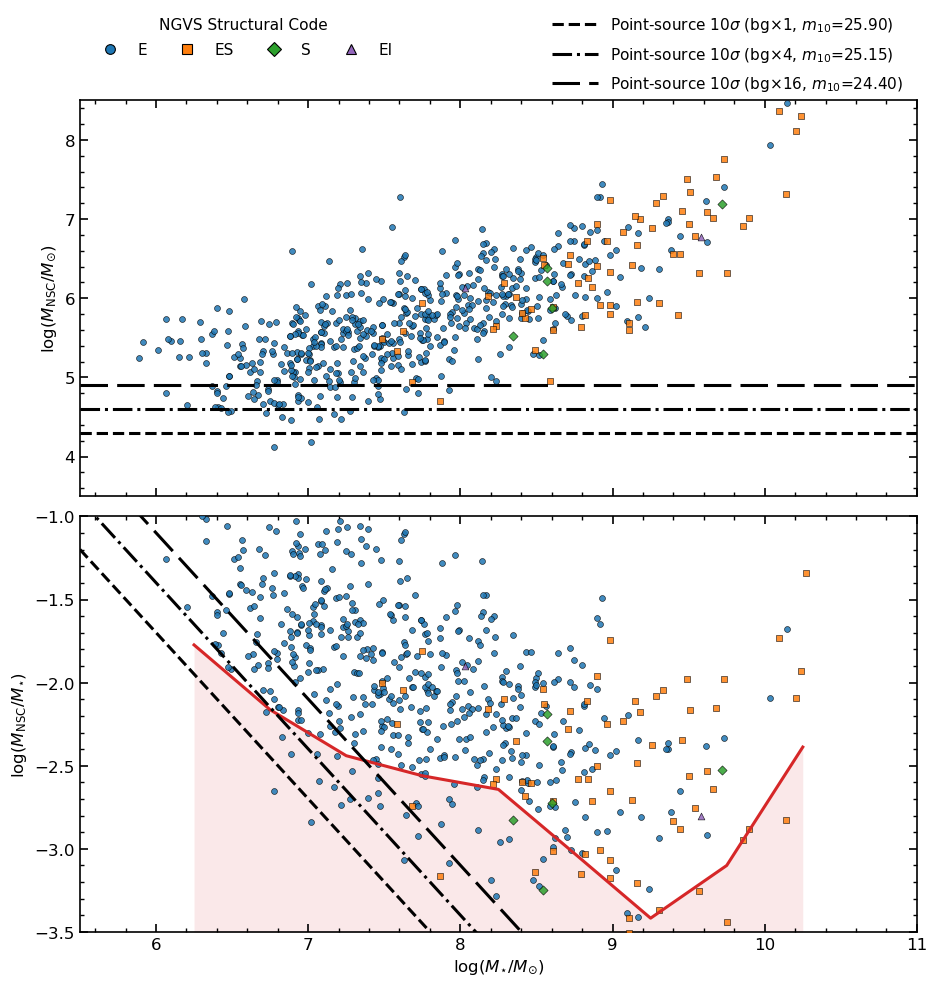}
    \caption{Incompleteness diagnostics for NSC measurements in the NGVS. (Upper) Stellar mass of the NSC as a function of host galaxy stellar mass for Sample~A galaxies. (Lower) The corresponding NSC mass fraction, $\log(M_{\rm NSC}/M_{\star})$, plotted as a function of $\log M_{\star}$. In both panels, the black lines indicate the detection limits corresponding to different $g$-band point-source depths relative to the nominal detection limit in dark time ($m_{g,\rm lim}=25.9$; heavy dashed line). Results are also shown for multiples of this dark-sky background (4$\times$, 16$\times$), converted to a limiting NSC mass 
    using the per-galaxy $M/L_g$ values in the NGVS catalog.  
    The red curve in the lower panel indicates an empirical lower envelope of the observed distribution, defined as the 10th percentile of $\log(M_{\rm  NSC}/M_{\star})$ in bins of $\log M_{\star}$ (bin width $\Delta\log M_{\star}=0.5$ dex). The empirical and depth-based guides illustrate the regime in which low-mass NSCs become increasingly difficult to detect in NGVS imaging.}
    \label{fig:nsc_completeness}
\end{figure*}

We conclude our discussion by reiterating that the detection of NSCs can be challenging for
galaxies with high central surface brightness, steeply rising profiles or complex nuclear structures. This is especially true when relying on ground-based images, as we have done here. 
The typical NGVS seeing of $0.75$\arcsec~in the $g$-band corresponds to a physical scale of $\sim60$~pc at the distance of Virgo. This is substantially larger than the half-light radii of most NSCs, which are typically only a few parsecs in size \citep[see, e.g.,][]{Neumayer2020}. NSCs in NGVS data are thus identified primarily through an excess in central surface brightness, rather than direct spatial resolution, which makes their detection sensitive to both nuclear-to-host contrast and the inner slope of the galaxy light profile \citep{Cote:2006aa,Turner:2012aa, Georgiev:2014aa}.

Figure~\ref{fig:nsc_completeness} provides an empirical assessment of these selection effects. In this figure, we plot NSC mass and mass fraction versus host-galaxy mass (upper and lower panels, respectively). In the lower panel, the red curve traces a lower envelope of the observed distribution of NSCs, defined as the 10th percentile of $\log(M_{\rm NSC}/M_{\star})$ in bins of $\log M_{\star}$. The parameter space below this envelope is sparsely populated, which may indicate a genuine paucity of NSCs, the onset of sample incompleteness, or a combination of both effects.

The dashed black lines in Figure~\ref{fig:nsc_completeness} show the detection boundary corresponding to the NGVS $g$-band point-source limit (25.9 at 10$\sigma$ during dark sky conditions; \citealt{Ferrarese2012}).We This limit is obtained by converting the point-source magnitude limit into a limiting NSC stellar mass using the per-galaxy stellar mass-to-light ratios in the NGVS catalog and the mean Virgo distance modulus.

To illustrate how galaxy surface brightness reduces these limits, we show the same curves after arbitrarily multiplying the background by factors of 4$\times$ and 16$\times$ (dot-dashed and long-dashed curves, respectively).

Lacking {\it a priori} knowledge of the intrinsic distribution of NSC masses, it is impossible to know if the apparent scarcity of NSCs with $\log ({M_{\rm NSC}}/M_{\odot}) \lesssim 5$ for galaxy masses below $\log ({M_{\ast}/M_ {\odot}}) \sim 7.5$ reflects a genuine dearth of low-mass NSCs, or is the result of reduced photometric sensitivity in this regime. A selection bias of this sort would certainly explain a curious feature, evident in this figure and noted earlier in \S\ref{sec:results}, in which NSCs appear to become increasingly rare among the lowest-mass galaxies, and yet, when present, can contribute $\sim10$\% or more of the total
galaxy mass. To definitively resolve this question, deep, high-resolution imaging from space will be needed to expand the census of NSCs in low-mass hosts. 


\section{Conclusions}
\label{sec:conclusions}

We have carried out an analysis of NSCs in the Virgo cluster using imaging and data products from the Next Generation Virgo Cluster Survey. Using a galaxy classification framework developed specifically for the NGVS \citep{Kurzner2025}, we have identified 1144 candidate nucleated galaxies among a sample of 3680 certain, likely or possible cluster members. All measurements and classifications for galaxies and NSCs used in this paper are available in the NGVS galaxy catalog \citep{Ferrarese+2026}. 

Using the full NGVS sample, which spans the mass range $4.5 \lesssim \log(M_\star/M_\odot) \lesssim 11.5$, we show that the nucleation fraction, $f_{\rm nuc}$, rises with increasing galaxy mass, peaks broadly at $\log(M_\star/M_\odot)\sim9$, and then declines towards higher masses. A key finding in this work, however, is that stellar mass alone does not uniquely determine a galaxy's nucleation probability. Instead, the observed $f_{\rm nuc}$ relation reflects the superposition of structurally distinct galaxy populations. Early-type ({\tt E}) galaxies dominate the primary peak at intermediate
masses, while late-type ({\tt S+SI}) and {\tt I}  systems contribute a secondary
enhancement at higher masses, and transition-type morphologies
({\tt ES}/{\tt EI}) exhibit an intermediate behavior. The global shape of the nucleation fraction is therefore not universal but emerges from the combined contributions of these populations.

The morphology dependence likely reflects a shift in the dominant balance of NSC formation and survival channels. Early-type galaxies reach their maximum nucleation efficiency near
$\log(M_\star/M_\odot)\sim9$, where GC inspiral is still efficient and the available cluster reservoir is large enough to build NSCs. In contrast, {\tt S}/{\tt SI} galaxies are typically more extended at fixed stellar mass, producing more extended GC systems and correspondingly longer dynamical-friction timescales. Their higher-mass nucleation peak should therefore not be interpreted as evidence for more efficient GC inspiral. Rather, increasing stellar mass provides a larger available cluster population, while their gas-rich disks may support secular inflows and central star formation in a regime where GC inspiral alone becomes increasingly inefficient. The higher-mass peak for late-type galaxies is therefore best interpreted as reflecting a changing balance between a larger cluster reservoir, dissipational growth, and reduced exposure to the dry-merger and core-scouring processes that suppress NSCs in the most massive early-type systems.
\color{black}


We find environmental trends within these subclasses to be comparatively weak. Nucleated galaxies are more spatially clustered than the overall galaxy population, but variations in nucleation fraction across the cluster are modest, at least within each morphology class. Apparent differences between environments may thus be attributable, in part, to variations in the underlying morphological mix; this reinforces the view that internal galaxy properties are the primary factor governing NSC formation (or destruction), with environment acting as a secondary modifier.

Simple calculations based on the properties of NGVS galaxies and their GC systems show that these empirical trends can be understood within a physically motivated framework that links galaxy structure, stellar mass, and dynamical processes. For low-mass galaxies, $\log(M_\star/M_\odot)\lesssim9$, NSC growth is likely dominated by GC infall driven by dynamical friction; although this process can be efficient in this regime, it is limited by the declining number of clusters per galaxy \citep[e.g.,][]{Peng:2008aa,Lim2025,DornanHarris2026}. At intermediate masses, where nucleation fractions peak around
$\log(M_\star/M_\odot)\sim9$, larger GC populations combine with
still-effective orbital decay, allowing GC inspiral to deliver
substantial stellar mass to the central regions. At higher masses, dynamical friction becomes increasingly inefficient as galaxies and their GC systems become more extended and characteristic velocity scales increase, preventing an increasing fraction of clusters from reaching the center within a Hubble time. At the highest masses, $\log(M_\star/M_\odot)\gtrsim 10.5$, NSC destruction due to mergers and core scouring by supermassive black hole binaries is probably an important factor as well. For galaxies in the mass range $9 \lesssim \log(M_\star/M_\odot) \lesssim 11$, the NGVS galaxy size-mass relation shows the well-known ``flaring" that correlates with morphology \citep[e.g.,][and see Figure~\ref{fig:size_mass}]{Cappellari2013a}. In this regime, additional processes such as dissipational gas inflows and {\it in situ} star formation are likely needed to explain the observed NSC masses, mass ratios and nucleation fractions \citep[e.g.,][]{Turner:2012aa,Antonini:2015aa}.

Several galaxy subpopulations identified within the NGVS sample appear to trace distinct evolutionary stages in NSC formation. We identify 97 dwarf-mass galaxies with early-type or transitional morphologies contain extended blue cores, a subset of which were noted in previous studies \citep{Lisker+06,Paudel:2011aa,Ulrich2017,Michea+2021}. These galaxies have nucleation fractions comparable to those of mass-matched control galaxies, while their blue central colors and nuclear properties indicate that recent or ongoing {\it in situ} star formation may contribute to NSC growth. An analysis of 79 galaxies with possible offset nuclei shows that these systems tend to be less centrally concentrated, modestly more flattened, and somewhat more extended at fixed stellar mass than centrally nucleated control galaxies. There is also some evidence that they favor lower-density environments. These structural differences are consistent with offset nuclei tracing transient or dynamically unsettled phases of NSC assembly. A sample of 17 eUCDs --- ultra-compact dwarf galaxies with faint envelopes --- from \citet{Wang+2026} falls along the extension of the scaling relation between NSC-galaxy luminosity ratio and host galaxy luminosity. This is consistent with the eUCD class being a very late evolutionary stage when nucleated galaxies have been almost entirely tidally stripped of their diffuse envelopes \citep{Bekki2001,Bekki+03,Seth:2014aa,Liu:2015aa,Wang2023}. Collectively, these diverse subpopulations provide snapshots of NSC evolutionary mechanisms operating across different morphological regimes, mass ranges and evolutionary timescales.

We conclude with an assessment of the potential impact of incompleteness on our results. Incompleteness at low mass likely affects the measured nucleation fractions and NSC mass distributions at the faint end of the survey. Such effects may play a role in explaining the curious trends for NSCs to become increasingly rare among the lowest-mass galaxies but, when present, to contribute an elevated fraction of the total mass ($\sim10$\% or more in some cases).

Deep, higher-resolution space-based imaging will be needed to fully characterize NSC populations at the lowest masses. In addition, spatially resolved spectroscopy for targeted samples, or fiber spectroscopy over wide fields, will provide direct constraints on stellar populations, ages, metallicities, and kinematics of both NSCs and their host galaxies, thereby constraining the {\it in situ} and migratory growth channels. Combined with machine-learning classification pipelines calibrated on visual benchmarks such as those presented here, future surveys will extend NSC demographics to larger volumes and a broader range of environments, helping to establish a comprehensive framework for understanding the origin and evolution of NSCs.


\begin{acknowledgments}

The authors acknowledge the use of the Canadian Advanced Network for Astronomy Research (CANFAR) Science Platform operated by the Canadian Astronomy Data Center (CADC) and the Digital Research Alliance of Canada (DRAC), with support from the National Research Council of Canada (NRC), the Canadian Space Agency (CSA), CANARIE, and the Canadian Foundation for Innovation (CFI) \citep{gaudet10}. This research has also made use of the NASA/IPAC Extragalactic Database (NED), which is operated by the Jet Propulsion Laboratory, California Institute of Technology, under contract with the National Aeronautics and Space Administration. PC and MMK gratefully acknowledge support provided by the Natural Sciences and Engineering Research Council of Canada (NSERC).
S.L. was supported by the National Research Foundation of Korea (NRF) grant funded by the Korea government (MSIT, No. 2710105275). PG acknowledges support from NSF grant AST-2206328. ET is thankful for the support from NSF-AST-2206498 grant. CL acknowledges support from the National Natural Science Foundation of China (NSFC, Grant No. 12173025), National Key R\&D Program of China(2023YFA1607800, 2023YFA1607804), 111 project (No. B20019), and Key Laboratory for Particle Physics, Astrophysics and Cosmology, Ministry of Education. MAT acknowledges the support of NSERC, 2023-03298.
\color{black}

\end{acknowledgments}

\facilities{CFHT(MegaCam)}

\software{
Astropy \citep{2013A&A...558A..33A,2018AJ....156..123A},
emcee \citep{Foreman-Mackey:2013aa},
matplotlib \citep{Matplotlib},
numpy \citep{numpy},
pandas \citep{McKinney2010,Reback2020},
photutils \citep{photutils},
scipy \citep{scipy}
}


\bibliographystyle{aasjournal}
\bibliography{references}

@article{Dressler:1980aa,
	Adsurl = {http://adsabs.harvard.edu/abs/1980ApJ...236..351D},
	Author = {{Dressler}, A.},
	Doi = {10.1086/157753},
	Journal = {\apj},
	Month = mar,
	Pages = {351-365},
	Title = {{Galaxy morphology in rich clusters - Implications for the formation and evolution of galaxies}},
	Volume = 236,
	Year = 1980}

@article{Peng:2008aa,
	Adsurl = {http://adsabs.harvard.edu/abs/2008ApJ...681..197P},
	Archiveprefix = {arXiv},
	Arxivurl = {http://arxiv.org/abs/0803.0330},
	Author = {{Peng}, E.~W. and {Jord{\'a}n}, A. and {C{\^o}t{\'e}}, P. and {Takamiya}, M. and {West}, M.~J. and {Blakeslee}, J.~P. and {Chen}, C.-W. and {Ferrarese}, L. and {Mei}, S. and {Tonry}, J.~L. and {West}, A.~A.},
	Doi = {10.1086/587951},
	Eprint = {0803.0330},
	Journal = {\apj},
	Month = jul,
	Pages = {197-224},
	Title = {{The ACS Virgo Cluster Survey. XV. The Formation Efficiencies of Globular Clusters in Early-Type Galaxies: The Effects of Mass and Environment}},
	Volume = 681,
	Year = 2008}

@article{Hartmann:2011aa,
	Adsurl = {http://adsabs.harvard.edu/abs/2011MNRAS.418.2697H},
	Archiveprefix = {arXiv},
	Arxivurl = {http://arxiv.org/abs/1103.5464},
	Author = {{Hartmann}, M. and {Debattista}, V.~P. and {Seth}, A. and {Cappellari}, M. and {Quinn}, T.~R.},
	Doi = {10.1111/j.1365-2966.2011.19659.x},
	Eprint = {1103.5464},
	Journal = {\mnras},
	Month = dec,
	Pages = {2697-2714},
	Title = {{Constraining the role of star cluster mergers in nuclear cluster formation: simulations confront integral-field data}},
	Volume = 418,
	Year = 2011}

@article{Babul:1992aa,
	Adsurl = {http://adsabs.harvard.edu/abs/1992MNRAS.255..346B},
	Author = {{Babul}, A. and {Rees}, M.~J.},
	Doi = {10.1093/mnras/255.2.346},
	Journal = {\mnras},
	Month = mar,
	Pages = {346-350},
	Title = {{On dwarf elliptical galaxies and the faint blue counts}},
	Volume = 255,
	Year = 1992}

@article{Lisker:2007aa,
	Adsurl = {http://adsabs.harvard.edu/abs/2007ApJ...660.1186L},
	Arxivurl = {http://arxiv.org/abs/astro-ph/0701429},
	Author = {{Lisker}, T. and {Grebel}, E.~K. and {Binggeli}, B. and {Glatt}, K.},
	Doi = {10.1086/513090},
	Eprint = {astro-ph/0701429},
	Journal = {\apj},
	Month = may,
	Pages = {1186-1197},
	Title = {{Virgo Cluster Early-Type Dwarf Galaxies with the Sloan Digital Sky Survey. III. Subpopulations: Distributions, Shapes, Origins}},
	Volume = 660,
	Year = 2007}

@article{Munoz:2015aa,
	Adsurl = {http://adsabs.harvard.edu/abs/2015ApJ...813L..15M},
	Archiveprefix = {arXiv},
	Arxivurl = {http://arxiv.org/abs/1510.02475},
	Author = {{Mu{\~n}oz}, R.~P. and {Eigenthaler}, P. and {Puzia}, T.~H. and {Taylor}, M.~A. and {Ordenes-Brice{\~n}o}, Y. and {Alamo-Mart{\'{\i}}nez}, K. and {Ribbeck}, K.~X. and {{\'A}ngel}, S. and {Capaccioli}, M. and {C{\^o}t{\'e}}, P. and {Ferrarese}, L. and {Galaz}, G. and {Hempel}, M. and {Hilker}, M. and {Jord{\'a}n}, A. and {Lan{\c c}on}, A. and {Mieske}, S. and {Paolillo}, M. and {Richtler}, T. and {S{\'a}nchez-Janssen}, R. and {Zhang}, H.},
	Doi = {10.1088/2041-8205/813/1/L15},
	Eid = {L15},
	Eprint = {1510.02475},
	Journal = {\apjl},
	Month = nov,
	Pages = {L15},
	Title = {{Unveiling a Rich System of Faint Dwarf Galaxies in the Next Generation Fornax Survey}},
	Volume = 813,
	Year = 2015}

@article{Eigenthaler:2018aa,
	Adsurl = {http://adsabs.harvard.edu/abs/2018ApJ...855..142E},
	Archiveprefix = {arXiv},
	Arxivurl = {http://arxiv.org/abs/1801.02633},
	Author = {{Eigenthaler}, P. and {Puzia}, T.~H. and {Taylor}, M.~A. and {Ordenes-Brice{\~n}o}, Y. and {Mu{\~n}oz}, R.~P. and {Ribbeck}, K.~X. and {Alamo-Mart{\'{\i}}nez}, K.~A. and {Zhang}, H. and {{\'A}ngel}, S. and {Capaccioli}, M. and {C{\^o}t{\'e}}, P. and {Ferrarese}, L. and {Galaz}, G. and {Grebel}, E.~K. and {Hempel}, M. and {Hilker}, M. and {Lan{\c c}on}, A. and {Mieske}, S. and {Miller}, B. and {Paolillo}, M. and {Powalka}, M. and {Richtler}, T. and {Roediger}, J. and {Rong}, Y. and {S{\'a}nchez-Janssen}, R. and {Spengler}, C.},
	Doi = {10.3847/1538-4357/aaab60},
	Eid = {142},
	Eprint = {1801.02633},
	Journal = {\apj},
	Month = mar,
	Pages = {142},
	Title = {{The Next Generation Fornax Survey (NGFS). II. The Central Dwarf Galaxy Population}},
	Volume = 855,
	Year = 2018}

@article{Ferguson:1989aa,
	Adsurl = {http://adsabs.harvard.edu/abs/1989ApJ...346L..53F},
	Author = {{Ferguson}, H.~C. and {Sandage}, A.},
	Doi = {10.1086/185577},
	Journal = {\apjl},
	Month = nov,
	Pages = {L53-L56},
	Title = {{The spatial distributions and intrinsic shapes of dwarf elliptical galaxies in the Virgo and Fornax Clusters}},
	Volume = 346,
	Year = 1989}

@article{Grant:2005aa,
	Adsurl = {http://adsabs.harvard.edu/abs/2005MNRAS.363.1019G},
	Author = {{Grant}, N.~I. and {Kuipers}, J.~A. and {Phillipps}, S.},
	Doi = {10.1111/j.1365-2966.2005.09518.x},
	Journal = {\mnras},
	Month = nov,
	Pages = {1019-1030},
	Title = {{Nucleated dwarf elliptical galaxies in the Virgo cluster}},
	Volume = 363,
	Year = 2005}

@article{Ordenes-Briceno:2018aa,
	Adsurl = {http://adsabs.harvard.edu/abs/2018ApJ...859...52O},
	Archiveprefix = {arXiv},
	Arxivurl = {http://arxiv.org/abs/1803.10784},
	Author = {{Ordenes-Brice{\~n}o}, Y. and {Eigenthaler}, P. and {Taylor}, M.~A. and {Puzia}, T.~H. and {Alamo-Mart{\'{\i}}nez}, K. and {Ribbeck}, K.~X. and {Mu{\~n}oz}, R.~P. and {Zhang}, H. and {Grebel}, E.~K. and {{\'A}ngel}, S. and {C{\^o}t{\'e}}, P. and {Ferrarese}, L. and {Hilker}, M. and {Lan{\c c}on}, A. and {Mieske}, S. and {Miller}, B.~W. and {Rong}, Y. and {S{\'a}nchez-Janssen}, R.},
	Doi = {10.3847/1538-4357/aaba70},
	Eid = {52},
	Eprint = {1803.10784},
	Journal = {\apj},
	Month = may,
	Pages = {52},
	Title = {{The Next Generation Fornax Survey (NGFS). III. Revealing the Spatial Substructure of the Dwarf Galaxy Population Inside Half of Fornax's Virial Radius}},
	Volume = 859,
	Year = 2018}

@article{Bell:2003aa,
	Adsurl = {http://adsabs.harvard.edu/abs/2003ApJS..149..289B},
	Arxivurl = {http://arxiv.org/abs/astro-ph/0302543},
	Author = {{Bell}, E.~F. and {McIntosh}, D.~H. and {Katz}, N. and {Weinberg}, M.~D.},
	Doi = {10.1086/378847},
	Eprint = {astro-ph/0302543},
	Journal = {\apjs},
	Month = dec,
	Pages = {289-312},
	Title = {{The Optical and Near-Infrared Properties of Galaxies. I. Luminosity and Stellar Mass Functions}},
	Volume = 149,
	Year = 2003}

@article{Ferrarese:2016aa,
	Adsurl = {http://adsabs.harvard.edu/abs/2016ApJ...824...10F},
	Archiveprefix = {arXiv},
	Arxivurl = {http://arxiv.org/abs/1604.06462},
	Author = {{Ferrarese}, L. and {C{\^o}t{\'e}}, P. and {S{\'a}nchez-Janssen}, R. and {Roediger}, J. and {McConnachie}, A.~W. and {Durrell}, P.~R. and {MacArthur}, L.~A. and {Blakeslee}, J.~P. and {Duc}, P.-A. and {Boissier}, S. and {Boselli}, A. and {Courteau}, S. and {Cuillandre}, J.-C. and {Emsellem}, E. and {Gwyn}, S.~D.~J. and {Guhathakurta}, P. and {Jord{\'a}n}, A. and {Lan{\c c}on}, A. and {Liu}, C. and {Mei}, S. and {Mihos}, J.~C. and {Navarro}, J.~F. and {Peng}, E.~W. and {Puzia}, T.~H. and {Taylor}, J.~E. and {Toloba}, E. and {Zhang}, H.},
	Doi = {10.3847/0004-637X/824/1/10},
	Eid = {10},
	Eprint = {1604.06462},
	Journal = {\apj},
	Month = jun,
	Pages = {10},
	Title = {{The Next Generation Virgo Cluster Survey (NGVS). XIII. The Luminosity and Mass Function of Galaxies in the Core of the Virgo Cluster and the Contribution from Disrupted Satellites}},
	Volume = 824,
	Year = 2016}

@article{Binggeli:1985aa,
	Adsurl = {http://adsabs.harvard.edu/abs/1985AJ.....90.1681B},
	Author = {{Binggeli}, B. and {Sandage}, A. and {Tammann}, G.~A.},
	Doi = {10.1086/113874},
	Journal = {\aj},
	Month = sep,
	Pages = {1681-1759},
	Title = {{Studies of the Virgo Cluster. II - A catalog of 2096 galaxies in the Virgo Cluster area.}},
	Volume = 90,
	Year = 1985}

@article{Binggeli:1987aa,
	Adsurl = {http://adsabs.harvard.edu/abs/1987AJ.....94..251B},
	Author = {{Binggeli}, B. and {Tammann}, G.~A. and {Sandage}, A.},
	Doi = {10.1086/114467},
	Journal = {\aj},
	Month = aug,
	Pages = {251-277},
	Title = {{Studies of the Virgo cluster. VI - Morphological and kinematical structure of the Virgo cluster}},
	Volume = 94,
	Year = 1987}

@article{Cantiello:2018aa,
	Adsurl = {http://adsabs.harvard.edu/abs/2018ApJ...856..126C},
	Archiveprefix = {arXiv},
	Arxivurl = {http://arxiv.org/abs/1802.05526},
	Author = {{Cantiello}, M. and {Blakeslee}, J.~P. and {Ferrarese}, L. and {C{\^o}t{\'e}}, P. and {Roediger}, J.~C. and {Raimondo}, G. and {Peng}, E.~W. and {Gwyn}, S. and {Durrell}, P.~R. and {Cuillandre}, J.-C.},
	Doi = {10.3847/1538-4357/aab043},
	Eid = {126},
	Eprint = {1802.05526},
	Journal = {\apj},
	Month = apr,
	Pages = {126},
	Title = {{The Next Generation Virgo Cluster Survey (NGVS). XVIII. Measurement and Calibration of Surface Brightness Fluctuation Distances for Bright Galaxies in Virgo (and Beyond)}},
	Volume = 856,
	Year = 2018}

@article{Cote:2004aa,
	Adsurl = {http://adsabs.harvard.edu/abs/2004ApJS..153..223C},
	Arxivurl = {http://arxiv.org/abs/astro-ph/0404138},
	Author = {{C{\^o}t{\'e}}, P. and {Blakeslee}, J.~P. and {Ferrarese}, L. and {Jord{\'a}n}, A. and {Mei}, S. and {Merritt}, D. and {Milosavljevi{\'c}}, M. and {Peng}, E.~W. and {Tonry}, J.~L. and {West}, M.~J.},
	Doi = {10.1086/421490},
	Eprint = {astro-ph/0404138},
	Journal = {\apjs},
	Month = jul,
	Pages = {223-242},
	Title = {{The ACS Virgo Cluster Survey. I. Introduction to the Survey}},
	Volume = 153,
	Year = 2004}

@article{Spengler:2017aa,
	Adsurl = {http://adsabs.harvard.edu/abs/2017ApJ...849...55S},
	Archiveprefix = {arXiv},
	Arxivurl = {http://arxiv.org/abs/1709.00406},
	Author = {{Spengler}, C. and {C{\^o}t{\'e}}, P. and {Roediger}, J. and {Ferrarese}, L. and {S{\'a}nchez-Janssen}, R. and {Toloba}, E. and {Liu}, Y. and {Guhathakurta}, P. and {Cuillandre}, J.-C. and {Gwyn}, S. and {Zirm}, A. and {Mu{\~n}oz}, R. and {Puzia}, T. and {Lan{\c c}on}, A. and {Peng}, E.~W. and {Mei}, S. and {Powalka}, M.},
	Doi = {10.3847/1538-4357/aa8a78},
	Eid = {55},
	Eprint = {1709.00406},
	Journal = {\apj},
	Month = nov,
	Pages = {55},
	Title = {{Virgo Redux: The Masses and Stellar Content of Nuclei in Early-type Galaxies from Multiband Photometry and Spectroscopy}},
	Volume = 849,
	Year = 2017}

@article{Cleveland:1979aa,
	Author = {Cleveland, William S},
	Journal = {Journal of the American Statistical Association},
	Number = {368},
	Pages = {829--836},
	Publisher = {Taylor \& Francis Group},
	Title = {Robust Locally Weighted Regression and Smoothing Scatterplots},
	Volume = {74},
	Year = {1979}}

@article{Lotz:2001aa,
	Adsurl = {http://adsabs.harvard.edu/abs/2001ApJ...552..572L},
	Arxivurl = {http://arxiv.org/abs/astro-ph/0102079},
	Author = {{Lotz}, J.~M. and {Telford}, R. and {Ferguson}, H.~C. and {Miller}, B.~W. and {Stiavelli}, M. and {Mack}, J.},
	Doi = {10.1086/320545},
	Eprint = {astro-ph/0102079},
	Journal = {\apj},
	Month = may,
	Pages = {572-581},
	Title = {{Dynamical Friction in DE Globular Cluster Systems}},
	Volume = 552,
	Year = 2001}

@article{Roediger:2017aa,
	Adsurl = {http://adsabs.harvard.edu/abs/2017ApJ...836..120R},
	Archiveprefix = {arXiv},
	Arxivurl = {http://arxiv.org/abs/1610.09377},
	Author = {{Roediger}, J.~C. and {Ferrarese}, L. and {C{\^o}t{\'e}}, P. and {MacArthur}, L.~A. and {S{\'a}nchez-Janssen}, R. and {Blakeslee}, J.~P. and {Peng}, E.~W. and {Liu}, C. and {Munoz}, R. and {Cuillandre}, J.-C. and {Gwyn}, S. and {Mei}, S. and {Boissier}, S. and {Boselli}, A. and {Cantiello}, M. and {Courteau}, S. and {Duc}, P.-A. and {Lan{\c c}on}, A. and {Mihos}, J.~C. and {Puzia}, T.~H. and {Taylor}, J.~E. and {Durrell}, P.~R. and {Toloba}, E. and {Guhathakurta}, P. and {Zhang}, H.},
	Doi = {10.3847/1538-4357/836/1/120},
	Eid = {120},
	Eprint = {1610.09377},
	Journal = {\apj},
	Month = feb,
	Pages = {120},
	Title = {{The Next Generation Virgo Cluster Survey (NGVS). XXIV. The Red Sequence to \~{}10$^{6}$ L $_{&sun;}$ and Comparisons with Galaxy Formation Models}},
	Volume = 836,
	Year = 2017}

@article{Pfeffer:2014aa,
	Adsurl = {http://adsabs.harvard.edu/abs/2014MNRAS.444.3670P},
	Archiveprefix = {arXiv},
	Author = {{Pfeffer}, J. and {Griffen}, B.~F. and {Baumgardt}, H. and {Hilker}, M.},
	Eprint = {1408.4467},
	Journal = {\mnras},
	Month = nov,
	Pages = {3670-3683},
	Title = {{Contribution of stripped nuclear clusters to globular cluster and ultracompact dwarf galaxy populations}},
	Volume = 444,
	Year = 2014}

@article{Mei:2007aa,
	Adsurl = {http://adsabs.harvard.edu/abs/2007ApJ...655..144M},
	Arxivurl = {http://arxiv.org/abs/astro-ph/0702510},
	Author = {{Mei}, S. and {Blakeslee}, J.~P. and {C{\^o}t{\'e}}, P. and {Tonry}, J.~L. and {West}, M.~J. and {Ferrarese}, L. and {Jord{\'a}n}, A. and {Peng}, E.~W. and {Anthony}, A. and {Merritt}, D.},
	Eprint = {astro-ph/0702510},
	Journal = {\apj},
	Month = jan,
	Pages = {144-162},
	Title = {{The ACS Virgo Cluster Survey. XIII. SBF Distance Catalog and the Three-dimensional Structure of the Virgo Cluster}},
	Volume = 655,
	Year = 2007}

@article{Seth:2008aa,
	Adsurl = {http://adsabs.harvard.edu/abs/2008ApJ...678..116S},
	Archiveprefix = {arXiv},
	Arxivurl = {http://arxiv.org/abs/0801.0439},
	Author = {{Seth}, A. and {Ag{\"u}eros}, M. and {Lee}, D. and {Basu-Zych}, A.},
	Eid = {116-130},
	Eprint = {0801.0439},
	Journal = {\apj},
	Month = may,
	Pages = {116-130},
	Title = {{The Coincidence of Nuclear Star Clusters and Active Galactic Nuclei}},
	Volume = 678,
	Year = 2008}

@article{Seth:2014aa,
	Adsurl = {http://adsabs.harvard.edu/abs/2014Natur.513..398S},
	Archiveprefix = {arXiv},
	Arxivurl = {http://arxiv.org/abs/1409.4769},
	Author = {{Seth}, A.~C. and {van den Bosch}, R. and {Mieske}, S. and {Baumgardt}, H. and {Brok}, M.~D. and {Strader}, J. and {Neumayer}, N. and {Chilingarian}, I. and {Hilker}, M. and {McDermid}, R. and {Spitler}, L. and {Brodie}, J. and {Frank}, M.~J. and {Walsh}, J.~L.},
	Eprint = {1409.4769},
	Journal = {\nat},
	Month = sep,
	Pages = {398-400},
	Title = {{A supermassive black hole in an ultra-compact dwarf galaxy}},
	Volume = 513,
	Year = 2014}

@article{Bekki:2007aa,
	Adsurl = {http://adsabs.harvard.edu/abs/2007PASA...24...77B},
	Author = {{Bekki}, K.},
	Journal = {\pasa},
	Month = jul,
	Pages = {77-94},
	Title = {{The Formation of Stellar Galactic Nuclei through Dissipative Gas Dynamics}},
	Volume = 24,
	Year = 2007}

@article{Taylor:2011aa,
	Adsurl = {http://adsabs.harvard.edu/abs/2011MNRAS.418.1587T},
	Archiveprefix = {arXiv},
	Arxivurl = {http://arxiv.org/abs/1108.0635},
	Author = {{Taylor}, E.~N. and {Hopkins}, A.~M. and {Baldry}, I.~K. and {Brown}, M.~J.~I. and {Driver}, S.~P. and {Kelvin}, L.~S. and {Hill}, D.~T. and {Robotham}, A.~S.~G. and {Bland-Hawthorn}, J. and {Jones}, D.~H. and {Sharp}, R.~G. and {Thomas}, D. and {Liske}, J. and {Loveday}, J. and {Norberg}, P. and {Peacock}, J.~A. and {Bamford}, S.~P. and {Brough}, S. and {Colless}, M. and {Cameron}, E. and {Conselice}, C.~J. and {Croom}, S.~M. and {Frenk}, C.~S. and {Gunawardhana}, M. and {Kuijken}, K. and {Nichol}, R.~C. and {Parkinson}, H.~R. and {Phillipps}, S. and {Pimbblet}, K.~A. and {Popescu}, C.~C. and {Prescott}, M. and {Sutherland}, W.~J. and {Tuffs}, R.~J. and {van Kampen}, E. and {Wijesinghe}, D.},
	Eprint = {1108.0635},
	Journal = {\mnras},
	Month = dec,
	Pages = {1587-1620},
	Primaryclass = {astro-ph.CO},
	Title = {{Galaxy And Mass Assembly (GAMA): stellar mass estimates}},
	Volume = 418,
	Year = 2011}

@article{Graham:2012aa,
	Adsurl = {http://adsabs.harvard.edu/abs/2012MNRAS.422.1586G},
	Archiveprefix = {arXiv},
	Arxivurl = {http://arxiv.org/abs/1202.4056},
	Author = {{Graham}, A.~W.},
	Eprint = {1202.4056},
	Journal = {\mnras},
	Month = may,
	Pages = {1586-1591},
	Title = {{Extending the M$_{bh}$-{$\sigma$} diagram with dense nuclear star clusters}},
	Volume = 422,
	Year = 2012}

@article{Carson:2015aa,
	Adsurl = {http://adsabs.harvard.edu/abs/2015AJ....149..170C},
	Archiveprefix = {arXiv},
	Arxivurl = {http://arxiv.org/abs/1501.05586},
	Author = {{Carson}, D.~J. and {Barth}, A.~J. and {Seth}, A.~C. and {den Brok}, M. and {Cappellari}, M. and {Greene}, J.~E. and {Ho}, L.~C. and {Neumayer}, N.},
	Eid = {170},
	Eprint = {1501.05586},
	Journal = {\aj},
	Month = may,
	Pages = {170},
	Title = {{The Structure of Nuclear Star Clusters in Nearby Late-type Spiral Galaxies from Hubble Space Telescope Wide Field Camera 3 Imaging}},
	Volume = 149,
	Year = 2015}

@article{Pfeffer:2013aa,
	Adsurl = {http://adsabs.harvard.edu/abs/2013MNRAS.433.1997P},
	Archiveprefix = {arXiv},
	Arxivurl = {http://arxiv.org/abs/1305.3656},
	Author = {{Pfeffer}, J. and {Baumgardt}, H.},
	Eprint = {1305.3656},
	Journal = {\mnras},
	Month = aug,
	Pages = {1997-2005},
	Title = {{Ultra-compact dwarf galaxy formation by tidal stripping of nucleated dwarf galaxies}},
	Volume = 433,
	Year = 2013}

@article{den-Brok:2014aa,
	Adsurl = {http://adsabs.harvard.edu/abs/2014MNRAS.445.2385D},
	Archiveprefix = {arXiv},
	Arxivurl = {http://arxiv.org/abs/1409.4766},
	Author = {{den Brok}, M. and {Peletier}, R.~F. and {Seth}, A. and {Balcells}, M. and {Dominguez}, L. and {Graham}, A.~W. and {Carter}, D. and {Erwin}, P. and {Ferguson}, H.~C. and {Goudfrooij}, P. and {Guzm{\'a}n}, R. and {Hoyos}, C. and {Jogee}, S. and {Lucey}, J. and {Phillipps}, S. and {Puzia}, T. and {Valentijn}, E. and {Kleijn}, G.~V. and {Weinzirl}, T.},
	Eprint = {1409.4766},
	Journal = {\mnras},
	Month = dec,
	Pages = {2385-2403},
	Title = {{The HST/ACS Coma Cluster Survey - X. Nuclear star clusters in low-mass early-type galaxies: scaling relations}},
	Volume = 445,
	Year = 2014}

@article{Georgiev:2016aa,
	Adsurl = {http://adsabs.harvard.edu/abs/2016MNRAS.457.2122G},
	Archiveprefix = {arXiv},
	Arxivurl = {http://arxiv.org/abs/1601.02613},
	Author = {{Georgiev}, I.~Y. and {B{\"o}ker}, T. and {Leigh}, N. and {L{\"u}tzgendorf}, N. and {Neumayer}, N.},
	Eprint = {1601.02613},
	Journal = {\mnras},
	Month = apr,
	Pages = {2122-2138},
	Title = {{Masses and scaling relations for nuclear star clusters, and their co-existence with central black holes}},
	Volume = 457,
	Year = 2016}

@article{Antonini:2015aa,
	Adsurl = {http://adsabs.harvard.edu/abs/2015ApJ...812...72A},
	Archiveprefix = {arXiv},
	Arxivurl = {http://arxiv.org/abs/1506.02050},
	Author = {{Antonini}, F. and {Barausse}, E. and {Silk}, J.},
	Eid = {72},
	Eprint = {1506.02050},
	Journal = {\apj},
	Month = oct,
	Pages = {72},
	Title = {{The Coevolution of Nuclear Star Clusters, Massive Black Holes, and Their Host Galaxies}},
	Volume = 812,
	Year = 2015}

@article{Gnedin:2014aa,
	Adsurl = {http://adsabs.harvard.edu/abs/2014ApJ...785...71G},
	Archiveprefix = {arXiv},
	Arxivurl = {http://arxiv.org/abs/1308.0021},
	Author = {{Gnedin}, O.~Y. and {Ostriker}, J.~P. and {Tremaine}, S.},
	Eid = {71},
	Eprint = {1308.0021},
	Journal = {\apj},
	Month = apr,
	Pages = {71},
	Title = {{Co-evolution of Galactic Nuclei and Globular Cluster Systems}},
	Volume = 785,
	Year = 2014}

@article{Guillard:2016aa,
	Adsurl = {http://adsabs.harvard.edu/abs/2016MNRAS.461.3620G},
	Archiveprefix = {arXiv},
	Arxivurl = {http://arxiv.org/abs/1606.09537},
	Author = {{Guillard}, N. and {Emsellem}, E. and {Renaud}, F.},
	Eprint = {1606.09537},
	Journal = {\mnras},
	Month = oct,
	Pages = {3620-3629},
	Title = {{New insights on the formation of nuclear star clusters}},
	Volume = 461,
	Year = 2016}

@article{Foreman-Mackey:2013aa,
	Adsurl = {http://adsabs.harvard.edu/abs/2013PASP..125..306F},
	Archiveprefix = {arXiv},
	Arxivurl = {http://arxiv.org/abs/1202.3665},
	Author = {{Foreman-Mackey}, D. and {Hogg}, D.~W. and {Lang}, D. and {Goodman}, J.},
	Eprint = {1202.3665},
	Journal = {\pasp},
	Month = mar,
	Pages = {306-312},
	Primaryclass = {astro-ph.IM},
	Title = {{emcee: The MCMC Hammer}},
	Volume = 125,
	Year = 2013}

@article{Bekki:2015aa,
	Adsurl = {http://adsabs.harvard.edu/abs/2015ApJ...812L..14B},
	Archiveprefix = {arXiv},
	Arxivurl = {http://arxiv.org/abs/1510.04388},
	Author = {{Bekki}, K.},
	Eid = {L14},
	Eprint = {1510.04388},
	Journal = {\apjl},
	Month = oct,
	Pages = {L14},
	Title = {{Formation of Ultra-compact Blue Dwarf Galaxies and Their Evolution into Nucleated Dwarfs}},
	Volume = 812,
	Year = 2015}

@article{Antonini:2012aa,
	Adsurl = {http://adsabs.harvard.edu/abs/2012ApJ...750..111A},
	Archiveprefix = {arXiv},
	Arxivurl = {http://arxiv.org/abs/1110.5937},
	Author = {{Antonini}, F. and {Capuzzo-Dolcetta}, R. and {Mastrobuono-Battisti}, A. and {Merritt}, D.},
	Eid = {111},
	Eprint = {1110.5937},
	Journal = {\apj},
	Month = may,
	Pages = {111},
	Title = {{Dissipationless Formation and Evolution of the Milky Way Nuclear Star Cluster}},
	Volume = 750,
	Year = 2012}

@article{Tremaine:1975aa,
	Adsurl = {http://adsabs.harvard.edu/abs/1975ApJ...196..407T},
	Author = {{Tremaine}, S.~D. and {Ostriker}, J.~P. and {Spitzer}, Jr., L.},
	Journal = {\apj},
	Month = mar,
	Pages = {407-411},
	Title = {{The formation of the nuclei of galaxies. I - M31}},
	Volume = 196,
	Year = 1975}

@article{Mihos:1994aa,
	Adsurl = {http://adsabs.harvard.edu/abs/1994ApJ...437L..47M},
	Author = {{Mihos}, J.~C. and {Hernquist}, L.},
	Journal = {\apjl},
	Month = dec,
	Pages = {L47-L50},
	Title = {{Dense stellar cores in merger remnants}},
	Volume = 437,
	Year = 1994}

@article{Zhang:2015aa,
	Adsurl = {http://adsabs.harvard.edu/abs/2015ApJ...802...30Z},
	Archiveprefix = {arXiv},
	Arxivurl = {http://arxiv.org/abs/1501.03167},
	Author = {{Zhang}, H.-X. and {Peng}, E.~W. and {C{\^o}t{\'e}}, P. and {Liu}, C. and {Ferrarese}, L. and {Cuillandre}, J.-C. and {Caldwell}, N. and {Gwyn}, S.~D.~J. and {Jord{\'a}n}, A. and {Lan{\c c}on}, A. and {Li}, B. and {Mu{\~n}oz}, R.~P. and {Puzia}, T.~H. and {Bekki}, K. and {Blakeslee}, J.~P. and {Boselli}, A. and {Drinkwater}, M.~J. and {Duc}, P.-A. and {Durrell}, P. and {Emsellem}, E. and {Firth}, P. and {S{\'a}nchez-Janssen}, R.},
	Eid = {30},
	Eprint = {1501.03167},
	Journal = {\apj},
	Month = mar,
	Pages = {30},
	Title = {{The Next Generation Virgo Cluster Survey. VI. The Kinematics of Ultra-compact Dwarfs and Globular Clusters in M87}},
	Volume = 802,
	Year = 2015}

@article{Liu:2015aa,
	Adsurl = {http://adsabs.harvard.edu/abs/2015ApJ...812...34L},
	Archiveprefix = {arXiv},
	Arxivurl = {http://arxiv.org/abs/1508.07334},
	Author = {{Liu}, C. and {Peng}, E.~W. and {C{\^o}t{\'e}}, P. and {Ferrarese}, L. and {Jord{\'a}n}, A. and {Mihos}, J.~C. and {Zhang}, H.-X. and {Mu{\~n}oz}, R.~P. and {Puzia}, T.~H. and {Lan{\c c}on}, A. and {Gwyn}, S. and {Cuillandre}, J.-C. and {Blakeslee}, J.~P. and {Boselli}, A. and {Durrell}, P.~R. and {Duc}, P.-A. and {Guhathakurta}, P. and {MacArthur}, L.~A. and {Mei}, S. and {S{\'a}nchez-Janssen}, R. and {Xu}, H.},
	Eid = {34},
	Eprint = {1508.07334},
	Journal = {\apj},
	Month = oct,
	Pages = {34},
	Title = {{The Next Generation Virgo Cluster Survey. X. Properties of Ultra-compact Dwarfs in the M87, M49, and M60 Regions.}},
	Volume = 812,
	Year = 2015}

@article{Wehner:2006aa,
	Adsurl = {http://adsabs.harvard.edu/abs/2006ApJ...644L..17W},
	Arxivurl = {http://arXiv.org/abs/astro-ph/0603801},
	Author = {{Wehner}, E.~H. and {Harris}, W.~E.},
	Eprint = {astro-ph/0603801},
	Journal = {\apjl},
	Month = jun,
	Pages = {L17-L20},
	Title = {{From Supermassive Black Holes to Dwarf Elliptical Nuclei: A Mass Continuum}},
	Volume = 644,
	Year = 2006}

@article{Walcher:2005aa,
	Adsurl = {http://adsabs.harvard.edu/abs/2005ApJ...618..237W},
	Arxivurl = {http://arXiv.org/abs/astro-ph/0409216},
	Author = {{Walcher}, C.~J. and {van der Marel}, R.~P. and {McLaughlin}, D. and {Rix}, H.-W. and {B{\"o}ker}, T. and {H{\"a}ring}, N. and {Ho}, L.~C. and {Sarzi}, M. and {Shields}, J.~C.},
	Eprint = {astro-ph/0409216},
	Journal = {\apj},
	Month = jan,
	Pages = {237-246},
	Title = {{Masses of Star Clusters in the Nuclei of Bulgeless Spiral Galaxies}},
	Volume = 618,
	Year = 2005}

@article{Walcher:2006aa,
	Adsurl = {http://adsabs.harvard.edu/abs/2006ApJ...649..692W},
	Arxivurl = {http://arXiv.org/abs/astro-ph/0604138},
	Author = {{Walcher}, C.~J. and {B{\"o}ker}, T. and {Charlot}, S. and {Ho}, L.~C. and {Rix}, H.-W. and {Rossa}, J. and {Shields}, J.~C. and {van der Marel}, R.~P.},
	Eprint = {astro-ph/0604138},
	Journal = {\apj},
	Month = oct,
	Pages = {692-708},
	Title = {{Stellar Populations in the Nuclei of Late-Type Spiral Galaxies}},
	Volume = 649,
	Year = 2006}

@article{Turner:2012aa,
	Adsurl = {http://adsabs.harvard.edu/abs/2012ApJS..203....5T},
	Archiveprefix = {arXiv},
	Arxivurl = {http://arXiv.org/abs/1208.0338},
	Author = {{Turner}, M.~L. and {C{\^o}t{\'e}}, P. and {Ferrarese}, L. and {Jord{\'a}n}, A. and {Blakeslee}, J.~P. and {Mei}, S. and {Peng}, E.~W. and {West}, M.~J.},
	Eid = {5},
	Eprint = {1208.0338},
	Journal = {\apjs},
	Month = nov,
	Pages = {5},
	Primaryclass = {astro-ph.CO},
	Title = {{The ACS Fornax Cluster Survey. VI. The Nuclei of Early-type Galaxies in the Fornax Cluster}},
	Volume = 203,
	Year = 2012}

@article{Scott:2013aa,
	Adsurl = {http://adsabs.harvard.edu/abs/2013ApJ...763...76S},
	Archiveprefix = {arXiv},
	Arxivurl = {http://arXiv.org/abs/1205.5338},
	Author = {{Scott}, N. and {Graham}, A.~W.},
	Eid = {76},
	Eprint = {1205.5338},
	Journal = {\apj},
	Month = feb,
	Pages = {76},
	Primaryclass = {astro-ph.CO},
	Title = {{Updated Mass Scaling Relations for Nuclear Star Clusters and a Comparison to Supermassive Black Holes}},
	Volume = 763,
	Year = 2013}

@article{Schinnerer:2008aa,
	Adsurl = {http://adsabs.harvard.edu/abs/2008ApJ...684L..21S},
	Archiveprefix = {arXiv},
	Arxivurl = {http://arXiv.org/abs/0808.0793},
	Author = {{Schinnerer}, E. and {B{\"o}ker}, T. and {Meier}, D.~S. and {Calzetti}, D.},
	Eprint = {0808.0793},
	Journal = {\apjl},
	Month = sep,
	Pages = {L21-L24},
	Title = {{Self-Regulated Fueling of Galaxy Centers: Evidence for Star Formation Feedback in IC 342's Nucleus}},
	Volume = 684,
	Year = 2008}

@article{Rossa:2006aa,
	Adsurl = {http://adsabs.harvard.edu/abs/2006AJ....132.1074R},
	Arxivurl = {http://arXiv.org/abs/astro-ph/0604140},
	Author = {{Rossa}, J. and {van der Marel}, R.~P. and {B{\"o}ker}, T. and {Gerssen}, J. and {Ho}, L.~C. and {Rix}, H.-W. and {Shields}, J.~C. and {Walcher}, C.-J.},
	Eprint = {astro-ph/0604140},
	Journal = {\aj},
	Month = sep,
	Pages = {1074-1099},
	Title = {{Hubble Space Telescope STIS Spectra of Nuclear Star Clusters in Spiral Galaxies: Dependence of Age and Mass on Hubble Type}},
	Volume = 132,
	Year = 2006}

@article{Peng:2010aa,
	Adsurl = {http://adsabs.harvard.edu/abs/2010AJ....139.2097P},
	Archiveprefix = {arXiv},
	Arxivurl = {http://arXiv.org/abs/0912.0731},
	Author = {{Peng}, C.~Y. and {Ho}, L.~C. and {Impey}, C.~D. and {Rix}, H.-W.},
	Eprint = {0912.0731},
	Journal = {\aj},
	Month = jun,
	Pages = {2097-2129},
	Primaryclass = {astro-ph.CO},
	Title = {{Detailed Decomposition of Galaxy Images. II. Beyond Axisymmetric Models}},
	Volume = 139,
	Year = 2010}

@article{Paudel:2011aa,
	Adsurl = {http://adsabs.harvard.edu/abs/2011MNRAS.413.1764P},
	Archiveprefix = {arXiv},
	Arxivurl = {http://arXiv.org/abs/1012.4092},
	Author = {{Paudel}, S. and {Lisker}, T. and {Kuntschner}, H.},
	Eprint = {1012.4092},
	Journal = {\mnras},
	Month = may,
	Pages = {1764-1776},
	Primaryclass = {astro-ph.GA},
	Title = {{Nuclei of early-type dwarf galaxies: insights from stellar populations}},
	Volume = 413,
	Year = 2011}

@article{Oh:2000aa,
	Adsurl = {http://adsabs.harvard.edu/abs/2000ApJ...543..620O},
	Author = {{Oh}, K.~S. and {Lin}, D.~N.~C.},
	Journal = {\apj},
	Month = nov,
	Pages = {620-633},
	Title = {{Nucleation of Dwarf Galaxies in the Virgo Cluster}},
	Volume = 543,
	Year = 2000}

@article{Milosavljevic:2004aa,
	Adsurl = {http://adsabs.harvard.edu/abs/2004ApJ...605L..13M},
	Arxivurl = {http://arXiv.org/abs/astro-ph/0310574},
	Author = {{Milosavljevi{\'c}}, M.},
	Eprint = {astro-ph/0310574},
	Journal = {\apjl},
	Month = apr,
	Pages = {L13-L16},
	Title = {{On the Origin of Nuclear Star Clusters in Late-Type Spiral Galaxies}},
	Volume = 605,
	Year = 2004}

@article{Lotz:2004aa,
	Adsurl = {http://adsabs.harvard.edu/abs/2004ApJ...613..262L},
	Arxivurl = {http://arXiv.org/abs/astro-ph/0406002},
	Author = {{Lotz}, J.~M. and {Miller}, B.~W. and {Ferguson}, H.~C.},
	Eprint = {astro-ph/0406002},
	Journal = {\apj},
	Month = sep,
	Pages = {262-278},
	Title = {{The Colors of Dwarf Elliptical Galaxy Globular Cluster Systems, Nuclei, and Stellar Halos}},
	Volume = 613,
	Year = 2004}

@article{Jordan:2007aa,
	Adsurl = {http://adsabs.harvard.edu/abs/2007ApJS..169..213J},
	Arxivurl = {http://arXiv.org/abs/astro-ph/0702320},
	Author = {{Jord{\'a}n}, A. and {Blakeslee}, J.~P. and {C{\^o}t{\'e}}, P. and {Ferrarese}, L. and {Infante}, L. and {Mei}, S. and {Merritt}, D. and {Peng}, E.~W. and {Tonry}, J.~L. and {West}, M.~J.},
	Eprint = {astro-ph/0702320},
	Journal = {\apjs},
	Month = apr,
	Pages = {213-224},
	Title = {{The ACS Fornax Cluster Survey. I. Introduction to the Survey and Data Reduction Procedures}},
	Volume = 169,
	Year = 2007}

@article{Georgiev:2014aa,
	Adsurl = {http://adsabs.harvard.edu/abs/2014MNRAS.441.3570G},
	Archiveprefix = {arXiv},
	Arxivurl = {http://arXiv.org/abs/1404.5956},
	Author = {{Georgiev}, I.~Y. and {B{\"o}ker}, T.},
	Eprint = {1404.5956},
	Journal = {\mnras},
	Month = jul,
	Pages = {3570-3590},
	Title = {{Nuclear star clusters in 228 spiral galaxies in the HST/WFPC2 archive: catalogue and comparison to other stellar systems}},
	Volume = 441,
	Year = 2014}

@article{Ferrarese:2006aa,
	Adsurl = {http://adsabs.harvard.edu/abs/2006ApJ...644L..21F},
	Arxivurl = {http://arXiv.org/abs/astro-ph/0603840},
	Author = {{Ferrarese}, L. and {C{\^o}t{\'e}}, P. and {Dalla Bont{\`a}}, E. and {Peng}, E.~W. and {Merritt}, D. and {Jord{\'a}n}, A. and {Blakeslee}, J.~P. and {Ha{\c s}egan}, M. and {Mei}, S. and {Piatek}, S. and {Tonry}, J.~L. and {West}, M.~J.},
	Eprint = {astro-ph/0603840},
	Journal = {\apjl},
	Month = jun,
	Pages = {L21-L24},
	Title = {{A Fundamental Relation between Compact Stellar Nuclei, Supermassive Black Holes, and Their Host Galaxies}},
	Volume = 644,
	Year = 2006}

@article{Ferrarese:2012aa,
	Adsurl = {http://adsabs.harvard.edu/abs/2012ApJS..200....4F},
	Author = {{Ferrarese}, L. and {C{\^o}t{\'e}}, P. and {Cuillandre}, J.-C. and {Gwyn}, S.~D.~J. and {Peng}, E.~W. and {MacArthur}, L.~A. and {Duc}, P.-A. and {Boselli}, A. and {Mei}, S. and {Erben}, T. and {McConnachie}, A.~W. and {Durrell}, P.~R. and {Mihos}, J.~C. and {Jord{\'a}n}, A. and {Lan{\c c}on}, A. and {Puzia}, T.~H. and {Emsellem}, E. and {Balogh}, M.~L. and {Blakeslee}, J.~P. and {van Waerbeke}, L. and {Gavazzi}, R. and {Vollmer}, B. and {Kavelaars}, J.~J. and {Woods}, D. and {Ball}, N.~M. and {Boissier}, S. and {Courteau}, S. and {Ferriere}, E. and {Gavazzi}, G. and {Hildebrandt}, H. and {Hudelot}, P. and {Huertas-Company}, M. and {Liu}, C. and {McLaughlin}, D. and {Mellier}, Y. and {Milkeraitis}, M. and {Schade}, D. and {Balkowski}, C. and {Bournaud}, F. and {Carlberg}, R.~G. and {Chapman}, S.~C. and {Hoekstra}, H. and {Peng}, C. and {Sawicki}, M. and {Simard}, L. and {Taylor}, J.~E. and {Tully}, R.~B. and {van Driel}, W. and {Wilson}, C.~D. and {Burdullis}, T. and {Mahoney}, B. and {Manset}, N.},
	Eid = {4},
	Journal = {\apjs},
	Month = may,
	Pages = {4},
	Title = {{The Next Generation Virgo Cluster Survey (NGVS). I. Introduction to the Survey}},
	Volume = 200,
	Year = 2012}

@article{Drinkwater:2003aa,
	Adsurl = {http://adsabs.harvard.edu/abs/2003Natur.423..519D},
	Arxivurl = {http://arXiv.org/abs/astro-ph/0306026},
	Author = {{Drinkwater}, M.~J. and {Gregg}, M.~D. and {Hilker}, M. and {Bekki}, K. and {Couch}, W.~J. and {Ferguson}, H.~C. and {Jones}, J.~B. and {Phillipps}, S.},
	Eprint = {astro-ph/0306026},
	Journal = {\nat},
	Month = may,
	Pages = {519-521},
	Title = {{A class of compact dwarf galaxies from disruptive processes in galaxy clusters}},
	Volume = 423,
	Year = 2003}

@article{Cote:2006aa,
	Adsurl = {http://adsabs.harvard.edu/abs/2006ApJS..165...57C},
	Arxivurl = {http://arXiv.org/abs/astro-ph/0603252},
	Author = {{C{\^o}t{\'e}}, P. and {Piatek}, S. and {Ferrarese}, L. and {Jord{\'a}n}, A. and {Merritt}, D. and {Peng}, E.~W. and {Ha{\c s}egan}, M. and {Blakeslee}, J.~P. and {Mei}, S. and {West}, M.~J. and {Milosavljevi{\'c}}, M. and {Tonry}, J.~L.},
	Eprint = {astro-ph/0603252},
	Journal = {\apjs},
	Month = jul,
	Pages = {57-94},
	Title = {{The ACS Virgo Cluster Survey. VIII. The Nuclei of Early-Type Galaxies}},
	Volume = 165,
	Year = 2006}

@article{Carollo:1997aa,
	Adsurl = {http://adsabs.harvard.edu/abs/1997AJ....114.2366C},
	Author = {{Carollo}, C.~M. and {Stiavelli}, M. and {de Zeeuw}, P.~T. and {Mack}, J.},
	Journal = {\aj},
	Month = dec,
	Pages = {2366},
	Title = {{Spiral Galaxies with WFPC2.I.Nuclear Morphology, Bulges, Star Clusters, and Surface Brightness Profiles}},
	Volume = 114,
	Year = 1997}

@article{Carollo:1998aa,
	Adsurl = {http://adsabs.harvard.edu/abs/1998AJ....116...68C},
	Arxivurl = {http://arXiv.org/abs/astro-ph/9804007},
	Author = {{Carollo}, C.~M. and {Stiavelli}, M. and {Mack}, J.},
	Eprint = {astro-ph/9804007},
	Journal = {\aj},
	Month = jul,
	Pages = {68-84},
	Title = {{Spiral Galaxies with WFPC2. II. The Nuclear Properties of 40 Objects}},
	Volume = 116,
	Year = 1998}

@article{Capuzzo-Dolcetta:2008aa,
	Adsurl = {http://adsabs.harvard.edu/abs/2008MNRAS.388L..69C},
	Archiveprefix = {arXiv},
	Arxivurl = {http://arXiv.org/abs/0804.4421},
	Author = {{Capuzzo-Dolcetta}, R. and {Miocchi}, P.},
	Eprint = {0804.4421},
	Journal = {\mnras},
	Month = jul,
	Pages = {L69-L73},
	Title = {{Self-consistent simulations of nuclear cluster formation through globular cluster orbital decay and merging}},
	Volume = 388,
	Year = 2008}

@article{Boker:2004aa,
	Adsurl = {http://adsabs.harvard.edu/abs/2004AJ....127..105B},
	Arxivurl = {http://arXiv.org/abs/astro-ph/0309761},
	Author = {{B{\"o}ker}, T. and {Sarzi}, M. and {McLaughlin}, D.~E. and {van der Marel}, R.~P. and {Rix}, H.-W. and {Ho}, L.~C. and {Shields}, J.~C.},
	Eprint = {astro-ph/0309761},
	Journal = {\aj},
	Month = jan,
	Pages = {105-118},
	Title = {{A Hubble Space Telescope Census of Nuclear Star Clusters in Late-Type Spiral Galaxies. II. Cluster Sizes and Structural Parameter Correlations}},
	Volume = 127,
	Year = 2004}

@article{Boker:2002aa,
	Adsurl = {http://adsabs.harvard.edu/abs/2002AJ....123.1389B},
	Arxivurl = {http://arXiv.org/abs/astro-ph/0112086},
	Author = {{B{\"o}ker}, T. and {Laine}, S. and {van der Marel}, R.~P. and {Sarzi}, M. and {Rix}, H.-W. and {Ho}, L.~C. and {Shields}, J.~C.},
	Eprint = {astro-ph/0112086},
	Journal = {\aj},
	Month = mar,
	Pages = {1389-1410},
	Title = {{A Hubble Space Telescope Census of Nuclear Star Clusters in Late-Type Spiral Galaxies. I. Observations and Image Analysis}},
	Volume = 123,
	Year = 2002}

@INPROCEEDINGS{gaudet10,
       author = {{Gaudet}, S{\'e}verin and {Hill}, Norman and {Armstrong}, Patrick and {Ball}, Nick and {Burke}, Jeff and {Chapel}, Brian and {Chapin}, Ed and {Damian}, Adrian and {Dowler}, Pat and {Gable}, Ian and {Goliath}, Sharon and {Ghiurea}, Isabella and {Fabbro}, S{\'e}bastien and {Gwyn}, Stephen and {Jenkins}, Dustin and {Kavelaars}, JJ and {Major}, Brian and {Ouellette}, John and {Paterson}, Mike and {Peddle}, Michael and {Penfold-Brown}, Duncan and {Pritchet}, Chris and {Schade}, David and {Sobie}, Randall and {Woods}, David and {Yeung}, Alinga and {Zhang}, Yuehai},
        title = "{CANFAR: the Canadian Advanced Network for Astronomical Research}",
    booktitle = {Software and Cyberinfrastructure for Astronomy},
         year = 2010,
       editor = {{Radziwill}, Nicole M. and {Bridger}, Alan},
       series = {Society of Photo-Optical Instrumentation Engineers (SPIE) Conference Series},
       volume = {7740},
        month = jul,
          eid = {77401I},
        pages = {77401I},
          doi = {10.1117/12.858026},
       adsurl = {https://ui.adsabs.harvard.edu/abs/2010SPIE.7740E..1IG}
}

@ARTICLE{2018AJ....156..123A,
       author = {{Astropy Collaboration} and {Price-Whelan}, A.~M. and {Sip{\H{o}}cz}, B.~M. and {G{\"u}nther}, H.~M. and {Lim}, P.~L. and {Crawford}, S.~M. and {Conseil}, S. and {Shupe}, D.~L. and {Craig}, M.~W. and {Dencheva}, N. and {Ginsburg}, A. and {VanderPlas}, J.~T. and {Bradley}, L.~D. and {P{\'e}rez-Su{\'a}rez}, D. and {de Val-Borro}, M. and {Aldcroft}, T.~L. and {Cruz}, K.~L. and {Robitaille}, T.~P. and {Tollerud}, E.~J. and {Ardelean}, C. and {Babej}, T. and {Bach}, Y.~P. and {Bachetti}, M. and {Bakanov}, A.~V. and {Bamford}, S.~P. and {Barentsen}, G. and {Barmby}, P. and {Baumbach}, A. and {Berry}, K.~L. and {Biscani}, F. and {Boquien}, M. and {Bostroem}, K.~A. and {Bouma}, L.~G. and {Brammer}, G.~B. and {Bray}, E.~M. and {Breytenbach}, H. and {Buddelmeijer}, H. and {Burke}, D.~J. and {Calderone}, G. and {Cano Rodr{\'\i}guez}, J.~L. and {Cara}, M. and {Cardoso}, J.~V.~M. and {Cheedella}, S. and {Copin}, Y. and {Corrales}, L. and {Crichton}, D. and {D'Avella}, D. and {Deil}, C. and {Depagne}, {\'E}. and {Dietrich}, J.~P. and {Donath}, A. and {Droettboom}, M. and {Earl}, N. and {Erben}, T. and {Fabbro}, S. and {Ferreira}, L.~A. and {Finethy}, T. and {Fox}, R.~T. and {Garrison}, L.~H. and {Gibbons}, S.~L.~J. and {Goldstein}, D.~A. and {Gommers}, R. and {Greco}, J.~P. and {Greenfield}, P. and {Groener}, A.~M. and {Grollier}, F. and {Hagen}, A. and {Hirst}, P. and {Homeier}, D. and {Horton}, A.~J. and {Hosseinzadeh}, G. and {Hu}, L. and {Hunkeler}, J.~S. and {Ivezi{\'c}}, {\v{Z}}. and {Jain}, A. and {Jenness}, T. and {Kanarek}, G. and {Kendrew}, S. and {Kern}, N.~S. and {Kerzendorf}, W.~E. and {Khvalko}, A. and {King}, J. and {Kirkby}, D. and {Kulkarni}, A.~M. and {Kumar}, A. and {Lee}, A. and {Lenz}, D. and {Littlefair}, S.~P. and {Ma}, Z. and {Macleod}, D.~M. and {Mastropietro}, M. and {McCully}, C. and {Montagnac}, S. and {Morris}, B.~M. and {Mueller}, M. and {Mumford}, S.~J. and {Muna}, D. and {Murphy}, N.~A. and {Nelson}, S. and {Nguyen}, G.~H. and {Ninan}, J.~P. and {N{\"o}the}, M. and {Ogaz}, S. and {Oh}, S. and {Parejko}, J.~K. and {Parley}, N. and {Pascual}, S. and {Patil}, R. and {Patil}, A.~A. and {Plunkett}, A.~L. and {Prochaska}, J.~X. and {Rastogi}, T. and {Reddy Janga}, V. and {Sabater}, J. and {Sakurikar}, P. and {Seifert}, M. and {Sherbert}, L.~E. and {Sherwood-Taylor}, H. and {Shih}, A.~Y. and {Sick}, J. and {Silbiger}, M.~T. and {Singanamalla}, S. and {Singer}, L.~P. and {Sladen}, P.~H. and {Sooley}, K.~A. and {Sornarajah}, S. and {Streicher}, O. and {Teuben}, P. and {Thomas}, S.~W. and {Tremblay}, G.~R. and {Turner}, J.~E.~H. and {Terr{\'o}n}, V. and {van Kerkwijk}, M.~H. and {de la Vega}, A. and {Watkins}, L.~L. and {Weaver}, B.~A. and {Whitmore}, J.~B. and {Woillez}, J. and {Zabalza}, V. and {Astropy Contributors}},
        title = "{The Astropy Project: Building an Open-science Project and Status of the v2.0 Core Package}",
      journal = {\aj},
         year = 2018,
        month = sep,
       volume = {156},
       number = {3},
          eid = {123},
        pages = {123},
          doi = {10.3847/1538-3881/aabc4f},
archivePrefix = {arXiv},
       eprint = {1801.02634},
 primaryClass = {astro-ph.IM},
       adsurl = {https://ui.adsabs.harvard.edu/abs/2018AJ....156..123A}
}

@ARTICLE{2013A&A...558A..33A,
       author = {{Astropy Collaboration} and {Robitaille}, Thomas P. and
         {Tollerud}, Erik J. and {Greenfield}, Perry and {Droettboom}, Michael and
         {Bray}, Erik and {Aldcroft}, Tom and {Davis}, Matt and
         {Ginsburg}, Adam and {Price-Whelan}, Adrian M. and
         {Kerzendorf}, Wolfgang E. and {Conley}, Alexander and {Crighton}, Neil and
         {Barbary}, Kyle and {Muna}, Demitri and {Ferguson}, Henry and
         {Grollier}, Fr{\'e}d{\'e}ric and {Parikh}, Madhura M. and
         {Nair}, Prasanth H. and {Unther}, Hans M. and {Deil}, Christoph and
         {Woillez}, Julien and {Conseil}, Simon and {Kramer}, Roban and
         {Turner}, James E.~H. and {Singer}, Leo and {Fox}, Ryan and
         {Weaver}, Benjamin A. and {Zabalza}, Victor and {Edwards}, Zachary I. and
         {Azalee Bostroem}, K. and {Burke}, D.~J. and {Casey}, Andrew R. and
         {Crawford}, Steven M. and {Dencheva}, Nadia and {Ely}, Justin and
         {Jenness}, Tim and {Labrie}, Kathleen and {Lim}, Pey Lian and
         {Pierfederici}, Francesco and {Pontzen}, Andrew and {Ptak}, Andy and
         {Refsdal}, Brian and {Servillat}, Mathieu and {Streicher}, Ole},
        title = "{Astropy: A community Python package for astronomy}",
      journal = {\aap},
         year = "2013",
        month = "Oct",
       volume = {558},
          eid = {A33},
        pages = {A33},
          doi = {10.1051/0004-6361/201322068},
archivePrefix = {arXiv},
       eprint = {1307.6212},
 primaryClass = {astro-ph.IM},
       adsurl = {https://ui.adsabs.harvard.edu/abs/2013A&A...558A..33A}
}

@ARTICLE{Durrell2014,
       author = {{Durrell}, Patrick R. and {C{\^o}t{\'e}}, Patrick and {Peng}, Eric W. and {Blakeslee}, John P. and {Ferrarese}, Laura and {Mihos}, J. Christopher and {Puzia}, Thomas H. and {Lan{\c{c}}on}, Ariane and {Liu}, Chengze and {Zhang}, Hongxin and {Cuillandre}, Jean-Charles and {McConnachie}, Alan and {Jord{\'a}n}, Andr{\'e}s and {Accetta}, Katharine and {Boissier}, Samuel and {Boselli}, Alessandro and {Courteau}, St{\'e}phane and {Duc}, Pierre-Alain and {Emsellem}, Eric and {Gwyn}, Stephen and {Mei}, Simona and {Taylor}, James E.},
        title = "{The Next Generation Virgo Cluster Survey. VIII. The Spatial Distribution of Globular Clusters in the Virgo Cluster}",
      journal = {\apj},
         year = 2014,
        month = oct,
       volume = {794},
       number = {2},
          eid = {103},
        pages = {103},
          doi = {10.1088/0004-637X/794/2/103},
archivePrefix = {arXiv},
       eprint = {1408.2821},
 primaryClass = {astro-ph.GA},
       adsurl = {https://ui.adsabs.harvard.edu/abs/2014ApJ...794..103D}
}

@ARTICLE{Lim2025,
       author = {{Lim}, Sungsoon and {Peng}, Eric W. and {C{\^o}t{\'e}}, Patrick and {Ferrarese}, Laura and {Roediger}, Joel C. and {Liu}, Chengze and {Spengler}, Chelsea and {Sola}, Elisabeth and {Duc}, Pierre-Alain and {Sales}, Laura V. and {Blakeslee}, John P. and {Cuillandre}, Jean-Charles and {Durrell}, Patrick R. and {Emsellem}, Eric and {Gwyn}, Stephen D.~J. and {Lan{\c{c}}on}, Ariane and {Marleau}, Francine R. and {Mihos}, J. Christopher and {M{\"u}ller}, Oliver and {Puzia}, Thomas H. and {S{\'a}nchez-Janssen}, Rub{\'e}n},
        title = "{The Spatial Distribution of Globular Cluster Systems in Early-type Galaxies: Estimation Procedure and Catalog of Properties for Globular Cluster Systems Observed with Deep Imaging Surveys}",
      journal = {\apjs},
         year = 2025,
        month = feb,
       volume = {276},
       number = {2},
          eid = {34},
        pages = {34},
          doi = {10.3847/1538-4365/ad97b7},
archivePrefix = {arXiv},
       eprint = {2411.17049},
 primaryClass = {astro-ph.GA},
       adsurl = {https://ui.adsabs.harvard.edu/abs/2025ApJS..276...34L}
}

@ARTICLE{Liu+20,
       author = {{Liu}, Chengze and {C{\^o}t{\'e}}, Patrick and {Peng}, Eric W. and {Roediger}, Joel and {Zhang}, Hongxin and {Ferrarese}, Laura and {S{\'a}nchez-Janssen}, Ruben and {Guhathakurta}, Puragra and {Yang}, Xiaohu and {Jing}, Yipeng and {Alamo-Mart{\'\i}nez}, Karla and {Blakeslee}, John P. and {Boselli}, Alessandro and {Cuilandre}, Jean-Charles and {Duc}, Pierre-Alain and {Durrell}, Patrick and {Gwyn}, Stephen and {Jord{\'a}n}, Andres and {Ko}, Youkyung and {Lan{\c{c}}on}, Ariane and {Lim}, Sungsoon and {Longobardi}, Alessia and {Mei}, Simona and {Mihos}, J. Christopher and {Mu{\~n}oz}, Roberto and {Powalka}, Mathieu and {Puzia}, Thomas and {Spengler}, Chelsea and {Toloba}, Elisa},
        title = "{The Next Generation Virgo Cluster Survey. XXXIV. Ultracompact Dwarf Galaxies in the Virgo Cluster}",
      journal = {\apjs},
         year = 2020,
        month = sep,
       volume = {250},
       number = {1},
          eid = {17},
        pages = {17},
          doi = {10.3847/1538-4365/abad91},
archivePrefix = {arXiv},
       eprint = {2007.15275},
 primaryClass = {astro-ph.GA},
       adsurl = {https://ui.adsabs.harvard.edu/abs/2020ApJS..250...17L}
}

@ARTICLE{Wang2023,
       author = {{Wang}, Kaixiang and {Peng}, Eric W. and {Liu}, Chengze and {Mihos}, J. Christopher and {C{\^o}t{\'e}}, Patrick and {Ferrarese}, Laura and {Taylor}, Matthew A. and {Blakeslee}, John P. and {Cuillandre}, Jean-Charles and {Duc}, Pierre-Alain and {Guhathakurta}, Puragra and {Gwyn}, Stephen and {Ko}, Youkyung and {Lan{\c{c}}on}, Ariane and {Lim}, Sungsoon and {MacArthur}, Lauren A. and {Puzia}, Thomas and {Roediger}, Joel and {Sales}, Laura V. and {S{\'a}nchez-Janssen}, Rub{\'e}n and {Spengler}, Chelsea and {Toloba}, Elisa and {Zhang}, Hongxin and {Zhu}, Mingcheng},
        title = "{An evolutionary continuum from nucleated dwarf galaxies to star clusters}",
      journal = {\nat},
         year = 2023,
        month = nov,
       volume = {623},
       number = {7986},
        pages = {296-300},
          doi = {10.1038/s41586-023-06650-z},
archivePrefix = {arXiv},
       eprint = {2311.05448},
 primaryClass = {astro-ph.GA},
       adsurl = {https://ui.adsabs.harvard.edu/abs/2023Natur.623..296W}
}

@ARTICLE{Mihos2005,
       author = {{Mihos}, J. Christopher and {Harding}, Paul and {Feldmeier}, John and {Morrison}, Heather},
        title = "{Diffuse Light in the Virgo Cluster}",
      journal = {\apjl},
         year = 2005,
        month = sep,
       volume = {631},
       number = {1},
        pages = {L41-L44},
          doi = {10.1086/497030},
archivePrefix = {arXiv},
       eprint = {astro-ph/0508217},
 primaryClass = {astro-ph},
       adsurl = {https://ui.adsabs.harvard.edu/abs/2005ApJ...631L..41M}
}

@ARTICLE{Mihos2015,
       author = {{Mihos}, J. Christopher and {Durrell}, Patrick R. and {Ferrarese}, Laura and {Feldmeier}, John J. and {C{\^o}t{\'e}}, Patrick and {Peng}, Eric W. and {Harding}, Paul and {Liu}, Chengze and {Gwyn}, Stephen and {Cuillandre}, Jean-Charles},
        title = "{Galaxies at the Extremes: Ultra-diffuse Galaxies in the Virgo Cluster}",
      journal = {\apjl},
         year = 2015,
        month = aug,
       volume = {809},
       number = {2},
          eid = {L21},
        pages = {L21},
          doi = {10.1088/2041-8205/809/2/L21},
archivePrefix = {arXiv},
       eprint = {1507.02270},
 primaryClass = {astro-ph.GA},
       adsurl = {https://ui.adsabs.harvard.edu/abs/2015ApJ...809L..21M}
}

@ARTICLE{Mihos2017,
       author = {{Mihos}, J. Christopher and {Harding}, Paul and {Feldmeier}, John J. and {Rudick}, Craig and {Janowiecki}, Steven and {Morrison}, Heather and {Slater}, Colin and {Watkins}, Aaron},
        title = "{The Burrell Schmidt Deep Virgo Survey: Tidal Debris, Galaxy Halos, and Diffuse Intracluster Light in the Virgo Cluster}",
      journal = {\apj},
         year = 2017,
        month = jan,
       volume = {834},
       number = {1},
          eid = {16},
        pages = {16},
          doi = {10.3847/1538-4357/834/1/16},
archivePrefix = {arXiv},
       eprint = {1611.04435},
 primaryClass = {astro-ph.GA},
       adsurl = {https://ui.adsabs.harvard.edu/abs/2017ApJ...834...16M}
}

@ARTICLE{Ferrarese2012,
       author = {{Ferrarese}, Laura and {C{\^o}t{\'e}}, Patrick and {Cuillandre}, Jean-Charles and {Gwyn}, S.~D.~J. and {Peng}, Eric W. and {MacArthur}, Lauren A. and {Duc}, Pierre-Alain and {Boselli}, A. and {Mei}, Simona and {Erben}, Thomas and {McConnachie}, Alan W. and {Durrell}, Patrick R. and {Mihos}, J. Christopher and {Jord{\'a}n}, Andr{\'e}s and {Lan{\c{c}}on}, Ariane and {Puzia}, Thomas H. and {Emsellem}, Eric and {Balogh}, Michael L. and {Blakeslee}, John P. and {van Waerbeke}, Ludovic and {Gavazzi}, Rapha{\"e}l and {Vollmer}, Bernd and {Kavelaars}, J.~J. and {Woods}, David and {Ball}, Nicholas M. and {Boissier}, S. and {Courteau}, St{\'e}phane and {Ferriere}, E. and {Gavazzi}, G. and {Hildebrandt}, Hendrik and {Hudelot}, P. and {Huertas-Company}, M. and {Liu}, Chengze and {McLaughlin}, Dean and {Mellier}, Y. and {Milkeraitis}, Martha and {Schade}, David and {Balkowski}, Chantal and {Bournaud}, Fr{\'e}d{\'e}ric and {Carlberg}, R.~G. and {Chapman}, S.~C. and {Hoekstra}, Henk and {Peng}, Chien and {Sawicki}, Marcin and {Simard}, Luc and {Taylor}, James E. and {Tully}, R. Brent and {van Driel}, Wim and {Wilson}, Christine D. and {Burdullis}, Todd and {Mahoney}, Billy and {Manset}, Nadine},
        title = "{The Next Generation Virgo Cluster Survey (NGVS). I. Introduction to the Survey}",
      journal = {\apjs},
         year = 2012,
        month = may,
       volume = {200},
       number = {1},
          eid = {4},
        pages = {4},
          doi = {10.1088/0067-0049/200/1/4},
       adsurl = {https://ui.adsabs.harvard.edu/abs/2012ApJS..200....4F}
}

@ARTICLE{Blakeslee2009,
       author = {{Blakeslee}, John P. and {Jord{\'a}n}, Andr{\'e}s and {Mei}, Simona and {C{\^o}t{\'e}}, Patrick and {Ferrarese}, Laura and {Infante}, Leopoldo and {Peng}, Eric W. and {Tonry}, John L. and {West}, Michael J.},
        title = "{The ACS Fornax Cluster Survey. V. Measurement and Recalibration of Surface Brightness Fluctuations and a Precise Value of the Fornax-Virgo Relative Distance}",
      journal = {\apj},
         year = 2009,
        month = mar,
       volume = {694},
       number = {1},
        pages = {556-572},
          doi = {10.1088/0004-637X/694/1/556},
archivePrefix = {arXiv},
       eprint = {0901.1138},
 primaryClass = {astro-ph.CO},
       adsurl = {https://ui.adsabs.harvard.edu/abs/2009ApJ...694..556B}
}

@article{Ferrarese2020,
doi = {10.3847/1538-4357/ab339f},
url = {https://dx.doi.org/10.3847/1538-4357/ab339f},
year = {2020},
month = {feb},
publisher = {The American Astronomical Society},
volume = {890},
number = {2},
pages = {128},
author = {Laura Ferrarese and Patrick Côté and Lauren A. MacArthur and Patrick R. Durrell and S. D. J. Gwyn and Pierre-Alain Duc and Rúben Sánchez-Janssen and Matthew Santos and John P. Blakeslee and Alessandro Boselli and Fred Boyer and Michele Cantiello and Stéphane Courteau and Jean-Charles Cuillandre and Eric Emsellem and Thomas Erben and Giuseppe Gavazzi and Puragra Guhathakurta and Marc Huertas-Company and Andrés Jordán and Ariane Lançon and Chengze Liu and Simona Mei and J. Christopher Mihos and Eric W. Peng and Thomas H. Puzia and Joel Roediger and David Schade and James E. Taylor and Elisa Toloba and Hongxin Zhang},
title = {The Next Generation Virgo Cluster Survey (NGVS). XIV. The Discovery of Low-mass Galaxies and a New Galaxy Catalog in the Core of the Virgo Cluster∗},
journal = {The Astrophysical Journal}
}

@ARTICLE{Elixir,
       author = {{Magnier}, E.~A. and {Cuillandre}, J. -C.},
        title = "{The Elixir System: Data Characterization and Calibration at the Canada-France-Hawaii Telescope}",
      journal = {\pasp},
         year = 2004,
        month = may,
       volume = {116},
       number = {819},
        pages = {449-464},
          doi = {10.1086/420756},
       adsurl = {https://ui.adsabs.harvard.edu/abs/2004PASP..116..449M}
}

@ARTICLE{Sanchez-Janssen2019,
       author = {{S{\'a}nchez-Janssen}, Rub{\'e}n and {C{\^o}t{\'e}}, Patrick and {Ferrarese}, Laura and {Peng}, Eric W. and {Roediger}, Joel and {Blakeslee}, John P. and {Emsellem}, Eric and {Puzia}, Thomas H. and {Spengler}, Chelsea and {Taylor}, James and {{\'A}lamo-Mart{\'\i}nez}, Karla A. and {Boselli}, Alessandro and {Cantiello}, Michele and {Cuillandre}, Jean-Charles and {Duc}, Pierre-Alain and {Durrell}, Patrick and {Gwyn}, Stephen and {MacArthur}, Lauren A. and {Lan{\c{c}}on}, Ariane and {Lim}, Sungsoon and {Liu}, Chengze and {Mei}, Simona and {Miller}, Bryan and {Mu{\~n}oz}, Roberto and {Mihos}, J. Christopher and {Paudel}, Sanjaya and {Powalka}, Mathieu and {Toloba}, Elisa},
        title = "{The Next Generation Virgo Cluster Survey. XXIII. Fundamentals of Nuclear Star Clusters over Seven Decades in Galaxy Mass}",
      journal = {\apj},
         year = 2019,
        month = jun,
       volume = {878},
       number = {1},
          eid = {18},
        pages = {18},
          doi = {10.3847/1538-4357/aaf4fd},
archivePrefix = {arXiv},
       eprint = {1812.01019},
 primaryClass = {astro-ph.GA},
       adsurl = {https://ui.adsabs.harvard.edu/abs/2019ApJ...878...18S}
}

@ARTICLE{MalinUnsharp,
       author = {{Malin}, D.~F.},
        title = "{Unsharp masking.}",
      journal = {AAS Photo Bulletin},
         year = 1977,
        month = jan,
       volume = {16},
        pages = {10-13},
       adsurl = {https://ui.adsabs.harvard.edu/abs/1977AASPB..16...10M}
}

@ARTICLE{Conroy2009,
       author = {{Conroy}, Charlie and {Gunn}, James E. and {White}, Martin},
        title = "{The Propagation of Uncertainties in Stellar Population Synthesis Modeling. I. The Relevance of Uncertain Aspects of Stellar Evolution and the Initial Mass Function to the Derived Physical Properties of Galaxies}",
      journal = {\apj},
         year = 2009,
        month = jul,
       volume = {699},
       number = {1},
        pages = {486-506},
          doi = {10.1088/0004-637X/699/1/486},
archivePrefix = {arXiv},
       eprint = {0809.4261},
 primaryClass = {astro-ph},
       adsurl = {https://ui.adsabs.harvard.edu/abs/2009ApJ...699..486C}
}

@ARTICLE{Su2022,
       author = {{Su}, Alan H. and {Salo}, Heikki and {Janz}, Joachim and {Venhola}, Aku and {Peletier}, Reynier F.},
        title = "{Photometric properties of nuclear star clusters and their host galaxies in the Fornax cluster}",
      journal = {\aap},
         year = 2022,
        month = aug,
       volume = {664},
          eid = {A167},
        pages = {A167},
          doi = {10.1051/0004-6361/202142593},
archivePrefix = {arXiv},
       eprint = {2206.01490},
 primaryClass = {astro-ph.GA},
       adsurl = {https://ui.adsabs.harvard.edu/abs/2022A&A...664A.167S}
}

@ARTICLE{Zanatta2021,
       author = {{Zanatta}, Em{\'\i}lio and {S{\'a}nchez-Janssen}, Rub{\'e}n and {Chies-Santos}, Ana L. and {de Souza}, Rafael S. and {Blakeslee}, John P.},
        title = "{A high occurrence of nuclear star clusters in faint Coma galaxies, and the roles of mass and environment}",
      journal = {\mnras},
         year = 2021,
        month = nov,
       volume = {508},
       number = {1},
        pages = {986-998},
          doi = {10.1093/mnras/stab2348},
archivePrefix = {arXiv},
       eprint = {2103.02123},
 primaryClass = {astro-ph.GA},
       adsurl = {https://ui.adsabs.harvard.edu/abs/2021MNRAS.508..986Z}
}

@ARTICLE{Hoyer2021,
       author = {{Hoyer}, Nils and {Neumayer}, Nadine and {Georgiev}, Iskren Y. and {Seth}, Anil C. and {Greene}, Jenny E.},
        title = "{The nucleation fraction of local volume galaxies}",
      journal = {\mnras},
         year = 2021,
        month = nov,
       volume = {507},
       number = {3},
        pages = {3246-3266},
          doi = {10.1093/mnras/stab2277},
archivePrefix = {arXiv},
       eprint = {2107.05313},
 primaryClass = {astro-ph.GA},
       adsurl = {https://ui.adsabs.harvard.edu/abs/2021MNRAS.507.3246H}
}

@ARTICLE{Ashok2023,
       author = {{Ashok}, Aishwarya and {Seth}, Anil and {Erwin}, Peter and {Debattista}, Victor P. and {de Lorenzo-C{\'a}ceres}, Adriana and {Gadotti}, Dmitri A. and {M{\'e}ndez-Abreu}, Jairo and {Beckman}, John E. and {Bender}, Ralf and {Drory}, Niv and {Fisher}, Deanne and {Hopp}, Ulrich and {Kluge}, Matthias and {Kolcu}, Tutku and {Maciejewski}, Witold and {Mehrgan}, Kianusch and {Parikh}, Taniya and {Saglia}, Roberto and {Seidel}, Marja and {Thomas}, Jens},
        title = "{Composite Bulges. III. A Study of Nuclear Star Clusters in Nearby Spiral Galaxies}",
      journal = {\apj},
         year = 2023,
        month = nov,
       volume = {958},
       number = {1},
          eid = {100},
        pages = {100},
          doi = {10.3847/1538-4357/ace341},
archivePrefix = {arXiv},
       eprint = {2308.03913},
 primaryClass = {astro-ph.GA},
       adsurl = {https://ui.adsabs.harvard.edu/abs/2023ApJ...958..100A}
}

@ARTICLE{Pechetti2020,
       author = {{Pechetti}, Renuka and {Seth}, Anil and {Neumayer}, Nadine and {Georgiev}, Iskren and {Kacharov}, Nikolay and {den Brok}, Mark},
        title = "{Luminosity Models and Density Profiles for Nuclear Star Clusters for a Nearby Volume-limited Sample of 29 Galaxies}",
      journal = {\apj},
         year = 2020,
        month = sep,
       volume = {900},
       number = {1},
          eid = {32},
        pages = {32},
          doi = {10.3847/1538-4357/abaaa7},
archivePrefix = {arXiv},
       eprint = {1911.09686},
 primaryClass = {astro-ph.GA},
       adsurl = {https://ui.adsabs.harvard.edu/abs/2020ApJ...900...32P}
}

@ARTICLE{Neumayer2020,
       author = {{Neumayer}, Nadine and {Seth}, Anil and {B{\"o}ker}, Torsten},
        title = "{Nuclear star clusters}",
      journal = {\aapr},
         year = 2020,
        month = jul,
       volume = {28},
       number = {1},
          eid = {4},
        pages = {4},
          doi = {10.1007/s00159-020-00125-0},
archivePrefix = {arXiv},
       eprint = {2001.03626},
 primaryClass = {astro-ph.GA},
       adsurl = {https://ui.adsabs.harvard.edu/abs/2020A&ARv..28....4N}
}

@ARTICLE{Marleau2021,
       author = {{Marleau}, Francine R. and {Habas}, Rebecca and {Poulain}, M{\'e}lina and {Duc}, Pierre-Alain and {M{\"u}ller}, Oliver and {Lim}, Sungsoon and {Durrell}, Patrick R. and {S{\'a}nchez-Janssen}, Rub{\'e}n and {Paudel}, Sanjaya and {Ahad}, Syeda Lammim and {Chougule}, Abhishek and {B{\'\i}lek}, Michal and {Fensch}, J{\'e}r{\'e}my},
        title = "{Ultra diffuse galaxies in the MATLAS low-to-moderate density fields}",
      journal = {\aap},
         year = 2021,
        month = oct,
       volume = {654},
          eid = {A105},
        pages = {A105},
          doi = {10.1051/0004-6361/202141432},
archivePrefix = {arXiv},
       eprint = {2109.13173},
 primaryClass = {astro-ph.GA},
       adsurl = {https://ui.adsabs.harvard.edu/abs/2021A&A...654A.105M}
}

@ARTICLE{Chabrier2003,
       author = {{Chabrier}, Gilles},
        title = "{Galactic Stellar and Substellar Initial Mass Function}",
      journal = {\pasp},
         year = 2003,
        month = jul,
       volume = {115},
       number = {809},
        pages = {763-795},
          doi = {10.1086/376392},
archivePrefix = {arXiv},
       eprint = {astro-ph/0304382},
 primaryClass = {astro-ph},
       adsurl = {https://ui.adsabs.harvard.edu/abs/2003PASP..115..763C}
}

@ARTICLE{Sanchez2019b,
       author = {{S{\'a}nchez-Janssen}, R. and {Puzia}, T.~H. and {Ferrarese}, L. and {C{\^o}t{\'e}}, P. and {Eigenthaler}, P. and {Miller}, B. and {Ordenes-Brice{\~n}o}, Y. and {Peng}, E.~W. and {Ribbeck}, K.~X. and {Roediger}, J. and {Spengler}, C. and {Taylor}, M.~A.},
        title = "{How nucleation and luminosity shape faint dwarf galaxies}",
      journal = {\mnras},
         year = 2019,
        month = jun,
       volume = {486},
       number = {1},
        pages = {L1-L5},
          doi = {10.1093/mnrasl/slz008},
archivePrefix = {arXiv},
       eprint = {1901.04509},
 primaryClass = {astro-ph.GA},
       adsurl = {https://ui.adsabs.harvard.edu/abs/2019MNRAS.486L...1S}
}

@ARTICLE{Ko2022,
       author = {{Ko}, Youkyung and {Peng}, Eric W. and {C{\^o}t{\'e}}, Patrick and {Ferrarese}, Laura and {Liu}, Chengze and {Longobardi}, Alessia and {Lan{\c{c}}on}, Ariane and {Mu{\~n}oz}, Roberto P. and {Puzia}, Thomas H. and {Alamo-Mart{\'\i}nez}, Karla A. and {Sales}, Laura V. and {Ramos-Almendares}, Felipe and {Abadi}, Mario G. and {Lee}, Myung Gyoon and {Hwang}, Ho Seong and {Caldwell}, Nelson and {Blakeslee}, John P. and {Boselli}, Alessandro and {Cuillandre}, Jean-Charles and {Duc}, Pierre-Alain and {Eyheramendy}, Susana and {Guhathakurta}, Puragra and {Gwyn}, Stephen and {Jord{\'a}n}, Andr{\'e}s and {Lim}, Sungsoon and {S{\'a}nchez-Janssen}, Rub{\'e}n and {Toloba}, Elisa},
        title = "{The Next Generation Virgo Cluster Survey. XXXIII. Stellar Population Gradients in the Virgo Cluster Core Globular Cluster System}",
      journal = {\apj},
         year = 2022,
        month = jun,
       volume = {931},
       number = {2},
          eid = {120},
        pages = {120},
          doi = {10.3847/1538-4357/ac63cf},
archivePrefix = {arXiv},
       eprint = {2204.05318},
 primaryClass = {astro-ph.GA},
       adsurl = {https://ui.adsabs.harvard.edu/abs/2022ApJ...931..120K}
}

@article{Cantiello2024,
doi = {10.3847/1538-4357/ad3453},
url = {https://dx.doi.org/10.3847/1538-4357/ad3453},
year = {2024},
month = {apr},
publisher = {The American Astronomical Society},
volume = {966},
number = {1},
pages = {145},
author = {Cantiello, Michele and Blakeslee, John P. and Ferrarese, Laura and Côté, Patrick and Raimondo, Gabriella and Cuillandre, Jean-Charles and Durrell, Patrick R. and Gwyn, Stephen and Hazra, Nandini and Peng, Eric W. and Roediger, Joel C. and Sánchez-Janssen, Rúben and Kurzner, Max},
title = {The Next Generation Virgo Cluster Survey (NGVS). III. A Catalog of Surface Brightness Fluctuation Distances and the Three-dimensional Distribution of Galaxies in the Virgo Cluster},
journal = {The Astrophysical Journal}
}

@ARTICLE{Hoffman1989,
       author = {{Hoffman}, G. Lyle and {Helou}, George and {Salpeter}, E.~E. and {Lewis}, B. Murray},
        title = "{Blue Compact Dwarf Galaxies in the Virgo Cluster: H i and IRAS Data and Upper Limits on Proto--Dwarf Galaxies}",
      journal = {\apj},
         year = 1989,
        month = apr,
       volume = {339},
        pages = {812},
          doi = {10.1086/167338},
       adsurl = {https://ui.adsabs.harvard.edu/abs/1989ApJ...339..812H}
}

@ARTICLE{Grossi2015,
       author = {{Grossi}, M. and {Hunt}, L.~K. and {Madden}, S.~C. and {Hughes}, T.~M. and {Auld}, R. and {Baes}, M. and {Bendo}, G.~J. and {Bianchi}, S. and {Bizzocchi}, L. and {Boquien}, M. and {Boselli}, A. and {Clemens}, M. and {Corbelli}, E. and {Cortese}, L. and {Davies}, J. and {De Looze}, I. and {di Serego Alighieri}, S. and {Fritz}, J. and {Pappalardo}, C. and {Pierini}, D. and {R{\'e}my-Ruyer}, A. and {Smith}, M.~W.~L. and {Verstappen}, J. and {Viaene}, S. and {Vlahakis}, C.},
        title = "{The Herschel Virgo Cluster Survey. XVIII. Star-forming dwarf galaxies in a cluster environment}",
      journal = {\aap},
         year = 2015,
        month = feb,
       volume = {574},
          eid = {A126},
        pages = {A126},
          doi = {10.1051/0004-6361/201424866},
archivePrefix = {arXiv},
       eprint = {1411.3960},
 primaryClass = {astro-ph.GA},
       adsurl = {https://ui.adsabs.harvard.edu/abs/2015A&A...574A.126G}
}

@ARTICLE{Schlegel1998,
       author = {{Schlegel}, David J. and {Finkbeiner}, Douglas P. and {Davis}, Marc},
        title = "{Maps of Dust Infrared Emission for Use in Estimation of Reddening and Cosmic Microwave Background Radiation Foregrounds}",
      journal = {\apj},
         year = 1998,
        month = jun,
       volume = {500},
       number = {2},
        pages = {525-553},
          doi = {10.1086/305772},
archivePrefix = {arXiv},
       eprint = {astro-ph/9710327},
 primaryClass = {astro-ph},
       adsurl = {https://ui.adsabs.harvard.edu/abs/1998ApJ...500..525S}
}

@ARTICLE{Lim2024,
       author = {{Lim}, Sungsoon and {Peng}, Eric W. and {C{\^o}t{\'e}}, Patrick and {Ferrarese}, Laura and {Roediger}, Joel C. and {Liu}, Chengze and {Spengler}, Chelsea and {Sola}, Elisabeth and {Duc}, Pierre-Alain and {Sales}, Laura V. and {Blakeslee}, John P. and {Cuillandre}, Jean-Charles and {Durrell}, Patrick R. and {Emsellem}, Eric and {Gwyn}, Stephen D.~J. and {Lan{\c{c}}on}, Ariane and {Marleau}, Francine R. and {Mihos}, J. Christopher and {M{\"u}ller}, Oliver and {Puzia}, Thomas H. and {S{\'a}nchez-Janssen}, Rub{\'e}n},
        title = "{The Next Generation Virgo Cluster Survey (NGVS). XXVII. The Size and Structure of Globular Cluster Systems and Their Connection to Dark Matter Halos}",
      journal = {\apj},
         year = 2024,
        month = may,
       volume = {966},
       number = {2},
          eid = {168},
        pages = {168},
          doi = {10.3847/1538-4357/ad3444},
archivePrefix = {arXiv},
       eprint = {2403.09926},
 primaryClass = {astro-ph.GA},
       adsurl = {https://ui.adsabs.harvard.edu/abs/2024ApJ...966..168L}
}

@article{Kurzner2025,
doi = {10.3847/1538-3881/ae1616},
url = {https://doi.org/10.3847/1538-3881/ae1616},
year = {2025},
month = {dec},
publisher = {The American Astronomical Society},
volume = {170},
number = {6},
pages = {363},
author = {Kurzner, Max M. and Côté, Patrick and Ferrarese, Laura and Wang, Kaixiang and Peng, Eric W. and Wilkinson, Scott and Roediger, Joel C. and Spengler, Chelsea and Brown, Toby and Liu, Chengze and Lim, Sungsoon and Sánchez-Janssen, Rubén and Toloba, Elisa and Guhathakurta, Puragra and Blakeslee, John P. and Durrell, Patrick R. and Lançon, Ariane and Mihos, J. Christopher and Taylor, Matthew A. and Woods, Tyrone E. and Thompson, Solveig and MacArthur, Lauren A.},
title = {The Next Generation Virgo Cluster Survey (NGVS). XL. The Morphological Classification of Virgo Cluster Galaxies},
journal = {The Astronomical Journal}
}

@ARTICLE{Carlsten2022a,
       author = {{Carlsten}, Scott G. and {Greene}, Jenny E. and {Beaton}, Rachael L. and {Danieli}, Shany and {Greco}, Johnny P.},
        title = "{The Exploration of Local VolumE Satellites (ELVES) Survey: A Nearly Volume-limited Sample of Nearby Dwarf Satellite Systems}",
      journal = {\apj},
         year = 2022,
        month = jul,
       volume = {933},
       number = {1},
          eid = {47},
        pages = {47},
          doi = {10.3847/1538-4357/ac6fd7},
archivePrefix = {arXiv},
       eprint = {2203.00014},
 primaryClass = {astro-ph.GA},
       adsurl = {https://ui.adsabs.harvard.edu/abs/2022ApJ...933...47C}
}

@ARTICLE{Carlsten2022b,
       author = {{Carlsten}, Scott G. and {Greene}, Jenny E. and {Beaton}, Rachael L. and {Greco}, Johnny P.},
        title = "{ELVES II: Globular Clusters and Nuclear Star Clusters of Dwarf Galaxies: the Importance of Environment}",
      journal = {\apj},
         year = 2022,
        month = mar,
       volume = {927},
       number = {1},
          eid = {44},
        pages = {44},
          doi = {10.3847/1538-4357/ac457e},
archivePrefix = {arXiv},
       eprint = {2105.03440},
 primaryClass = {astro-ph.GA},
       adsurl = {https://ui.adsabs.harvard.edu/abs/2022ApJ...927...44C}
}

@ARTICLE{Leaman2022,
       author = {{Leaman}, Ryan and {van de Ven}, Glenn},
        title = "{On the link between nuclear star cluster and globular cluster system mass, nucleation fraction, and environment}",
      journal = {\mnras},
         year = 2022,
        month = nov,
       volume = {516},
       number = {4},
        pages = {4691-4715},
          doi = {10.1093/mnras/stab1966},
archivePrefix = {arXiv},
       eprint = {2112.05573},
 primaryClass = {astro-ph.GA},
       adsurl = {https://ui.adsabs.harvard.edu/abs/2022MNRAS.516.4691L}
}

@ARTICLE{Fahrion2021,
       author = {{Fahrion}, K. and {Lyubenova}, M. and {van de Ven}, G. and {Hilker}, M. and {Leaman}, R. and {Falc{\'o}n-Barroso}, J. and {Bittner}, A. and {Coccato}, L. and {Corsini}, E.~M. and {Gadotti}, D.~A. and {Iodice}, E. and {McDermid}, R.~M. and {Mart{\'\i}n-Navarro}, I. and {Pinna}, F. and {Poci}, A. and {Sarzi}, M. and {de Zeeuw}, P.~T. and {Zhu}, L.},
        title = "{Diversity of nuclear star cluster formation mechanisms revealed by their star formation histories}",
      journal = {\aap},
         year = 2021,
        month = jun,
       volume = {650},
          eid = {A137},
        pages = {A137},
          doi = {10.1051/0004-6361/202140644},
archivePrefix = {arXiv},
       eprint = {2104.06412},
 primaryClass = {astro-ph.GA},
       adsurl = {https://ui.adsabs.harvard.edu/abs/2021A&A...650A.137F}
}

@article{Akaike1974,
  title={A new look at the statistical model identification},
  author={Akaike, Hirotugu},
  journal={IEEE transactions on automatic control},
  volume={19},
  number={6},
  pages={716--723},
  year={1974},
  publisher={Ieee}
}

@ARTICLE{Behroozi2013  ,
       author = {{Behroozi}, Peter S. and {Wechsler}, Risa H. and {Conroy}, Charlie},
        title = "{The Average Star Formation Histories of Galaxies in Dark Matter Halos from z = 0-8}",
      journal = {\apj},
         year = 2013,
        month = jun,
       volume = {770},
       number = {1},
          eid = {57},
        pages = {57},
          doi = {10.1088/0004-637X/770/1/57},
archivePrefix = {arXiv},
       eprint = {1207.6105},
 primaryClass = {astro-ph.CO},
       adsurl = {https://ui.adsabs.harvard.edu/abs/2013ApJ...770...57B}
}

@ARTICLE{Moster2013,
       author = {{Moster}, Benjamin P. and {Naab}, Thorsten and {White}, Simon D.~M.},
        title = "{Galactic star formation and accretion histories from matching galaxies to dark matter haloes}",
      journal = {\mnras},
         year = 2013,
        month = feb,
       volume = {428},
       number = {4},
        pages = {3121-3138},
          doi = {10.1093/mnras/sts261},
archivePrefix = {arXiv},
       eprint = {1205.5807},
 primaryClass = {astro-ph.CO},
       adsurl = {https://ui.adsabs.harvard.edu/abs/2013MNRAS.428.3121M}
}

@ARTICLE{CapuzzoDolcetta1993,
       author = {{Capuzzo-Dolcetta}, Roberto},
        title = "{The Evolution of the Globular Cluster System in a Triaxial Galaxy: Can a Galactic Nucleus Form by Globular Cluster Capture?}",
      journal = {\apj},
         year = 1993,
        month = oct,
       volume = {415},
        pages = {616},
          doi = {10.1086/173189},
archivePrefix = {arXiv},
       eprint = {astro-ph/9301006},
 primaryClass = {astro-ph},
       adsurl = {https://ui.adsabs.harvard.edu/abs/1993ApJ...415..616C}
}

@article{Graham2009,
  author = {Graham, A. W. and Spitler, L. R.},
  title = {Quantifying the Coexistence of Massive Black Holes and Dense Nuclear Star Clusters},
  journal = {Monthly Notices of the Royal Astronomical Society},
  year = {2009},
  volume = {397},
  pages = {2148-2162},
  doi = {10.1111/j.1365-2966.2009.15118.x},
  archivePrefix = {arXiv},
  eprint = {0907.5250}
}

@ARTICLE{Cappellari2016,
       author = {{Cappellari}, Michele},
        title = "{Structure and Kinematics of Early-Type Galaxies from Integral Field Spectroscopy}",
      journal = {\araa},
         year = 2016,
        month = sep,
       volume = {54},
        pages = {597-665},
          doi = {10.1146/annurev-astro-082214-122432},
archivePrefix = {arXiv},
       eprint = {1602.04267},
 primaryClass = {astro-ph.GA},
       adsurl = {https://ui.adsabs.harvard.edu/abs/2016ARA&A..54..597C}
}

@ARTICLE{Boselli2006,
       author = {{Boselli}, Alessandro and {Gavazzi}, Giuseppe},
        title = "{Environmental Effects on Late-Type Galaxies in Nearby Clusters}",
      journal = {\pasp},
         year = 2006,
        month = apr,
       volume = {118},
       number = {842},
        pages = {517-559},
          doi = {10.1086/500691},
archivePrefix = {arXiv},
       eprint = {astro-ph/0601108},
 primaryClass = {astro-ph},
       adsurl = {https://ui.adsabs.harvard.edu/abs/2006PASP..118..517B}
}

@ARTICLE{Zanatta2024,
       author = {{Zanatta}, Em{\'\i}lio and {S{\'a}nchez-Janssen}, Rub{\'e}n and {de Souza}, Rafael S. and {Chies-Santos}, Ana L. and {Blakeslee}, John P.},
        title = "{NSCs from groups to clusters: a catalogue of dwarf galaxies in the Shapley supercluster and the role of environment in galaxy nucleation}",
      journal = {\mnras},
         year = 2024,
        month = may,
       volume = {530},
       number = {3},
        pages = {2670-2687},
          doi = {10.1093/mnras/stae849},
archivePrefix = {arXiv},
       eprint = {2403.14847},
 primaryClass = {astro-ph.GA},
       adsurl = {https://ui.adsabs.harvard.edu/abs/2024MNRAS.530.2670Z}
}

@ARTICLE{Ulrich2017,
       author = {{Urich}, Linda and {Lisker}, Thorsten and {Janz}, Joachim and {van de Ven}, Glenn and {Leaman}, Ryan and {Boselli}, Alessandro and {Paudel}, Sanjaya and {Sybilska}, Agnieszka and {Peletier}, Reynier F. and {den Brok}, Mark and {Hensler}, Gerhard and {Toloba}, Elisa and {Falc{\'o}n-Barroso}, Jes{\'u}s and {Niemi}, Sami-Matias},
        title = "{Young, metal-enriched cores in early-type dwarf galaxies in the Virgo cluster based on colour gradients}",
      journal = {\aap},
         year = 2017,
        month = oct,
       volume = {606},
          eid = {A135},
        pages = {A135},
          doi = {10.1051/0004-6361/201730897},
archivePrefix = {arXiv},
       eprint = {1707.01228},
 primaryClass = {astro-ph.GA},
       adsurl = {https://ui.adsabs.harvard.edu/abs/2017A&A...606A.135U}
}

@ARTICLE{Cote:2007,
       author = {{C{\^o}t{\'e}}, Patrick and {Ferrarese}, Laura and {Jord{\'a}n}, Andr{\'e}s and {Blakeslee}, John P. and {Chen}, Chin-Wei and {Infante}, Leopoldo and {Merritt}, David and {Mei}, Simona and {Peng}, Eric W. and {Tonry}, John L. and {West}, Andrew A. and {West}, Michael J.},
        title = "{The ACS Fornax Cluster Survey. II. The Central Brightness Profiles of Early-Type Galaxies: A Characteristic Radius on Nuclear Scales and the Transition from Central Luminosity Deficit to Excess}",
      journal = {\apj},
         year = 2007,
        month = dec,
       volume = {671},
       number = {2},
        pages = {1456-1465},
          doi = {10.1086/522822},
archivePrefix = {arXiv},
       eprint = {0711.1358},
 primaryClass = {astro-ph},
       adsurl = {https://ui.adsabs.harvard.edu/abs/2007ApJ...671.1456C}
}

@ARTICLE{Binggeli2000,
       author = {{Binggeli}, B. and {Barazza}, F. and {Jerjen}, H.},
        title = "{Off-center nuclei in dwarf elliptical galaxies}",
      journal = {\aap},
         year = 2000,
        month = jul,
       volume = {359},
        pages = {447-456},
       adsurl = {https://ui.adsabs.harvard.edu/abs/2000A&A...359..447B}
}

@ARTICLE{Bekki2001,
       author = {{Bekki}, Kenji and {Couch}, Warrick J. and {Drinkwater}, Michael J.},
        title = "{Galaxy Threshing and the Formation of Ultracompact Dwarf Galaxies}",
      journal = {\apjl},
         year = 2001,
        month = may,
       volume = {552},
       number = {2},
        pages = {L105-L108},
          doi = {10.1086/320339},
archivePrefix = {arXiv},
       eprint = {astro-ph/0106402},
 primaryClass = {astro-ph},
       adsurl = {https://ui.adsabs.harvard.edu/abs/2001ApJ...552L.105B}
}

@ARTICLE{Postman1984,
       author = {{Postman}, M. and {Geller}, M.~J.},
        title = "{The morphology-density relation - The group connection.}",
      journal = {\apj},
         year = 1984,
        month = jun,
       volume = {281},
        pages = {95-99},
          doi = {10.1086/162078},
       adsurl = {https://ui.adsabs.harvard.edu/abs/1984ApJ...281...95P}
}

@ARTICLE{Whitemore1993,
       author = {{Whitmore}, Bradley C. and {Gilmore}, Diane M. and {Jones}, Christine},
        title = "{What Determines the Morphological Fractions in Clusters of Galaxies?}",
      journal = {\apj},
         year = 1993,
        month = apr,
       volume = {407},
        pages = {489},
          doi = {10.1086/172531},
       adsurl = {https://ui.adsabs.harvard.edu/abs/1993ApJ...407..489W}
}

@ARTICLE{Blanton2009,
       author = {{Blanton}, Michael R. and {Moustakas}, John},
        title = "{Physical Properties and Environments of Nearby Galaxies}",
      journal = {\araa},
         year = 2009,
        month = sep,
       volume = {47},
       number = {1},
        pages = {159-210},
          doi = {10.1146/annurev-astro-082708-101734},
archivePrefix = {arXiv},
       eprint = {0908.3017},
 primaryClass = {astro-ph.GA},
       adsurl = {https://ui.adsabs.harvard.edu/abs/2009ARA&A..47..159B}
}

@ARTICLE{Clark1954,
       author = {{Clark}, P.~J. and {Evans}, F.~C.},
        title = "{Distances to nearest neighbor as a measure of spatial relationships in populations}",
      journal = {Ecology},
         year = 1954,
        month = jan,
       volume = {35},
        pages = {445-453},
          doi = {10.2307/1931034},
       adsurl = {https://ui.adsabs.harvard.edu/abs/1954Ecol...35..445C}
}

@ARTICLE{Bekki+03,
       author = {{Bekki}, K. and {Couch}, W.~J. and {Drinkwater}, M.~J. and {Shioya}, Y.},
        title = "{Galaxy threshing and the origin of ultra-compact dwarf galaxies in the Fornax cluster}",
      journal = {\mnras},
         year = 2003,
        month = sep,
       volume = {344},
       number = {2},
        pages = {399-411},
          doi = {10.1046/j.1365-8711.2003.06916.x},
archivePrefix = {arXiv},
       eprint = {astro-ph/0308243},
 primaryClass = {astro-ph},
       adsurl = {https://ui.adsabs.harvard.edu/abs/2003MNRAS.344..399B}
}

@ARTICLE{Drinkwater+00,
       author = {{Drinkwater}, M.~J. and {Jones}, J.~B. and {Gregg}, M.~D. and {Phillipps}, S.},
        title = "{Compact Stellar Systems in the Fornax Cluster: Super-massive Star Clusters or Extremely Compact Dwarf Galaxies?}",
      journal = {\pasa},
         year = 2000,
        month = dec,
       volume = {17},
       number = {3},
        pages = {227-233},
          doi = {10.1071/AS00034},
archivePrefix = {arXiv},
       eprint = {astro-ph/0002003},
 primaryClass = {astro-ph},
       adsurl = {https://ui.adsabs.harvard.edu/abs/2000PASA...17..227D}
}

@ARTICLE{Hilker+99,
       author = {{Hilker}, M. and {Infante}, L. and {Vieira}, G. and {Kissler-Patig}, M. and {Richtler}, T.},
        title = "{The central region of the Fornax cluster. II. Spectroscopy and radial velocities of member and background galaxies}",
      journal = {\aaps},
         year = 1999,
        month = jan,
       volume = {134},
        pages = {75-86},
          doi = {10.1051/aas:1999434},
archivePrefix = {arXiv},
       eprint = {astro-ph/9807144},
 primaryClass = {astro-ph},
       adsurl = {https://ui.adsabs.harvard.edu/abs/1999A&AS..134...75H}
}

@ARTICLE{Taylor+25,
       author = {{Taylor}, Matthew A. and {Tahmasebzadeh}, Behzad and {Thompson}, Solveig and {Vasiliev}, Eugene and {Valluri}, Monica and {Drinkwater}, Michael J. and {C{\^o}t{\'e}}, Patrick and {Ferrarese}, Laura and {Roediger}, Joel and {Baumgardt}, Holger and {Bentz}, Misty C. and {Dage}, Kristen and {Peng}, Eric W. and {Lapeer}, Drew and {Liu}, Chengze and {Sumners}, Zach and {Wang}, Kaixiang and {Baldassare}, Vivienne and {Blakeslee}, John P. and {Ko}, Youkyung and {Woods}, Tyrone E.},
        title = "{A Supermassive Black Hole in a Diminutive Ultracompact Dwarf Galaxy Discovered with JWST/NIRSpec+IFU}",
      journal = {\apjl},
         year = 2025,
        month = sep,
       volume = {991},
       number = {1},
          eid = {L24},
        pages = {L24},
          doi = {10.3847/2041-8213/ae028e},
archivePrefix = {arXiv},
       eprint = {2503.00113},
 primaryClass = {astro-ph.GA},
       adsurl = {https://ui.adsabs.harvard.edu/abs/2025ApJ...991L..24T}
}

@ARTICLE{Lisker+06,
       author = {{Lisker}, Thorsten and {Glatt}, Katharina and {Westera}, Pieter and {Grebel}, Eva K.},
        title = "{Virgo Cluster Early-Type Dwarf Galaxies with the Sloan Digital Sky Survey. II. Early-Type Dwarfs with Central Star Formation}",
      journal = {\aj},
         year = 2006,
        month = dec,
       volume = {132},
       number = {6},
        pages = {2432-2452},
          doi = {10.1086/508414},
archivePrefix = {arXiv},
       eprint = {astro-ph/0608294},
 primaryClass = {astro-ph},
       adsurl = {https://ui.adsabs.harvard.edu/abs/2006AJ....132.2432L}
}

@ARTICLE{Mayes2021b,
       author = {{Mayes}, Rebecca J. and {Drinkwater}, Michael J. and {Pfeffer}, Joel and {Baumgardt}, Holger and {Liu}, Chengze and {Ferrarese}, Laura and {C{\^o}t{\'e}}, Patrick and {Peng}, Eric W.},
        title = "{Contribution of stripped nuclei to the ultracompact dwarf galaxy population in the Virgo cluster}",
      journal = {\mnras},
         year = 2021,
        month = feb,
       volume = {501},
       number = {2},
        pages = {1852-1867},
          doi = {10.1093/mnras/staa3731},
archivePrefix = {arXiv},
       eprint = {2107.03555},
 primaryClass = {astro-ph.GA},
       adsurl = {https://ui.adsabs.harvard.edu/abs/2021MNRAS.501.1852M}
}

@ARTICLE{Cappellari2013b,
       author = {{Cappellari}, Michele and {Scott}, Nicholas and {Alatalo}, Katherine and {Blitz}, Leo and {Bois}, Maxime and {Bournaud}, Fr{\'e}d{\'e}ric and {Bureau}, M. and {Crocker}, Alison F. and {Davies}, Roger L. and {Davis}, Timothy A. and {de Zeeuw}, P.~T. and {Duc}, Pierre-Alain and {Emsellem}, Eric and {Khochfar}, Sadegh and {Krajnovi{\'c}}, Davor and {Kuntschner}, Harald and {McDermid}, Richard M. and {Morganti}, Raffaella and {Naab}, Thorsten and {Oosterloo}, Tom and {Sarzi}, Marc and {Serra}, Paolo and {Weijmans}, Anne-Marie and {Young}, Lisa M.},
        title = "{The ATLAS$^{3D}$ project - XV. Benchmark for early-type galaxies scaling relations from 260 dynamical models: mass-to-light ratio, dark matter, Fundamental Plane and Mass Plane}",
      journal = {\mnras},
         year = 2013,
        month = jul,
       volume = {432},
       number = {3},
        pages = {1709-1741},
          doi = {10.1093/mnras/stt562},
archivePrefix = {arXiv},
       eprint = {1208.3522},
 primaryClass = {astro-ph.CO},
       adsurl = {https://ui.adsabs.harvard.edu/abs/2013MNRAS.432.1709C}
}

@ARTICLE{Cappellari2013a,
       author = {{Cappellari}, Michele and {McDermid}, Richard M. and {Alatalo}, Katherine and {Blitz}, Leo and {Bois}, Maxime and {Bournaud}, Fr{\'e}d{\'e}ric and {Bureau}, M. and {Crocker}, Alison F. and {Davies}, Roger L. and {Davis}, Timothy A. and {de Zeeuw}, P.~T. and {Duc}, Pierre-Alain and {Emsellem}, Eric and {Khochfar}, Sadegh and {Krajnovi{\'c}}, Davor and {Kuntschner}, Harald and {Morganti}, Raffaella and {Naab}, Thorsten and {Oosterloo}, Tom and {Sarzi}, Marc and {Scott}, Nicholas and {Serra}, Paolo and {Weijmans}, Anne-Marie and {Young}, Lisa M.},
        title = "{The ATLAS$^{3D}$ project - XX. Mass-size and mass-{\ensuremath{\sigma}} distributions of early-type galaxies: bulge fraction drives kinematics, mass-to-light ratio, molecular gas fraction and stellar initial mass function}",
      journal = {\mnras},
         year = 2013,
        month = jul,
       volume = {432},
       number = {3},
        pages = {1862-1893},
          doi = {10.1093/mnras/stt644},
archivePrefix = {arXiv},
       eprint = {1208.3523},
 primaryClass = {astro-ph.CO},
       adsurl = {https://ui.adsabs.harvard.edu/abs/2013MNRAS.432.1862C}
}

@ARTICLE{Cappellari2006,
       author = {{Cappellari}, Michele and {Bacon}, R. and {Bureau}, M. and {Damen}, M.~C. and {Davies}, Roger L. and {de Zeeuw}, P.~T. and {Emsellem}, Eric and {Falc{\'o}n-Barroso}, Jes{\'u}s and {Krajnovi{\'c}}, Davor and {Kuntschner}, Harald and {McDermid}, Richard M. and {Peletier}, Reynier F. and {Sarzi}, Marc and {van den Bosch}, Remco C.~E. and {van de Ven}, Glenn},
        title = "{The SAURON project - IV. The mass-to-light ratio, the virial mass estimator and the Fundamental Plane of elliptical and lenticular galaxies}",
      journal = {\mnras},
         year = 2006,
        month = mar,
       volume = {366},
       number = {4},
        pages = {1126-1150},
          doi = {10.1111/j.1365-2966.2005.09981.x},
archivePrefix = {arXiv},
       eprint = {astro-ph/0505042},
 primaryClass = {astro-ph},
       adsurl = {https://ui.adsabs.harvard.edu/abs/2006MNRAS.366.1126C}
}

@ARTICLE{Bertin2002,
       author = {{Bertin}, G. and {Ciotti}, L. and {Del Principe}, M.},
        title = "{Weak homology of elliptical galaxies.}",
      journal = {\aap},
         year = 2002,
        month = apr,
       volume = {386},
        pages = {149-168},
          doi = {10.1051/0004-6361:20020248},
archivePrefix = {arXiv},
       eprint = {astro-ph/0202208},
 primaryClass = {astro-ph},
       adsurl = {https://ui.adsabs.harvard.edu/abs/2002A&A...386..149B}
}

@ARTICLE{DornanHarris2026,
       author = {{Dornan}, Veronika and {Harris}, William E.},
        title = "{Globular Cluster Systems in Dwarf Galaxies: Catalogs and Comparisons}",
      journal = {\apj},
         year = 2026,
        month = feb,
       volume = {998},
       number = {1},
          eid = {41},
        pages = {41},
          doi = {10.3847/1538-4357/ae2742},
archivePrefix = {arXiv},
       eprint = {2512.08453},
 primaryClass = {astro-ph.GA},
       adsurl = {https://ui.adsabs.harvard.edu/abs/2026ApJ...998...41D}
}

@ARTICLE{Faber1997,
       author = {{Faber}, S.~M. and {Tremaine}, Scott and {Ajhar}, Edward A. and {Byun}, Yong-Ik and {Dressler}, Alan and {Gebhardt}, Karl and {Grillmair}, Carl and {Kormendy}, John and {Lauer}, Tod R. and {Richstone}, Douglas},
        title = "{The Centers of Early-Type Galaxies with HST. IV. Central Parameter Relations.}",
      journal = {\aj},
         year = 1997,
        month = nov,
       volume = {114},
        pages = {1771},
          doi = {10.1086/118606},
archivePrefix = {arXiv},
       eprint = {astro-ph/9610055},
 primaryClass = {astro-ph},
       adsurl = {https://ui.adsabs.harvard.edu/abs/1997AJ....114.1771F}
}

@ARTICLE{Cappellari+2011,
       author = {{Cappellari}, Michele and {Emsellem}, Eric and {Krajnovi{\'c}}, Davor and {McDermid}, Richard M. and {Scott}, Nicholas and {Verdoes Kleijn}, G.~A. and {Young}, Lisa M. and {Alatalo}, Katherine and {Bacon}, R. and {Blitz}, Leo and {Bois}, Maxime and {Bournaud}, Fr{\'e}d{\'e}ric and {Bureau}, M. and {Davies}, Roger L. and {Davis}, Timothy A. and {de Zeeuw}, P.~T. and {Duc}, Pierre-Alain and {Khochfar}, Sadegh and {Kuntschner}, Harald and {Lablanche}, Pierre-Yves and {Morganti}, Raffaella and {Naab}, Thorsten and {Oosterloo}, Tom and {Sarzi}, Marc and {Serra}, Paolo and {Weijmans}, Anne-Marie},
        title = "{The ATLAS$^{3D}$ project - I. A volume-limited sample of 260 nearby early-type galaxies: science goals and selection criteria}",
      journal = {\mnras},
         year = 2011,
        month = may,
       volume = {413},
       number = {2},
        pages = {813-836},
          doi = {10.1111/j.1365-2966.2010.18174.x},
archivePrefix = {arXiv},
       eprint = {1012.1551},
 primaryClass = {astro-ph.CO},
       adsurl = {https://ui.adsabs.harvard.edu/abs/2011MNRAS.413..813C}
}

@ARTICLE{Milosavljevic+2001, 
       author = {{Milosavljevi{\'c}}, Milo{\v{s}} and {Merritt}, David},
        title = "{Formation of Galactic Nuclei}",
      journal = {\apj},
         year = 2001,
        month = dec,
       volume = {563},
       number = {1},
        pages = {34-62},
          doi = {10.1086/323830},
archivePrefix = {arXiv},
       eprint = {astro-ph/0103350},
 primaryClass = {astro-ph},
       adsurl = {https://ui.adsabs.harvard.edu/abs/2001ApJ...563...34M}
}

@ARTICLE{Graham2004,
       author = {{Graham}, Alister W.},
        title = "{Core Depletion from Coalescing Supermassive Black Holes}",
      journal = {\apjl},
         year = 2004,
        month = sep,
       volume = {613},
       number = {1},
        pages = {L33-L36},
          doi = {10.1086/424928},
archivePrefix = {arXiv},
       eprint = {astro-ph/0503177},
 primaryClass = {astro-ph},
       adsurl = {https://ui.adsabs.harvard.edu/abs/2004ApJ...613L..33G}
}

@ARTICLE{Paudel2010,
       author = {{Paudel}, Sanjaya and {Lisker}, Thorsten and {Kuntschner}, Harald and {Grebel}, Eva K. and {Glatt}, Katharina},
        title = "{Stellar populations of Virgo cluster early-type dwarf galaxies with and without discs: a dichotomy in age?}",
      journal = {\mnras},
         year = 2010,
        month = jun,
       volume = {405},
       number = {2},
        pages = {800-820},
          doi = {10.1111/j.1365-2966.2010.16507.x},
archivePrefix = {arXiv},
       eprint = {1002.2301},
 primaryClass = {astro-ph.CO},
       adsurl = {https://ui.adsabs.harvard.edu/abs/2010MNRAS.405..800P}
}

@ARTICLE{Burstein+1997,
       author = {{Burstein}, David and {Bender}, Ralf and {Faber}, S. and {Nolthenius}, R.},
        title = "{Global Relationships Among the Physical Properties of Stellar Systems.}",
      journal = {\aj},
         year = 1997,
        month = oct,
       volume = {114},
        pages = {1365},
          doi = {10.1086/118570},
archivePrefix = {arXiv},
       eprint = {astro-ph/9707037},
 primaryClass = {astro-ph},
       adsurl = {https://ui.adsabs.harvard.edu/abs/1997AJ....114.1365B}
}

@ARTICLE{Gray+2025,
       author = {{Gray}, Emily I. and {Read}, Justin I. and {Taylor}, Ethan and {Orkney}, Matthew D.~A. and {Rey}, Martin P. and {Yates}, Robert M. and {Kim}, Stacy Y. and {No{\"e}l}, Noelia E.~D. and {Agertz}, Oscar and {Andersson}, Eric P. and {Pontzen}, Andrew},
        title = "{EDGE: a new model for nuclear star cluster formation in dwarf galaxies}",
      journal = {\mnras},
         year = 2025,
        month = may,
       volume = {539},
       number = {2},
        pages = {1167-1179},
          doi = {10.1093/mnras/staf521},
archivePrefix = {arXiv},
       eprint = {2405.19286},
 primaryClass = {astro-ph.GA},
       adsurl = {https://ui.adsabs.harvard.edu/abs/2025MNRAS.539.1167G}
}

@ARTICLE{Chandrasekhar1943,
       author = {{Chandrasekhar}, S.},
        title = "{Dynamical Friction. I. General Considerations: the Coefficient of Dynamical Friction.}",
      journal = {\apj},
         year = 1943,
        month = mar,
       volume = {97},
        pages = {255},
          doi = {10.1086/144517},
       adsurl = {https://ui.adsabs.harvard.edu/abs/1943ApJ....97..255C}
}

@ARTICLE{Jo2026,
       author = {{Jo}, Yongseok and {Jung}, Minyong and {Bryan}, Greg L. and {Kim}, Seoyoung and {Kim}, Ji-hoon and {Lee}, Ahram},
        title = "{Evolution of Nuclear Star Cluster in Dwarf Galaxy through Mergers and In-Situ Star Formation}",
      journal = {arXiv e-prints},
         year = 2026,
        month = mar,
          eid = {arXiv:2603.18242},
        pages = {arXiv:2603.18242},
          doi = {10.48550/arXiv.2603.18242},
archivePrefix = {arXiv},
       eprint = {2603.18242},
 primaryClass = {astro-ph.GA},
       adsurl = {https://ui.adsabs.harvard.edu/abs/2026arXiv260318242J}
}

@ARTICLE{Michea+2021,
       author = {{Michea}, Josefina and {Pasquali}, Anna and {Smith}, Rory and {Kraljic}, Katarina and {Grebel}, Eva K. and {Calder{\'o}n-Castillo}, Paula and {Lisker}, Thorsten},
        title = "{Brought to Light. I. Quantification of Disk Substructure in Dwarf Early-type Galaxies}",
      journal = {\aj},
         year = 2021,
        month = jun,
       volume = {161},
       number = {6},
          eid = {268},
        pages = {268},
          doi = {10.3847/1538-3881/abf24b},
archivePrefix = {arXiv},
       eprint = {2103.16579},
 primaryClass = {astro-ph.GA},
       adsurl = {https://ui.adsabs.harvard.edu/abs/2021AJ....161..268M}
}

@ARTICLE{Forbes2018,
       author = {{Forbes}, Duncan A. and {Bastian}, Nate and {Gieles}, Mark and {Crain}, Robert A. and {Kruijssen}, J.~M. Diederik and {Larsen}, S{\o}ren S. and {Ploeckinger}, Sylvia and {Agertz}, Oscar and {Trenti}, Michele and {Ferguson}, Annette M.~N. and {Pfeffer}, Joel and {Gnedin}, Oleg Y.},
        title = "{Globular cluster formation and evolution in the context of cosmological galaxy assembly: open questions}",
      journal = {Proceedings of the Royal Society of London Series A},
         year = 2018,
        month = feb,
       volume = {474},
       number = {2210},
          eid = {20170616},
        pages = {20170616},
          doi = {10.1098/rspa.2017.0616},
archivePrefix = {arXiv},
       eprint = {1801.05818},
 primaryClass = {astro-ph.GA},
       adsurl = {https://ui.adsabs.harvard.edu/abs/2018RSPSA.47470616F}
}

@ARTICLE{Harris2013,
       author = {{Harris}, William E. and {Harris}, Gretchen L.~H. and {Alessi}, Matthew},
        title = "{A Catalog of Globular Cluster Systems: What Determines the Size of a Galaxy's Globular Cluster Population?}",
      journal = {\apj},
         year = 2013,
        month = aug,
       volume = {772},
       number = {2},
          eid = {82},
        pages = {82},
          doi = {10.1088/0004-637X/772/2/82},
archivePrefix = {arXiv},
       eprint = {1306.2247},
 primaryClass = {astro-ph.GA},
       adsurl = {https://ui.adsabs.harvard.edu/abs/2013ApJ...772...82H}
}

@ARTICLE{Hudson2014,
       author = {{Hudson}, Michael J. and {Harris}, Gretchen L. and {Harris}, William E.},
        title = "{Dark Matter Halos in Galaxies and Globular Cluster Populations}",
      journal = {\apjl},
         year = 2014,
        month = may,
       volume = {787},
       number = {1},
          eid = {L5},
        pages = {L5},
          doi = {10.1088/2041-8205/787/1/L5},
archivePrefix = {arXiv},
       eprint = {1404.1920},
 primaryClass = {astro-ph.GA},
       adsurl = {https://ui.adsabs.harvard.edu/abs/2014ApJ...787L...5H}
}

@ARTICLE{Kormendy2004,
       author = {{Kormendy}, John and {Kennicutt}, Jr., Robert C.},
        title = "{Secular Evolution and the Formation of Pseudobulges in Disk Galaxies}",
      journal = {\araa},
         year = 2004,
        month = sep,
       volume = {42},
       number = {1},
        pages = {603-683},
          doi = {10.1146/annurev.astro.42.053102.134024},
archivePrefix = {arXiv},
       eprint = {astro-ph/0407343},
 primaryClass = {astro-ph},
       adsurl = {https://ui.adsabs.harvard.edu/abs/2004ARA&A..42..603K}
}

@ARTICLE{Fisher2010,
       author = {{Fisher}, David B. and {Drory}, Niv},
        title = "{Bulges of Nearby Galaxies with Spitzer: Scaling Relations in Pseudobulges and Classical Bulges}",
      journal = {\apj},
         year = 2010,
        month = jun,
       volume = {716},
       number = {2},
        pages = {942-969},
          doi = {10.1088/0004-637X/716/2/942},
archivePrefix = {arXiv},
       eprint = {1004.5393},
 primaryClass = {astro-ph.CO},
       adsurl = {https://ui.adsabs.harvard.edu/abs/2010ApJ...716..942F}
}

@ARTICLE{Coelho2011,
       author = {{Coelho}, P. and {Gadotti}, D.~A.},
        title = "{Bars Rejuvenating Bulges? Evidence from Stellar Population Analysis}",
      journal = {\apjl},
         year = 2011,
        month = dec,
       volume = {743},
       number = {1},
          eid = {L13},
        pages = {L13},
          doi = {10.1088/2041-8205/743/1/L13},
archivePrefix = {arXiv},
       eprint = {1111.1736},
 primaryClass = {astro-ph.CO},
       adsurl = {https://ui.adsabs.harvard.edu/abs/2011ApJ...743L..13C}
}

@ARTICLE{Ahn2017,
       author = {{Ahn}, Christopher P. and {Seth}, Anil C. and {den Brok}, Mark and {Strader}, Jay and {Baumgardt}, Holger and {van den Bosch}, Remco and {Chilingarian}, Igor and {Frank}, Matthias and {Hilker}, Michael and {McDermid}, Richard and {Mieske}, Steffen and {Romanowsky}, Aaron J. and {Spitler}, Lee and {Brodie}, Jean and {Neumayer}, Nadine and {Walsh}, Jonelle L.},
        title = "{Detection of Supermassive Black Holes in Two Virgo Ultracompact Dwarf Galaxies}",
      journal = {\apj},
         year = 2017,
        month = apr,
       volume = {839},
       number = {2},
          eid = {72},
        pages = {72},
          doi = {10.3847/1538-4357/aa6972},
archivePrefix = {arXiv},
       eprint = {1703.09221},
 primaryClass = {astro-ph.GA},
       adsurl = {https://ui.adsabs.harvard.edu/abs/2017ApJ...839...72A}
}

@ARTICLE{Ahn2018,
       author = {{Ahn}, Christopher P. and {Seth}, Anil C. and {Cappellari}, Michele and {Krajnovi{\'c}}, Davor and {Strader}, Jay and {Voggel}, Karina T. and {Walsh}, Jonelle L. and {Bahramian}, Arash and {Baumgardt}, Holger and {Brodie}, Jean and {Chilingarian}, Igor and {Chomiuk}, Laura and {den Brok}, Mark and {Frank}, Matthias and {Hilker}, Michael and {McDermid}, Richard M. and {Mieske}, Steffen and {Neumayer}, Nadine and {Nguyen}, Dieu D. and {Pechetti}, Renuka and {Romanowsky}, Aaron J. and {Spitler}, Lee},
        title = "{The Black Hole in the Most Massive Ultracompact Dwarf Galaxy M59-UCD3}",
      journal = {\apj},
         year = 2018,
        month = may,
       volume = {858},
       number = {2},
          eid = {102},
        pages = {102},
          doi = {10.3847/1538-4357/aabc57},
archivePrefix = {arXiv},
       eprint = {1804.02399},
 primaryClass = {astro-ph.GA},
       adsurl = {https://ui.adsabs.harvard.edu/abs/2018ApJ...858..102A}
}

@ARTICLE{scipy,
  author  = {Virtanen, Pauli and Gommers, Ralf and Oliphant, Travis E. and
            Haberland, Matt and Reddy, Tyler and Cournapeau, David and
            Burovski, Evgeni and Peterson, Pearu and Weckesser, Warren and
            Bright, Jonathan and {van der Walt}, St{\'e}fan J. and
            Brett, Matthew and Wilson, Joshua and Millman, K. Jarrod and
            Mayorov, Nikolay and Nelson, Andrew R. J. and Jones, Eric and
            Kern, Robert and Larson, Eric and Carey, C J and
            Polat, {\.I}lhan and Feng, Yu and Moore, Eric W. and
            {VanderPlas}, Jake and Laxalde, Denis and Perktold, Josef and
            Cimrman, Robert and Henriksen, Ian and Quintero, E. A. and
            Harris, Charles R. and Archibald, Anne M. and
            Ribeiro, Ant{\^o}nio H. and Pedregosa, Fabian and
            {van Mulbregt}, Paul and {SciPy 1.0 Contributors}},
  title   = {{{SciPy} 1.0: Fundamental Algorithms for Scientific
            Computing in Python}},
  journal = {Nature Methods},
  year    = {2020},
  volume  = {17},
  pages   = {261--272},
  adsurl  = {https://rdcu.be/b08Wh},
  doi     = {10.1038/s41592-019-0686-2},
}

@ARTICLE{Matplotlib,
       author = {{Hunter}, John D.},
        title = "{Matplotlib: A 2D Graphics Environment}",
      journal = {Computing in Science and Engineering},
         year = 2007,
        month = may,
       volume = {9},
       number = {3},
        pages = {90-95},
          doi = {10.1109/MCSE.2007.55},
       adsurl = {https://ui.adsabs.harvard.edu/abs/2007CSE.....9...90H}
}

@ARTICLE{numpy,
       author = {{Harris}, Charles R. and {Millman}, K. Jarrod and {van der Walt}, St{\'e}fan J. and {Gommers}, Ralf and {Virtanen}, Pauli and {Cournapeau}, David and {Wieser}, Eric and {Taylor}, Julian and {Berg}, Sebastian and {Smith}, Nathaniel J. and {Kern}, Robert and {Picus}, Matti and {Hoyer}, Stephan and {van Kerkwijk}, Marten H. and {Brett}, Matthew and {Haldane}, Allan and {del R{\'\i}o}, Jaime Fern{\'a}ndez and {Wiebe}, Mark and {Peterson}, Pearu and {G{\'e}rard-Marchant}, Pierre and {Sheppard}, Kevin and {Reddy}, Tyler and {Weckesser}, Warren and {Abbasi}, Hameer and {Gohlke}, Christoph and {Oliphant}, Travis E.},
        title = "{Array programming with NumPy}",
      journal = {\nat},
         year = 2020,
        month = sep,
       volume = {585},
       number = {7825},
        pages = {357-362},
          doi = {10.1038/s41586-020-2649-2},
archivePrefix = {arXiv},
       eprint = {2006.10256},
 primaryClass = {cs.MS},
       adsurl = {https://ui.adsabs.harvard.edu/abs/2020Natur.585..357H}
}

@misc{Reback2020,
  author       = {Reback, J. and McKinney, W. and others},
  title        = {pandas-dev/pandas: Pandas},
  year         = {2020},
  howpublished = {\url{https://pandas.pydata.org/}},
  doi          = {10.5281/zenodo.3509134},
  note         = {Version 1.x}
}

@InProceedings{McKinney2010,
  author    = { {W}es {M}c{K}inney },
  title     = { {D}ata {S}tructures for {S}tatistical {C}omputing in {P}ython },
  booktitle = { {P}roceedings of the 9th {P}ython in {S}cience {C}onference },
  pages     = { 56 - 61 },
  year      = { 2010 },
  editor    = { {S}t\'efan van der {W}alt and {J}arrod {M}illman },
  doi       = { 10.25080/Majora-92bf1922-00a }
}

@misc{photutils,
  author       = {Bradley, L. and others},
  title        = {photutils: Photometry Tools},
  year         = {2023},
  howpublished = {\url{https://photutils.readthedocs.io/}},
  doi          = {10.5281/zenodo.596036},
  note         = {Astropy project, Version X.Y}
}

@ARTICLE{Lyu2025,
       author = {{Lyu}, Wenhe and {Zhang}, Hong-Xin and {Paudel}, Sanjaya and {Li}, Tie and {Tang}, Yimeng and {Chen}, Guangwen and {Kong}, Xu and {Peng}, Eric W.},
        title = "{Dependence of Metal Enrichment of Nuclear Star Clusters on Galaxy Stellar Mass}",
      journal = {\apj},
         year = 2025,
        month = jan,
       volume = {979},
       number = {1},
          eid = {85},
        pages = {85},
          doi = {10.3847/1538-4357/ad9b0d},
archivePrefix = {arXiv},
       eprint = {2412.03132},
 primaryClass = {astro-ph.GA},
       adsurl = {https://ui.adsabs.harvard.edu/abs/2025ApJ...979...85L}
}

@ARTICLE{Fahrion2022,
       author = {{Fahrion}, Katja and {Leaman}, Ryan and {Lyubenova}, Mariya and {van de Ven}, Glenn},
        title = "{Disentangling the formation mechanisms of nuclear star clusters}",
      journal = {\aap},
         year = 2022,
        month = feb,
       volume = {658},
          eid = {A172},
        pages = {A172},
          doi = {10.1051/0004-6361/202039778},
archivePrefix = {arXiv},
       eprint = {2112.05610},
 primaryClass = {astro-ph.GA},
       adsurl = {https://ui.adsabs.harvard.edu/abs/2022A&A...658A.172F}
}

@ARTICLE{Barazza2002,
       author = {{Barazza}, F.~D. and {Binggeli}, B. and {Jerjen}, H.},
        title = "{More evidence for hidden spiral and bar features in bright early-type dwarf galaxies}",
      journal = {\aap},
         year = 2002,
        month = sep,
       volume = {391},
        pages = {823-831},
          doi = {10.1051/0004-6361:20020875},
archivePrefix = {arXiv},
       eprint = {astro-ph/0206275},
 primaryClass = {astro-ph},
       adsurl = {https://ui.adsabs.harvard.edu/abs/2002A&A...391..823B}
}

@ARTICLE{Jerjen2000,
       author = {{Jerjen}, H. and {Kalnajs}, A. and {Binggeli}, B.},
        title = "{IC3328: A ``dwarf elliptical galaxy'' with spiral structure}",
      journal = {\aap},
         year = 2000,
        month = jun,
       volume = {358},
        pages = {845-849},
          doi = {10.48550/arXiv.astro-ph/0004248},
archivePrefix = {arXiv},
       eprint = {astro-ph/0004248},
 primaryClass = {astro-ph},
       adsurl = {https://ui.adsabs.harvard.edu/abs/2000A&A...358..845J}
}

@ARTICLE{Poulain+2021,
       author = {{Poulain}, M{\'e}lina and {Marleau}, Francine R. and {Habas}, Rebecca and {Duc}, Pierre-Alain and {S{\'a}nchez-Janssen}, Rub{\'e}n and {Durrell}, Patrick R. and {Paudel}, Sanjaya and {Ahad}, Syeda Lammim and {Chougule}, Abhishek and {M{\"u}ller}, Oliver and {Lim}, Sungsoon and {B{\'\i}lek}, Michal and {Fensch}, J{\'e}r{\'e}my},
        title = "{Structure and morphology of the MATLAS dwarf galaxies and their central nuclei}",
      journal = {\mnras},
         year = 2021,
        month = oct,
       volume = {506},
       number = {4},
        pages = {5494-5511},
          doi = {10.1093/mnras/stab2092},
archivePrefix = {arXiv},
       eprint = {2107.09379},
 primaryClass = {astro-ph.GA},
       adsurl = {https://ui.adsabs.harvard.edu/abs/2021MNRAS.506.5494P}
}

@ARTICLE{Ashok+2023,
       author = {{Ashok}, Aishwarya and {Seth}, Anil and {Erwin}, Peter and {Debattista}, Victor P. and {de Lorenzo-C{\'a}ceres}, Adriana and {Gadotti}, Dmitri A. and {M{\'e}ndez-Abreu}, Jairo and {Beckman}, John E. and {Bender}, Ralf and {Drory}, Niv and {Fisher}, Deanne and {Hopp}, Ulrich and {Kluge}, Matthias and {Kolcu}, Tutku and {Maciejewski}, Witold and {Mehrgan}, Kianusch and {Parikh}, Taniya and {Saglia}, Roberto and {Seidel}, Marja and {Thomas}, Jens},
        title = "{Composite Bulges. III. A Study of Nuclear Star Clusters in Nearby Spiral Galaxies}",
      journal = {\apj},
         year = 2023,
        month = nov,
       volume = {958},
       number = {1},
          eid = {100},
        pages = {100},
          doi = {10.3847/1538-4357/ace341},
archivePrefix = {arXiv},
       eprint = {2308.03913},
 primaryClass = {astro-ph.GA},
       adsurl = {https://ui.adsabs.harvard.edu/abs/2023ApJ...958..100A}
}

@ARTICLE{vanDonkelaar+2023,
       author = {{van Donkelaar}, Floor and {Mayer}, Lucio and {Capelo}, Pedro R. and {Tamfal}, Tomas and {Quinn}, Thomas R. and {Madau}, Piero},
        title = "{Stellar cluster formation in a Milky Way-sized galaxy at z > 4 - I. The proto-globular cluster population and the imposter amongst us}",
      journal = {\mnras},
         year = 2023,
        month = jun,
       volume = {522},
       number = {2},
        pages = {1726-1735},
          doi = {10.1093/mnras/stad946},
archivePrefix = {arXiv},
       eprint = {2210.04915},
 primaryClass = {astro-ph.GA},
       adsurl = {https://ui.adsabs.harvard.edu/abs/2023MNRAS.522.1726V}
}

@ARTICLE{Hoyer+2023,
       author = {{Hoyer}, Nils and {Neumayer}, Nadine and {Seth}, Anil C. and {Georgiev}, Iskren Y. and {Greene}, Jenny E.},
        title = "{Photometric and structural parameters of newly discovered nuclear star clusters in Local Volume galaxies}",
      journal = {\mnras},
         year = 2023,
        month = apr,
       volume = {520},
       number = {3},
        pages = {4664-4682},
          doi = {10.1093/mnras/stad220},
archivePrefix = {arXiv},
       eprint = {2212.04151},
 primaryClass = {astro-ph.GA},
       adsurl = {https://ui.adsabs.harvard.edu/abs/2023MNRAS.520.4664H}
}

@ARTICLE{Roman+2023,
       author = {{Rom{\'a}n}, J. and {S{\'a}nchez-Alarc{\'o}n}, P.~M. and {Knapen}, J.~H. and {Peletier}, R.},
        title = "{Evidence for globular cluster collapse after a dwarf-dwarf merger: A potential nuclear star cluster in formation}",
      journal = {\aap},
         year = 2023,
        month = mar,
       volume = {671},
          eid = {L7},
        pages = {L7},
          doi = {10.1051/0004-6361/202345928},
archivePrefix = {arXiv},
       eprint = {2302.08516},
 primaryClass = {astro-ph.GA},
       adsurl = {https://ui.adsabs.harvard.edu/abs/2023A&A...671L...7R}
}

@ARTICLE{Lambert+2024,
       author = {{Lambert}, Mika and {Khim}, Donghyeon J. and {Zaritsky}, Dennis and {Donnerstein}, Richard},
        title = "{Systematically Measuring Ultra-diffuse Galaxies (SMUDGes). VI. Nuclear Star Clusters}",
      journal = {\aj},
         year = 2024,
        month = feb,
       volume = {167},
       number = {2},
          eid = {61},
        pages = {61},
          doi = {10.3847/1538-3881/ad0f25},
archivePrefix = {arXiv},
       eprint = {2311.12795},
 primaryClass = {astro-ph.GA},
       adsurl = {https://ui.adsabs.harvard.edu/abs/2024AJ....167...61L}
}

@ARTICLE{Khim+2024,
       author = {{Khim}, Donghyeon J. and {Zaritsky}, Dennis and {Lambert}, Mika and {Donnerstein}, Richard},
        title = "{Properties of Nuclear Star Clusters in Low Surface Brightness Galaxies}",
      journal = {\aj},
         year = 2024,
        month = jul,
       volume = {168},
       number = {1},
          eid = {45},
        pages = {45},
          doi = {10.3847/1538-3881/ad4ed3},
archivePrefix = {arXiv},
       eprint = {2405.14948},
 primaryClass = {astro-ph.GA},
       adsurl = {https://ui.adsabs.harvard.edu/abs/2024AJ....168...45K}
}

@ARTICLE{Cao+2025,
       author = {{Cao}, Tian-Wen and {Chen}, Zi-Qi and {Li}, Zi-Jian and {Cheng}, Cheng and {Galaz}, Gaspar and {Kalari}, Venu M. and {Wang}, Jun-feng and {Zhang}, Chun-Yi and {Chen}, Pei-Bin and {Shen}, Meng-Ting and {Wu}, Hong},
        title = "{Formation of a Nuclear Star Cluster through a Merger Event in the Low-surface-brightness Galaxy AGC 223218}",
      journal = {\apj},
         year = 2025,
        month = sep,
       volume = {990},
       number = {2},
          eid = {131},
        pages = {131},
          doi = {10.3847/1538-4357/adf559},
archivePrefix = {arXiv},
       eprint = {2510.06668},
 primaryClass = {astro-ph.GA},
       adsurl = {https://ui.adsabs.harvard.edu/abs/2025ApJ...990..131C}
}

@ARTICLE{Poulain+2025b,
       author = {{Poulain}, M{\'e}lina and {Marleau}, Francine R. and {Duc}, Pierre-Alain and {S{\'a}nchez-Janssen}, Rub{\'e}n and {Durrell}, Patrick R. and {Paudel}, Sanjaya and {Habas}, Rebecca and {M{\"u}ller}, Oliver and {Lim}, Sungsoon and {Heesters}, Nick and {Fensch}, J{\'e}r{\'e}my},
        title = "{Dwarf galaxies in the MATLAS survey: Hubble Space Telescope observations of nuclear star clusters}",
      journal = {\aap},
         year = 2025,
        month = dec,
       volume = {704},
          eid = {A113},
        pages = {A113},
          doi = {10.1051/0004-6361/202554845},
archivePrefix = {arXiv},
       eprint = {2509.19068},
 primaryClass = {astro-ph.GA},
       adsurl = {https://ui.adsabs.harvard.edu/abs/2025A&A...704A.113P}
}

@ARTICLE{Poulain+2025a,
       author = {{Poulain}, M{\'e}lina and {Smith}, Rory and {Duc}, Pierre-Alain and {Marleau}, Francine R. and {Habas}, Rebecca and {Durrell}, Patrick R. and {Fensch}, J{\'e}r{\'e}my and {Lim}, Sungsoon and {M{\"u}ller}, Oliver and {Paudel}, Sanjaya and {S{\'a}nchez-Janssen}, Rub{\'e}n},
        title = "{Evidence of star cluster migration and merger in dwarf galaxies}",
      journal = {\nat},
         year = 2025,
        month = apr,
       volume = {640},
       number = {8060},
        pages = {902-906},
          doi = {10.1038/s41586-025-08783-9},
archivePrefix = {arXiv},
       eprint = {2504.06749},
 primaryClass = {astro-ph.GA},
       adsurl = {https://ui.adsabs.harvard.edu/abs/2025Natur.640..902P}
}

@ARTICLE{Baldassare+2022,
       author = {{Baldassare}, Vivienne F. and {Stone}, Nicholas C. and {Foord}, Adi and {Gallo}, Elena and {Ostriker}, Jeremiah P.},
        title = "{Massive Black Hole Formation in Dense Stellar Environments: Enhanced X-Ray Detection Rates in High-velocity Dispersion Nuclear Star Clusters}",
      journal = {\apj},
         year = 2022,
        month = apr,
       volume = {929},
       number = {1},
          eid = {84},
        pages = {84},
          doi = {10.3847/1538-4357/ac5f51},
archivePrefix = {arXiv},
       eprint = {2203.02517},
 primaryClass = {astro-ph.HE},
       adsurl = {https://ui.adsabs.harvard.edu/abs/2022ApJ...929...84B}
}

@ARTICLE{Hoyer+2024,
       author = {{Hoyer}, N. and {Arcodia}, R. and {Bonoli}, S. and {Merloni}, A. and {Neumayer}, N. and {Zhang}, Y. and {Comparat}, J.},
        title = "{Massive black holes in nuclear star clusters. Investigation with SRG/eROSITA X-ray data}",
      journal = {\aap},
         year = 2024,
        month = feb,
       volume = {682},
          eid = {A36},
        pages = {A36},
          doi = {10.1051/0004-6361/202347665},
archivePrefix = {arXiv},
       eprint = {2401.17288},
 primaryClass = {astro-ph.GA},
       adsurl = {https://ui.adsabs.harvard.edu/abs/2024A&A...682A..36H}
}

@ARTICLE{Dullo+2024,
       author = {{Dullo}, B.~T. and {Knapen}, J.~H. and {Baldi}, R.~D. and {Williams}, D.~R.~A. and {Beswick}, R.~J. and {McHardy}, I.~M. and {Green}, D.~A. and {Gil de Paz}, A. and {Aalto}, S. and {Alberdi}, A. and {Argo}, M.~K. and {Gallagher}, J.~S. and {Kl{\"o}ckner}, H.-R. and {Marcaide}, J.~M. and {Mutie}, I.~M. and {Saikia}, D.~J. and {Saikia}, P. and {Stevens}, I.~R. and {Torrej{\'o}n}, S.},
        title = "{LeMMINGs. Multiwavelength constraints on the co-existence of nuclear star clusters and AGN in nucleated galaxies}",
      journal = {\mnras},
         year = 2024,
        month = aug,
       volume = {532},
       number = {4},
        pages = {4729-4751},
          doi = {10.1093/mnras/stae1732},
archivePrefix = {arXiv},
       eprint = {2407.10850},
 primaryClass = {astro-ph.GA},
       adsurl = {https://ui.adsabs.harvard.edu/abs/2024MNRAS.532.4729D}
}

@ARTICLE{Decarli+2007,
       author = {{Decarli}, R. and {Gavazzi}, G. and {Arosio}, I. and {Cortese}, L. and {Boselli}, A. and {Bonfanti}, C. and {Colpi}, M.},
        title = "{The census of nuclear activity of late-type galaxies in the Virgo cluster}",
      journal = {\mnras},
         year = 2007,
        month = oct,
       volume = {381},
       number = {1},
        pages = {136-150},
          doi = {10.1111/j.1365-2966.2007.12208.x},
archivePrefix = {arXiv},
       eprint = {0707.0999},
 primaryClass = {astro-ph},
       adsurl = {https://ui.adsabs.harvard.edu/abs/2007MNRAS.381..136D}
}

@ARTICLE{Edler+2023,
       author = {{Edler}, H.~W. and {de Gasperin}, F. and {Shimwell}, T.~W. and {Hardcastle}, M.~J. and {Boselli}, A. and {Heesen}, V. and {McCall}, H. and {Bomans}, D.~J. and {Br{\"u}ggen}, M. and {Bulbul}, E. and {Chy{\.z}y}, K.~T. and {Ignesti}, A. and {Merloni}, A. and {Pacaud}, F. and {Reiprich}, T.~H. and {Roberts}, I.~D. and {Rottgering}, H.~J.~A. and {van Weeren}, R.~J.},
        title = "{VICTORIA project: The LOFAR HBA Virgo Cluster Survey}",
      journal = {\aap},
         year = 2023,
        month = aug,
       volume = {676},
          eid = {A24},
        pages = {A24},
          doi = {10.1051/0004-6361/202346458},
archivePrefix = {arXiv},
       eprint = {2306.04513},
 primaryClass = {astro-ph.GA},
       adsurl = {https://ui.adsabs.harvard.edu/abs/2023A&A...676A..24E}
}

@ARTICLE{Hou+2024,
       author = {{Hou}, Meicun and {Hu}, Zhensong and {Li}, Zhiyuan},
        title = "{An X-Ray Census of Active Galactic Nuclei in the Virgo and Fornax Clusters of Galaxies with SRG/eROSITA}",
      journal = {\apjl},
         year = 2024,
        month = apr,
       volume = {965},
       number = {2},
          eid = {L24},
        pages = {L24},
          doi = {10.3847/2041-8213/ad3965},
archivePrefix = {arXiv},
       eprint = {2402.07275},
 primaryClass = {astro-ph.GA},
       adsurl = {https://ui.adsabs.harvard.edu/abs/2024ApJ...965L..24H}
}

@ARTICLE{Wang+2026,
       author = {{Wang}, Kaixiang and {Peng}, Eric W. and {Liu}, Chengze and {Ferrarese}, Laura and {C{\^o}t{\'e}}, Patrick and {Mihos}, J. Christopher and {Taylor}, Matthew A. and {Blakeslee}, John P. and {Cuillandre}, Jean-Charles and {Duc}, Pierre-Alain and {Guhathakurta}, Puragra and {Gwyn}, Stephen and {Ko}, Youkyung and {Lan{\c{c}}on}, Ariane and {Lim}, Sungsoon and {MacArthur}, Lauren A. and {Puzia}, Thomas and {Roediger}, Joel and {Sales}, Laura V. and {S{\'a}nchez-Janssen}, Rub{\'e}n and {Spengler}, Chelsea and {Toloba}, Elisa and {Zhang}, Hong-Xin and {Zhu}, Mingcheng},
        title = "{The Next Generation Virgo Cluster Survey (NGVS). XXXVI. Caught in the Act of Transformation: Morphological Decomposition and Classification of Ultracompact Dwarf Galaxies in the Virgo Cluster}",
      journal = {\apjs},
         year = 2026,
        month = jul,
       volume = {285},
       number = {1},
          eid = {27},
        pages = {27},
          doi = {10.3847/1538-4365/ae5231},
       adsurl = {https://ui.adsabs.harvard.edu/abs/2026ApJS..285...27W}
}

@ARTICLE{Ferrarese+2026,
       author = {{Ferrarese}, Laura and {C{\^o}t{\'e}}, Patrick and {MacArthur}, Lauren A. and {Roediger}, Joel C. and {Blakeslee}, John P. and {Cantiello}, Michele and {Cuillandre}, Jean-Charles and {Guhathakurta}, Puragra and {Gwyn}, Stephen and {Kurzner}, Max M. and {Peng}, Eric W. and {Santos}, Matthew and {Todd}, Eleanore B. and {Toloba}, Elisa and {Duc}, Pierre-Alain and {Durrell}, Patrick R. and {Fantin}, Nicholas and {Feng}, Yuting and {Lan{\c{c}}on}, Ariane and {Lim}, Sungsoon and {Liu}, Chengze and {Lokhorst}, Deborah and {Longobardi}, Alessia and {Mei}, Simona and {Mihos}, J. Christopher and {Paudel}, Sanjaya and {Puzia}, Thomas H. and {Raichoor}, Anand and {S{\'a}nchez-Janssen}, Rub{\'e}n and {Wang}, Kaixiang and {Zhang}, Hongxin and {Boselli}, Alessandro and {Balogh}, Michael L. and {Boissier}, Samuel and {Emsellem}, Eric and {Kavelaars}, J.~J. and {McConnachie}, Alan W. and {Morgan}, Cameron R. and {Spengler}, Chelsea and {Taylor}, James E. and {Taylor}, Matthew A. and {Brown}, Toby and {Fabbro}, S{\'e}bastien and {Gavazzi}, Raphael and {Hildebrandt}, Hendrik and {Jain}, Rashi and {Jord{\'a}n}, Andr{\'e}s and {Smith}, Rory and {Thompson}, Solveig and {Yoshino}, Haruka and {van Driel}, Wim and {van Waerbeke}, Ludovic and {Woods}, Tyrone E. and {Yeung}, Jade},
        title = "{The Next Generation Virgo Cluster Survey (NGVS). II. A Catalog of Galaxies in the Virgo Cluster}",
      journal = {\apjs},
         year = 2026,
        month = sep,
       volume = {286},
       number = {1},
          eid = {36},
        pages = {36},
          doi = {10.3847/1538-4365/ae8fc7},
       adsurl = {https://ui.adsabs.harvard.edu/abs/2026ApJS..286...36F}
}

\end{document}